\documentclass{emulateapj}
\usepackage{natbib}
\slugcomment{The Astrophysical Journal Supplement Series, in press, 17 July 2012}

\def\simlt{\mathrel{\hbox{\rlap{\hbox{\lower4pt\hbox{$\sim$}}}\hbox{$<$}}}}
\def\simgt{\mathrel{\hbox{\rlap{\hbox{\lower4pt\hbox{$\sim$}}}\hbox{$>$}}}}

\def\ale{\mathrel{\hbox{\rlap{\hbox{\lower4pt\hbox{$\sim$}}}\hbox{$<$}}}}
\def\age{\mathrel{\hbox{\rlap{\hbox{\lower4pt\hbox{$\sim$}}}\hbox{$>$}}}}

\def\nodata{---}

\def\kms{km~s$^{-1}$}

\def\g1256{1255--0}
\def\Reff{$R_{\rm e}$}
\def\aeff{$a_{\rm e}$}

\def\Rd{R_{\rm d}}

\def\vc{v_{\rm c}}
\def\v0{v_0}
\def\vs{v_s}

\def\vrotpx{$v_{\rm rot,p}(x)$}

\def\jp{$j_{\rm p}$}
\def\jt{$j_{\rm t}$}
\def\js{$j_\star$}
\def\jv{$j_{\rm vir}$}

\def\Ms{$M_\star$}
\def\Mv{$M_{\rm vir}$}
\def\tjt{$\tilde{j_{\rm t}}$}
\def\tjp{$\tilde{j_{\rm p}}$}
\def\MLs{$\Upsilon_\star$}
\def\atlas3d{ATLAS$^{\rm 3D}$}
\def\Dlj{$\Delta \log j_\star$}

\def\spose#1{\hbox to 0pt{#1\hss}}
\newcommand\lsim{\mathrel{\spose{\lower 3pt\hbox{$\mathchar''218$}}
     \raise 2.0pt\hbox{$\mathchar''13C$}}}
\newcommand\gsim{\mathrel{\spose{\lower 3pt\hbox{$\mathchar''218$}}
     \raise 2.0pt\hbox{$\mathchar''13E$}}}

\begin{document}

\title{Angular momentum and galaxy formation revisited}

\author{
Aaron J. Romanowsky\altaffilmark{1}\footnote{Current address: Department 
of Physics and Astronomy, San Jos\'e State University, San Jose, CA 95192, USA}, 
S. Michael Fall\altaffilmark{2}}

\affil{
\altaffilmark{1}University of California Observatories, Santa Cruz, CA 95064, USA\\
\altaffilmark{2}Space Telescope Science Institute, 3700 San Martin Drive, Baltimore, MD 21218, USA\\
}

\shorttitle{Angular momentum and galaxy formation revisited}
\shortauthors{Romanowsky \& Fall}

\begin{abstract}
Motivated by a new wave of kinematical tracers in the outer regions of 
early-type galaxies (ellipticals and lenticulars), we 
re-examine the role of angular momentum 
in galaxies of all types.
We present new methods for quantifying the specific angular momentum $j$, focusing mainly
on the more challenging case of early-type galaxies, in order to 
derive firm empirical relations between stellar \js\ and mass \Ms\
(thus extending the work by \citealt{1983IAUS..100..391F}).
We carry out detailed analyses of eight galaxies with kinematical data extending as
far out as ten effective radii, and find that data at two effective radii are generally
sufficient to estimate total \js\ reliably.
Our results
contravene suggestions that ellipticals could harbor large reservoirs of hidden \js\
in their outer regions owing to angular momentum transport in major mergers.
We then carry out a comprehensive analysis of extended kinematic data from
the literature for a sample of $\sim$~100 nearby bright galaxies of all types,
placing them on a diagram of \js\ versus \Ms.
The ellipticals and spirals form two parallel \js--\Ms\ tracks, with log-slopes
of $\sim$~0.6, which for the spirals is closely related to the Tully-Fisher relation,
but for the ellipticals derives from a remarkable conspiracy between masses,
sizes, and rotation velocities.
We find that on average, the ellipticals contain roughly 3--4 times
less angular momentum than spirals of equal mass.
We decompose the spirals into disks and bulges and find that these subcomponents
follow similar \js--\Ms\ trends to the overall ones for spirals and ellipticals.
The lenticulars have an intermediate trend, and we propose that the morphological
types of galaxies reflect disk and bulge subcomponents that follow
separate, fundamental \js--\Ms\ scaling relations.
This provides a physical motivation for characterizing galaxies most basically
with two parameters: mass and bulge-to-disk ratio.
Next, in an approach complementary to numerical simulations,
we construct idealized models of angular momentum content in a
cosmological context, using estimates of dark matter halo spin and mass
from theoretical and empirical studies.
We find that the width of the halo spin distribution cannot account for the differences
between spiral and elliptical \js, but that the observations are
reproduced well if these galaxies simply
retained different fractions of their initial $j$ complement
($\sim$~60\% and $\sim$~10\%, respectively).
We consider various physical mechanisms for the simultaneous evolution
of \js\ and \Ms\ (including outflows, stripping, collapse bias, and merging), 
emphasizing that the vector sum of all such processes 
must produce the observed \js--\Ms\ relations.
We suggest that a combination of early collapse and
multiple mergers (major or minor) 
may account naturally for the trend for ellipticals.
More generally, the observed variations in angular momentum represent
simple but fundamental constraints for any model of galaxy formation.

\end{abstract}

\vspace{2cm}

\section{Introduction}\label{sec:intro}

Many schemes for classifying galaxies have been presented over the years,
focusing on somewhat ephemeral properties such as morphology and color.
Alternatively, one may consider three fundamental physical parameters:
mass $M$, energy $E$, and angular momentum $J$.
Qualitatively, these are related to the amount of material in a galaxy,
to the linear size, and to the rotation velocity.

An important advantage of these parameters is that they may be related back to 
the earlier states of galaxies without having to unravel all of the
messy intervening details
such as baryonic dissipation, star formation, and morphological transformation.
As an example, the simple assumption that $J$ is approximately conserved during the
collapse of gas within hierarchically-forming dark matter halos
naturally explains the
observed basic scaling relations of disk galaxies
\citep{1980MNRAS.193..189F,1997ApJ...482..659D,1998MNRAS.295..319M}.

Here ``conserved'' means that the initial $J_\star$ is retained at a factor of $\sim$~2 level,
unlike $E$, which can be readily lost by factors of $\sim$~10 through dissipative collapse and radiation.
Note that the ``weak'' conservation 
of {\it total} $J$ is less restrictive and more plausible than the ``strong''
conservation of
the {\it internal} distribution of $J$ with radius, which 
could be readily altered
by secular processes within disks while still preserving total $J$
(e.g., \citealt{2004ARA&A..42..603K};
see \citealt{2002ASPC..275..389F}
and \citealt{2002ARA&A..40..487F} for further discussion).

In this vein,
\citet[hereafter F83]{1983IAUS..100..391F} introduced a general diagram of
\js\ versus stellar mass \Ms, where $j_\star \equiv J_\star/M_\star$ is the 
stellar specific angular momentum.
This diagram has the important advantages that it deals
with conservable physical quantities, and that the axes represent independent
variables.  The \Ms\ axis embodies a {\it mass} scale, while the \js\ axis
represents a {\it length} scale times a {\it rotation-velocity} scale.
On the contrary, the standard relations between \Ms\ and circular velocity $v_{\rm c}$
(e.g., \citealt{1977A&A....54..661T,2010MNRAS.407....2D,2011ApJ...742...16T}) 
involve correlated variables,
since $v_{\rm c}$ may be directly connected to \Ms.
Another related parameter is the {\it spin} ($\lambda$), which is useful for
characterizing dark matter halo rotation, and which we will discuss later in this paper.

\begin{figure}
\centering{
\includegraphics[width=3.5in]{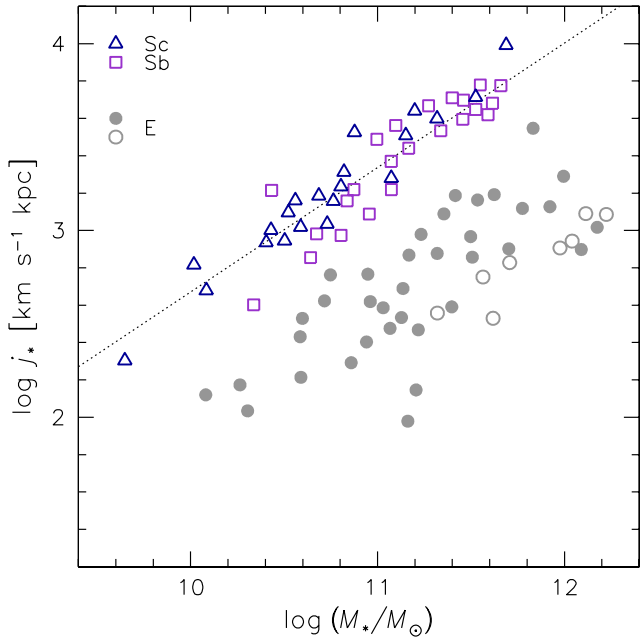} 
}
\caption{The total intrinsic stellar specific angular momentum of galaxies
plotted against their total stellar mass, reproduced from
\citet{1983IAUS..100..391F}, 
with corrections from a Hubble constant of $h=0.5$ to $0.7$.  
The symbols show galaxy types according to the legend at the upper left;
for the ellipticals (E), open circles show galaxies with an upper-limit estimate of \js.
The dotted line shows a trend of $j_\star \propto M_\star^{2/3}$.
The logarithms plotted here and used throughout the paper are in base~10.
These \js--\Ms\ scaling relations are the focus of this paper, and will 
eventually be updated
in Figure~\ref{fig:JMM0}.
\label{fig:JMM00}
}
\end{figure}

The simple \js--\Ms\ diagram is still charged with useful
information for understanding galaxies, and to orient the remainder of our
discussion, we begin by reproducing the original version from F83
here in Figure~\ref{fig:JMM00}.
The only change is to rescale the data for
a Hubble constant of $h=0.7$ rather than $h=0.5$.
These data were for late-type spirals (Sb and Sc) based on extended optical
rotation curves, and for elliptical galaxies based on observations from their
inner half-light radii, as feasible in that era.

The first key feature to note from Figure~\ref{fig:JMM00} is that
the spirals follow a fairly tight scaling relation of $j_\star \propto M_\star^\alpha$,
where $\alpha\sim0.7$
(see also \citealt{1967PASJ...19..409T,1969ApL.....3..153H,1970ApJ...160..811F,1973ApJ...184..735N}),
which is a phenomenology that is now understood 
to provide a remarkable link between visible galaxies
and their invisible dark matter halos.
F83 provided a simple theoretical framework in which the gaseous baryons of galaxies
are initially mixed with the dark matter and share in the same $j$.
The baryons then cool and decouple from the dark matter, collapsing into star-forming disks.
If the baryonic $j$ is approximately conserved in this process,
both the {\it zeropoint} and the {\it slope} of the observed spiral-galaxy \js--\Ms\ relation 
are reproduced.

The formation of disk galaxies can thus be explained at a basic level through this
long-standing picture of (weak) $j$ conservation.
To provide further understanding, hydrodynamical simulations of galaxy
formation have been pursued for decades, with the \js--\Ms\ observational diagram 
from F83 as a key benchmark for theory.  
Attaining that benchmark has turned out to be a major challenge, with early
studies finding catastrophic $j$ loss (e.g., \citealt{1991ApJ...377..365K,1991ApJ...380..320N,1995MNRAS.275...56N,1997ApJ...478...13N}).

This angular momentum ``catastrophe'' can be attributed partially to numerical limitations,
and partially to uncertainties in modeling baryonic processes such as feedback
following star formation, as reviewed by \citet{2002ASPC..275..389F}.
Over the years, the simulations have improved and 
can now come close to reproducing the \js--\Ms\ observations
(e.g., \citealt{2007MNRAS.374.1479G,2011MNRAS.410.1391A,2011ApJ...742...76G}),
although much work still remains in understanding both the numerics and the physics.

Besides the angular momentum benchmark from F83 which has become a standard ingredient in
modeling the formation of disk galaxies, there is another aspect 
of the original \js--\Ms\ diagram 
that has received relatively little attention:
the inclusion of elliptical galaxies along with the spirals.
The diagram thereby provides
a fundamental diagnostic of scaling relations for {\it all} galaxies, which
is important because
there is still not a full explanation for such a basic property
as the \citet{1926ApJ....64..321H} sequence of galaxy morphologies.

Star formation considerations aside,
there is an obvious {\it dynamical} distinction between galaxy
disks and spheroids, which are characterized
by cold, ordered rotation versus random motions with fairly low net rotation,
respectively.
Differences in the conservation and distribution of $j$ may very well
be pivotal to explaining these differences and to governing the fates of
galaxies.

As shown in Figure~\ref{fig:JMM00},
F83 found that ellipticals followed a \js--\Ms\ trend roughly parallel to the spirals,
but lower by a factor of $\sim$~6, and with more apparent scatter
(see also \citealt{1975ApJ...200..439B}).
There are several potential explanations for such a
difference between spirals and ellipticals, but the most plausible one is
traced to a violent, clumpy genesis for spheroids.
For example, mergers
could naturally redistribute angular momentum from the central regions of a galaxy to its
outer parts by dynamical friction 
(e.g., \citealt{1980ApJ...236...43A,1981MNRAS.197..179G,1987ApJ...319..575B,1988ApJ...330..519Z,1992ApJ...393..484B,1992ApJ...400..460H,1994MNRAS.267..401N,1996ApJ...463...69H,2007MNRAS.380L..58D,2008MNRAS.387..364Z}).
Thus, $j$ should be basically conserved but inconveniently locked up in unobservable components such
as the dark halo and the faint outer stars.

With this theoretical sketch in hand, the \js\ disparity between spirals
and ellipticals has received little further attention over the years.
However, the scenario of angular momentum redistribution has not yet been directly
tested by observations---a
situation that may now finally be remedied via the advent of new
techniques for optical spectroscopy in galaxy halos
(with preliminary results along these lines
reported in \citealt{2004IAUS..220..165R}).

\begin{figure}
\centering{
\includegraphics[width=3.5in]{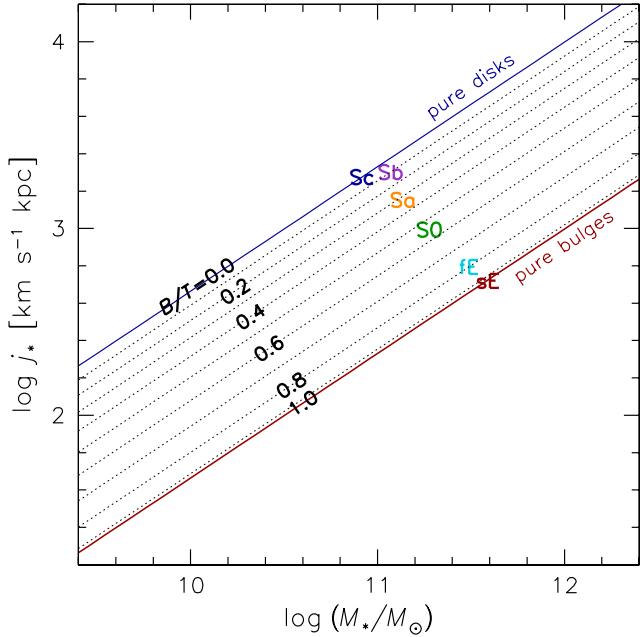} 
}
\caption{
Physically-motivated classification diagram of galaxies, using the parameter
space of stellar mass and specific angular momentum.
The solid blue and red lines show parallel scaling relations for disks and bulges,
which are based loosely on our observational results to be
presented in Section~\ref{sec:obsres}.
Approximate positions are also shown for different galaxy types:
Sc, Sb, Sa, S0, fE, and sE (the latter two
being fast and slow-rotating ellipticals).
}
\label{fig:schem1}
\end{figure}

In this paper we re-open various questions about angular momentum in 
all types of bright galaxies,
following and extending the treatment of F83.
Are the \js--\Ms\ slopes, zeropoints, and scatter in Figure~\ref{fig:JMM00}
supported upon re-examination?
Does the ``missing'' \js\ in ellipticals emerge in large-radius data?
Can the \js\ variations be associated with the natural dispersion in spin expected for 
standard dark matter halos, or is it necessary to invoke additional baryonic $j$ evolution?

F83 also proposed that the Hubble sequence may be understood as a systematic variation
in \js\ at a fixed \Ms\ 
(or equivalently, variation in \Ms\ at fixed \js), 
but could not test this idea owing to the lack of 
adequate data for the crucial, intermediate cases of Sa and S0 galaxies.
Here we will pursue this theme, and advance a framework where
every galaxy can be considered basically as a linear combination of a disk and a bulge,
with each of these components following a characteristic \js--\Ms\ scaling relation.
In this idealized model, the \js--\Ms\ parameter space maps uniquely to a space
of \Ms\ and bulge fraction $B/T$.

Figure~\ref{fig:schem1} provides a schematic overview of this framework, showing
decompositions of the Hubble sequence in \js--\Ms\ parameter space.
One of our goals in this paper will be to include observational results for 
Sa and S0 galaxies in this diagram for the first time, to see if such 
systems fill in the gap (if any) between earlier and later types,
and if bulges and disks are homologous enough to explain the
\js--\Ms\ trends as primarily reflecting a $B/T$ sequence.

The \js--\Ms\ diagram does not simply provide a basic {\it description} of galaxies
and their subcomponents, but also permits a novel approach to modeling the
{\it evolution} of galaxies which is complementary to numerical simulations.
As mentioned previously, there are simple models for the formation of disk galaxies
that relate their \js\ and \Ms\ values to the initial conditions of their host halos.
More generally, {\it any} stage in the evolution of a galaxy will involve
a vector of change in the $j$-$M$ diagram
that is not arbitrary, since in real physical processes, changes in
$j$ and $M$ will be linked in characteristic ways.
Therefore the empirical offsets between the \js--\Ms\ sequences of different galaxy types,
and of their subcomponents including bulges, disks, and dark matter halos,
can reveal the evolutionary connections among them.

We set out to explore the preceding questions and issues as follows.
In Section~\ref{sec:gen} we present a methodology for careful estimation of \js\ in
various types of galaxies and observations, with most of the details of its derivation
given in Appendix~\ref{sec:form}.
Section~\ref{sec:examp} uses detailed models of a handful of real galaxies
to examine a simplified procedure for \js\ estimation.
Our updated analysis
of the observed \js\ trends in a large sample of galaxies
follows, with the observational ingredients and their inter-correlations
described in Section~\ref{sec:data},
and the full results presented in Section~\ref{sec:obsres}
including a definitive confirmation of the large offset between spirals and ellipticals.
These empirical \js\ trends can be considered as fundamental, enduring
tools for constraining theories of galaxy evolution.
In Section~\ref{sec:theory} we go on to connect the observations to generalized theoretical
predictions for angular momentum
in a modern cosmological context.
We summarize in Section~\ref{sec:concl}.

In addition, Appendix~\ref{sec:form} is an important part of this paper, providing an extended presentation
of new content relating to the derivation of \js, which has been split off from
the main text for the sake of readability.
Appendices~\ref{sec:obsapp}--\ref{sec:decomp} provide data tables of \js\ and other properties of observed galaxies,
along with detailed discussion of the observations and data analysis for a subsample
of these galaxies.

The reader looking for immediate answers to the questions above may wish to skip
ahead to the results of Sections~\ref{sec:obsresults} and onwards.

\section{Basic formulae: disks and spheroids}\label{sec:gen}

The foundation for this paper is a revised, general observational analysis of
specific stellar angular momentum \js\ for bright galaxies in the nearby universe.
This quantity is most generally calculated 
by the following expression:
\begin{equation}\label{eqn:j3d}
{\bf j}_{\rm t} \equiv \frac{{\bf J}_{\rm t}}{M_\star} = \frac{\int_{\bf r} {\bf r}\times \bar{\bf v} \rho \, d^3{\bf r}}{\int_{\bf r} \rho \, d^3{\bf r}} ,
\end{equation}
where the subscript ``t'' denotes the ``true'' angular momentum in three-dimensional space,
${\bf r}$ and $\bar{\bf v}({\bf r})$ are the position and mean-velocity vectors (with respect to the center of mass of the galaxy),
and $\rho({\bf r})$ is the three-dimensional density of the population under study
(generally assumed to be stars in this project).

For spiral galaxies, we approximate the density distribution as an infinitely-thin,
axisymmetric disk with an exponential surface density profile.
Assuming also a radially-constant rotation curve,
Equation~(\ref{eqn:j3d}) yields the simple expression:
\begin{equation}\label{eqn:F83eq1}
j_{\rm t} = 2 \, \vc \, \Rd\ ,
\end{equation}
where $\vc$ is the intrinsic circular rotation velocity,
and $\Rd$ is the intrinsic exponential-disk scale length.
These deprojected quantities are relatively easy to infer from observations
because it is straightforward to estimate disk galaxy inclinations.
Equation~({\ref{eqn:F83eq1}) is widely used in the literature (including in F83), 
but we will demonstrate explicitly that it provides an excellent approximation
to real galaxies whose rotation curves vary with radius.

For more general cases including elliptical galaxies,\footnote{We use the term
``spheroid'' to mean a pressure-dominated stellar system
(which may also rotate).
A ``bulge'' is the spheroidal component of a spiral galaxy.
An ``elliptical'' is a galaxy with only a spheroidal component, 
although many galaxies commonly classified as ellipticals probably have
embedded disklike components, similar to those in lenticulars but less obvious.  
We consider jointly the ellipticals and lenticulars under the general rubric of
``early-type'' galaxies.}
there is no established recipe equivalent to Equation~(\ref{eqn:F83eq1}).
For multiple reasons, estimating \jt\ for these galaxies is much harder
than for spirals.  Not only are their inclinations and intrinsic shapes
uncertain, but large-radius rotation measurements are both more difficult and
more critical.

We illustrate the last point with some basic galaxy models.
Adopting the simple assumption of an axisymmetric system with cylindrical
rotation that is constant with respect to the intrinsic radius $R$, we consider both
a disk galaxy with an exponential surface density profile, and an elliptical
galaxy with a standard \citet{1948AnAp...11..247D} $R^{1/4}$ profile.
Although ellipticals are in general triaxial systems,
the axisymmetric model is sufficiently accurate for our purposes.

Figure~\ref{fig:fR} then shows the cumulative distribution of angular momentum
(both total and specific) with radius.  
For the disk galaxy, the specific angular momentum reaches roughly half of its
total value at the effective radius \Reff\ that encloses half of the stellar light.
This implies that observational estimates of \jt\ will be relatively easy for disk galaxies.

\begin{figure}
\centering{
\includegraphics[width=3.5in]{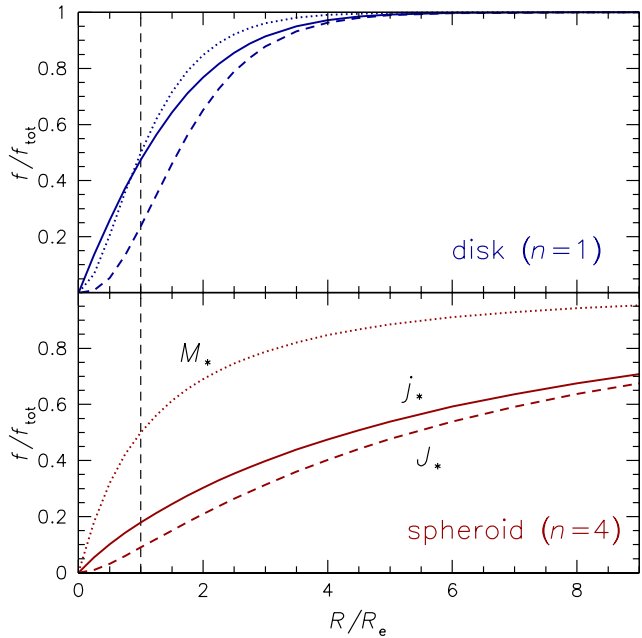} 
}
\caption{Fraction of enclosed cumulative quantities vs.\
cylindrical galactocentric radius (normalized by
the effective radius \Reff) for model galaxies with an exponential profile
($n=1$ disk, {\it top}) and a de Vaucouleurs profile ($n=4$ spheroid, {\it bottom}).
A constant, cylindrical rotation field is assumed.
The quantities are projected stellar mass \Ms\ ({\it dotted curve}),
angular momentum $J_\star$ ({\it dashed}), and specific angular momentum \js\
({\it solid}).
The latter quantity is computed using the cumulative values of both $J_\star$
and \Ms\ within the radius $R$.
The vertical dashed line marks 1~\Reff.
To capture half of \js, the observations must extend to $\sim$~1~\Reff\ in a disk
galaxy, and to $\sim$~(4--5)~\Reff\ in a spheroid.
\label{fig:fR}
}
\end{figure}

For the elliptical galaxy on the other hand, the halfway mark for \jt\ is reached
at 4.5~\Reff.
This is because ellipticals contain a fairly large fraction of their light in their
outer regions where the radius lever-arm in ${\bf r}\times\bar{\bf v}$ is large.
The implication is that observations of elliptical galaxies need to
extend to much larger radii than for spirals, in order to be confident of 
capturing the total \jt.

Typical stellar kinematics observations in 1983 extended to $\sim$~1~\Reff, 
and even today, only a small handful of galaxies have been observed 
kinematically out to $\sim$~5~\Reff,
which means the positions of the ellipticals in the original \js--\Ms\ diagram 
(Figure~\ref{fig:JMM00}) were highly uncertain, and continue to be challenging
to determine with surety. 
Fortunately, after a great deal of experimentation, which we will discuss below,
we find that there is a heuristic approach where
observations around $\sim$~2~\Reff\ can be used to
estimate the total \jt\ of ellipticals with reasonable accuracy.

Returning to a general framework for estimating \jt\ from observations,
there is not only the challenge of extending the data to large radii, but
also of having only three of the six phase-space quantities in
Equation~(\ref{eqn:j3d}) accessible (i.e., the projected positions and line-of-sight velocity).
Even the projection of ${\bf j}_{\rm t}$ on the sky involves unobservable
velocity components tangential to the line of sight, and requires
additional modeling assumptions.

To cope with these issues, we will model the observed rotation and luminosity profiles of
galaxies and convert these to \jt\ estimates using approximate deprojection factors.
Although these factors are based on highly simplified models,
the dominant source of uncertainty is still the limited extent of the data to
large radii.

We derive in Appendix~\ref{sec:form} two alternative expressions for estimating \jt\
from observations, both of them based again on the simplifying assumption of cylindrical rotation.
The first expression starts with a detailed calculation of a ``projected''
specific angular momentum proxy that can be estimated directly from observations:
\begin{equation}
j_{\rm p} = \frac{\int v_{\rm rot,p}(x) \, \Sigma(x) \, x^2 \, dx}{\int \Sigma(x) \, x \, dx} \, .
\label{eqn:JMp}
\end{equation}
Here \vrotpx\ is the observed profile of rotation velocity along the projected
semi-major axis $x$, and
$\Sigma(x)$ is the surface density profile, again along the semi-major axis.

The quantity \jp\ is related to \jt\ through a ``deprojection'' factor $C_i$:
\begin{equation}\label{eqn:proxy}
j_{\rm t} = C_i \, j_{\rm p} .
\end{equation}
Therefore the problem of estimating \jt\ separates into two parts:
the calculation of \jp\ from observations, and the factor $C_i$ which can be
calibrated from theoretical models.

As we describe in Appendix~\ref{sec:form},
this latter factor has some dependence on the detailed density--velocity structure
of the galaxy, but is primarily a function of the inclination $i$ relative to the
line of sight.  For thin-disk galaxies, it is simply $C_i=(\sin i)^{-1}$.
With spheroidal galaxies, there is an additional dilution effect that comes from
the line-of-sight intersecting the rotation field at non-tangent points.
In principle, this effect is dependent on the detailed shape of the
rotation profile, but we have found with simplified test models that such
variations can be neglected in practice.
We also find that as long as the major-axis radius $x$, rather than a circularized
radius $R$, is used in Equation~(\ref{eqn:JMp}), then $C_i$ is insensitive to galaxy flattening.

A general approximation to $C_i$ as a function of inclination
is provided by Equation~(\ref{eqn:Cform}).
It is normally difficult to determine $i$ for spheroidal galaxies,
and we will when needed adopt inclination-averaged values.

Equation~(\ref{eqn:JMp}) yields accurate results that are commensurate with the quality
of modern observations, but involves numerical integration,
and careful compilation of $\Sigma(x)$ and \vrotpx\ profiles along with
extrapolation beyond the bounds of the data.

We could in principle simplify the problem further by using parametric
models for \vrotpx\ and $\Sigma(x)$.  Unfortunately, the diversity of
observed rotation profiles (when non-spiral galaxies are considered)
 defies parametrization.
We can at least adopt for the surface density
the general \citet{1968adga.book.....S} law
which accurately represents a wide range of galaxy types:
\begin{equation}\label{eqn:sersic}
\Sigma(x) \propto \exp\left[-b_n(x/a_{\rm e})^{1/n}\right],
\end{equation}
where \aeff\ is the effective radius along the {\it semi-major axis}, and
the shape index $n$ determines the steepness of the outer density profile
(higher values are shallower: e.g., an exponential disk profile has $n=1$
and the de Vaucouleurs law for ellipticals has $n=4$),
while $b_n$ is a numerical function of $n$ 
[Equation~(\ref{eqn:bn})].

We use this $\Sigma(x)$ simplification in practice when deriving \jp\ from
a detailed \vrotpx\ profile in expression (\ref{eqn:JMp}).
We also generally base our $\Sigma(x)$ profiles on observations of stellar
surface brightness profiles $I(x)$, assuming for simplicity 
that there are no variations of stellar mass-to-light ratio with radius
(e.g., due to dust).

Our second method 
is a quick-and-dirty shortcut for estimating \jt,
as needed to generate an initial overview of the trends for a large
sample of galaxies.
We simply calculate the following linear scalar expression
[derived in Appendix~\ref{sec:form} from Equation~(\ref{eqn:JMp})]:
\begin{equation}\label{eqn:jCK0}
\tilde{j_{\rm p}} = k_n \, \vs \, a_{\rm e} ,
\end{equation}
where \tjp\ means an approximation for \jp,
$\vs$ is the {\it observed} rotation velocity at some arbitrary measurement
location $x_s$, and $k_n\sim$~1--5 is a numerical coefficient that depends on the
S\'ersic index $n$ of the galaxy
[see Equation~(\ref{eqn:kn})].
As in Equation~(\ref{eqn:proxy}), \tjp\ is multiplied by $C_i$ to provide an approximate \tjt.
Here the basic idea is that a galaxy can be represented by a characteristic
observed rotation velocity scale $\vs$, a length scale \aeff,
and a factor $k_n$ that relates to the moment of inertia (discussed further below).

The heuristic approximation that we make here is to
select $\vs$ at $x_s\sim$~2~\aeff\ for
all galaxies. We will show in the next section that this choice
allows us to estimate \jp\ with an accuracy of $\sim \pm$~0.1~dex,
which is good enough to start making some interesting inferences about
trends in \jt.

For $n=4$ spheroids, the expression equivalent to
Equation~(\ref{eqn:F83eq1}) for spirals is:
\begin{equation}\label{eqn:eqn4}
\tilde{j_{\rm t}} = 3.03 \, \vs \, R_{\rm e} ,
\end{equation}
for a median, unknown inclination (Equation~(\ref{eqn:j1})).
An important concept with the more general expression (\ref{eqn:jCK0})
is that $k_n$ increases strongly with $n$;
for fixed galaxy size and rotation velocity, a
more extended luminosity profile implies a higher \jp\
owing to the large fraction of mass residing at large radii. 
This also means that a spheroidal ($n\sim4$) galaxy 
with the {\it same} observed rotation $\vs$ and size \aeff\ as a spiral
has a {\it larger} specific angular momentum.
Late-type and early-type galaxies near the $L^*$ characteristic luminosity
{\it do} have similar sizes for the same stellar mass 
(e.g., \citealt{2003MNRAS.343..978S}). 
Therefore we can already make the basic prediction that if
\jp\ at a fixed mass is independent of morphology,
then the early-types should have $\vs$ values relative to late-types of
$\sim k_1/k_4$, i.e., lower by a factor of $\sim$~2.

The \js\ formalism that we have outlined here represents a modest extension of the simpler
methods in F83.  The improvements introduced here
include allowance for a range of luminosity profiles (not only $n=1$ and $n=4$),
and better treatment of elliptical galaxies where rotation at large radii is 
critically important.
It also becomes more straightforward to understand the interplay between
observations and uncertainties in the \js\ estimates, as explored in the
next section.

\section{Observations: analysis methods}\label{sec:examp}

Before we move on to \js--\Ms\ analyses of a large sample of galaxies,
we examine a small sample in more detail.
The goals here are to illustrate the nature of the available data,
to demonstrate that the simplified Equations~(\ref{eqn:F83eq1}) and (\ref{eqn:jCK0}) 
are good approximations to a full treatment with Equation~(\ref{eqn:JMp}),
and to understand some systematic effects in the \js\ and \Ms\ determinations.

Because this paper is concerned with the angular momentum bound up in the stellar
components of galaxies, 
the preferred kinematic tracer comes from integrated-light absorption-line spectroscopy.
In many cases, such data do not extend to large enough radii,
so we make use of additional tracers as proxies for the field stars:
cold and warm gas, planetary nebulae (PNe), and metal-rich globular clusters (GCs).

We consider disk- and bulge-dominated galaxies in Sections~\ref{sec:disk}
and \ref{sec:bulge}, respectively.
We evaluate our simplified \tjp\ estimate (\ref{eqn:jCK0}) in Section~\ref{sec:simp},
describe our mass estimates in Section~\ref{sec:mass}, and then consider
systematic uncertainties in Section~\ref{sec:sys}.

\subsection{Disk-dominated galaxies}\label{sec:disk}

\vspace{0.2cm}

\noindent
The most straightforward galaxies for estimating angular momentum are the
gas-rich spirals, since the stellar rotation profile, which cannot
always be measured directly, follows the gas rotation profile to a good approximation.
Also, the observed rotation can easily be
corrected for projection effects in order to recover the intrinsic value
(see Appendix~\ref{sec:thin}).
The detailed analysis below is overkill for these galaxies, whose \jt\ can be
readily estimated through Equation~(\ref{eqn:F83eq1}), but we wish to illustrate
how our more general treatment works for them, before moving on to the spheroids.

We consider two real  galaxies:
NGC~3054 and NGC~3200, which are well-studied disk-dominated spirals from
the classic optical rotation curve analyses of \citet{1982ApJ...261..439R}.
These cases are chosen to bracket the typical range of inner 
rotation profile shapes for spirals (slowly and rapidly rising, respectively).

We take the long-slit major-axis ionized-gas kinematics data from \citet{2004A&A...424..447P},
shown in Figure~\ref{fig:spirals} after a modest amount of re-binning.
These rotation profiles have
high-frequency bumps and wiggles that
are presumably caused by local perturbations such as spiral arms.
Fortunately, these features tend to average out when calculating a cumulative $j$
and are not important in this context.

To calculate the projected specific angular momentum \jp,
we carry out a piecewise integration of Equation~(\ref{eqn:JMp}),
using the major-axis rotation-velocity data \vrotpx\ up to $\sim$~2~\aeff,
along with simple power-law extrapolations at larger radii,
as shown in Figure~\ref{fig:spirals}.
For $\Sigma(x)$, we use an exponential model
[$n=1$ in Equation~(\ref{eqn:sersic})], with the disk scale-lengths $\Rd$
taken from $r$-band photometry as we will discuss in the next section.
Note that $a_{\rm e} = 1.68 \Rd$ for a pure exponential disk.

\begin{figure}
\centering{
\includegraphics[width=3.5in]{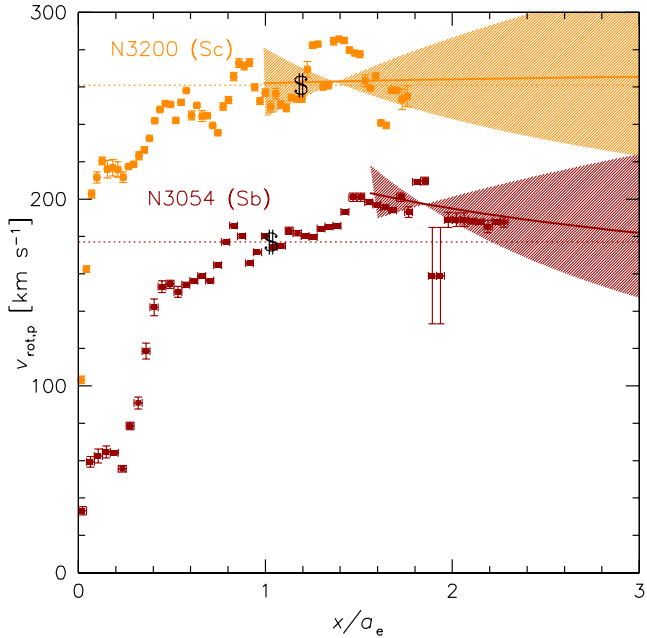} 
}
\caption{Observed rotation-velocity profiles of two spiral galaxies (NGC~3054 and NGC~3200)
vs.\ semi-major axis radius
(renormalized by the effective radius).
Each galaxy is labeled with its Hubble type.
The data are ionized gas velocities from \citet{2004A&A...424..447P}.
The solid curves with shaded regions show power-law fits (with uncertainties)
used to extrapolate the rotation velocity to larger radii.
See main text and Appendices~\ref{sec:form}~and~\ref{sec:obsapp} for further details.
Dotted horizontal lines show the characteristic rotation velocity $\vs$ for each galaxy;
the approximate intersection with the corresponding rotation-velocity profile is
marked with a \$\ symbol and defines the radius $x_s$
(see Section~\ref{sec:simp}).
\vskip 0.1cm
\label{fig:spirals}
}
\end{figure}

\begin{figure}
\centering{
\includegraphics[width=3.5in]{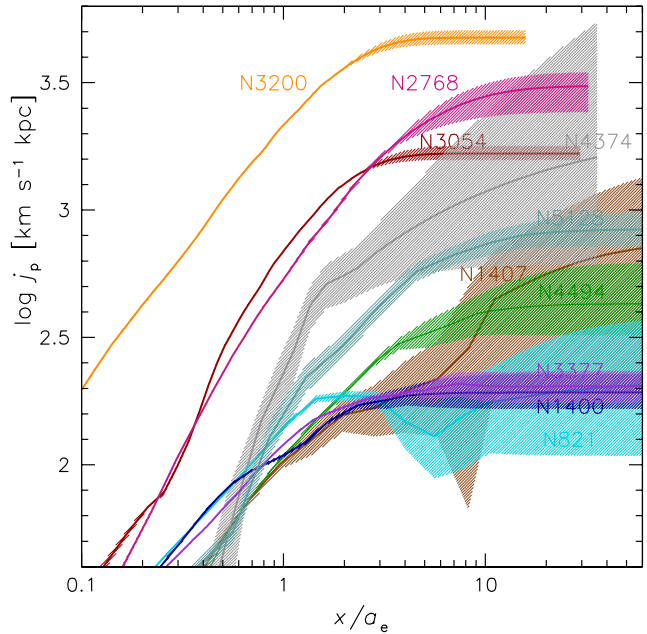} 
}
\caption{The cumulative projected specific angular momentum, $j_{\rm p}(<x)$,
of several nearby galaxies as a function of semi-major axis radius
(with log axes), based on modeling of kinematic observations.
Solid curves show the best-fit models, with shaded regions illustrating the 
uncertainties (including those due to extrapolations at large radii).
See Table~B\ref{tab:gal} for the distances and \aeff\ values adopted.
For most of the galaxies, \jp\ has nearly reached its asymptotic value
by $x \sim$~5~\aeff.
\label{fig:JMcum1}
}
\end{figure}

The resulting cumulative $j_{\rm p}(\leq x)$ profiles with radius for these galaxies are shown in
Figure~\ref{fig:JMcum1}. Here it would be trivial to convert $j_{\rm p}(\leq x)$ immediately to 
$j_{\rm t}(\leq R)$ using the known inclinations of these galaxies, but our general strategy
is to focus first on the direct modeling of the observations for all galaxies,
and later apply the deprojection factors $C_i$, which involve different systematics.

It can be seen that \jp\ hardly changes outside $\sim$~3~\aeff,
and that the large-radius extrapolations
make very little difference: 
the regions outside $\sim$~2--2.5~\aeff\ ($\sim$~3--4~$\Rd$)
contain only $\sim$~8\%--15\%
of the total luminosity, and contribute only $\sim$~15\%--25\% of the total \jp\
(half of \jp\ is enclosed within $\sim$~1.2~\aeff~$\sim$~$2\Rd$; 
Figure~\ref{fig:fR}).
Given reasonable extrapolations of the data, the total \jp\ for these 
two galaxies, using our basic modeling assumptions, is constrained to $\sim$~5\%
($\sim$~0.02~dex).

Thus the kinematics is not a major source of uncertainty for \jt\ estimation 
in disk-dominated galaxies.  
Additional complications that we have not considered here are
deviations of the disk surface density profile from 
a simple constant mass-to-light ratio exponential model,
and inclusion of a bulge (to be discussed later).
We will examine more general systematic uncertainties
in Section~\ref{sec:sys}.  

\subsection{Bulge-dominated galaxies}\label{sec:bulge}

We now turn to the novel component of this paper, which is the careful treatment
of \jt\ in early-type, bulge-dominated galaxies.
Figure~\ref{fig:fR} demonstrated that traditional observations within 1~\aeff\
provide little assurance about the total angular momentum content of these
systems, while even current cutting-edge observations out to
$\sim$~5~\aeff\ might in principle not be adequate.

\begin{figure*}
\centering{
\includegraphics[width=5in]{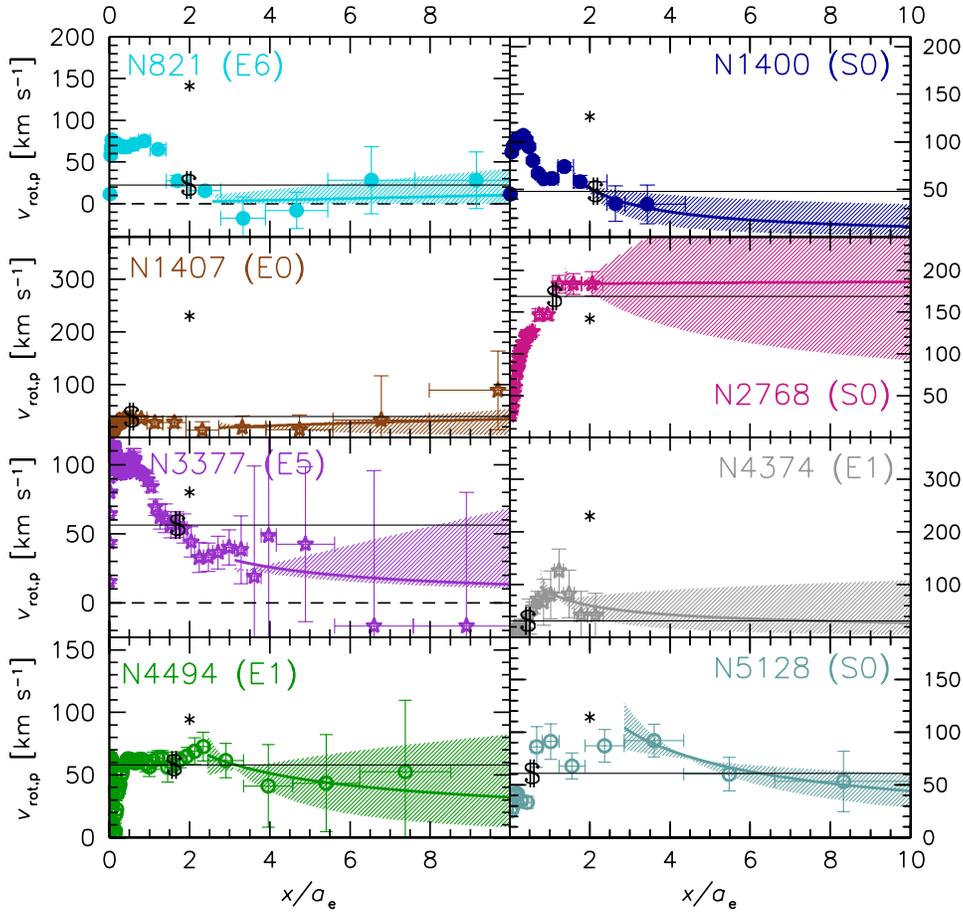} 
}
\caption{Rotation-velocity profiles for eight early-type galaxies.
See Figure~\ref{fig:spirals} for further details, including 
an explanation of the shaded uncertainty regions.
For ease of inter-comparisons, the vertical axis
of each panel has been scaled according to the velocity
dispersion of the galaxy at at 2~\aeff, which is marked in each panel by a $*$ symbol.
Note the dashed lines at zero rotation velocity in some cases.
The galaxies show a diversity of rotation-velocity trends with radius.
\label{fig:etg}
}
\end{figure*}

Here we analyze a sample of eight real galaxies in detail in
order to characterize the accuracy of \jt\ estimations.
Seven of these galaxies were chosen because of the availability of
high-quality extended kinematic data
using integrated stellar light spectroscopy
from two recent papers \citep{2009MNRAS.394.1249C,2009MNRAS.398...91P}.
Both papers represent the first installments of systematic surveys of
early-type galaxies in the local universe, and there is no obvious selection bias for
the seven galaxies.
Five of them are ``ordinary'' near-$L^*$ early-types 
with central ``fast-rotator'' 
kinematics as is typical for such galaxies \citep{1996ApJ...464L.119K,2011MNRAS.414..888E}.  
The other two (NGC~1407 and NGC~4374 $=$ M84) are examples of round, bright ``slow rotators''
that are common in high-density environments \citep{2011MNRAS.416.1680C}.

Five of these galaxies also have PN or GC kinematics data available
\citep{2009MNRAS.394.1249C,2009AJ....137.4956R}, 
which we incorporate into our analysis
in order to extend the range of galactocentric radii probed.
We include an eighth galaxy in our sample, NGC~5128 (Cen~A), because it has the most
extended (PN) kinematics data of any early-type galaxy in the literature
\citep{2004ApJ...602..685P}.  It may also be the remnant of a recent
major merger (e.g., \citealt{2006MNRAS.370.1737B}), 
which as discussed in Section~\ref{sec:intro} is expected to
generally transfer angular momentum into the outer regions.
Analysis of this galaxy thus provides a golden opportunity to search for
the ``missing'' angular momentum, and to see if any clear \jt\ difference
emerges with respect to the other galaxies in the sample.

The use of PNe and GCs to provide proxies for stellar kinematics 
may seem risky, given the considerable uncertainties that remain about the 
parent stellar populations of these tracers.
However, in most galaxies studied to date,
both the density and kinematical profiles of PN and metal-rich GC
 systems have been found to correspond well
to those of the full stellar population in the regions of overlap
(e.g., \citealt{2009MNRAS.394.1249C,2010A&A...518A..44M,2011MNRAS.415.1244D,2012A&A...539A..11M,Cortesi12,Pota12}).
We have also verified that this is generally the case for the galaxies in our sample.

Further details of the observations as well as of the kinematical modeling are
provided in Appendix~\ref{sec:obsapp}, along with the resulting rotation and angular momentum
profiles.  It should be emphasized that the careful, homogeneous construction of these
profiles is laborious, which is why the current sample of galaxies that we consider
in detail is relatively small.

The rotation-velocity profiles of these eight galaxies are summarized
in Figure~\ref{fig:etg}.
Unlike the spirals (Figure~\ref{fig:spirals}), the early-types show great diversity
in the characteristic shapes of their profiles.
Some are fairly constant with radius, 
others plummet rapidly from a central high value, and
one continues increasing to the limits of the data.
This diversity is {\it not} simply a matter of inclination, as can be seen
by the divergent cases of NGC~821 and NGC~2768, which are both highly flattened 
and probably close to edge-on.
We thus find that the central rotation properties of early-type galaxies cannot be used 
to reliably estimate the total angular momentum content, and there is probably no
simple function that universally characterizes their full rotation-velocity profiles.

As with the spirals, we fit power laws to the outer
regions of the rotation data in order to extrapolate to larger radii
(see Appendix~\ref{sec:obsapp} for further details).
We then use Equation~(\ref{eqn:JMp}) to calculate profiles of cumulative 
\jp\ with radius, which we plot in Figure~\ref{fig:JMcum1}.
Even though the data do not reach the total asymptotic value for \jp, the requirement
of a smooth power-law extrapolation for the rotation-velocity profile does in most cases
strongly limit the total \jp, which is typically determined at the $\pm$~15\% level
($\pm$~0.06~dex).
The radius enclosing half of the total \jp\ varies from galaxy to galaxy
depending on the shape of its rotation-velocity profile: 0.7--3~\aeff\
(for the two spirals, it is 1~\aeff).

The exceptions to these findings are the
two bright, round ellipticals NGC~1407 and NGC~4374.
Figure~\ref{fig:JMcum1} shows that
much of the angular momentum in these galaxies is found 
at very large radii (half of \jp\ within 9~\aeff\ and 4~\aeff, respectively), 
as expected from their fairly high S\'ersic indices of $n\sim$~4--8
(the ordinary early-types have $n\sim$~2--4).
However, beyond the usual uncertainties introduced by extrapolating the rotation velocity,
there are a couple of other practical considerations.

One issue is that although these particular galaxies have relatively well studied
surface brightness profiles, many such massive ellipticals do not,
with their $n$ and \aeff\ values poorly known.  This situation produces
``double jeopardy'' for angular momentum estimation, since both the luminosity and
the rotation-velocity profiles at very large radii are important yet poorly constrained.

The other issue demonstrated with NGC~4374 is that its cumulative \jp\ has
not yet converged at the (estimated total) virial radius of 
$\sim$~35~\aeff, so it is not clear how its angular momentum should even be defined.
This class of high-$n$ galaxies is clearly problematic, and we will consider
any \jt\ results on them to be tentative for now.

Figure~\ref{fig:JMcum1} also reveals a first glimpse of the basic result of this paper.
For most of the early-types in the sample, there is relatively little angular momentum
hidden beyond $\sim$~1--2~\aeff, and their total values of \jp\ are lower than those of
the spirals.  We will make more detailed comparisons later in this paper.

\subsection{Simple $J/M$ approximations}\label{sec:simp}

We now arrive at a question that is critical for the wider survey
of angular momentum in the rest of this paper:
how accurate is the simplified Equation (\ref{eqn:jCK0})?
As a reminder, this \tjp-estimator would replace the detailed calculations
based on Equation~(\ref{eqn:JMp}) that we have carried out in the preceding subsections,
but which are time-consuming to carry out for a larger sample of galaxies, and
are not even possible for cases without very extended kinematic data.

In Appendix~\ref{sec:form},
we have motivated the construction of Equation~(\ref{eqn:jCK0})
via toy models of galaxies, and calculated the corresponding coefficient $k_n$.
We will now apply this formula to the set of ten real galaxies just discussed
(both late- and early-types), and find an optimum radial location $x_s$ for measuring the
characteristic rotation velocity $\vs$.

For each galaxy, it is straightforward to find the constant value of $\vs$
which when substituted in Equation~(\ref{eqn:JMp}) yields the same \jp\
as with the full observed rotation-velocity profile.
These results are listed in Table~B\ref{tab:gal} and
 shown in Figures~\ref{fig:spirals} and \ref{fig:etg}, where the intersection of
$\vs$ with the rotation-velocity profile determines the characteristic measurement
radius $x_s$. As an example,
for NGC~821, it is clear that $x_s\sim$~2\aeff.
For NGC~4494 on the other hand, a broad range of choices for $x_s$ would work,
owing to its nearly constant rotation-velocity profile.

Considering this issue in more detail, we
calculate \tjp\ using Equation~(\ref{eqn:jCK0}) 
with an arbitrary choice for $x_s$ (which in turn
determines a guess for $\vs$ from the observed rotation velocity at this radius).
The results for $x_s/a_{\rm e} = (1,2,3)$
are shown in Figure~\ref{fig:guess}, plotted against \jp\ calculated in full
from Equation~(\ref{eqn:JMp}).
It can be seen that $x_s/a_{\rm e}=2$ provides a reasonably good match between
\tjp\ and \jp\ for all of the galaxies in this sample.
The other radius choices fare worse, owing to 
galaxies like NGC~821 that have rotation-velocity profiles
with a distinct transition between the inner and outer regions near 2~\aeff,
and thus $\vs$ measurement elsewhere would be biased.

\begin{figure}
\centering{
\includegraphics[width=3.5in]{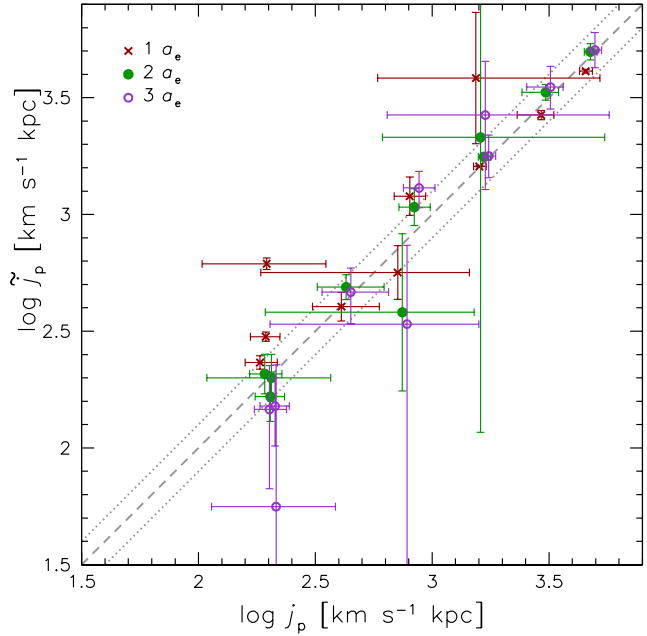} 
}
\caption{Comparison of a simple projected specific angular momentum estimate
[\tjp; Equation~(\ref{eqn:jCK0})] with the more accurate value (\jp).
Results are shown for ten different galaxies, each with a choice of
three reference radii: $x_s/a_{\rm e}=$~1 (red crosses),
2 (green filled circles), and 3 (purple open circles).
Some of the points are given a 0.02~dex horizontal offset for visibility.
The dashed and dotted lines mark the one-to-one relation with a $\pm$~0.1~dex scatter.
The optimal choice here for $x_s$ is $2$~\aeff.
}\label{fig:guess}
\end{figure}

Now to home in more finely on a choice for $x_s$, in Figure~\ref{fig:jrat}
we present the ratio of estimated and ``correct'' \jp, as a function of
the chosen $x_s$, for each galaxy.
Some of the galaxies permit a broad range of choices for $x_s$, while
others do not.  Especially noteworthy again are the galaxies like NGC~821 and NGC~3377
which have sharp drops in their rotation-velocity profiles, so $\vs$ measured
at small radii would overestimate \jp\ by factors of $\sim$~2--3.

We do not find a strong correlation between $n$ and optimal $x_s$ as
expected from the simple models we constructed in Appendix~\ref{sec:simpapp};
the dominant effect on $x_s$ with the real galaxy sample is the scatter in
the shapes of the rotation-velocity profiles.  
Future detailed analyses of a larger sample of galaxies may reveal systematic
trends with $n$ that motivate improved \jp\ estimation methods, but for now
we stick with our simple \tjp\ approach.

Because the real galaxies so far do not show strongly rising outer rotation-velocity profiles,
and if anything the reverse,
$x_s\sim$~2~\aeff\ appears to be a good overall choice for the rotation-velocity
measurement radius. This minimizes
the galaxy-to-galaxy scatter in the \tjp\ approximation ($\sim \pm$~0.1~dex),
and appears to produce little systematic bias ($<\sim$~0.1~dex).
Such ``errors'' are comparable to the uncertainties from carrying out the
full \jp\ calculations, and are therefore acceptable
for our purposes in this paper.

\begin{figure}
\centering{
\includegraphics[width=3.5in]{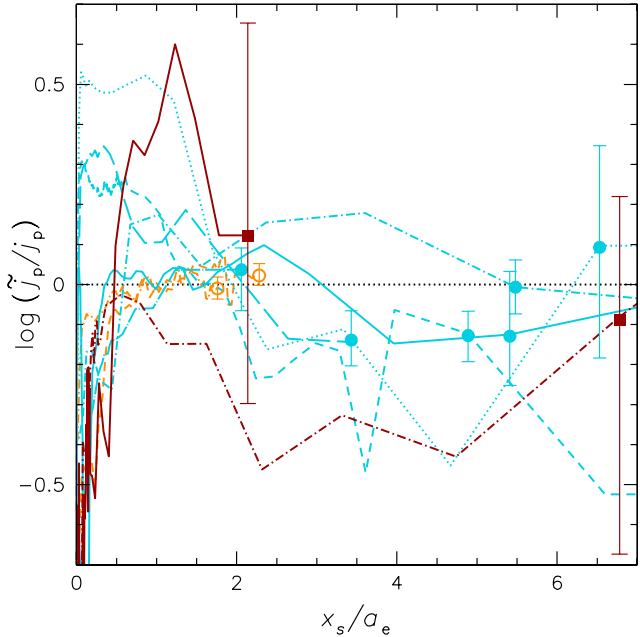} 
}
\caption{The logarithmic ratio between 
simple estimates of projected specific angular momentum [Equation~(\ref{eqn:jCK0})]
and more accurate values [Equation~(\ref{eqn:JMp})],
vs.\ the rotation-measurement radius $x_s$ in units of the effective radius.
Each point indicates a sample ratio for an individual galaxy,
with error bars indicating the kinematics-driven uncertainties in total \jp\ 
from the detailed models.
Results are plotted for 10 galaxies: two spirals (orange profiles with open circles),
six ordinary early-types (blue profiles with filled circles),
and two giant ellipticals (red profiles with filled squares).
As in Figure~\ref{fig:guess}, $x_s \sim 2 a_{\rm e}$ provides a good
measurement location, resulting in minimal scatter and bias 
for the angular momentum estimates.
\label{fig:jrat}
}
\end{figure}

One caveat here is that this sample of galaxies is still small, and we cannot
yet be sure of the universal validity of our approximation, e.g., for the
larger sample of galaxies that we will study in the remainder of this paper.
However, we will show that there is no apparent systematic bias, i.e.,
the overall scientific conclusions are consistent with the
subset of detailed \jp\ profiles.

\subsection{Stellar mass estimates}\label{sec:mass}

So far we have focused on estimating \js, but the other key component
in constructing the \js--\Ms\ diagram is of course the stellar mass \Ms.
Assuming that we have a well-determined surface brightness profile $I(x)$
or total luminosity, 
we then need to know the stellar mass-to-light ratio \MLs.
In this paper, we assume for simplicity that \MLs\ is constant 
throughout each galaxy, which also means that its value is not relevant
in our \js\ calculations.

Estimating \MLs\ in galaxies is a classic and not fully resolved problem.
One standard approach is to use theoretical models for stellar
populations in combination with observations of the stellar light
(e.g., broad-band colors, or spectroscopic line indices).
Although there are well-known degeneracies between the ages and metallicities
inferred for the stars, fortunately \MLs\ can be estimated with
more certainty (e.g., \citealt{2009MNRAS.396.1132T}), modulo
the initial mass function (IMF) of the stellar populations.

The IMF affects the overall normalization of \MLs\
via the mass contributions of late-type dwarf stars or compact stellar remnants,
which are observationally difficult to tally.
If all galaxies have the same IMF, then our analyses of the
{\it relative} differences between galaxies in the \js--\Ms\ plane will be secure.
There are also recent, indirect claims for possible galaxy-to-galaxy IMF variations
(e.g., \citealt{2008MNRAS.385..147D,2010ApJ...709.1195T,2010ApJ...721L...1T,2011ApJ...735L..13V,2012MNRAS.422L..33D,2012arXiv1206.1594F,2012arXiv1206.4311S}).
However, even in this case we do not expect a major impact on our conclusions.

As an example, the recent analysis of \citet{2012Natur.484..485C}
implies that strong IMF variations tend to occur in only the most
massive, and relatively rare, early-type galaxies, 
which would have $\log\,(M_\star/M_\odot) \ga 11.3$ in our plots
(based on a standard IMF).
Such galaxies might have masses larger than our estimates by factors of
$\sim$~2, but given the relatively small numbers of such galaxies and the 
weak constraints on their \js\ values, they will have little effect on
our estimated \js--\Ms\ trends.

It is outside the scope of this paper to estimate \MLs\ for each galaxy in detail.
Instead, we adopt the simplification that all galaxies have $\Upsilon_{\star,K}=1.0$.
The near infrared (NIR) $K$-band 
is only mildly affected by internal and foreground extinction, is
thought to be fairly insensitive to variations in stellar populations, 
and has uniform photometry available from the 2MASS survey \citep{2006AJ....131.1163S}.
The systematic variation in $\Upsilon_{\star,K}$ across our entire sample of late
and early-type galaxies is conventionally expected to
be no more than $\sim$~0.1~dex, based on $B-V$ colors \citep{2003ApJS..149..289B},
although there are recent suggestions of variations at the level of $\sim$~0.4 dex
(\citealt{2009MNRAS.400.1181Z,2011ApJ...739L..47B}).
Our adopted value of $\Upsilon_{\star,K}=1.0$ is motivated by dynamical estimates
in both spirals and lenticulars from
\citet[figure~9]{2009MNRAS.400.1665W}, and corresponds to an IMF midway
between \citet{2001MNRAS.322..231K} and \citet{1955ApJ...121..161S}.

Our calculations of \Ms\ also require estimates of total luminosity, $L_K$.
However, we do {\it not}
simply adopt the total magnitudes provided by the 2MASS archive.
These values are not reliable for early-type galaxies 
(e.g., \citealt{2007MNRAS.381.1463N,2009ApJ...702..955D,2009MNRAS.400.1665W,2011arXiv1107.1728S}),
particularly the variety with extended high-$n$ envelopes,
where the 2MASS values could be too faint by as much as 1 mag.

Instead, we construct our own ``aperture corrections''.
We adopt the 2MASS magnitudes within the 20$^{\rm th}$-mag isophote, $K_{20}$,
and use the best available optical photometry for each galaxy along with a
S\'ersic model fit to estimate the fraction of the galaxy light residing
beyond $K_{20}$.

This procedure neglects any bandpass-dependence in the light profiles $I(x)$,
which are often more radially extended in bluer bands (e.g., 
\citealt{1961ApJS....5..233D,1990AJ....100.1091P,2011MNRAS.416.1983R}).
Such differences imply \MLs\ variations with radius \citep{2011MNRAS.418.1557T},
which is a reminder of the limitations of our constant-\MLs\ approximation.
Given our reliance on optical profiles $I(x)$ to derive $\Sigma(x)$ and estimate \jp,
as in Equation~(\ref{eqn:JMp}), for consistency we do need to use
the optical data to extrapolate the $K$-band photometry in estimating \Ms.
However, the scale-lengths \aeff\ of the stellar {\it mass} distributions
are probably smaller on average
than the \aeff\ values that we use based on optical {\it luminosity} distributions,
leading us to overestimate both \jp\ and \Ms.  Improvement on this point
could be made in the future by analysis of deep $I(x)$ data at NIR wavelengths.
NIR spectroscopy would then also be needed for full consistency of both
\jp\ and \Ms\ estimates (e.g., 
\citealt{2003AJ....125.2809S,2008ApJ...674..194S,2011MNRAS.412.2017V}).

\subsection{The \js--\Ms\ diagram}\label{sec:sys}

Here we focus on the \js--\Ms\ plane, our ultimate destination in this paper,
but for now considering the projected specific angular momentum \jp\
rather than the true \jt\ in
order to isolate various effects that are disjoint from inclination uncertainties.
Figure~\ref{fig:JMM1} shows our detailed galaxy sample where cumulative 
$j_{\rm p}(<R) = J_{\rm p}(<R)/M_\star(<R)$ is plotted
not as a function of radius (as in Figure~\ref{fig:JMcum1})
but of {\it enclosed projected stellar mass}, \Ms.

\begin{figure}
\centering{
\includegraphics[width=3.5in]{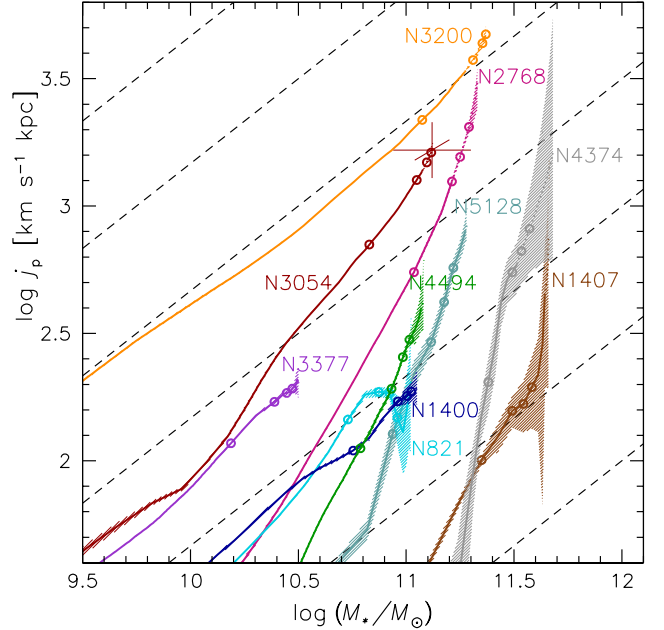} 
}
\caption{The
cumulative projected specific angular momentum of nearby galaxies
(as in Figure~\ref{fig:JMcum1}), now plotted
vs.\ cumulative projected stellar mass.
The curves are {\it solid} where constrained by the data,
and {\it dotted} for extrapolations.
Circles show intervals of 1~\aeff, up to 4~\aeff.
Error bars at the end of the NGC~3054 curve
illustrate the effects of systematic uncertainties (see text for details):
diagonal for the distance, vertical for scale-length, and horizontal for \MLs.
Diagonal dashed lines show tracks of $j_{\rm p} \propto M_\star^{2/3}$,
which represent constant halo spin.
\label{fig:JMM1}
}
\end{figure}

For reference, we show dashed lines corresponding to
$j_{\rm p} \propto M_\star^{\alpha}$, with $\alpha=2/3$.
This value for $\alpha$ is motivated by previous observations 
(Section~\ref{sec:intro}), and by theoretical predictions for \jt-\Ms,
given
constant values of an initial halo spin parameter $\lambda$, as we will see
in Section~\ref{sec:theory1}.
We are most concerned with the locations of galaxies relative to these tracks,
and with any systematic effects that could shift the data in a direction
{\it perpendicular} to them.

The shaded regions of the curves in Figure~\ref{fig:JMM1}
indicate the uncertainties due to the kinematic data, 
including the extrapolations to large radii.  For most of the galaxies,
the asymptotic position in the \jp-\Ms\ diagram is relatively well determined.
The main exceptions are NGC~1407 and NGC~4374, 
which as discussed before are extended giant ellipticals
whose total \jp\ is very difficult to determine.
The early-type galaxy NGC~2768 is also a concern even though the formal \jp\ uncertainties are small,
since there are large contributions to the total \jp\ estimate from the region
of extrapolation.

An offset in total \jp\ between the late-types and most of the early-types as in
Figure~\ref{fig:JMcum1} is also apparent in Figure~\ref{fig:JMM1}.
However, the mass dimension brings the relative positions into sharper focus.
For example, NGC~4374 and NGC~5128 have similar \jp\ values to NGC~3054, but also have
larger stellar masses, which means that their 
inferred halo spins will be
{\it lower} (considering distances perpendicular to the dashed tracks).

We next consider some systematic uncertainties that apply even if the
rotation-velocity profiles are perfectly measured.
First, there is a typical distance uncertainty of $\sim$~10\%.
This affects \jp\ linearly and \Ms\ quadratically, moving
the position of the data by a very small amount
nearly parallel to the $\lambda$ 
tracks (see sample error bars marked for NGC~3054 in the Figure).

Next we consider an uncertainty of $\sim$~30\% ($\sim$~0.11~dex) in the scale lengths \aeff, 
which translates into a similar uncertainty in \jp\ [see Equation~(\ref{eqn:F83eq1})].\footnote{
In practice, the \aeff\ uncertainty is correlated with an
uncertainty in the galaxy luminosity and thus in \Ms, but this is a relatively
weak effect.}
Also, in some cases the surface brightness profile is well constrained
and the associated \jp\ uncertainty is very small (e.g., $\sim$~5\% or $\sim$~0.02~dex
in the case of the $n\sim3$ elliptical NGC~4494).

Finally there is \MLs, which may be uncertain by a factor of $\sim$~50\% 
($\sim$~0.2~dex) and would affect \Ms\ by the same amount.  
For spiral galaxies, this is probably the limiting factor for inferring
their $\lambda$ values.
For the early-types, the inclination is generally unknown and may be a
significant source of uncertainty for estimating \jt, even when \jp\ is well constrained.
We will return to this theme in Section~\ref{sec:less}.

\section{Observations: scaling relations and derivations of $J/M$ for the full sample}\label{sec:data}

Having carried out detailed analyses of \js\ for a handful of galaxies in the previous section,
we now derive \js\ for a much larger galaxy sample, using simpler methods.
Besides these derivations, in this section we also examine some basic scaling
relations for galaxies, in order to understand the observational underpinnings of
the \js--\Ms\ results in the next section, and to verify that our results
are consistent with some well-known properties of galaxies.
We also introduce a novel, generalized version of the Tully-Fisher relation
for galaxies of all types.
Those who are keen to get straight to the angular momentum results
may wish to skip to Section~\ref{sec:obsresults}.

In order to populate the observational \js--\Ms\ diagram,
we will use the \tjp\ approximation of Equation~(\ref{eqn:jCK0})
which we have found to be generally accurate at the $\sim$~0.1~dex 
($\sim$~25\%) level.
The basic parameters that we then need for all of the galaxies are:
the total stellar mass (\Ms) and its
scale-length ($\Rd$ or \aeff), the S\'ersic index $n$, and
the characteristic rotation velocity $\vs$.

The distances to the galaxies are estimated from redshifts and surface brightness fluctuations.
As discussed in Section~\ref{sec:mass}, 
\Ms\ is derived from aperture-corrected 2MASS magnitudes $m_K$, assuming
$\Upsilon_{\star,K}=1.0$.

The other parameters are derived differently for the late-type and early-type samples,
as we will discuss in Sections~\ref{sec:ltgdata} and \ref{sec:etgdata}, respectively.
Section~\ref{sec:scale} brings the data together in an examination of
basic scaling relations, before proceeding to the final \js--\Ms\ analyses of 
Section~\ref{sec:obsres}.

\subsection{Late-types}\label{sec:ltgdata}

Because spiral galaxies are dominated by their disk components, whose
photometric and kinematic properties are relatively straightforward to measure,
past studies of their angular momenta have generally treated them as
pure disks, e.g., using Equation~(\ref{eqn:F83eq1}) to calculate \jt.
However, this approximation may be inadequate for the spirals with relatively
large bulges (Sa and some Sb), and it is one of the goals of this paper to
consider these components.

With Equation~(\ref{eqn:jCK0}) in mind, we could use values for the parameters
$n$, \aeff, and $v_{\rm s}$ that characterize the composite bulge--disk systems
(e.g., with an overall $n$ somewhat larger than $1$).
However, the required stellar photometry and kinematic data are not available
for a large sample of galaxies.
Instead, we analyze disk and bulge components separately, make some
simple assumptions for the bulges to compensate for the missing data,
and then combine the disks and bulges into global \js\ analyses.

We focus on the classic spiral galaxy data set assembled
by \citet{1986AJ.....91.1301K,1987AJ.....93..816K,1988AJ.....96..514K},
comprising 64 galaxies from type Sa to Sm, at distances ranging from 1 to 100~Mpc.
These data include $r$-band CCD photometry along with 
bulge/disk decompositions, and inclination-corrected gas-disk rotation curves 
from both optical emission-lines 
(e.g., \citealt{1980ApJ...238..471R,1982ApJ...261..439R,1985ApJ...289...81R}) 
and HI radio emission (based on various sources in the literature).
Most of Kent's sample comes from the Rubin et~al.\ surveys, which selected for
spiral galaxies with high inclinations, 
spanning a wide range of luminosities, scale-lengths, and Hubble types,
and without strong bars.
Despite advances in observational resources in the intervening decades,
we know of no comparable, publicly-available sample that includes both rotation curves and
photometry with detailed bulge/disk decompositions for a wide range of disk-galaxy types.

We estimate the disk and bulge scale-lengths ($\Rd$ and $a_{\rm e, b}$)
by modeling the nonparametric Kent decompositions
with simple exponential and de Vaucouleurs profiles ($n=1$ and $n=4$, 
respectively).
Our models thereby treat all bulges as ``classical'', with $n\sim4$,
neglecting some variations in their detailed properties, such as
the $n \sim$~1--2 indices of ``pseudo'' bulges \citep{2004ARA&A..42..603K}.
The latter bulges tend to be much less massive, and make only minor
contributions to the total \js\ for spirals, which is insensitive to the 
details of the adopted bulge density and rotation profiles.\footnote{More
extensive observations and modeling in the future could
be used to establish the \js--\Ms\ trends for morphologically different bulges,
and thereby provide physically-based information as to whether or not
there are genuinely distinct subtypes.}

For 34 of these sample galaxies (type Sb to Sc), independent decompositions 
were carried out on the {\it same data set}
by \citet{1994MNRAS.267..283A}, using parametric fits to the raw surface
brightness profiles.
Our $\Rd$ values agree with theirs at the $\sim$~10\% level,
while the bulge results are highly variable, both between our
analyses and theirs, and between different model fits by these authors.
Most of these galaxies are very disk dominated ($B/T \la 0.1$),
so it is not surprising that the bulge parameters would be very uncertain.
Fortunately the bulges in such cases turn out to be only very minor
contributors to the total \js\ of their host galaxies.
Other parameters and their sources are listed in Table~C\ref{tab:spirals}.

For $\vs$ of the stellar disk components of these galaxies, we assume that they rotate
with the same velocities as their gas disks.
We derive $v_{\rm c}$ based on the rotation curves
over the range (2--3)~$\Rd$, re-projecting this intrinsic value to
the observed $\vs$ according to the inclination 
($\vs = v_{\rm c} \sin i$).

The final and most challenging parameter to estimate is 
the characteristic rotation velocity $\vs$ for the bulges.
Direct estimates of bulge rotation-velocity profiles over a large
range in radius require extensive spectroscopic data combined with careful bulge--disk 
kinematic decomposition.
As far as we know, this has only been done for {\it one} spiral galaxy to date
\citep{2012ApJ...752..147D}.
Thus we are much worse off with estimating \js\ for spiral bulges than for
early-type galaxies, and must make even stronger simplifying assumptions than in the
original F83 analysis of ellipticals.  Fortunately, because the spirals are disk-dominated,
we will find that their total \js\ estimates
are only mildly sensitive to the assumptions about bulge kinematics.

Our strategy for the bulge $\vs$ values is to estimate these
indirectly, based on other observables:
the ellipticity $\epsilon\equiv 1-q$ 
and the central velocity dispersion $\sigma_0$.
These three parameters may be related together through the following model:
\begin{equation}\label{eqn:binney}
\vs = \left(\frac{v}{\sigma}\right)^* \sigma_0 \left(\frac{\epsilon}{1-\epsilon}\right)^{1/2} ,
\end{equation}
where $(v/\sigma)^*$ is a parameter describing the relative dynamical importance of rotation
and pressure.
In an edge-on galaxy, $(v/\sigma)^* \simeq 1$ represents an oblate isotropic system where the
observed ellipticity is supported by rotation, and this model also turns out to work
well at other inclinations
\citep{1982modg.proc..113K}.

The standard lore is that spiral bulges and low-luminosity ellipticals are
near oblate-isotropic, with typical $(v/\sigma)^*\sim$~0.9 
\citep{1982ApJ...256..460K,1983ApJ...266...41D,1998gaas.book.....B,2008gady.book.....B}.
However, some concerns about these conclusions were raised early on
\citep{1984ApJ...287...66W,1986ApJ...302..208F}
and modern integral-field analysis of early-types has revealed that their
rotation velocities tend to be significantly {\it lower} than in the oblate isotropic model 
\citep{2007MNRAS.379..418C,2011MNRAS.414..888E}.
The rotation of spiral bulges, on the other hand,
has not seen systematic investigation in decades
(some new work has just appeared in \citealt{2012ApJ...754...67F}),
and here we attempt only a quick look at the implications of recent papers that have 
reported bulge kinematics for a handful of cases.

\begin{figure}
\centering{
\includegraphics[width=3.5in]{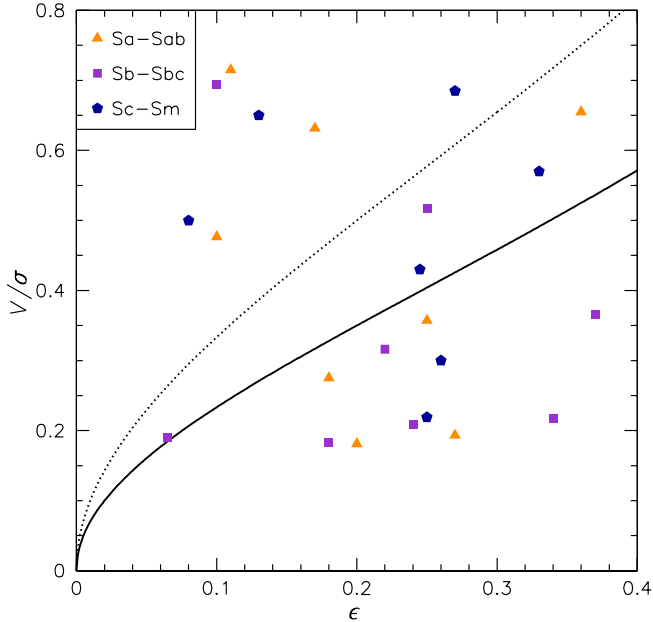} 
}
\caption{Relation between bulge rotation velocity and velocity dispersion
as a function of ellipticity.
The points show data for 26 spiral galaxies from the literature,
with symbol shapes and colors corresponding to different Hubble types as in the legend.
The curves show Equation~(\ref{eqn:binney}) with
$(v/\sigma)^* = 1$ and $(v/\sigma)^*=0.7$ for the dotted and solid curves, respectively.
We adopt $(v/\sigma)^*=0.7$ as our default model.
}
\label{fig:vsigb}
\end{figure}

We take results on $(v/\sigma)$ and $\epsilon$ from
\citet{2007MNRAS.381..401L}, \citet{2008MNRAS.389..341M}, and \citet{2009MNRAS.395...28M},
and plot them in Figure~\ref{fig:vsigb}.
We see that the oblate isotropic model is {\it not} a good representation of most of the data,
nor is any other simple value of $(v/\sigma)^*$.
However, in order to have a simplified framework for bulge
rotation, we characterize this data set as having $(v/\sigma)^*=0.7\pm0.4$
(median and 68\% scatter).

We therefore adopt the following procedure for estimating bulge \js.
We use the observational values for $\epsilon$ and $\sigma_0$, and then estimate
$\vs$ using Equation~(\ref{eqn:binney}) with $(v/\sigma)^*=0.7$ representing
a typical value for bulges.
We test the impact of the latter assumption on the results by also using
$(v/\sigma)^*=0.3$ and $1.1$ to bracket the possible range of average
bulge rotation.  We thereby explore the systematic uncertainty in bulge rotation
but not the intrinsic scatter, keeping in mind also that this bulge model
is based on the central regions and does not account for the uncertainties
in extrapolating the rotation to large radii,
as discussed in detail for the early-type galaxies.

The $\epsilon$ values are taken from the Kent derivations.
We take the $\sigma_0$ measurements in most cases from HyperLeda \citep{2003A&A...412...45P},
and also from \citet{1999A&A...342..671C} and \citet{2004A&A...424..447P}.
For some of the later-type galaxies, there are no $\sigma_0$
measurements available, and for these we use an empirical relation 
(which we infer from other galaxies in these studies)
that $\sigma_0$ is approximately equal to the gas-disk rotation velocity.
Such cases all have $B/T < 0.15$, so this approximation is not of major importance for
the total \js\ estimates, but any inferences for these particular bulges 
will be relatively uncertain.

We now have enough information to proceed with the specific angular momentum
calculations for the spiral galaxies.
Again, our basic approach is to
estimate separately the bulge and disk angular momenta $j_{\rm b}$ and $j_{\rm d}$.
Given a bulge stellar mass fraction quantified as $f_{\rm b}$, we can then estimate the
total specific angular momentum by:
\begin{equation}\label{eqn:jtot}
j = f_{\rm b} j_{\rm b} + (1-f_{\rm b}) j_{\rm d} .
\end{equation}
In practice, we use the bulge-to-total $r$-band luminosity ratio $B/T$ (from
the series of Kent papers) as a proxy for $f_{\rm b}$.

To calculate the projected values of $j_{\rm b}$ and $j_{\rm d}$, we use
Equation~(\ref{eqn:jCK0}).  For the intrinsic values,
we assume that both the bulge and the disk in a given galaxy have the same inclination $i$, 
which is estimated from the observed disk ellipticity.
We then use the deprojection factor $C_i$ to convert projected to intrinsic values
[see Equation~(\ref{eqn:proxy})].
For the disk, this is a simple factor of $(\sin i)^{-1}$, and the calculation reduces
to Equation~(\ref{eqn:F83eq1}).
For the bulge, we calculate $C_i$ from Equation~(\ref{eqn:Cform}).

Using these procedures, 
we construct a catalog of spiral galaxies with
characteristic masses, scale-lengths, and rotation velocities for both their bulge
and disk components.  We report these values in Table~C\ref{tab:spirals}, along with
the total galactic specific angular momenta (bulge and disk combined), 
both projected and intrinsic.
When we vary the assumed bulge rotation systematically across the bracketing range,
the total \js\ is changed by no more than $\sim$~0.03~dex ($\sim$~7\%) for the vast majority 
of the galaxies, and up to $\sim$~0.1~dex ($\sim$~25\%) for a few of the Sa--Sab galaxies.
Therefore the details of the bulge modeling are of only very mild importance
to the overall \js\ results for the spirals.
These data will be used in later sections to examine various scaling relations for
these galaxies and for their subcomponents.

\subsection{Early-types}\label{sec:etgdata}

For the gas-poor early-type galaxies (lenticulars and ellipticals), 
the challenge is to assemble a large sample with all of the ingredients
that we need to calculate \js\ (i.e., $\vs$, \aeff, $n$).
The information is scarcest for $\vs$, and therefore we have scoured
the literature for kinematic data sets extending to radii of at least
$\sim$~2~\aeff, assembling a sample that, although not exhaustive,
is unprecedented in its size and scope.
The sources include integrated-starlight absorption-line spectroscopy,
and velocities of GCs and PNe.  
To estimate approximate values for $\vs$,
we simply read off the major-axis rotation velocity at 2~\aeff\ 
(as explained in Section~\ref{sec:simp}).
We thereby assemble a total sample of 40 early-type galaxies, including the 8 galaxies
that we modeled in detail in Section~\ref{sec:examp}.

Table~C\ref{tab:etg} provides a summary of our sample,
along with the sources of kinematic data.
Given that the data are drawn from a variety of literature sources with
complex selection effects, it is important to check whether or not the sample
is a fair representation of early-types in the nearby universe.
We have done so in Appendix~\ref{sec:obsfull}, using the \atlas3d\ volume-limited sample
of nearby galaxies as a reference,
and focusing on the
masses \Ms\ and central rotation parameters $(v/\sigma)^*$.

We find that the distribution of our sample galaxies in the $(v/\sigma)^*$--\Ms\
parameter space is fairly similar to that of an unbiased sample over a similar mass range.
The median galaxy mass in our sample is $\log\,(M_\star/M_\odot)=10.8$, which is near the 
characteristic mass $M_\star^*$ of nearby galaxies \citep{2010MNRAS.404.1111G}.
We thus conclude that our observational results should be representative of low-redshift
ordinary early-type galaxies.  
The only caveat here is that our sample is biased toward ellipticals at the
expense of lenticulars, which we must take into account later when drawing 
conclusions about the overall population of early-type galaxies.

An alternative scheme for classifying early-types is as ``fast rotators''
(including almost all lenticulars) and ``slow rotators'', 
based on their central kinematics \citep{2007MNRAS.379..401E}.
The central rotation is known to correlate with many other galaxy properties
\citep{1983ApJ...266...41D,1996ApJ...464L.119K}, 
and the fast and slow rotators have been interpreted as having
different formation histories.
Therefore it is important that we investigate to what extent
the {\it global} specific angular momentum \js\
correlates with the central rotation classification.  
Our sample includes three slow-rotators, which is consistent with the
fraction of such galaxies in the nearby universe \citep{2011MNRAS.414..888E}, 
and will provide a rough initial idea of any systematic differences between
fast and slow rotators.

Returning to the remaining observational parameters,
for each early-type density profile, we need both the S\'ersic index $n$ and
the corresponding scale-length \aeff\ (which can differ significantly from
the value obtained with a classic $n=4$ fit,
e.g., in the RC3 catalog of \citealt{1991trcb.book.....D}).
Unfortunately there is no comprehensive source available for such measurements,
and we resort to a medley of literature data.

For 34 of the galaxies in our sample, there are published S\'ersic fits,
and we take the $(a_{\rm e}, n)$ values according to the following priority:
detailed photometric analysis in individual galaxy papers
(e.g., \citealt{2009MNRAS.393..329N});
the \citet{2009ApJS..182..216K} tabulation for Virgo galaxies;
\citet{2009ApJS..181..135H,2009ApJS..181..486H,2001MNRAS.326.1517D}.

For the remaining 6 galaxies, 
we have as a starting point the RC3 value for the effective radius.
Then we use the well-established observation that there are strong correlations between
early-type galaxy size and luminosity, and the S\'ersic index $n$
(e.g., \citealt{1993MNRAS.265.1013C,1997A&A...321..111P,2003AJ....125.2936G,2003ApJ...594..186B,2009ApJS..182..216K}).  
This allows us to estimate a most-probable $n$ value for each galaxy
(see Appendix~\ref{sec:obsfull} for details).

Note that if we were simply to approximate all of the early-types as $n=4$ spheroids,
the $k_n$ values in Equation~(\ref{eqn:jCK0}) would be too high on average by
$\sim$~30\% ($\sim$~0.15~dex, given a median index value of $n\sim2.5$).
This would translate to an equivalent systematic error on \js.
We could adjust for this effect by adopting $n=2.5$ in all cases, but $n$ also
has a systematic dependence on galaxy mass, and ignoring this fact would produce
a spurious mass-dependent trend in \js\ of $\sim$~50\% ($\sim$~0.2~dex) over the 
full range in mass.

In Table~C\ref{tab:etg}, we compile the observed parameters $\vs$, \aeff, and $n$ for 
our full early-type galaxy sample.
We use these to calculate \jp\ approximately from Equation~(\ref{eqn:jCK0}),
and tabulate these values as well.
For some of the very extended galaxies like NGC~4374, 
the total luminosity and angular momentum
(via the factor $k_n$) are integrated out only to the estimated virial radius.

In order to convert projected \jp\ to intrinsic \jt\ for analysis in later sections, 
we must apply a deprojection factor $C_i$ which depends on the inclination $i$.
Unfortunately, 
the individual inclinations are not generally known, but neither are they completely random,
because of an inclination-bias in galaxy classification.
As discussed in Appendix~\ref{sec:spheroid}, we therefore apply median deprojection factors of
$C_{\rm med}=1.21$ ($+0.08$~dex) to the lenticulars, 
and $C_{\rm med}=1.65$ ($+0.22$~dex) to the ellipticals.

Since one of our eventual goals will be to quantify the intrinsic scatter in the observed
\js--\Ms\ relations, it is important to be clear about the error budget
in our analyses.  Again, the basic parameters that go into our \js\ calculations are
$C_i$, \aeff, $n$, and $\vs$.  For early-type galaxies with an assumed $n=4$ profile,
the typical uncertainties in \aeff\ are $\sim$~25\% ($\sim$~0.1~dex;
\citealt{2011MNRAS.413..813C}).  If we allow for a more general $n$, which
for some galaxies is measured directly and in other cases is derived statistically
(Appendix~\ref{sec:obsfull}), then we estimate a combined uncertainty on \js\ from
\aeff\ and $n$ of $\sim$~40\% ($\sim$~0.15~dex).
The uncertainty on $\vs$ from our simplified measurement and extrapolation approach is
$\sim$~25\% ($\sim$~0.1~dex; Section~\ref{sec:simp}).

Table~\ref{tab:err} summarizes the uncertainties introduced by a
number of different ingredients in the \js--\Ms\ calculations.
The separate uncertainties for \js\ and \Ms\ are
mapped to the direction
perpendicular to a $j_\star \propto M_\star^{2/3}$ 
trend, as discussed in Section~\ref{sec:sys}.
This net uncertainty is designated $\Delta \lambda$, owing to the
connection with spin-based theoretical models.

The total uncertainty in $\lambda$ for late-type galaxies is 
typically $\sim$~30\% ($\sim$~0.1 dex), and is driven by the 
estimate of \Ms\ (via \MLs) rather than \js.
For the vast majority of the early-types (apart from the special
class of massive, extended ellipticals), the uncertainty is $\sim$~60\%
($\sim$~0.2~dex), and is driven by the four parameters
mentioned above that enter into the \js\ calculation.

\begin{table}
\begin{center}
\caption{Uncertainty budget}\label{tab:err}
\noindent{\smallskip}\\
\begin{tabular}{l c c c c c c c c}
\hline
Galaxy & & & & $\Delta \lambda$ (dex) \\
type & $D$ & $C_i$ & $v_s$ & $\tilde{v}_s$ & $n,a_{\rm e}$ & bulge & \MLs\ & total\\
\hline
Sb--Sm & 0.01 & 0.01 & 0.02 & 0.03 & 0.05 & 0.03 & 0.07  & 0.09 \\
Sa--Sab & 0.01 & 0.01 & 0.02 & 0.03 & 0.05 & 0.1 & 0.07 & 0.13 \\
S0 & 0.01 & 0.05 & 0.06 & 0.1 & 0.15 & 0 & 0.07 & 0.18 \\
fE & 0.01 & 0.15 & 0.06 & 0.1 & 0.15 & 0 & 0.07 & 0.22 \\
sE & 0.01 & 0.12 & 0.35 & 0.35 & 0.2 & 0 & 0.2 & 0.40 \\
\hline
\hline
\end{tabular}
\\
\tablecomments{
The uncertainties on \js\ and \Ms\ have been converted into equivalent
uncertainties on $\lambda$.
The different galaxy types include 
fast- and slow-rotating ellipticals (fE and sE).
The listed sources of potential error are
distance ($D$),
corrections for projection effects including inclination ($C_i$),
the rotation velocity scale calculated in detail ($v_s$),
the alternative approximate rotation velocity scale ($\tilde{v}_s$),
the stellar density profile S\'ersic index ($n$) and 
scale radius ($a_{\rm e}$),
the incorporation of bulge contributions,
and the stellar mass-to-light ratio including IMF variations (\MLs).
}
\end{center}
\end{table}

This full \js--\Ms\ dataset is assembled from
a generally unbiased $\sim M_\star^*$ galaxy sample that we can use to
investigate differences in angular momentum not only between early-types and spirals, 
but also between ellipticals and lenticulars, and between fast and slow rotators.

\subsection{Size and rotation-velocity scaling relations}\label{sec:scale}

Before considering specific angular momenta
and their correlations in the next section, we examine some trends among the
raw ingredients that go into these analyses, \aeff, $\vs$, and \Ms.
Doing so provides a check that our results are consistent with the familiar size--mass
and mass--rotation~velocity (Tully-Fisher) 
relations that have been established for nearby galaxies.
We also introduce novel relations involving rotation, and explore some
preliminary indications about angular momentum.

We first consider the standard scaling relation of galaxy size versus mass,
or $a_{\rm e}$ versus \Ms\ in our notation, showing the results in Figure~\ref{fig:size},
where we again compare our results to the
volume-limited \atlas3d\ sample as a baseline check.
We find that in both samples, late- and early-type galaxies have {\it roughly}
 the same sizes 
at a given mass
(cf.\ \citealt{2003MNRAS.343..978S,2007MNRAS.379..400S}), but
there is a clear systematic trend for the more bulge-dominated galaxies to be more compact
(see also \citealt{2004MNRAS.355.1155D,2009MNRAS.393.1531G,2010MNRAS.402..282M,2011MNRAS.414.2055M,2011MNRAS.416..322D}).
Given the many different assumptions and data sources that went into our
sizes and masses, these parameters match the \atlas3d\ results remarkably well overall
(with some nuances discussed further in Appendix~\ref{sec:obsfull}).
This suggests that our size and mass data are representative and reliable at the $\sim$~0.1~dex level.

\begin{figure}
\centering{
\includegraphics[width=3.5in]{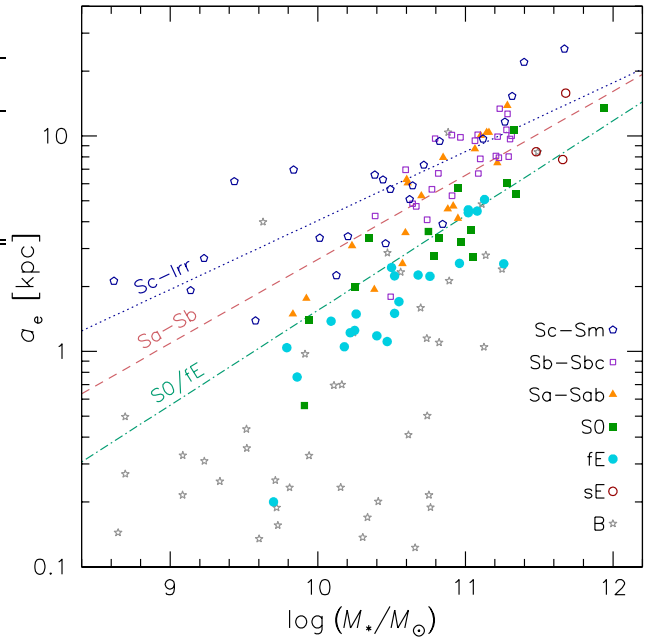} 
}
\caption{Relation between size and stellar mass for our galaxy sample.
The former is the semi-major axis effective radius, and the latter
is based on $K$-band total luminosities with an adopted mass-to-light ratio of
$M_\star/L_K=1$ in solar units.
Different symbols denote different galaxy types as shown in the legend;
for the spirals,
the disk and bulge (``B'') components are shown separately.
The range of the plot is restricted in order to better see the
main trends in the data; the bulge data extend 
to radii as small as $a_{\rm e}\sim 0.01$~kpc (note also that the most compact elliptical
shown is NGC~4486B, which is considered a rare, highly-stripped galaxy).
For comparison, diagonal lines show
power-law model fits to the data from the \atlas3d\ survey
(i.e., {\it independent} from our data set):
lenticulars and fast-rotator ellipticals (dot-dashed),
Sa--Sb spirals (dashed), and Sc--Irr spirals (dotted).
For both data sets, the late-type galaxies are systematically larger than the early-types
at a given stellar mass.  The absolute normalizations of the trends are similar 
between the \atlas3d\ sample and ours,
with some small differences as discussed in the text.
} \label{fig:size}
\end{figure}

We can also consider separately the spiral {\it bulges}, plotting their sizes and masses
for our sample
in Figure~\ref{fig:size}.
Although the full range of sizes is not visible in this plot, the bulges
follow a roughly parallel size--mass relation to the elliptical galaxies,
but smaller on average by a factor of $\sim$~4 ($\sim$~0.6~dex) 
and with a great deal of scatter
(possibly because of the approximate nature of these size measurements).
Other studies have also found that bulges are more compact than ellipticals
\citep{2008MNRAS.388.1708G,2009MNRAS.393.1531G,2010MNRAS.405.1089L,2011MNRAS.416..322D},
but the quantitative details vary considerably, and we therefore regard
our bulge scaling relations as provisional.  

\begin{figure*}
\centering{
\includegraphics[width=6.6in]{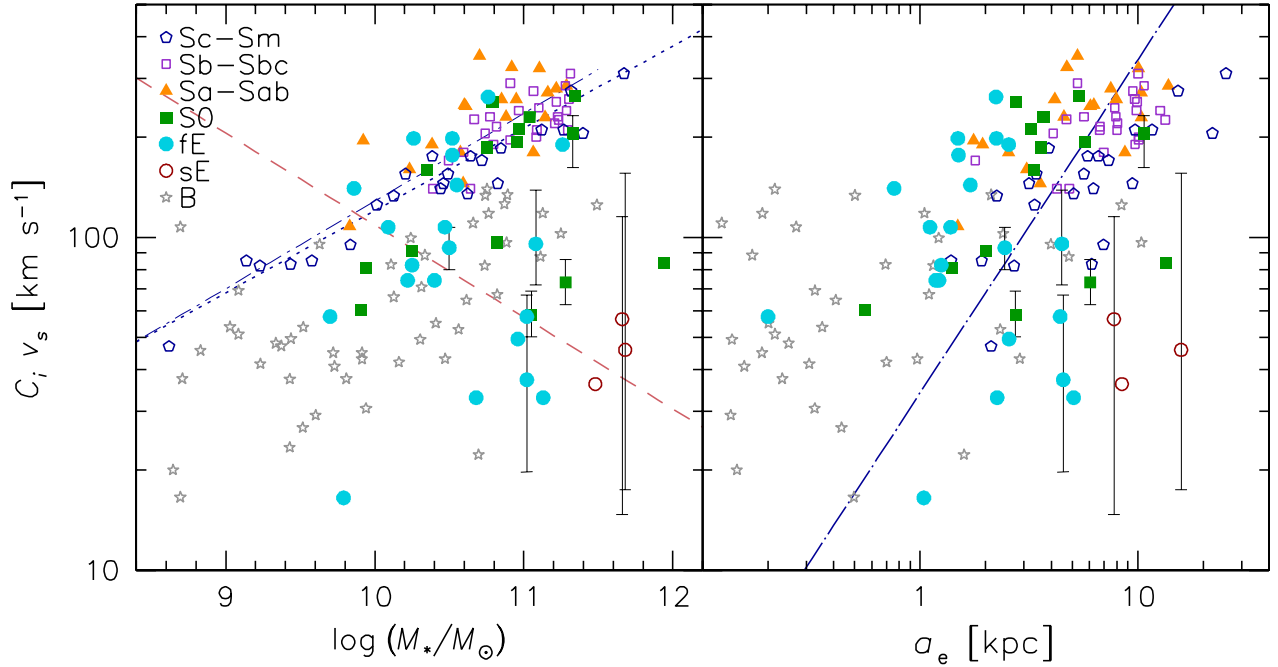} 
}
\caption{Relations between characteristic rotation velocity $C_i\,\vs$, stellar mass
(left-hand panel), and size (right-hand panel) for our
full galaxy sample, using the same data sources and symbols as in
Figure~\ref{fig:size}.
For the spiral disks, $\vs$ is the outer gas-disk rotation velocity.
For the lenticulars and ellipticals, $\vs$ is the stellar rotation velocity measured along
the semi-major axis at 2~\aeff, except for the points with error bars,
which are the eight cases studied in detail in Section~\ref{sec:examp}, with
$\vs$ derived from full modeling of the rotation-velocity profiles.
For the bulges, $\vs$ is estimated indirectly using flattening and velocity
dispersion observations (Section~\ref{sec:ltgdata}).
In all cases, the rotation velocity has been deprojected for both inclination and ``dilution''
effects, using the factor $C_i$ (see text for details).
In the left-hand panel, the dotted blue line shows a least-square fit
to the Sb--Sc disks,
a dashed red line shows a proposed inverse trend for a subset of the E/S0s, and
the blue dot-dashed line shows the baryonic Tully-Fisher relation
for late-type galaxies from \citet{2011ApJ...742...16T} for comparison.
In the right-hand panel, the diagonal line shows a prediction for the spiral
disks based on $\Lambda$CDM models
(see Section~\ref{sec:theory2}).
Overall, the spiral and elliptical galaxies follow mass--rotation~velocity and 
size--rotation~velocity trends that have remarkably opposite slopes.
The trends for the lenticulars are between the spirals and ellipticals.
} \label{fig:rot}
\end{figure*}

The next scaling relation that we consider is rotation velocity
versus mass.  For spiral
galaxies, this is the Tully-Fisher relation, but it has to
our knowledge never been constructed previously for all galaxy types.
We can already generate a broad expectation for what we will find,
given the observed size--mass relations along with the assumption that \js\ is independent
of galaxy type.
As mentioned in Section~\ref{sec:gen}, we can then use
Equation~(\ref{eqn:jCK0}) to predict the ratio of characteristic rotation velocities for
ellipticals and spirals:
\begin{equation}
\frac{v_{s,{\rm E}}}{v_{s,{\rm Sp}}} \sim \frac{k_1}{k_4} \frac{a_{\rm e,Sp}}{a_{\rm e,E}} ,
\end{equation}
where we are approximating the spiral galaxy parameters as dominated by the disk component.
With $k_1/k_4=0.5$, and $a_{\rm e,Sp}/a_{\rm e,E} \sim 2$ for our sample,
we therefore predict $v_{s,{\rm E}}/v_{s,{\rm Sp}} \sim 1$. Thus,
{\it ellipticals should rotate at roughly the same velocity as spirals if they have
the same specific angular momenta at a given mass.}

Without proceeding any further, this scaling analysis already suggests that
ellipticals have lower \js\ than spirals, or else they would be extremely flattened
by rotation, similarly to the spiral disks which have near-maximal rotational support
(modulo possible differences in dynamical mass between
spiral and elliptical galaxies at the same stellar mass).
The same argument applies even more strongly to the spiral bulges, since they
are far more compact than the disks at a given mass.
If the bulges had the same \js\ as the disks, then 
they would have to rotate much {\it faster}, which is impossible.

We now examine what our new collection of observations tells us directly about the
rotation scaling relations. The left-hand panel of
Figure~\ref{fig:rot} shows the characteristic rotation velocity $\vs$
for the elliptical and lenticular galaxies, and the spiral disk and bulge subcomponents,
in our sample.
Here we are plotting the {\it intrinsic} rotation velocity,
multiplying by the deprojection factor $C_i$, which is just $(\sin i)^{-1}$ 
for disks (see Appendix~\ref{sec:thin}), 
and Equation (\ref{eqn:Cform}) for bulges.
For the early-type galaxies, the inclinations are unknown, and we have adopted
median factors for $C_i$ as discussed in Section~\ref{sec:etgdata}.

We see that the disks follow a fairly tight relation of approximately
$C_i\,v_{\rm s} \propto M_\star^{0.25}$,
with a residual trend for the later-type disks to rotate more slowly.
This is equivalent to the familiar Tully-Fisher relation, and in the Figure 
we include a recent result from the literature \citep{2011ApJ...742...16T}, 
which matches our data very well
(cf.\ the type-dependence among spirals found by \citealt{2008AJ....135.1738M}).
We also show in the right-hand panel of Figure~\ref{fig:rot} the relation between
size and rotation velocity, which are strongly correlated parameters for disk galaxies.

The elliptical galaxies are completely different,
showing an {\it anti-correlation} between rotation velocity and mass,\footnote{This
echoes a similar trend in the {\it central} rotation properties of
early-type galaxies in general (shown in Figure~\ref{fig:atlas1}).
The eight galaxies studied in detail (points with error bars in Figure~\ref{fig:rot})
are consistent with this trend but do not include enough lower-luminosity ellipticals
to distinguish between $v_{\rm s}$ being constant or decreasing with mass.}
with $C_i\,v_{\rm s} \propto M_\star^{-0.1}$.
This result also contrasts markedly with standard relations
for ellipticals involving the velocity dispersion $\sigma_0$ or the dynamical mass 
(e.g., $\sigma_0 \propto M_\star^{0.25}$;
\citealt{1976ApJ...204..668F,2011ApJ...742...16T}).
In galaxy disks, the rotation velocity traces the dynamical mass,
so the Tully-Fisher relation is a measure of both mass and angular momentum.
In elliptical galaxies, on the other hand,
the mass and angular momentum relations are decoupled.
We also find an anti-correlation between rotation velocity and size (right-hand panel)
that we will discuss later in this paper.

The behavior of the lenticulars in the mass--rotation~velocity diagram is difficult to 
discern in detail owing to the small sample size, but in general it appears
intermediate to the other galaxy types.  We also notice
an interesting pattern when considering the lenticulars and ellipticals together:
there may be a {\it bimodal} mass--rotation~velocity relation,\footnote{This
pattern may be partially an artifact of inclination effects.
In particular, some of the edge-on lenticulars
were observed with long-slit spectroscopy
directly along their embedded disks, which may not provide an accurate measurement
of the overall rotation.  However, for the ellipticals we find no correlation
between apparent rotation velocity and ellipticity.
An additional issue is that the occasional extremely low-inclination
galaxy will not be treated well by our median-deprojection method
(cf.\ the right-hand panel of Figure~\ref{fig:Cr}), so in any fits to the data,
we will discard outliers with very low $\vs$ or \js\ (e.g., NGC~1419).
}
with some galaxies following
the trend for spirals, and others following a steep reverse relation,
$C_i\,v_{\rm s} \propto M_\star^{-0.3}$.
The implication is that there may be two distinct populations of early-type galaxies,
one of which is closely related to spirals, and which are not equivalent to
standard E and S0 classifications.

The bulge rotation velocities appear to follow a similar trend to the spirals, 
at about half the amplitude.
Here it should be remembered that the bulge ``data'' points are {\it indirect}
estimates constructed in order to provide plausible adjustments to the total
angular momenta of the spiral galaxies (Section~\ref{sec:ltgdata}).
The results so far suggest that bulges are different from
ellipticals in their mass--size--rotation~velocity relations, and we will see in
the next section how their angular momenta compare.

Since both the sizes and the rotation velocities of elliptical galaxies are systematically
lower than for spiral disks, we can already predict that the ellipticals will 
on average have much lower \js.
Note that although this conclusion has already been widely adopted for decades,
only now have the kinematic data reached large enough radii to confirm it
with confidence.

To see that the low characteristic rotation velocities for ellipticals
are not a mathematical sleight of hand,
one may consider the specific cases of NGC~821 and NGC~3377 in Figure~\ref{fig:etg}.
The rotation-velocity profiles of these galaxies decline dramatically outside $x\sim$~(1--2)~\aeff,
which may be contrasted with the spiral galaxies in Figure~\ref{fig:spirals}.
Preliminary analysis of additional {\it edge-on} cases, where the deprojection
uncertainties are minimized, indicates that such declines are a {\it generic
feature} of $\sim M^*$ early-type galaxies (A. Romanowsky et al., in preparation).

This conclusion includes NGC~2768, which from the current data
 appears consistent with a constant
or rising outer rotation velocity, but which with more extensive new PN data may have
a declining outer profile.
Even the cases of strongly rising rotation-velocity profiles out to $x\sim$~2~\aeff\ 
found by \citet{1999ApJ...513L..25R} 
appear upon closer inspection to turn over at larger radii.
These results all contrast with early claims of high outer rotation
in some early-types, which were recently overturned with improved observations
(e.g., \citealt{1994Msngr..76...40A,1998AJ....116.2237K,2006pnbm.conf..294R,2010A&A...518A..44M,2011ApJS..197...33S}).

We can also begin making some interesting inferences
about the relations among other galaxy types, based on both
size and rotation-velocity trends (Figures~\ref{fig:size} and \ref{fig:rot}).
As discussed, the lenticulars share similar properties to spirals
in some cases, and to ellipticals in others.
The distinction between ``fast'' and ``slow'' rotator ellipticals
based on their inner regions does not appear to hold up when considering
their global rotation properties.

This overview of the observable scaling relations between mass, size, and rotation velocity
gives us a preview of some of our overall conclusions about angular momentum, and
provides more confidence in the solidity of those conclusions.
We construct a novel mass--rotation~velocity relation for ellipticals, which is the
analogue of the Tully-Fisher relation for spirals, but with the remarkable difference
of having a negative slope.
The data also imply  that both elliptical galaxies and spiral bulges must have 
lower specific angular momenta than spiral disks of the same mass.
We address this issue more quantitatively in the next section, incorporating
the additional mass-dependent factor $k_n$ in calculating \js.

\section{Observations: angular momenta of the full sample}\label{sec:obsres}

Having derived estimates of the \js\ and \Ms\ parameters for our
full galaxy sample, we now examine the resulting observational trends,
which constitute the key results of this paper.
We begin by focusing on the late-type galaxies in Section~\ref{sec:less}, and
combine these with the early-types in Section~\ref{sec:obsresults}.
We discuss our proposed replacement for the Hubble sequence in Section~\ref{sec:replace},
which we test by examining systematic residuals from the \js--\Ms\ trends in Section~\ref{sec:resid}.
We further convert the \js--\Ms\ data into one-dimensional histograms in Section~\ref{sec:hist}.

\subsection{Lessons from spirals}\label{sec:less}

Although the main novelty of this paper is our careful consideration of
early type galaxies, we also include the oft-studied category of spirals in order
to provide an integrated analysis of bright galaxies of all types.
Furthermore,
the well-constrained angular momenta of the spirals also permit us to better understand
systematic issues such as inclination corrections that are trickier to handle
for early-types.

\begin{figure}
\centering{
\includegraphics[width=3.3in]{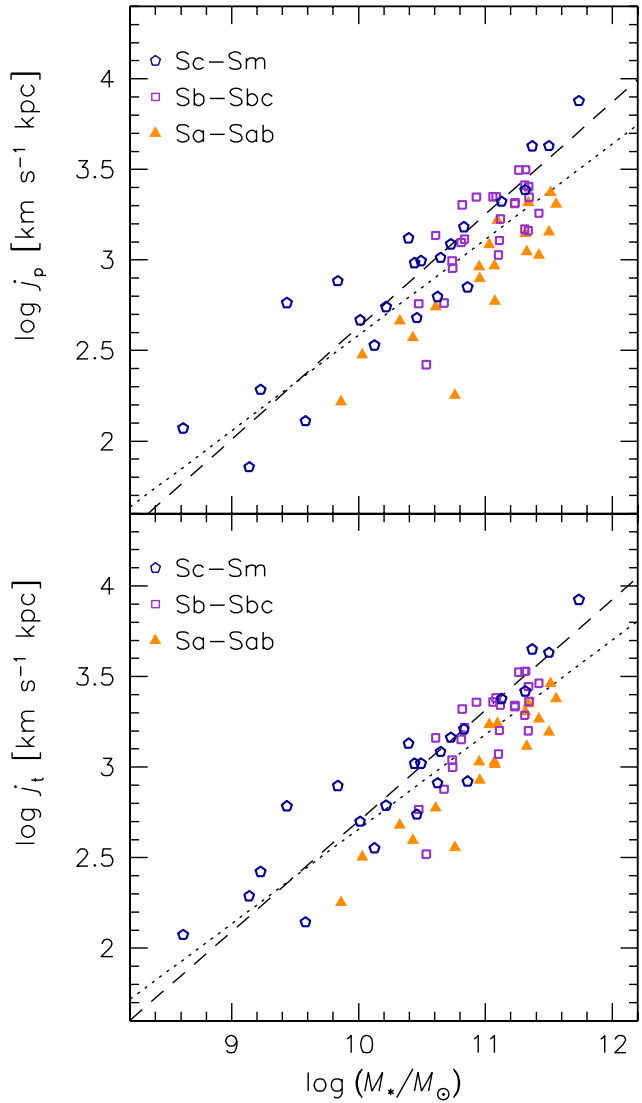} 
}
\caption{The total (disk plus bulge) stellar specific angular momentum of nearby spiral galaxies
plotted against total stellar mass.
The top and bottom panels show estimates of projected and intrinsic \js, 
respectively;
the uncertainty in \js\ for each galaxy is in almost all cases
smaller than the plotted symbols.
Different symbols denote galaxy sub-types as specified in the legends.
The dotted lines show fits to the data in each panel,
while the dashed lines show fits to the disk components alone (data not shown).
The spiral galaxies follow a universal \js--\Ms\ relation, with some
dependence on Hubble type.  The projected relation is very similar to the
intrinsic relation, but with a small offset, and slightly increased scatter, in \js.
\label{fig:JMsp}
}
\end{figure}

We plot the total (disk+bulge) \js--\Ms\ data for the spirals from
Table~C\ref{tab:spirals} in
Figure~\ref{fig:JMsp}.  In the top panel, we show the projected value, \jp,
and in the bottom panel, the intrinsic value, \jt.
These are related trivially by the disk inclination, but we wish to investigate
how well the trends in projection reflect the intrinsic trends, since 
deprojection for the early-type galaxies will be more difficult.

\begin{table}
\begin{center}
\caption{Mass--angular momentum fits to data}\label{tab:loglog}
\noindent{\smallskip}\\
\begin{tabular}{l c c c}
\hline
Sample & $\log j_0$ & $\alpha$ & $\sigma_{\log j_\star}$ \\
\hline
All spirals, total, projected & $3.11\pm0.03$ & $0.53\pm0.05$ & 0.22 \\ 
All spirals, total, intrinsic & $3.18\pm0.03$ & $0.52\pm0.04$ & 0.19 \\ 
Sa--Sab, total, projected & $2.93\pm0.05$ & $0.60\pm0.06$ & 0.17 \\
Sa--Sab, total, intrinsic & $3.02\pm0.04$ & $0.64\pm0.07$ & 0.12 \\
Sb--Sbc, total, projected & $3.15\pm0.03$ & $0.65\pm0.14$ & 0.16 \\
Sb--Sbc, total, intrinsic & $3.21\pm0.03$ & $0.68\pm0.13$ & 0.15 \\
Sc--Sm, total, projected & $3.25\pm0.04$ & $0.58\pm0.06$ & 0.20 \\
Sc--Sm, total, intrinsic & $3.29\pm0.04$ & $0.55\pm0.05$ & 0.18 \\
\hline
All spirals, disks, projected & $3.25\pm0.02$ & $0.62\pm0.05$ & 0.20 \\
All spirals, disks, intrinsic & $3.31\pm0.02$ & $0.61\pm0.04$ & 0.17 \\
Sa--Sab, disks, projected & $3.25\pm0.05$ & $0.76\pm0.09$ & 0.21 \\
Sa--Sab, disks, intrinsic & $3.34\pm0.04$ & $0.82\pm0.08$ & 0.17 \\
Sb--Sbc, disks, projected & $3.24\pm0.03$ & $0.71\pm0.14$ & 0.16 \\
Sb--Sbc, disks, intrinsic & $3.30\pm0.03$ & $0.75\pm0.12$ & 0.13 \\
Sc--Sm, disks, projected & $3.29\pm0.05$ & $0.61\pm0.07$ & 0.21 \\
Sc--Sm, disks, intrinsic & $3.33\pm0.05$ & $0.57\pm0.05$ & 0.19 \\
\hline
All spirals, bulges, projected & $2.20\pm0.31$ & $0.69\pm0.11$ & 0.58 \\
All spirals, bulges, intrinsic & $2.32\pm0.31$ & $0.69\pm0.10$ & 0.57 \\
Sa--Sab, bulges, projected & $2.30\pm0.32$ & $0.99\pm0.15$ & 0.47 \\
Sa--Sab, bulges, intrinsic & $2.44\pm0.32$ & $0.99\pm0.15$ & 0.46 \\
Sb--Sbc, bulges, projected & $1.89\pm0.34$ & $0.34\pm0.20$ & 0.58 \\
Sb--Sbc, bulges, intrinsic & $2.01\pm0.33$ & $0.34\pm0.19$ & 0.56 \\
Sc--Sm, bulges, projected & $2.21\pm0.57$ & $0.64\pm0.27$ & 0.60 \\
Sc--Sm, bulges, intrinsic & $2.30\pm0.58$ & $0.63\pm0.28$ & 0.60 \\
\hline
Lenticulars, projected & $2.97\pm0.08$ & $0.80\pm0.14$ & 0.29 \\
Lenticulars, intrinsic & $3.05\pm0.08$ & $0.80\pm0.14$ & 0.29 \\
Ellipticals, projected & $2.52\pm0.05$ & $0.60\pm0.09$ & 0.24 \\
Ellipticals, intrinsic & $2.73\pm0.05$ & $0.60\pm0.09$ & 0.24 \\
\hline
Sb--Sm, intrinsic, fixed $\alpha=2/3$ & $3.28\pm0.03$ & 0.67 & 0.19 \\
Ellipticals, intrinsic, fixed $\alpha=2/3$ & $2.75\pm0.05$ & 0.67 & 0.24 \\
{\it $\Lambda$CDM halos} & {\it 2.50} & {\it 0.67} & {\it 0.23} \\
\hline
\end{tabular}
\\
\end{center}
\end{table}

Overall, the spiral galaxies appear to follow fairly tight 
\js--\Ms\ trends, with similar slopes, regardless of Hubble sub-type.
In more detail, we carry out least-square fits to \js\ as a function of \Ms\
in log-log space:
\begin{equation}\label{eqn:loglog}
\log j_{\rm mod} = \log j_0 + \alpha \left[ \log (M_\star/M_\odot) -11 \right] ,
\end{equation}
with a residual rms scatter that we parameterize as $\sigma_{\log j_\star}$.
The uncertainties in the fit parameters $j_0$ and $\alpha$ are
estimated by bootstrap resampling.

Our fitting results for various spiral subsamples are reported in Table~\ref{tab:loglog}.
For total \js, the systematic uncertainties from the bulge rotation
(see Section~\ref{sec:ltgdata}) turn out to be smaller than or equal to
the statistical fitting uncertainties, even for the Sa--Sab galaxies,
and in the Table we have combined both uncertainties in quadrature.

The data are basically consistent with a universal \js--\Ms\ slope for spiral
galaxies of all types, with $\alpha$~$\sim$~0.6 and an rms scatter of 
$\sigma_{\log j}$~$\sim$~0.2~dex.
There is also a clear residual trend with Hubble type: 
the Sa--Sab galaxies have systematically lower \js\ than the Sb--Sm galaxies.
These conclusions hold for both \jp\ and \jt, although the uncertainties and
the scatter are smaller for \jt, as expected if there are genuine, underlying 
physical correlations that become clearer after deprojection.

The multi-component nature of our model galaxies allows
us to look further at disk and bulge properties separately.
We will take up this issue in Section~\ref{sec:obsresults}, and for now provide the fits to 
the $j_{\rm d}$--$M_{\rm d}$ and $j_{\rm b}$--$M_{\rm b}$ relations
in Table~\ref{tab:loglog}.
It should be remembered that the bulge results depend on
model assumptions, although as discussed, we have plausibly bracketed their
upper and lower limits for \js.

\begin{figure*}
\centering{
\includegraphics[width=3.5in]{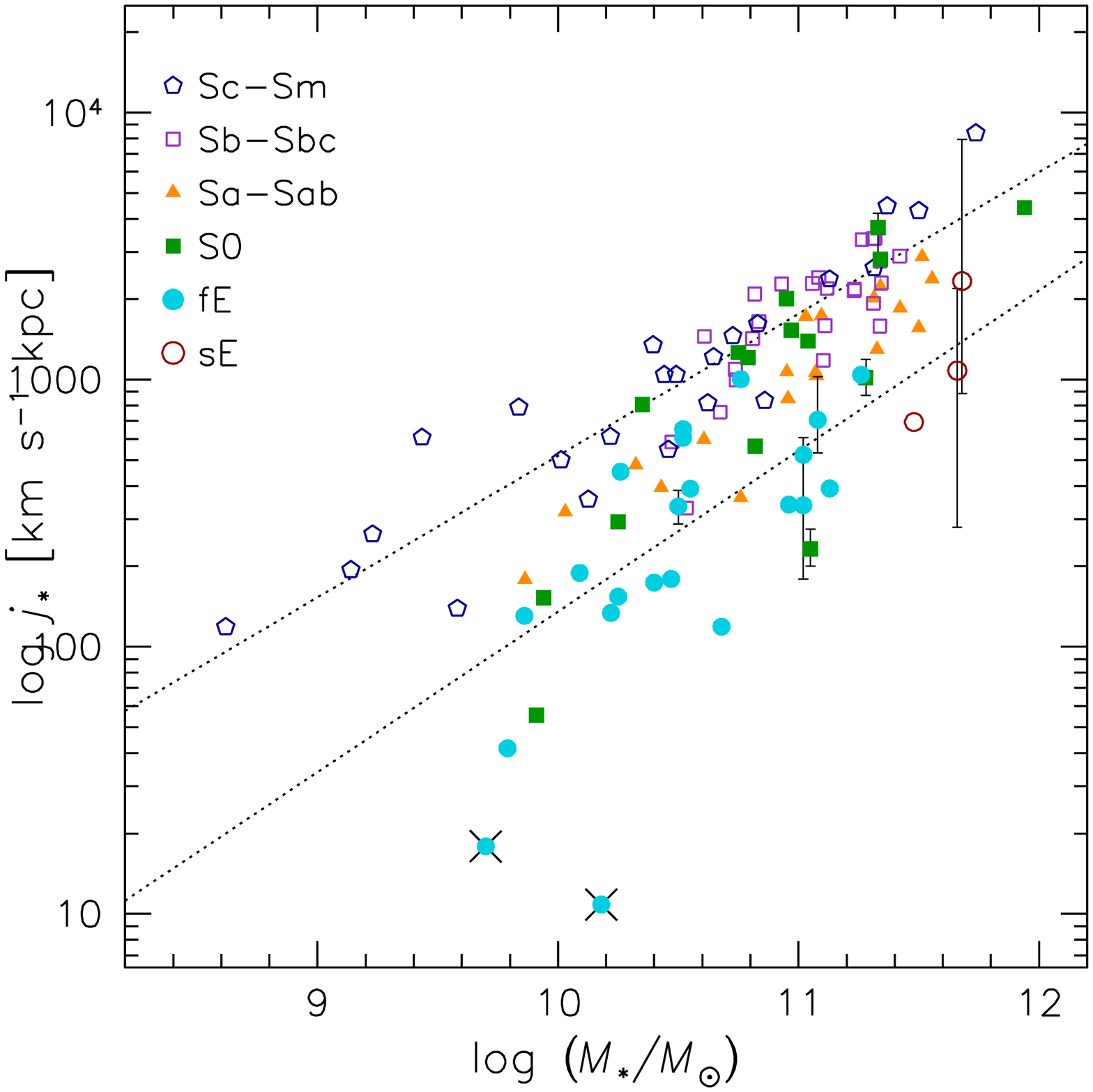} 
\includegraphics[width=3.5in]{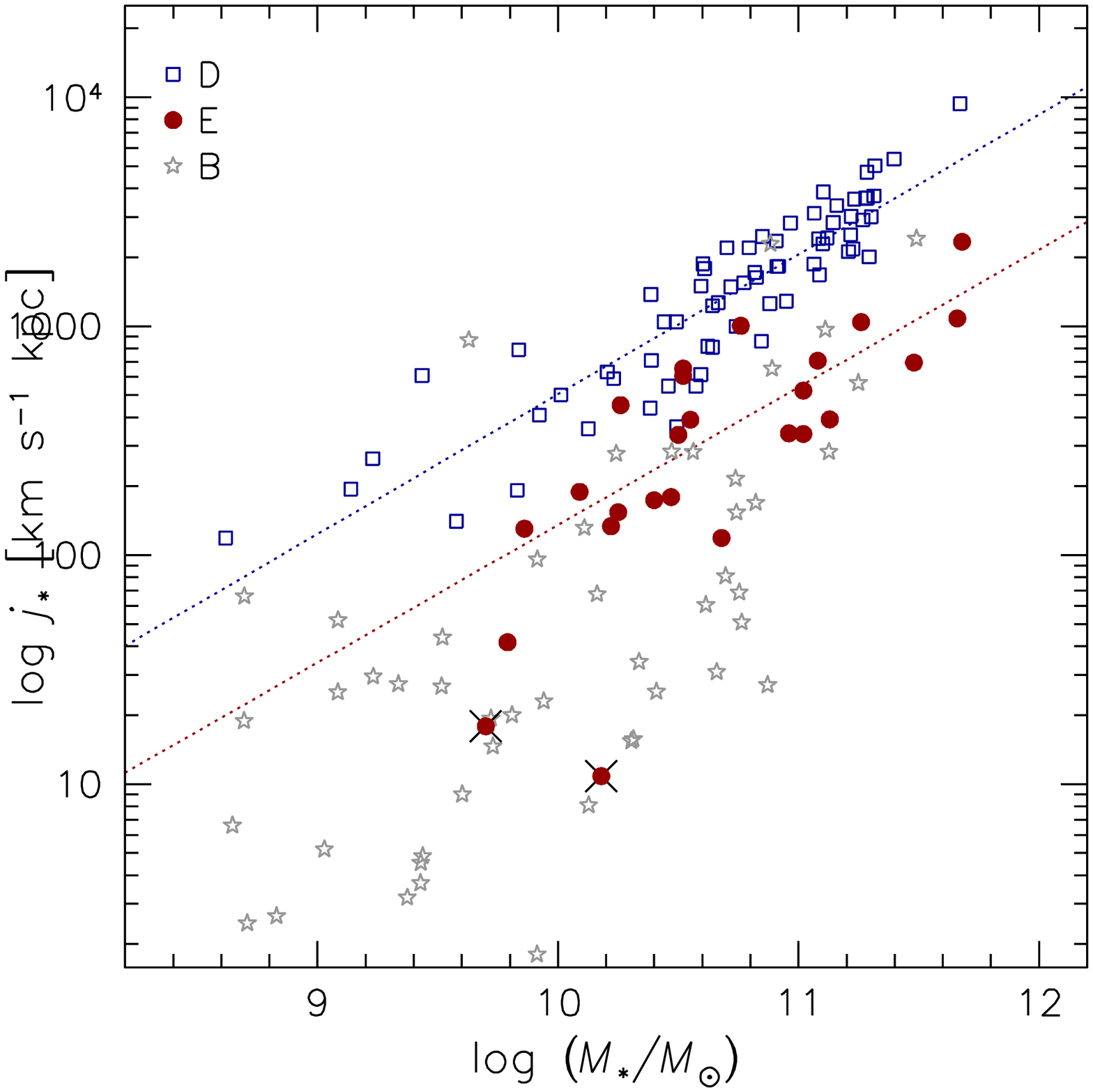} 
}
\caption{{\it Left-hand panel:} The total intrinsic specific angular momentum of galaxies
plotted against their total stellar mass.
Symbols show galaxy types according to the legend at the upper left.
The points with error bars shown are based on the
more detailed \js\ estimator [Equation~(\ref{eqn:JMp})];
for the remainder of the galaxies, 
the approximate \js\ estimator [Equation~(\ref{eqn:jCK0})] was used.
The uncertainties are similar in both cases.
The deprojection from observed \jp\ to intrinsic \jt\ was accomplished using
individual inclinations for the spirals, and median deprojection factors
for the lenticulars and ellipticals (see main text).
The least massive early-type galaxy in the sample is
the compact elliptical NGC~4486B, which is probably
in the process of being tidally stripped by the giant galaxy M87;
the other low-\js\ outlier is NGC~1419. Both are
marked with black $\times$ symbols and excluded from all fits in this paper.
Dotted lines show the best fits for the Sb--Sm and elliptical galaxies:
these two galaxy types follow \js--\Ms\ trends that are parallel but separated in
\js\ by $\sim$~0.5~dex.
{\it Right-hand panel:}
As left-hand panel, but now plotting spiral disks and bulges alone,
along with elliptical galaxies, as indicated by the legend.
The upper line is now the fit to the disks (for all spiral types)
rather than to the whole galaxies.
Note that the slopes of the lines in this panel and the left-hand one 
should not be compared by eye,
owing to the different axis ranges.
The uncertainties in \js\ for the disks are typically $\sim$~0.04 dex,
and for the bulges at least $\sim$~0.2~dex;
the \Ms\ uncertainties are systematic (see main text).
Many of the most massive spiral bulges appear to a follow a similar \js--\Ms\ 
relation to the ellipticals.
\label{fig:JMM0}
}
\end{figure*}

As anticipated, the bulges turn out to have little impact on the total \js\ trends for
the Sb--Sm galaxies,
which are dominated by the disk components.
For the Sa--Sab galaxies, the bulges are responsible for the systematic
offset with respect to the later types;
this offset changes slightly but persists when
adopting the upper or lower limits to the bulge rotation.
The {\it disks} of all the galaxy types
turn out to follow nearly the same \js--\Ms\ relations.

This analysis demonstrates that inclination effects are not expected to have a major 
impact on our overall results, since for both disks and bulges,
the intrinsic and projected \js--\Ms\ trends as well as their scatter are very similar.
There is an overall offset between disk \jt\ and \jp\ of $\sim$~0.07~dex,  
which is comparable to the range of 0.04--0.06~dex that we would expect,
given the median inclination $i=67^\circ$ of our sample, and
depending on whether the \js--\Ms\ trend represents a median or an average fit
(see Appendix~\ref{sec:thin} for further discussion).

For our ensuing study of early-type galaxies, we will therefore simply adopt median 
deprojection values for all of the galaxies, which we estimated in Section~\ref{sec:etgdata}
to mean adding offsets of 0.08~dex and 0.22~dex to \jp\ to derive \jt,
for lenticulars and ellipticals, respectively.
We can also in general drop the usage of \jp\ in the rest of this paper,
in favor of the more physically meaningful \jt\ which we now adopt as
our estimate for \js.

\subsection{Combined observational results}\label{sec:obsresults}

We are now ready to include the early-type galaxies in our analysis,
and thereby address most of the key science questions raised in Section~\ref{sec:intro}.
As a reminder, our starting point is the \js--\Ms\ diagram from F83 that
we have reproduced in Figure~\ref{fig:JMM00}.
Do we find the same \js--\Ms\ trends with an updated and expanded dataset,
and more detailed analysis?
Do ellipticals still appear to have systematically low \js\ relative to spirals,
or do we discover large reservoirs of additional \js\ at large galactocentric radii,
using modern data?
Do Sa and S0 galaxies fill in any ``gap'' between spirals and ellipticals,
and can we then connect the Hubble sequence to a sequence in \js?
Can we characterize all galaxies as combinations of disks and bulges that follow
universal scaling relations?
(The main remaining question that connects to galaxy formation theory will be pursued
in the next section.)

Taking our early-type galaxy \js\ and \Ms\ estimates from Table~C\ref{tab:etg}
(after statistically correcting projected to intrinsic quantities;
see Table~\ref{tab:err} for an error analysis),
we plot them in Figure~\ref{fig:JMM0} (left), along with the spirals results discussed
in Section~\ref{sec:less}.
This new Figure is the centerpiece of our paper.
{\it Focusing first on the elliptical galaxies, our basic finding is that 
they follow a \js--\Ms\ trend which is
roughly parallel to the spirals but with a large systematic offset to lower \js.}

We thereby confirm the conclusions of F83, finding from 
a new synthesis of modern photometric and kinematic data
that the ``missing'' angular momentum in ellipticals does {\it not} emerge at large radii,
as had been expected from some theoretical studies.
As discussed in Section~\ref{sec:scale}, the new observations tend to show outer rotation
profiles that {\it decline} rather than rise.
Even the nearby galaxy NGC~5128 (Cen~A), 
which is often considered to be an elliptical formed through a recent major merger, 
shows a relatively low \js\ when compared to spirals of the same stellar mass.
Whether or not these observations pose a genuine problem to major-merger explanations 
for forming ellipticals will require renewed theoretical analysis, but
as discussed in Section~\ref{sec:scale}, there seems to be 
a pattern in the literature
of misdiagnoses of high outer rotation from early, sparse data -- which led to 
premature claims of evidence for major mergers.\footnote{\citet{2012MNRAS.421.1485N}
also recently noted an emerging trend for low rotation in elliptical-galaxy halos,
at odds with major-merger expectations.
One possible counter-example is the S0 galaxy NGC~1316, which is generally thought to be
a major-merger remnant.  Based on the new PN kinematics results from 
\citet{2012A&A...539A..11M},
we confirm the finding of \citet{1998ApJ...507..759A} that the \js--\Ms\ values
for this galaxy are close to the mean trend for spirals.
However, we caution that our photometric parameters and \MLs\ value
are particularly insecure for this galaxy.
}

The specific angular momentum difference between spirals and ellipticals is
also apparent from a simple,
direct consideration of the data in Section~\ref{sec:scale},
where the smaller sizes and rotation velocities for ellipticals suggested that they 
have lower \js.
As an arbitrary benchmark, we use the median \js\ at the $L^*$ 
characteristic luminosity, which is $\log\,(L^*_K/L_{K,\odot}) \sim 11$,
corresponding to $\log\,(M_\star/M_\odot) \sim 11$.
For ellipticals and Sb--Sm spirals,
we find projected values of $j_{\rm p} \sim 330$~\kms~kpc
and $\sim 1600$~\kms~kpc, respectively, and true values
of $j_\star = j_{\rm t} \sim 540$~\kms~kpc and $\sim 1800$~\kms~kpc.

In more detail, we report fits to the \js--\Ms\ data 
toward the end of Table~\ref{tab:loglog}.
The fitted slope for the ellipticals is consistent with that for the Sb--Sm spirals,
but is significantly offset to lower \js\ by a factor of $\sim$~3.4 ($\sim$~0.5 dex).
These findings are consistent with F83, except that the gap has narrowed from
a factor of  $\sim$~6 ($\sim$~0.8~dex).\footnote{Our
revised Sb--Sm relation is 
$\sim$~0.1~dex lower than in F83, partly owing to the inclusion of bulges, and partly to
new estimates for disk sizes and mass-to-light ratios.
Our revised ellipticals relation is 
$\sim$~0.2~dex higher than in F83;
this difference appears to arise not so much from the rotation
data (the extrapolations to large radius by F83 turn out very good on average),
but from a refined treatment of the total angular momentum calculation for spheroids.
Our slopes of $\alpha=0.53\pm0.04$ and $0.60\pm0.09$ for the
Sb--Sm and elliptical galaxies are shallower than the $\alpha=0.75$ slope 
suggested by F83; for the Sb--Sm galaxies, this difference is driven mostly by
our inclusion of bulges and of lower-mass galaxies
[$\log\,(M_\star/M_\odot)\sim$~9];
while for the ellipticals, a shallower slope was already apparent in F83.
}
Note that if the $K$-band \MLs\ for the ellipticals were systematically higher than
for the spirals by a factor of $\sim$~2 (perhaps owing to age or IMF differences;
cf.\ Section~\ref{sec:mass}), 
then the \js\ offset would increase to a factor of $\sim$~5 ($\sim$~0.7~dex).

The scatter of $\sigma_{\log j_\star}=$~0.24~dex for the ellipticals
is similar to the \jp\ scatter for the spirals.
We also note that the general trends for the ellipticals
are supported by the small sample of galaxies that we modeled in detail
(see points with error bars in Figure~\ref{fig:JMM0}, left).
Although one might still have concerns that large formal
uncertainties in \js\ remain for most of the sample after extrapolating
their rotation-velocity profiles beyond 2~\Reff, in order to
close the \js\ gap between spirals and ellipticals, 
the rotation velocity would have to rise rapidly by a factor of $\sim$~4 outside these radii,
which seems implausible (cf.\ Figure~\ref{fig:etg}).

The parallel nature of the spiral and elliptical trends
is an interesting and non-trivial result, since Figure~\ref{fig:rot} showed
that the slopes of the rotation-velocity scaling relations for these galaxies 
have opposite signs.
Some mass-dependent conspiracy of size, rotation velocity, and S\'ersic index must be at work
in order for the \js--\Ms\ slopes to turn out the same.

The few ``slow rotator'' ellipticals in our sample show no indication of deviating 
systematically from the overall \js--\Ms\ trend for ellipticals, which
disagrees with earlier findings of much lower \js\ for such galaxies
\citep{1990A&A...239...97B}.
Although their outer regions, like their central parts, rotate slowly relative
to most of the fast rotators (Figure~\ref{fig:rot}), we find that 
this is compensated for by their larger scale radii and S\'ersic indices
(keeping in mind that the results for these galaxies are the most uncertain).
Thus the global \js\ measurements suggest that the slow and fast rotators may
have more in common than was previously suspected.

Having confirmed the basic observational findings of F83, we now move on to
fresh territory, beginning with the
inclusion of Sa and S0 galaxies in Figure~\ref{fig:JMM0} (left).
F83 suggested that these would fill the gap in \js--\Ms\ space between ellipticals
and late-type spirals, which is confirmed by our sample.
Both of these galaxy types are on average offset to lower \js\ from the Sb--Sm spirals trend
by a factor of $\sim$~1.8 ($\sim$~0.25 dex; we will discuss variations about
the average in Section~\ref{sec:resid}).

One natural interpretation of this new finding is that the Hubble classifications are related to
an underlying physical structure, where all galaxies are composed of
some combination of two basic components: a disk and a spheroid
(as illustrated schematically in Figure~\ref{fig:schem1} of Section~\ref{sec:intro}).
These components would define two distinct sequences in the \js--\Ms\ plane, which
in combination would move the total values of galaxies to intermediate regions in this plane,
depending on the bulge-to-total mass ratios, $B/T$.

To explore this idea, we plot the \js--\Ms\ data
separately for elliptical galaxies, and for spiral disk and bulge subcomponents, 
in the right-hand panel of Figure~\ref{fig:JMM0}.
The disks follow a similar relation to spiral galaxies overall, since these are
dominated by their disks.
More remarkably, the \js--\Ms\ trend for {\it bulges} 
is fairly similar to the trend for ellipticals over the mass range where they 
overlap.\footnote{At lower bulge masses, the apparent tendency to
relatively low \js\ values
should be viewed as speculative,
since it is based on classical bulges rather than the pseudo-bulges
that may predominate in this regime.}
This is a surprising result, because as shown in Figure~\ref{fig:size}, the bulge
{\it sizes} are systematically smaller than the ellipticals, and thus their
rotation velocities (Figure~\ref{fig:rot}) must be higher, in an apparent
conspiracy to produce roughly the same \js.

A similar analysis could in principle be carried out for the fast-rotator ellipticals,
since they 
are widely considered to host hidden, embedded disk-like components.
Do the disk and bulge subcomponents of ellipticals 
follow the same \js--\Ms\ relations as those of the spirals?
We have investigated this question
in Appendix~\ref{sec:decomp} using decompositions from
the literature, but the results are somewhat ambiguous.
Thus, although we have been able to address all of the major questions
raised initially about empirical \js--\Ms\ trends, we flag the trends
for the subcomponents in ellipticals (and lenticulars) as an important
aspect remaining in need of clarification.

\subsection{Replacing the Hubble diagram}\label{sec:replace}

The foregoing discussion brings us to the diagram that
we have already introduced schematically with Figure~\ref{fig:schem1}, which
constitutes our own, physically-motivated, substitute for the classic Hubble tuning fork,
and which could provide the underlying explanation for the observational trends
found in Figure~\ref{fig:JMM0}.
In this scheme, all galaxies are composed of a disk and a bulge, each adhering to a
distinct and parallel \js--\Ms\ scaling relation.
If the disk and bulge relations are universal 
(which we will further test in Section~\ref{sec:resid}), then the
location of a galaxy in \js--\Ms\ space can immediately be used to infer its $B/T$ value uniquely,
and vice-versa 
(i.e., there is a coordinate transformation between the two parameter spaces).
Elliptical galaxies would then be the cases with $B/T \sim 1$,
and bulges could be thought of as mini-ellipticals.

\begin{figure*}
\centering{
\includegraphics[width=3.5in]{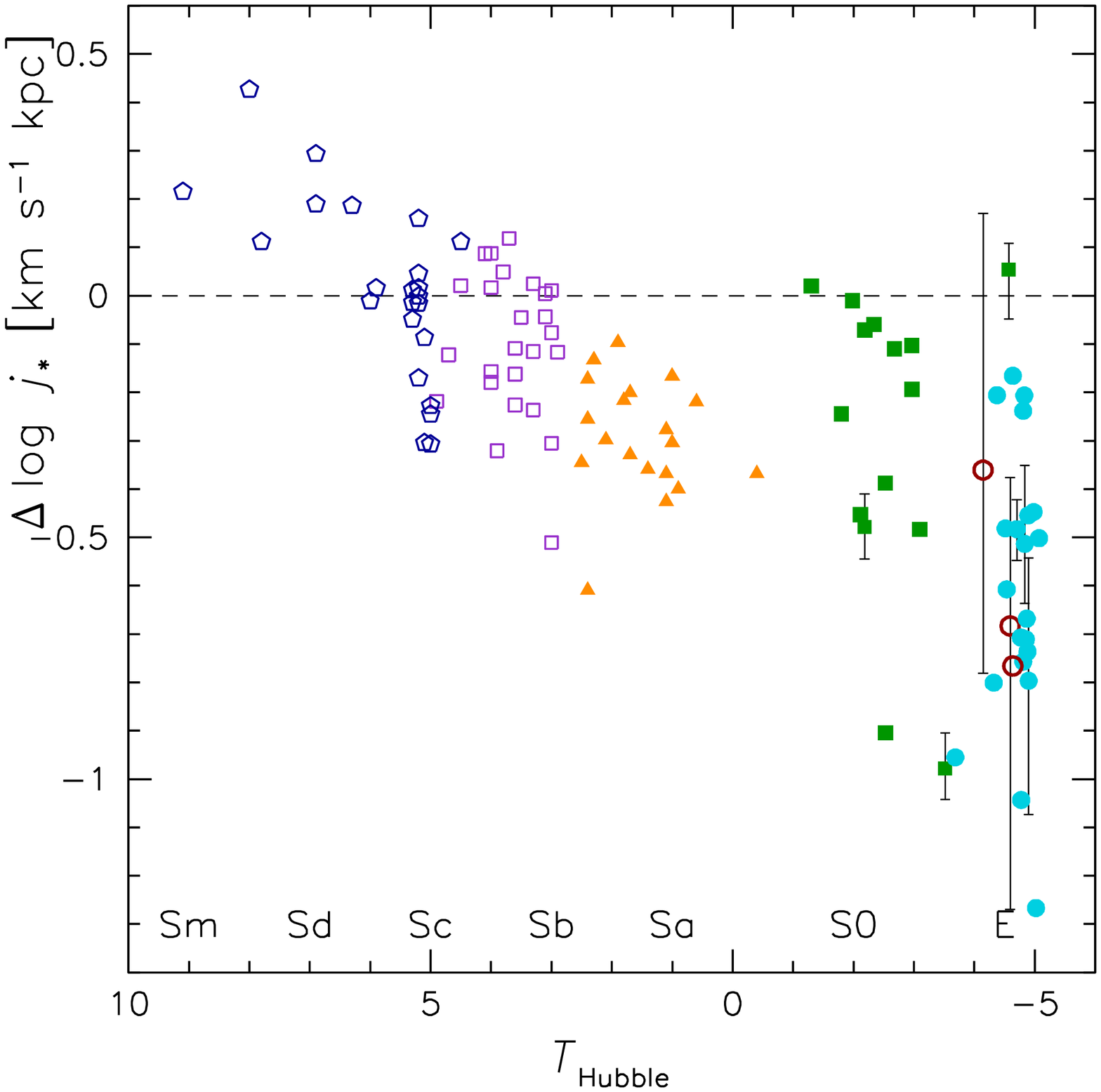} 
\includegraphics[width=3.5in]{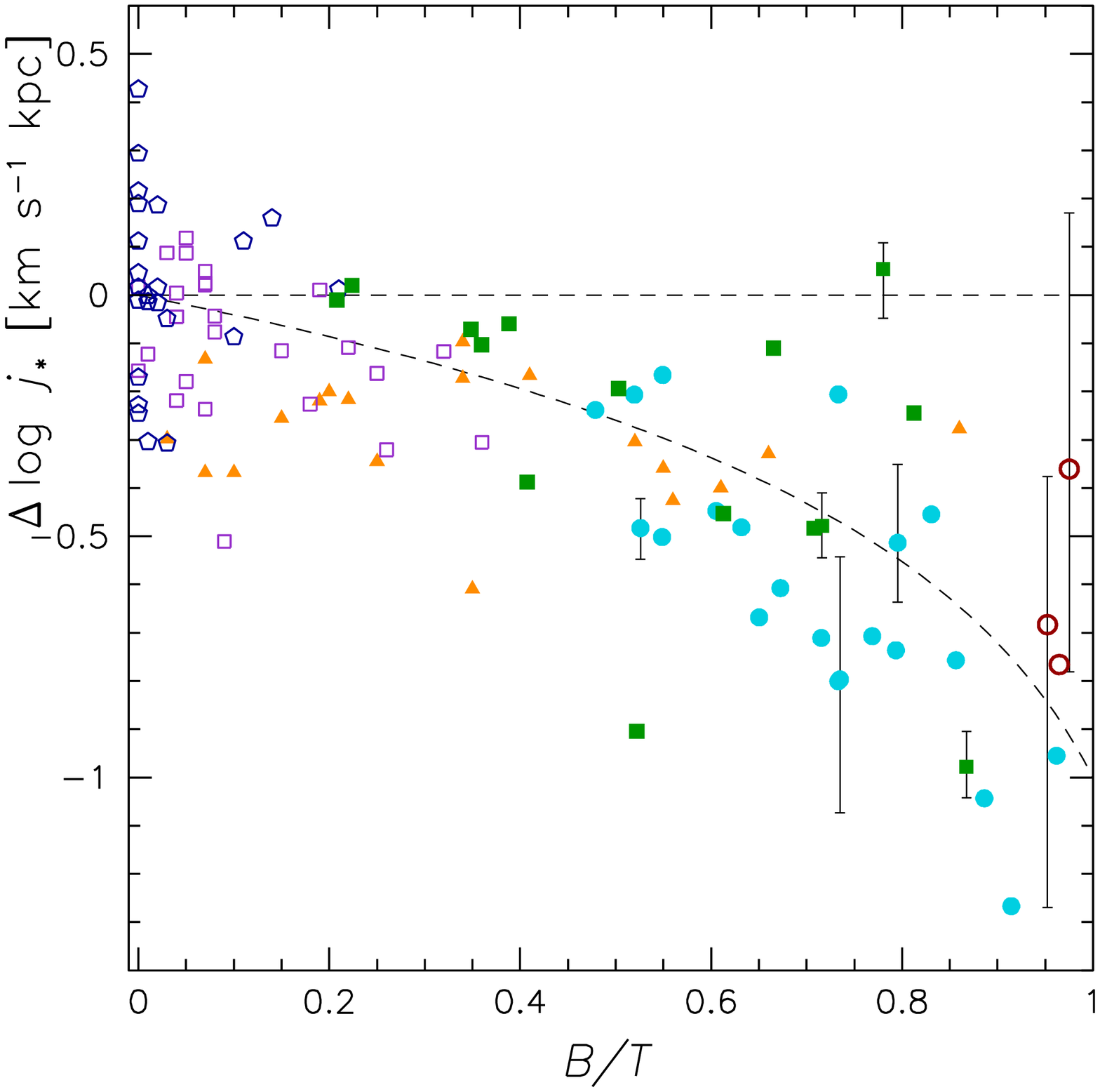} 
}
\caption{Specific angular momentum relative to the best-fitted
trend for spiral disks.
In the {\it left-hand panel}, these residuals are plotted vs.\ Hubble stage.
For clarity, small random offsets have been added in the horizontal direction
for the early-type galaxies.
In the {\it right-hand panel}, the residuals are plotted vs.\ bulge-to-total mass ratio.
The curved line shows a sample model prediction 
(not a fit to the data; see text for details).
There are strong systematic trends of the \js\ residuals with respect to both
Hubble type and to bulge-fraction, and the relative smoothness of
this trend (particularly for the E/S0s)
suggests that bulge-fraction is the more fundamental
driving parameter.
\label{fig:diffj}
}
\end{figure*}

As with the original Hubble diagram, our \js--\Ms\ diagram 
provides a simple {\it description}
of galaxies, along with the temptation to interpret it as some kind of
{\it evolutionary sequence}.
However, our diagram differs, since the parameters used are physical
quantities that may in principle be  conserved, and thus it is actually justified to begin
using the diagram directly as a tool to motivate and test some evolutionary scenarios
for galaxies.  This will be the objective of Section~\ref{sec:theory}.

A key feature of our diagram is that it views galaxies as fundamentally populating
a space of {\it two parameters}, angular momentum and mass, which are nearly
equivalent to the more observationally accessible properties of bulge fraction and luminosity.
In this framework, galaxies {\it cannot} be fruitfully reduced to a one-dimensional family
controlled by a single parameter (e.g., \citealt{2008Natur.455.1082D}).

Our diagram may also be contrasted with another currently fashionable way to understand
galaxies: as color-magnitude sequences that are generally related to
star formation histories (e.g., \citealt{2004ApJ...600..681B,2007ApJ...665..265F}).
These properties are loosely related to \js--\Ms\ space if star formation generally
occurs in high-\js\ disks. However, our framework is
less astronomical and more astrophysical in nature, and
we expect it to provide novel insights to galaxy formation that are complementary to
other classifications, and perhaps more fundamental.

Another recently-introduced classification for galaxies is also based loosely
on specific angular momentum concepts:
$\lambda_R$ \citep{2007MNRAS.379..401E},
which measures the rotational dominance in the central regions
(typically inside $\sim R_{\rm e}/2$) and is similar to a $v/\sigma$ metric.
Applied to early-type galaxies, a host of interesting patterns and
correlations have emerged \citep{2011MNRAS.414..888E}.
However, this metric in practice is not only very scale dependent,
but also misses exactly those scales that are
most important for measuring true, physical angular momentum (recall Figure~\ref{fig:fR}).
In fact, we have seen evidence that \js\ and the central $\lambda_R$ 
are disjoint properties:
the slow rotators (low-$\lambda_R$ galaxies) do not appear to deviate from
the \js--\Ms\ trend for fast rotators.

A final related diagram to mention is \js--$v_{\rm c}$, where $v_{\rm c}$ is the 
circular velocity, tracing the dynamical mass of a galaxy within some characteristic radius
(e.g., \citealt{2000ApJ...538..477N,2012MNRAS.424..502K}).
There are complications with using this parameter space, since for spiral galaxies
both \js\ and $v_{\rm c}$ are normally based on the same rotation-velocity measurements,
which causes a built-in correlation.  
Unlike \Ms, $v_{\rm c}$ is not a physical quantity subject to
straightforward conservation laws.
In addition, a critical point for our goal of analyzing all types of galaxies 
in a unified manner is that
it is very hard to estimate $v_{\rm c}$ for a large sample of early-types
since they rarely host extended gas disks.  Instead, extensive data are
required from other tracers such as stellar kinematics
(as needed for \js\ estimation), as well as
 grueling dynamical modeling which even with the state-of-the art techniques
can still leave considerable uncertainties \citep{2009MNRAS.395...76D}.
Similar problems apply to a \js--\Mv\ (virial mass) diagram, where
the masses can be estimated only on a statistical rather than on an individual basis
(e.g., \citealt{2012MNRAS.421..608D}).

\subsection{Examining the residuals}\label{sec:resid}

Our bulge--disk framework, although rather compelling, is not a unique
explanation for the systematic trends in the left-hand panel of Figure~\ref{fig:JMM0}.
It is possible that the vertical displacements of \js\ in this diagram are
somehow more directly related to Hubble morphology than to $B/T$
(although one should keep in mind that $B/T$ is one of the main factors
in the morphological classifications, along with spiral arm winding and
clumpiness).

To consider this point more clearly, and to
better see the relative trends in the data, we flatten the \js--\Ms\
relations into one dimension, dividing by the mean 
trend for the spiral disks and thus generating the quantity:
\begin{equation}\label{eqn:Dj}
\Delta \log j_\star \equiv \log j_\star - \log\,j_{\rm mod}(M_\star) ,
\end{equation}
where $j_{\rm mod}$ is given by Equation~(\ref{eqn:loglog}).
We plot \Dlj\ versus the Hubble stage parameter $T_{\rm Hubble}$ in Figure~\ref{fig:diffj}
(left-hand panel).
There is clearly a strong positive correlation between $T_{\rm Hubble}$ and the \js--\Ms\ residuals.
Among the spirals, this trend is clearest when considering the Sa--Sab versus Sb--Sc galaxies.
The Scd--Sm galaxies appear to continue the trend, but they inhabit the lowest-mass
area of the \js--\Ms\ diagram, where the mean relation is not defined well enough
to be certain of the residuals.

The S0s break the smooth trend of
\Dlj\ decreasing for smaller $T_{\rm Hubble}$.
Many of them appear to have comparable specific angular momenta to typical 
Sb--Sc galaxies, which was foreshadowed by the rotation scaling relations
of Figure~\ref{fig:rot}.
The implication is that lenticulars and spirals are overall dynamically similar,
differing more in their finer morphological features which
may be related to star formation activity.
We can thus think of these lenticulars as faded spirals, or of
the spirals as rejuvenated lenticulars, although
they differ in average $B/T$ values, and
more nuanced comparisons will require analysis of \MLs\
(cf.\ \citealt{2010MNRAS.409.1330W}).
As for the subset of lenticulars with low \Dlj, they may either be
very close to face-on, or else belong to a different family of objects
that are related to the ellipticals.

Returning to our original hypothesis that $B/T$ is the key parameter
affecting the \js--\Ms\ trends, we consider its correlation with
the residuals \Dlj.
Since we do not actually 
have bulge/disk decompositions for the early-type galaxies in our sample,
we introduce a novel technique that uses
the degree of central rotational support as a rough proxy for $B/T$.
The idea here is that the bulge is to a first approximation non-rotating,
so any observed rotation is from the disk:
objects with higher $(v/\sigma)$ imply higher disk fractions and lower $B/T$.
Appendix~\ref{sec:decomp} describes our methods for early-type $B/T$ estimation in
more detail.
For the late-types, we already have $B/T$ estimates based on decompositions in
the literature, as discussed earlier.

We show the results in the right-hand panel of Figure~\ref{fig:diffj}.
The residuals {\it do} correlate clearly with $B/T$, in a fairly smooth trend
that is followed equally well by all of the galaxy types, and
which contrasts with the $T_{\rm Hubble}$ trend.
We have marked a simple expectation for the $B/T$ trend with the curved line, given 
the summation of Equation~(\ref{eqn:jtot}), along with an arbitrarily assumed 
$j_{\rm b}=0.1 \times j_{\rm d}$.
This model mimics the data remarkably well, although it should be remembered
that the agreement is somewhat built-in already, since
correlated rotational properties
were used both to estimate $B/T$ and to calculate \js.

Recalling that we also had to make strong modeling assumptions for
the spiral bulges when calculating \js,
the better connection of the residuals to $B/T$ 
rather than $T_{\rm Hubble}$
should be considered preliminary.
It is also difficult to tell how much of the scatter in \js\ at 
fixed $B/T$ is due to observational error, and how much is due to intrinsic
variations, i.e., with bulges and/or disks not following perfectly standardized
\js--\Ms\ relations.
Definitive resolution of these issues
 will require more detailed bulge--disk decompositions of all
types of galaxies, including spectroscopic information 
(cf.\ \citealt{2011MNRAS.414..642C,2012MNRAS.422.2590J,2012ApJ...752..147D,Forbes12}}),
and allowances for \MLs\ variations.  

We would however like to advance the proposition that bulge fraction is the
fundamental driving parameter behind \js\ variations, and 
is responsible for many of the observed variations in galaxy properties
(see discussion in previous subsection).
Not only does this make sense from a physical standpoint, but the
agreements between ellipticals and spiral bulges in Figure~\ref{fig:JMM0} (right), 
and between model and data in Figure~\ref{fig:diffj} (right), 
provide provisional but strongly suggestive observational support.
The radially-declining rotation-velocity profiles of galaxies like NGC~821 and NGC~3377
in Figure~\ref{fig:etg} could also be naturally explained by central disk components
embedded in non-rotating bulges.
Furthermore, we will see from consideration of a cosmological context in Section~\ref{sec:theory2} 
that the distribution of \js\ is more naturally reconciled with distinct disk and spheroid
subpopulations than with a simple continuum of galaxy \js.

\subsection{Histograms of stellar $j$ residuals}\label{sec:hist}

\begin{figure}
\centering{
\includegraphics[width=3.5in]{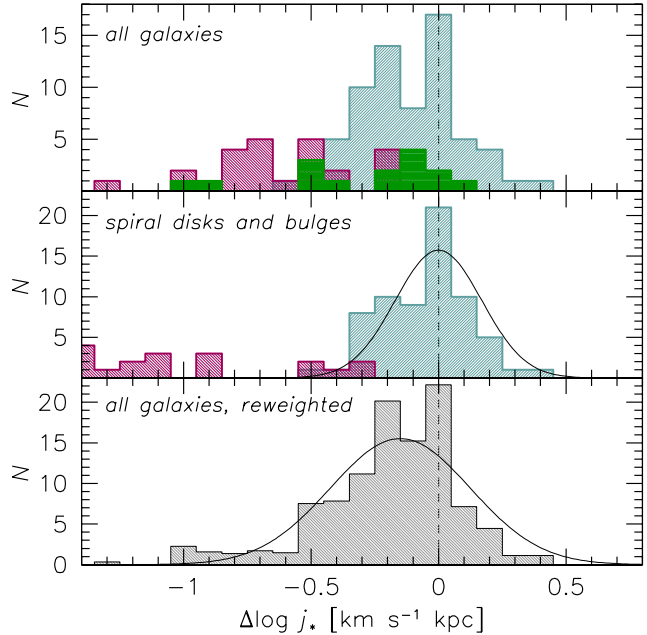} 
}
\caption{Histogram of specific angular momentum relative to the 
mean observed trend for spiral disks.
In two of the panels, curves show example lognormal distributions for
comparison to the data.
In the upper panel, the red, green, and blue
histograms show data from Figure~\ref{fig:diffj} for spirals, lenticulars,
and ellipticals, respectively.
The middle panel shows the bulge and disk subcomponents of spiral galaxies,
with red and blue histograms, respectively.
The lower panel is a summation of the data from the upper panel,
after renormalizing each galaxy sub-type type by its frequency in the nearby universe
(see main text).
The specific angular momentum does not appear to have a simple lognormal distribution,
and may even be bimodal.
\label{fig:histdiff}
}
\end{figure}

Before moving on to theoretical analyses, 
we construct one more representation of the data 
whose relevance will become particularly clear in the next section.
We compress
the preceding \js--\Ms\ information into a histogram of residuals from the 
spiral disk relation, showing the results
in Figure~\ref{fig:histdiff} (upper panel).
Here it is apparent that the spiral galaxy data comprise 
a roughly lognormal distribution in $\Delta j_\star$, with an rms dispersion of $\sim$~0.2 dex.
The ellipticals have a less well-defined distribution that
partially overlaps the spirals but is offset $\sim$~0.5 dex lower,
while the small sample of lenticulars spans almost the full range of residuals.

In the middle panel of Figure~\ref{fig:histdiff}, we look instead at the
disk and bulge subcomponents of the spiral galaxies,
where we have also
overplotted a Gaussian with a width of $\sigma_{\log j_\star} = 0.17$~dex for reference.  
Given the uncertainties and possible selection bias in our analysis,
we consider the disks to be reasonably consistent with a lognormal distribution.

The \Dlj\ distribution for the spiral bulges resembles that of the ellipticals
in the sense that both are systematically offset to lower values, as we have previously seen.
The bulges apparently extend to much lower \Dlj\ than the ellipticals, but as discussed in
Section~\ref{sec:obsresults},
this is not a secure result, given the uncertainties in the
bulge calculations.

Returning to the overall results, we would like to know whether or not galaxies follow
a bimodal distribution in \Dlj\ as the top panel of Figure~\ref{fig:histdiff}
suggests. The complication here is possible bias in the galaxy sample:
if we were to study {\it all} bright galaxies in a volume-limited sample,
the \Dlj\ distribution might look very different.
To investigate this issue, we must 
re-weight the distribution of \js\ in our sample by galaxy type.

The simplest approach is to renormalize by frequency or number density.
We use the \atlas3d\ results that 70\%, 22\%, and 8\% of the
galaxies in the nearby universe are spirals, lenticulars, and ellipticals 
(over a stellar mass range similar to our observational sample;
\citealt{2011MNRAS.413..813C}).
The fractions in our sample are 63\%, 14\%, and 23\%, demonstrating a strong bias
toward ellipticals at the expense of lenticulars.

We plot the re-weighted results in the lower panel of Figure~\ref{fig:histdiff},
showing also for reference a lognormal 
curve with $\sigma_{\log j_\star}=0.27$~dex
(a width that will be motivated in Section~\ref{sec:theory2}).
The total distribution of $\log\,j_\star$ residuals appears slightly non-Gaussian,
with a tail extending to low values.
This feature may not be significant if one allows for systematic
uncertainties in the selection effects,
but the skewness will become clearer when compared to
theory in Section~\ref{sec:theory2}.

An alternative scheme would be to re-weight by the stellar mass density of the
different galaxy types.  This would bring us closer to a total distribution function
for stellar $j$ in the universe, rather than a distribution of galaxies with given \js.
It is beyond the scope of this paper to carry out such an exercise in detail, but the
basic outcome is clear.
The high end of the mass distribution is dominated by early-types (cf.\ lower panel
of Figure~\ref{fig:atlas1}), which means that the mass weighting would enhance the 
contributions of these galaxies relative to number weighting.
The universal distribution of \js\ would then appear {\it more non-Gaussian} than in the
lower panel of Figure~\ref{fig:histdiff}.

We therefore find evidence that the residuals of the specific angular momenta of
galaxies from the mean relation are not simply lognormal.
The best match to a lognormal model is provided by the disk components of spirals,
while the bulges and the ellipticals may comprise a distinct second 
population.\footnote{\citet{2007MNRAS.375..163H} used a large photometric survey
to estimate \js\ indirectly, with results that are less accurate than those
presented here, but which similarly imply a bimodal distribution
for ellipticals and spirals.}
Again, a natural interpretation of this finding is that
all galaxies are composed of some combination of high-
and low-\js\ material, which may be identified with disks and bulges, respectively.

Some implications of these results for galaxy formation in a modern cosmological
context will be discussed in the next section.
It should be remembered, however, that our empirical findings---of specific,
strong correlations between galactic angular momentum, mass, morphology, and bulge 
fraction---stand on their
own and must be explicable by any successful theory of galaxy formation,
whether now or in the future.

\section{Connecting to theory}\label{sec:theory}

We are now ready to present a fresh theoretical way of looking at galaxies, 
using the \js--\Ms\ diagram, which was introduced in F83, and which may now be
reinvigorated by populating it with observational data for galaxies of all types.
Our general approach is to take a step back from galactic {\it details},
whether these be spiral arms and dust lanes in observations, or unresolved
gas physics and star formation recipes in simulations,
and return to some simple physical parameters and conservation rules
that may provide robust constraints and insights to galaxy formation.

We have shown in Sections~\ref{sec:obsresults} and \ref{sec:resid}
that the specific stellar angular momenta
of observed galaxies follow remarkably tight correlations with their masses and bulge fractions.
Such patterns in Nature demand theoretical explanations, as they could
be tracing fundamental physical processes.
Indeed, the \js--\Ms\ relation for spiral galaxies is well known in some circles,
and provides a crucial benchmark for models of galaxy formation.
However, the correlation for elliptical galaxies (already shown in a preliminary version
by F83) is less well known and addressed with theoretical models.
Our goal is to advance a general, physical framework for integrating these
observational constraints into models of galaxy formation and evolution.

Our approach here is different from, and complementary to, the
active field of hydrodynamical simulations of galaxy formation.
Although such simulations have made notable progress toward the ultimate goal
of reproducing realistic galaxies,
they still have a long way to go,
with recent work highlighting large differences in the basic properties of
simulated galaxies, depending on what code, resolution, and physical recipes are used
\citep{2012MNRAS.423.1726S,2011arXiv1110.5635T}.

Historically, such methods missed reproducing observed
\js\ trends by factors of up to $\sim$~30, and even the most recent work shows
variations at the factor of $\sim$~2 level.
The general concern is that many of the large-scale properties of galaxies could
well depend strongly on transport processes at the scales of molecular clouds,
which are not yet modeled satisfactorily in cosmological simulations.
Therefore some caution is still needed in assuming that the simulations
are providing an adequate representation of reality.

In this context, simplified ``toy'' models continue to play a key role in
defining the broad but solid outlines of the galaxy formation theory that
is required to match the observational constraints.
These models may also prove useful in physical understanding of the output
of numerical hydrodynamical simulations.

We frame our analysis in the context
of the current standard cosmological model for structure formation:
cold dark matter with a cosmological constant ($\Lambda$CDM; \citealt{2011ApJS..192...18K}).
This model makes specific, robust predictions for the angular momenta of DM halos.
Because the visible galaxies, consisting of stars and gas,
are presumed to reside in these DM halos, we may then
ask whether or not the observed stellar angular momenta bear any resemblance to
the predictions for DM halos.

We begin with the properties of $\Lambda$CDM halos as our ``initial conditions''
for galaxy formation, which we map to our observable space: \js--\Ms\ for 
the stellar components of galaxies.
We do this by parameterizing the retention
of mass and angular momentum during galaxy formation,
and 
then by introducing a menu of \js--\Ms\ vectors of change that correspond
to plausible physical processes (outflows, mergers, etc.).

We emphasize that the primary aim of this paper is {\it not} to concoct a new theory
of galaxy formation, nor to weigh in on competing models
by vetting specific simulation outputs against the \js--\Ms\ diagram.
Instead, we wish to lay out a generalized framework that can
both constrain and explain the models.
The methodology and merits of this approach should become clearer as we develop the
ideas throughout this section,
and as we eventually work through some practical examples.

We develop general theoretical predictions and make basic inferences
about $j$ retention in Section~\ref{sec:theory1}.
In Section~\ref{sec:theory2} we investigate two possible explanations for
the observed $j_\star$ dichotomy between spirals and ellipticals.
In Section~\ref{sec:theory3} we consider coupling between changes in mass and 
angular momentum,
and connect these to evolutionary scenarios for galaxies.

\subsection{Basic constraints}\label{sec:theory1}

The overdense regions in an expanding universe are not spherically symmetric
and exert tidal torques on each other, inducing a net angular 
momentum in each collapsing galaxy \citep{1951pca..conf..195H}.
This rotational behavior is usually specified in terms of a dimensionless spin
parameter that 
quantifies the dynamical importance of rotation, and
is a combination of fundamental physical quantities:
\begin{equation}\label{eqn:lambda}
\lambda \equiv \frac{J |E|^{1/2}}{G M^{5/2}} ,
\end{equation}
where $J$ is the angular momentum, $E$ is the energy (kinetic and potential),
$G$ is the gravitational constant, and $M$ is the mass
\citep{1969ApJ...155..393P}.\footnote{Recall
that the parameters $(J,E,M)$ can be translated roughly into a more observationally oriented
basis set of rotation velocity, effective radius, and luminosity $(v_{\rm rot}, R_{\rm e}, L)$, 
where in approximate terms:
$M \propto L$, $E \propto L^2 R_{\rm e}^{-1}$, and $J \propto v_{\rm rot} L R_{\rm e}$.}
Whether analyzed through linear tidal torque theory, or through $N$-body simulations
of galaxy assembly, $\lambda$ is predicted to follow an almost lognormal distribution
that is relatively insensitive to cosmological parameters, time, galaxy mass,
and environment (e.g., \citealt{1987ApJ...319..575B,1988ApJ...330..519Z,1995MNRAS.272..570S,1996MNRAS.281..716C,2007MNRAS.378...55M}).

The spin parameter provides a convenient way to characterize DM halos, but
it is not straightforward to connect
$\lambda$ to baryonic galaxies
because it is not a physically conserved quantity (as energy is dissipated).
We instead conduct our theoretical analysis in terms of 
the specific angular momentum parameter $j$, as we have done
with the observations.
Along with the mass $M$, $j$ is a quantity that is potentially conserved 
at some approximate level during the evolutionary history of a galaxy.

\begin{figure*}
\centering{
\includegraphics[width=7in]{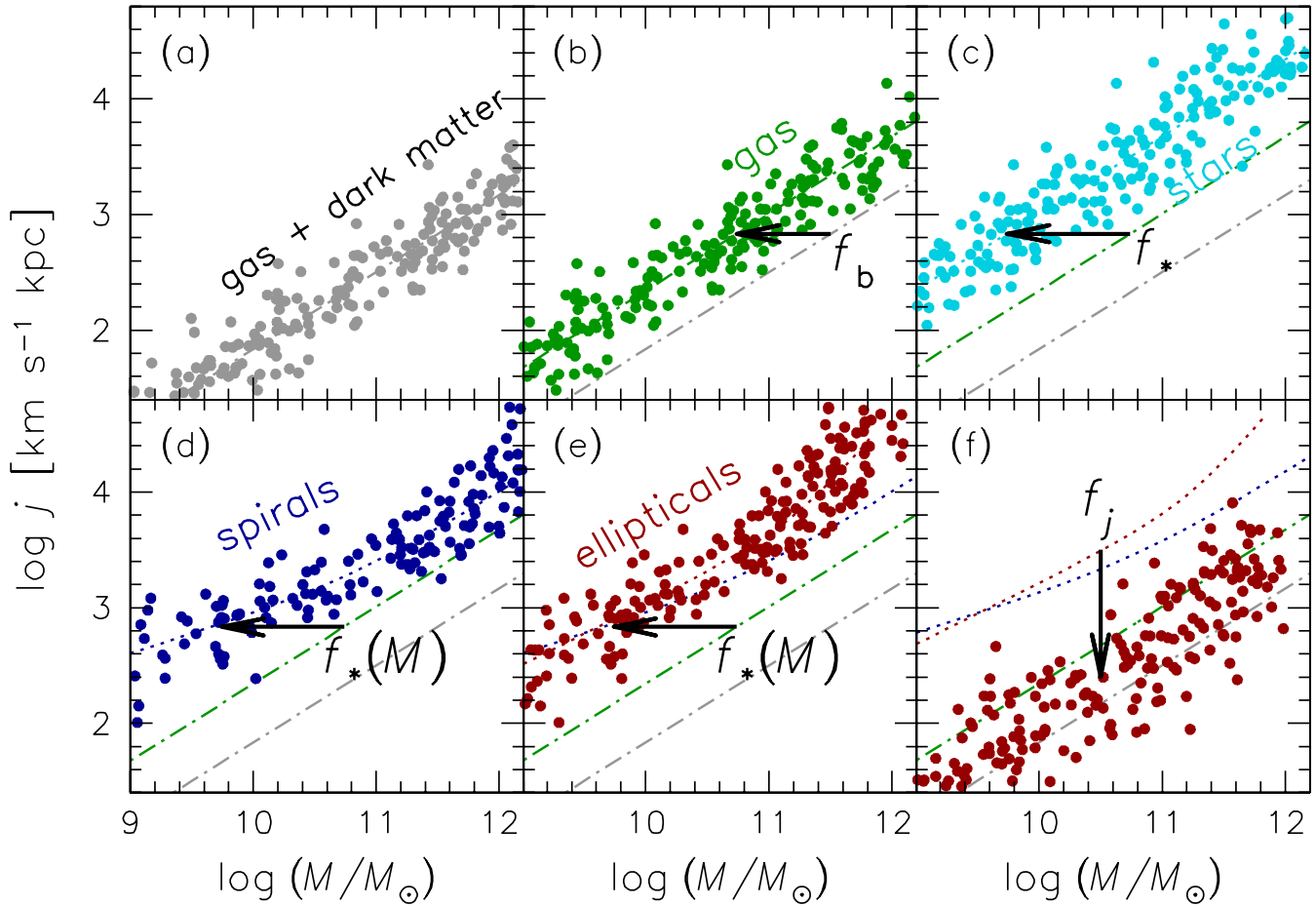} 
}
\caption{
Schematic evolution of galaxies in the space of specific angular momentum and mass.
Each point shows a galaxy randomly selected from a simple model (see main text).
Panel (a) shows the initial galactic halos of gas and DM.
Panel (b) shows the gas component only, adopting a baryon fraction of $f_{\rm b}=0.17$, 
with an arrow illustrating the direction that a single galaxy takes in this diagram.
Panel (c) shows the stellar component after forming from the gas
with an average relative fraction of $\langle f_\star \rangle=0.1$.
Panels (d) and (e) show the stars of spiral and elliptical galaxies, respectively, after
adopting more realistic variations of $\langle f_\star \rangle$ with mass.
Panel (f) shows the effect of angular momentum loss, with a factor of 
$\langle f_j \rangle=0.1$.
Note that these are simple, idealized models, and not every aspect should be
taken literally; e.g., spiral galaxies probably do not exist at masses of
$M_\star \ga 10^{12} M_\odot$.
}
\label{fig:schem2}
\end{figure*}

To re-cast $\lambda$ to $j$, we adopt a $\Lambda$CDM-based 
spherically-symmetric halo profile from \citet{1996ApJ...462..563N},
truncated at the virial radius.\footnote{The virial radius is defined
as bounding a region
inside which the mean halo density is a factor of $\Delta_{\rm vir}$
times the critical density $\rho_{\rm crit} \equiv 3 H^2 / (8 \pi G)$.
We adopt a WMAP5 cosmology, with
$H=$~72~\kms~Mpc$^{-1}$ and $\Delta_{\rm vir}=95.3$ at $z=0$
\citep{2008MNRAS.391.1940M}.
To calculate $E$ for this halo, we use an expression from 
\citet{1998MNRAS.295..319M} with a fixed concentration of $c_{\rm vir}=9.7$; 
and we ignore variations due to concentration which affect $\lambda$ 
at the $\sim$~5\% level.
A related spin-proxy parameter, $\lambda^\prime$, is based on a singular
isothermal sphere \citep{2001ApJ...555..240B},
and is $\simeq$~11\% smaller than $\lambda$.}
We then obtain:
\begin{equation}\label{eqn:jlam}
j_{\rm vir} = 4.23\times10^4 \mbox{ } \lambda \left(\frac{M_{\rm vir}}{10^{12} M_\odot}\right)^{2/3} \mbox{\kms~kpc} .
\end{equation}
We adopt a characteristic value\footnote{This is
based on the average value of $\log\,\lambda$, but throughout this
paper we use shorthand such as $\langle \lambda \rangle$
and $\langle j \rangle$ for log-averages.}
of $\langle\lambda\rangle=0.035$, 
along with a 1~$\sigma$ log dispersion of 0.23~dex,
based on a study of relaxed halos in a cosmological simulation
with WMAP5 parameters, by \citet{2008MNRAS.391.1940M}.
The log-averaged numerical coefficient in Equation~(\ref{eqn:jlam}) then becomes
1460~\kms~kpc.
Other recent studies are generally consistent with these results 
at the level of $\sim$~10\%.
The $\alpha=2/3$ exponent is also an explicit prediction of tidal torque theory
\citep{1984ApJ...281...56S,1988MNRAS.232..339H}, and
provides a reasonable approximation to the trends from {\it direct} calculations 
of \jv\ and \Mv\ in $N$-body simulations \citep{2010MNRAS.407.1338A}.

Equation~(\ref{eqn:jlam}) can be considered as setting firm ``initial conditions''
for galaxies, characterizing their angular momenta near the
time of virialization.
This is shown schematically in panel (a) of Figure~\ref{fig:schem2},
which we have populated with toy-model ``galaxies'' consisting of primordial halos of gas and
DM.  Their masses are drawn from a uniform logarithmic distribution, and their
angular momenta from a lognormal distribution using $\langle j_{\rm vir} \rangle$
and $\sigma_{\log j_{\rm vir}}$ as above.

We next consider a series of idealized evolutionary steps that allow us to
parameterize evolution in the $j$--$M$ diagram.
We assume that the baryons consist initially of gas
that is well mixed with the dark matter of its parent halo, and that does not
collapse within the halo until after the linear and translinear
regimes of tidal torque when most of the angular momentum is acquired.
The gas may then be 
assumed to have the same value of $j$ as the halo, which
we show in panel (b) as a simple shift of the points to the left,
according to a cosmological baryon fraction of $f_{\rm b}=0.17$
\citep{2011ApJS..192...18K}.  

In panel (c) we show what happens in a simple case
where a fraction of the baryons form into stars, with a particular value
of $\langle f_\star\rangle=0.1$, and a dispersion of $\sigma_{\log f_\star}=0.15$~dex.
Again, $j$ is assumed to be conserved, and the galaxies shift to the left.
It is also usually assumed, though not required by the diagram,
that this process involves the formation of a thin
stellar disk whose collapse was halted by the balance between gravity and centrifugal
force.

Our analysis does however assume that the baryon collapse extends all the way out
to the halo virial radius. This conventional assumption is at some level
implausible since DM collapse and gas cooling are governed by different physical
scales in space and time.  A more generalized approach where the baryon collapse
radius is allowed to vary 
will be considered in Section~\ref{sec:bias}.

Note that the $f_\star$ parameter can take on a more general meaning
of {\it net} stellar mass fraction relative to initial gas mass, which
allows for stars that are accreted by or ejected from the galaxy.
We will shortly discuss a more refined model where $f_\star$ varies systematically
with mass, but for now we continue with our very simplified constant-$f_\star$
model in order to consider its basic implications.

Our next model ingredient is an idealized process of angular momentum loss, with
no concomitant change in mass, which we quantify by a fractional $j$ net retention
factor of $f_j$. An example of such a process would be internal $j$ transfer from
the stars to the DM halo.
Given the parameters $f_\star$ and $f_j$, we may then translate
the $j$--$M$ relation~(\ref{eqn:jlam}) for DM halos to an equivalent one for the
stellar components of galaxies:
\begin{equation}\label{eqn:jslam}
j_\star = 2.92\times10^4 \, f_j \, f_\star^{-2/3} \, \lambda \left(\frac{M_\star}{10^{11} M_\odot}\right)^{2/3} \mbox{\kms~kpc} ,
\end{equation}
where again using the prediction for 
$\langle\lambda\rangle$, 
the numerical coefficient for $\langle j_\star \rangle$
becomes 1010~\kms~kpc.

This relation is identical to our parameterized fit to the
observational data with Equation~(\ref{eqn:loglog}),
modulo the numerical factors and the value for the exponent $\alpha$.
Since the observed \js--\Ms\ relation can be approximated with $\alpha=2/3$
and a normalization $j_0$, then we
can express the difference between observation and theory through
a simple combination of the parameters $f_j$ and $f_\star$:
\begin{equation}\label{eqn:comb}
j_0 = 1010 \, \langle f_j \, f_\star^{-2/3} \rangle \, {\rm km \, s^{-1} \, kpc} .
\end{equation}

Equations~(\ref{eqn:jlam})--(\ref{eqn:comb})
are simple but powerful, allowing us
to connect the visible properties of galaxies
to their invisible DM halos, using some simple parameters and assumptions.
They also provide robust observational constraints on some basic
characteristics of galaxy formation that are still far beyond the ability
of raw theory to predict reliably.
{\it The average value of $f_j \, f_\star^{-2/3}$ for a population of galaxies
can be determined by observations as a strict constraint on theory.}

We can immediately use Equation~(\ref{eqn:comb}) in combination with the observational
results for $j_0$ from Table~\ref{tab:loglog} for fixed $\alpha=2/3$.
We find that $\langle f_j \, f_\star^{-2/3}\rangle \simeq$~1.9 for Sb--Sm spirals
and $\simeq$~0.5 for ellipticals.
For example, if we assumed an arbitrary $\langle f_\star \rangle=$~0.2 for both 
types of galaxies, then we would infer $\langle f_j\rangle\simeq$~0.65 for spirals
and $\simeq$~0.1 for ellipticals.
This means a systematic difference in net angular momentum retention between
the two galaxy types which, although there are many further details to work through
below, will hold up as a basic result of this paper.

To derive firmer constraints on $f_j$, we need to break the $f_\star$--$f_j$
degeneracy by introducing well-motivated values for $f_\star$,
for both spirals and ellipticals.
We also need to consider the complication that $f_\star$ cannot in reality
have a simple, constant value, even on average.  
This is because the observed luminosity function of galaxies has
a dramatically different shape from the predicted mass function of DM halos
(e.g., \citealt{1978MNRAS.183..341W,1991ApJ...379...52W,2002ApJ...569..101M,2003MNRAS.339.1057Y,2010ApJ...710..903M}).
Below the characteristic ``knee'' luminosity $L^*$,
the galaxies are observed to follow a shallower slope than the DM mass function
$dN/dM \propto M^{-2}$, while at higher luminosities, the observations are
{\it steeper} than the predictions.
The implication is that the fraction of luminous-to-dark matter 
declines rapidly for galaxies fainter and brighter than $L^*$;
i.e., assuming a constant $f_{\rm b}$, the function $\langle f_\star\rangle(M_{\rm vir})$ has a
characteristic inverted U shape.

This empirical trend is thought to be caused physically
by various feedback effects that inhibit
star formation and become increasingly important in the low- and high-mass regimes
(such as stellar and supermassive black hole feedback, respectively; e.g., 
\citealt{1993ApJ...402...15L,1994MNRAS.271..781C,1999MNRAS.310.1087S,2006MNRAS.370..645B,2006MNRAS.365...11C}). 
Regardless of the explanation, any self-consistent
$\Lambda$CDM-based model must incorporate a strong,
systematic mass dependence on star formation efficiency, $\langle f_\star\rangle(M_{\rm vir})$.

One might be concerned that such a mass dependence would transform an underlying 
$j \propto M^{2/3}$ 
relation for DM halos into something very different for 
the stellar components of galaxies, and quite unlike our observational results.
To check this, we will modify our simple model above to allow for a varying
function $\langle f_*\rangle(M_{\rm vir})$.
Since this function is a tracer of undetermined baryonic physics during galaxy
evolution, there is not yet any robust theoretical prediction for it,
but fortunately it can be estimated empirically.
This is done in an {\it average} sense through various techniques such
as weak gravitational lensing, stacked satellite kinematics, and
matching up the mass and luminosity functions mentioned above.

There have been many studies that estimated $\langle f_\star\rangle(M_{\rm vir})$, but
few that did so separately for different galaxy types, which is important for our analysis.
We therefore adopt the relations for $\langle f_\star\rangle(M_\star)$ derived
by \citet{2010MNRAS.407....2D}.
For the spiral galaxies, we use their relation for ``late-type'' galaxies:
\begin{equation}\label{eqn:fsMltg}
\langle f_\star\rangle(M_\star) = \frac{f_0 \left(M_\star/M_0\right)^{1/2}}{\left[1+\left(M_\star/M_0\right)\right]^{1/2}} .
\end{equation}
Below a characteristic mass $\log\,(M_0/M_\odot) \simeq 10.8$, 
this relation has a dependence
$\langle f_\star\rangle \propto M_\star^{1/2}$. At higher masses, it
approaches a constant, $f_0 \simeq 0.33$.
Here we have converted the Dutton~et~al.\ results to our definition of the virial mass
and to our adopted stellar IMF, while using $h=0.72$.

For elliptical galaxies, we adopt the Dutton et al.\ relation
for ``early-type'' galaxies:\footnote{There has been very little work along these
lines for elliptical and lenticular galaxies separately, but there is some
recent evidence that the halo masses for these types are the same
\citep{2011ApJ...742...16T}.
Note also that the Dutton~et~al.\ relations were derived for somewhat smaller mass ranges
than covered by our data, 
and that their stellar mass determinations 
may not be fully consistent with our methods.}
\begin{equation}\label{eqn:fsMetg}
\langle f_\star \rangle (M_\star) = \frac{f_0 \left(M_\star/M_0\right)^{0.15}}{\left[1+\left(M_\star/M_0\right)^2\right]^{1/2}} ,
\end{equation}
where $\log\,(M_0/M_\odot)\simeq11.2$, $f_0 \simeq 0.14$,
and the asymptotic behaviors at low and high masses are
$\langle f_\star\rangle \sim M_\star^{\,0.15}$ and 
$\langle f_\star\rangle \sim M_\star^{-0.85}$, respectively.
One of the key features to notice here is that an elliptical galaxy typically
has a much lower value of $f_\star$ than a spiral with the same stellar mass:
i.e., ellipticals inhabit systematically more massive DM halos, which in many
cases extend up to ``group'' masses of $M_{\rm vir}\sim10^{13} M_\odot$ and beyond
(see also \citealt{2011A&A...534A..14V}).

These $\langle f_\star\rangle(M_\star)$ relations can be uniquely transformed to 
$\langle f_\star\rangle(M_{\rm vir})$,
and taken together define an inverted U-shaped trend as discussed above.
The relations were constructed using a compilation of different literature
results, 
which showed an encouraging degree of mutual consistency,
so we conclude that the average trends above are probably reliable at the
$\sim$~50\% ($\sim$~0.2~dex) level.
There may also be non-zero galaxy-to-galaxy variations in $f_\star$
at a fixed mass and type; the value of this scatter is less well established, but
recent analyses suggest that it may be $\sim$~0.15~dex 
\citep{2010ApJ...717..379B,2011MNRAS.410..210M}.
We adopt this as our default value, which fortunately is smaller than
the expected dispersion in halo spin of $\simeq$~0.23~dex and so will not
have much impact on our conclusions.

Using these variable $\langle f_\star\rangle(M_{\rm vir})$ relations to construct mock \js--\Ms\
data sets as before, we plot the results in panels (d) and (e) of Figure~\ref{fig:schem2}.
For both spirals and ellipticals, we can see that the curvature in 
$\langle f_\star\rangle (M_{\rm vir})$
translates to systematic deviations in the \js--\Ms\ relation from a simple 
$\alpha=2/3$ power-law.
We will investigate how these deviations compare to real observations
in the next subsection.

Panels (d) and (e) of Figure~\ref{fig:schem2}
also demonstrate that at masses of $M_\star \ga 10^{11} M_\odot$,
the ellipticals are predicted to have {\it higher} \js\ than the spirals of the same mass,
owing to their differences in $f_\star$.  The more massive DM halos of
ellipticals ought to provide larger virial-radius lever arms that lead
to larger $j_{\rm vir}$, and therefore larger \js---{\it if}
they retain as much fractional angular momentum as spiral galaxies do.
Therefore the observed offset in \js--\Ms\ between spirals and ellipticals
implies an even larger difference in $\langle f_j\rangle$ than in the simple example above
with fixed $\langle f_\star\rangle=0.2$.
We will examine this apparent $f_j$ dichotomy further in the next
subsection.

As a final illustrative exercise,
we generate a mock data set for elliptical galaxies as in panel (e), then
adopt $\langle f_j\rangle=0.1$,
with an assumed dispersion of $\sigma_{\log f_j}=0.15$~dex.
The results are plotted in panel (f), where we see that the galaxies
have coincidentally returned to nearly the original
$j$--$M$ sequence for halos, modulo a little curvature and increased scatter.

Figure~\ref{fig:schem2} thus shows how one could map the observed
\js--\Ms\ properties of a population of galaxies (panel f) to a theoretical
prediction for their halos (panel a), and recover some basic parameters describing
galaxy formation (see Equation~(\ref{eqn:comb})).
This formulation is closely related to a classic theoretical framework
for the formation of spiral galaxy disks, whose observed sizes and rotation velocities
are generally
 consistent with the approximate conservation of primordial specific angular momentum
($f_j \sim 1$; e.g., \citealt{1980MNRAS.193..189F,1997ApJ...482..659D,1998MNRAS.295..319M}).
However, our formulation is more general by including also the early-type galaxies,
as well as the bulge components within spiral galaxies (which we will discuss below).

\subsection{Investigating the spread in \js}\label{sec:theory2}

As just discussed, the observed dichotomy between the \js--\Ms\
relations of spirals and ellipticals may imply differences in
their specific angular momentum retention, expressed here by the factor $f_j$.
This interpretation is based on an implicit assumption that the parent
halos of both galaxy types had the same average $\lambda$.
However, a natural halo-to-halo scatter in $\lambda$ is expected,
and one could instead imagine the other extreme case, in which
$f_j$ is the same for the two galaxy types, while
their halo $\lambda$ values are systematically different 
(e.g., \citealt{1982MNRAS.200..585K,1984Natur.311..517B,1996MNRAS.282..436C}).
In other words, spirals and ellipticals are drawn from the high-
and low-spin tails of the $\lambda$ distribution, respectively.

We call these two alternatives the ``variable $f_j$''
and ``spin bias'' scenarios.
In reality, a mixture of both scenarios may be present, which would be
difficult to disentangle,
but we can begin by investigating these two limiting cases in detail.
Thus the aim of this section is to test how consistent each of
these cases is with the data.

The reason we can make headway on this issue is that
there are predictions from $\Lambda$CDM not only for the average value of $\lambda$,
but also for its probability distribution, i.e., a lognormal
with a characteristic dispersion as discussed in Section~\ref{sec:theory1}.
We continue to focus on the spirals and ellipticals as 
the two interesting extremes of the observed \js\ range (at fixed \Ms), 
and consider the lenticulars as intermediate either in
$f_j$ or in $\lambda$.

We begin with the spin-bias scenario.  If correct, adopting a constant $f_j$ value for
a complete, unbiased galaxy sample would allow us to work backwards to infer
the underlying $\lambda$ distribution, which could then be
compared to the theoretical prediction.
One might think that we have already implicitly carried out this test
by examining the residuals from the observed \js--\Ms\ relation in
Section~\ref{sec:hist} and Figure~\ref{fig:histdiff}.
However, that analysis did not account for the differences in $f_\star$
between different galaxy types.

We therefore proceed with a more direct comparison to theory by
generating \js--\Ms\ model predictions for each galaxy type, and calculating
the observed residuals with respect to these models.
We use Equation~(\ref{eqn:jslam}) with $\lambda=\langle \lambda \rangle=0.035$,
along with  the empirical
$\langle f_\star\rangle(M_\star)$ relations (\ref{eqn:fsMltg}) and (\ref{eqn:fsMetg}), 
and an ad-hoc $\langle f_j\rangle=0.55$,
to predict a mean \js--\Ms\ relation for each galaxy type.
We then derive the residuals $\Delta \log j_\star$ by subtracting the model from
the observations as in Equation~(\ref{eqn:Dj}).
If the spin bias scenario is correct, then the properly reweighted distribution
of these residuals ought to follow a lognormal with dispersion
$\sigma_{\log j_\star}\simeq0.27$ (which accounts for observational
errors and the intrinsic scatter in $f_\star$).

\begin{figure}
\centering{
\includegraphics[width=3.5in]{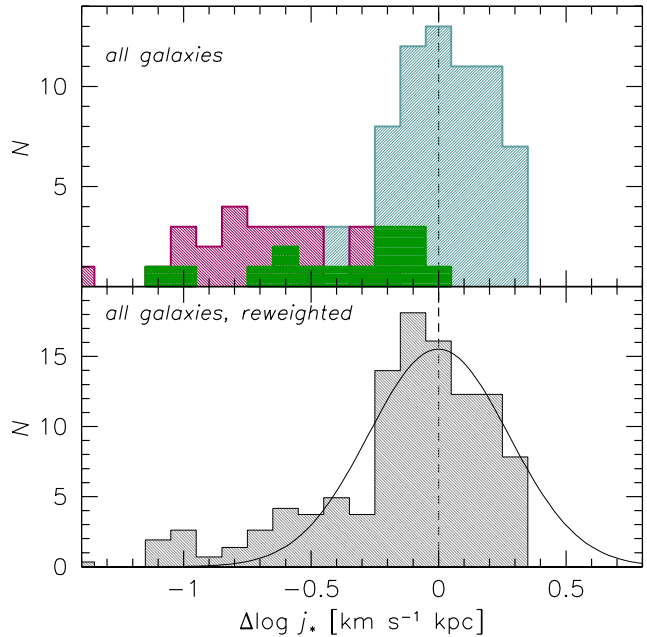} 
}
\caption{Distributions of residuals in the
observed stellar specific angular momentum,
with respect to the mean theoretical prediction for $\Lambda$CDM halos,
after assuming a fixed $j$-retention parameter, $f_j=0.55$.
As in Figure~\ref{fig:histdiff}, red, green, and blue histograms in the top
panels show the residuals for elliptical, lenticular, and spiral galaxies, 
respectively.  The bottom panel shows the same distribution, renormalized
for the relative frequencies of galaxies in the nearby universe.
The curve shows a predicted lognormal distribution for comparison.
The distribution of residuals for spiral galaxies is narrower than
expected from the distribution of halo spins, while the overall galaxy distribution
shows clear departures from the lognormal model
(with an excess at low \js\ and a deficit at high \js).
}\label{fig:histdiff3}
\end{figure}

\begin{figure*}
\centering{
\includegraphics[width=7.0in]{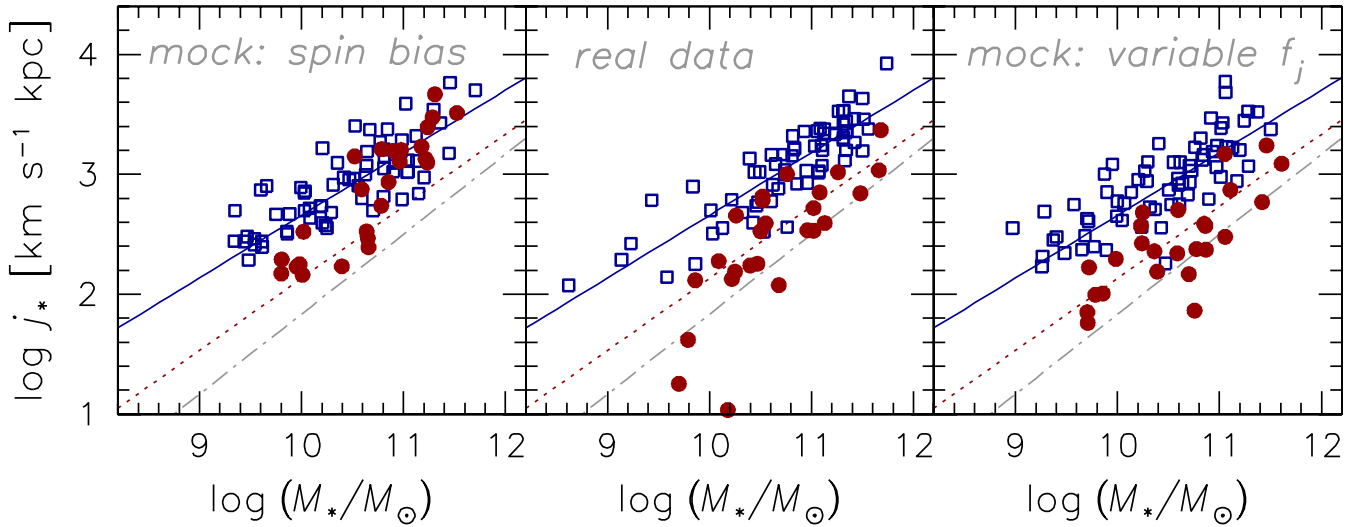} 
}
\caption{Stellar specific angular momentum vs.\ stellar mass, comparing
mock data generated from $\Lambda$CDM-based models (left- and right-hand panels)
to real data (middle panel).
The model on the left includes halo spin bias, while the model on the right
assumes systematic differences in angular momentum retention
between spirals and ellipticals.
Blue open squares and red filled circles show spirals and ellipticals, respectively,
with the solid blue and dotted red lines showing the best-fit power-laws 
for the real data.
The relation for halos is also shown for reference as a gray dot-dashed line.
The mock data sets include intrinsic scatter in the parameters $\lambda$
and $f_\star$ at a given mass, but {\it not} observational errors.
The simple variable-$f_j$ mock data on the right resemble the real data,
while the spin-biased model does not.
}\label{fig:mock2}
\end{figure*}

Figure~\ref{fig:histdiff3} presents histograms of these residuals, both
by separate galaxy types (top panel), and in combination (bottom panel), which
uses a renormalization
by frequency of galaxy types from the
\atlas3d\ survey, as in Section~\ref{sec:hist},
We find that overall, the total distribution of $\Delta j_\star$ 
has approximately the predicted width.
However, the distribution in detail appears significantly different from a lognormal:
there is an excess of low-$\Delta j_\star$ galaxies,
and a missing tail at high-$\Delta j_\star$.
In particular, there are too many elliptical galaxies in the nearby universe to
be explained by the tail of low-spin halos.\footnote{\citet{2007MNRAS.375..163H}
also found in attempting to infer halo $\lambda$ values for spirals and ellipticals
that an ad-hoc rescaling of the elliptical values was required in order
to avoid a double-peaked $\lambda$ distribution.}

This histogram analysis appears to exclude a simple spin-bias scenario,
but there are some caveats,
such as small sample sizes and the assumption of perfect lognormality 
for the distribution of halo spins.
We can make further progress by recognizing that the scenario makes predictions
for the \js\ residuals not only for all galaxies combined, but also as a function
of mass.  This is because $\lambda$ is not predicted to depend on halo mass,
while the relative frequencies of different galaxy types are
observed to vary strongly.
One can then immediately see a serious problem with the spin-bias scenario:
at high masses, almost all of the galaxies are ellipticals, which should thus
be an unbiased population representing the full range of halo spins
(\citealt{2012MNRAS.421..608D} made a similar point for low-mass disk galaxies).

We investigate this issue in more detail by constructing
a mock data set as in
Figure~\ref{fig:schem2}, while this time incorporating a schematic model for spin bias.
We now assume that all galaxies have $f_j=0.45$, with
the late-types inhabiting the high-spin halos, and the early-types the low-spin ones.
Using the number densities of early- and late-types as a function of \Ms\ from \atlas3d,
we use the $\langle f_\star\rangle(M_\star)$ relations to
translate this to the relative fractions at fixed halo mass
(which can be quite different from the fractions at fixed \Ms).
We then randomly draw a distribution of biased spin parameters for each galaxy type; 
e.g., if spirals comprise 25\% of galaxies at a given mass,
we draw mock spirals from the top quarter of the spin distribution.
We also adopt a similar mass range and total number of galaxies as in our
real data sets.

We show the resulting \js--\Ms\ mock data set in the left-hand panel
of Figure~\ref{fig:mock2}, which can be compared to the real data in the middle panel.
We see that the low-mass ellipticals could indeed be drawn from only the low-spin tail
because of their rarity.  However, at high masses the ellipticals are common
and their predicted \js\ values are similar to the spirals.
To salvage the spin-bias scenario would thus seem to require a mass-dependent
bias, which seems epicyclic and therefore not appealing.\footnote{There
may be reasons of stability for ellipticals to be dominant at high masses (e.g.,
\citealt{1997ApJ...482..659D,1998ApJ...507..601V,2012MNRAS.421..608D}), 
but this ostensibly changes the {\it morphology} and not \js.}

The biasing idea can also be discredited by environmental considerations:  
there are strong observational correlations between environmental
density and galaxy morphology, but as mentioned earlier,
halo spins in theory depend only weakly on environment
(which has some observational support in the case of disk galaxies;
\citealt{2008MNRAS.388..863C,2008MNRAS.391..197B}).
In addition, if we consider disks and bulges to be manifestations of the same
\js--\Ms\ trends as spiral and elliptical galaxies, then the coexistence of
these subcomponents within the same galaxies provides a clear argument against halo
spin bias.  

We next turn to the variable-$f_j$ scenario, where spirals and ellipticals
are drawn from the same underlying distribution of halo spins, but their
baryonic components have systematic differences in retaining $j$.
Given that we know $\langle f_\star\rangle$ for each galaxy type, Equation~(\ref{eqn:comb}) 
suggests that we can immediately use the observed $j_0$ normalization to infer $\langle f_j\rangle$.
However, the situation is more complicated since $\langle f_\star\rangle$ varies with mass
and therefore one does not expect an exact
$\alpha=2/3$ for fixed $f_j$ (recall Figure~\ref{fig:schem2}(d,e)).

As we did for the spin-bias scenario, we again construct mean \js--\Ms\
relations for each galaxy type, while now leaving $f_j$ as a free parameter.
Carrying out least-square fits to the data, we find
values of $\langle f_j\rangle=0.56\pm0.03$ and $\langle f_j\rangle=0.12\pm0.01$ for
the spiral and elliptical galaxies, respectively.
The difference in $\langle f_j\rangle$ of a factor of $4.7\pm0.8$ is slightly larger than
the observed \js--\Ms\ relative offset, as anticipated in the 
previous section because of the differences in $\langle f_\star\rangle$ 
(e.g., Equation~(\ref{eqn:comb})).\footnote{Given the degeneracy
between $f_j$ and $f_\star$, in principle the inferred $f_j$ dichotomy
could be an artifact of errors in our adopted values for $\langle f_\star\rangle$.
However, these errors would have to amount to a combined factor of $\sim$~5: 
e.g., with true $\langle f_\star\rangle \sim$~0.1 for the spirals
along with $\sim$~0.2 for the ellipticals,
rather than $\sim$~0.25 and $\sim$~0.1.}

The next step is to verify that these best-fit models 
provide reasonable representations of the data.
We again construct mock data sets, using the new $f_j$ models
(with 0.15~dex of scatter in $f_\star$),
and show the results in the right-hand panel of Figure~\ref{fig:mock2}.
Here we see that, unlike the spin bias model, these variable-$f_j$ models provide
a remarkably good match to the data.  The curvature of the predicted \js--\Ms\
relation turns out to be imperceptible, once we account for
observational errors, small-number statistics, and a limited mass range.\footnote{
Future empirical estimates of \js\ and \Ms\ over a larger dynamic range 
could provide a strong test of constant-$f_j$ scenarios.
Given the observational difficulty of measuring \js\ at high masses where the
underlying halos pertain to entire galaxy groups and clusters, the
best prospect for improvement would be to study lower-mass galaxies,
with $\log\,(M_\star/M_\odot) \la$~9.}
Furthermore, the observed slope for the spirals is shallower than
$\alpha=2/3$, which is predicted by the model.

This comparison does not entirely succeed in accounting for
the {\it scatter} about the \js--\Ms\ relations.
As can be seen in Figure~\ref{fig:mock2},
the real observations appear to follow {\it tighter} trends than
predicted by our simple model, for both spirals and ellipticals.
The model fits give rms scatters of $\sigma_{\log f_j}=$~0.18~dex and
0.25~dex for the spirals and ellipticals,
which is already {\it less} than the expected scatter
of 0.27~dex from $\lambda$ and $f_\star$,
even without allowing for measurement errors, and scatter in $f_j$
(see also the histogram of spirals in the
top panel of Figure~\ref{fig:histdiff3}, compared to the curve in the lower panel).

One possible explanation for this reduced scatter
is that the baryonic processes responsible for $j$-loss could act
as some kind of ``attractor'' to specific values of $f_j$
(cf.\ \citealt{2000ApJ...545..781D}).
Alternatively, halo spin bias could be at work in a secondary role,
even while $f_j$ variation is the primary effect.\footnote{It has
been suggested that later type galaxies are biased to {\it lower}
spin halos \citep{2004ApJ...612L..13D}. If correct,
the net impact on the \js\ scatter is unclear, but one implication
is that the $f_j$ dichotomy between spirals and ellipticals would
be even larger than in our no-bias scenario.}

Our overall conclusion is that the variable-$f_j$ model
reproduces the \js--\Ms\ observations well in general,
is fairly insensitive to the exact trend of $\langle f_\star\rangle$ with mass,
and does not require any additional variation of $\langle f_j\rangle$ with mass.
The spirals appear to 
have been fairly efficient in preserving the
specific angular momentum imprints of their parent halos, while
ellipticals have lost the vast majority of theirs.

This is a plausible scenario from a physical standpoint
if we return to our proposed framework where all galaxies are composed of bulges and disks
(Figure~\ref{fig:schem1} and Section~\ref{sec:replace}).
Unfortunately, we do not have $\langle f_\star\rangle(M_\star)$ relations for the bulges and disks 
themselves in order to
directly derive their $\langle f_j\rangle$ trends. However, given the similarities in \js--\Ms\
that we found between these subcomponents and the galaxies overall,
it seems reasonable to suppose that bulges and disks have $\langle f_j\rangle\sim$~0.1
and $\sim$~0.6, respectively, and that these values are characteristic
of two distinct modes of galaxy evolution.\footnote{One concern here is that for
more bulge-dominated galaxies, one might expect the disk-only $\langle f_\star\rangle$ to be
relatively low, and thus the disk \js\ to appear relatively high.
However, the observations are somewhat suggestive of the {\it opposite} trend,
i.e., disk \js\ anti-correlating with $B/T$.}
We will return to this topic in the next section.

Our conclusions about {\it spiral} galaxies echo similar findings in the literature, 
which have typically inferred $\langle f_j\rangle \sim$~0.5--0.6 overall
(e.g., \citealt{2000ApJ...538..477N,2007ApJ...654...27D,2009ASPC..419....3B,2012MNRAS.421..608D,2012MNRAS.424..502K}).
In particular, \citet{2012MNRAS.421..608D} used a
model parametrization similar to our $(f_\star,f_j)$,
and found that $\langle f_j\rangle$ is fairly constant over a wide mass range.
Note that these authors used  a
parametrized mass model
to fit the Tully-Fisher relation, which was then converted to
an average $j_{\rm vir}$--$M_{\rm vir}$ relation.
Our approach works instead in the space of observables, \js--\Ms, which is more direct
and transparent while also allowing us to analyze galaxy-to-galaxy 
variations.\footnote{As 
a consistency check, we also take a slightly different approach
and make a model prediction for the mean relation
between size and rotation~velocity for spirals (cf.\ \citealt{1998MNRAS.295..319M,2004ASSL..319..341B}).
We adopt a value of $\langle f_\star \rangle=0.56$, and rather than assuming
some function $\langle f_\star\rangle(M_\star)$, we relate the disk rotation and the virial
circular velocity by $\vs \simeq 1.2 v_{\rm vir}$.  Given $\langle \lambda \rangle=0.035$,
there is a linear relation predicted between $\vs$ and \aeff, which we show in the
right-hand panel of Figure~\ref{fig:rot}. To zeroth order, this prediction agrees
well with the spiral data.}

Our finding for the {\it ellipticals} is novel,
as neither the predictions for \js--\Ms\ of ellipticals, nor their
subsequent $f_j$ inferences, have been well-studied before now.
We have not carried out a comparable analysis on {\it lenticulars}
since the constraints on them are less certain.
Qualitatively speaking, their observed $\log j_\star$ normalization is between the
other two galaxy types, which for plausible values of $\langle f_\star\rangle$
implies $\langle f_j \rangle$ values that are intermediate to
those for the spirals and ellipticals.
In addition, there may be two subpopulations of lenticulars
as discussed in Section~\ref{sec:resid}, with low and high $\langle f_j\rangle$.

There are two interesting implications about these findings.  
One is that that we now have a remarkably simple and successful framework for 
describing and connecting some of the most fundamental
properties of galaxies.
The observable galaxies may be connected to their unobservable host halos
using \js\ and \Ms\ along with some relatively basic parameters $f_j$ and $f_\star$.
Such a model may appear implausibly oversimplified in the light of our
ever-expanding awareness of the complexities of galaxy formation physics,
but for some reason it seems to work.
 
\begin{figure}
\centering{
\includegraphics[width=3.5in]{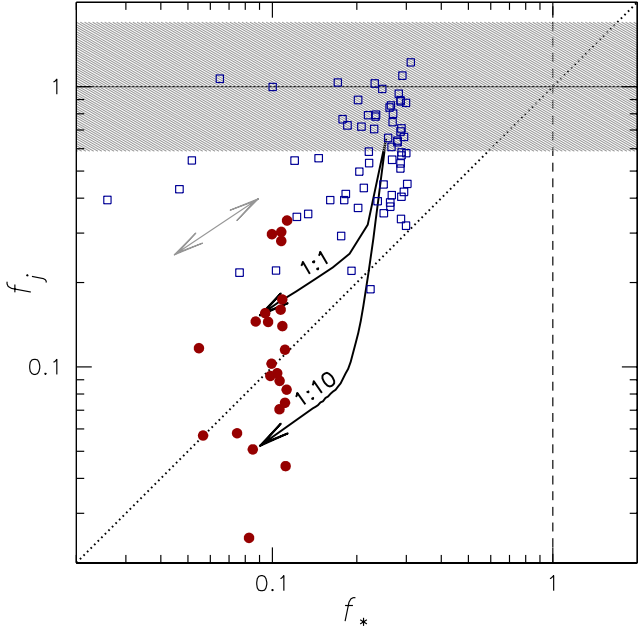} 
}
\caption{Specific angular momentum retention fraction plotted against
stellar mass fraction,
as inferred for individual galaxies, with symbols as in
Figure~\ref{fig:mock2}.
The dotted diagonal line is the one-to-one relation,
and the gray double-arrow shows the direction of the
uncertainties as driven by the $f_j \propto f_\star^{2/3}$ degeneracy.
The width of the shaded region around $f_j=1$ corresponds to the
scatter in spin expected for $\Lambda$CDM halos.
The black arrows show schematic vectors from 1:1 and 1:10 mergers,
as discussed in Section~\ref{sec:theory3}.
The spiral and elliptical galaxies occupy distinct regions of the diagram,
while a simple model implies that converting spirals into ellipticals would
require a very large amount of growth through $\sim$~1:3 mergers.
}\label{fig:fvsf}
\end{figure}

The other implication is that these parameters may give us insight into the
formation of disks and bulges, and into the origins of the Hubble sequence.
To illustrate this point, we use our modeling procedures as described
above to work backwards and estimate
$f_\star$ and $f_j$ values for {\it individual} galaxies.
The outcome is shown in Figure~\ref{fig:fvsf}, where one should focus on
the {\it average} results for each galaxy type, since
no attempt was made to model the scatter in $f_\star$ and $\lambda$.

The general picture that we obtain is that
spiral and elliptical galaxies are clumped
around two regions of parameter space:
$(f_\star,f_j) \sim (0.25,0.55)$, and $\sim (0.1,0.1)$, respectively.
{\it Whatever processes formed and shaped these galaxies were
efficient at both forming stars and retaining total specific angular momentum
for the spirals, and inefficient for the ellipticals.}

As discussed in Section~\ref{sec:intro}, early cosmologically-based simulations 
struggled to reproduce such high $f_j$ values for spirals, finding typically
$f_j \sim$~0.01--0.1, which
was later realized to be due in part to numerical artifacts,
and in part to inadequate feedback recipes.
Feedback could be particularly important for slowing down gas collapse and star formation
so that the baryons are not affected by torque-driven $j$ transfer during early mergers
\citep{1998MNRAS.300..773W,2003ApJ...596...47S,2012ApJ...749..140H,2012MNRAS.423.1726S}.
However, whatever physical processes are now invoked to explain the $f_j$ values
of spirals must simultaneously allow for much lower $f_j$ in ellipticals
(e.g., by having less efficient feedback; \citealt{2008MNRAS.387..364Z,2008MNRAS.389.1137S}).

\subsection{Physically-motivated models for galaxy evolution}\label{sec:theory3}

\begin{figure*}
\centering{
\includegraphics[width=7.1in]{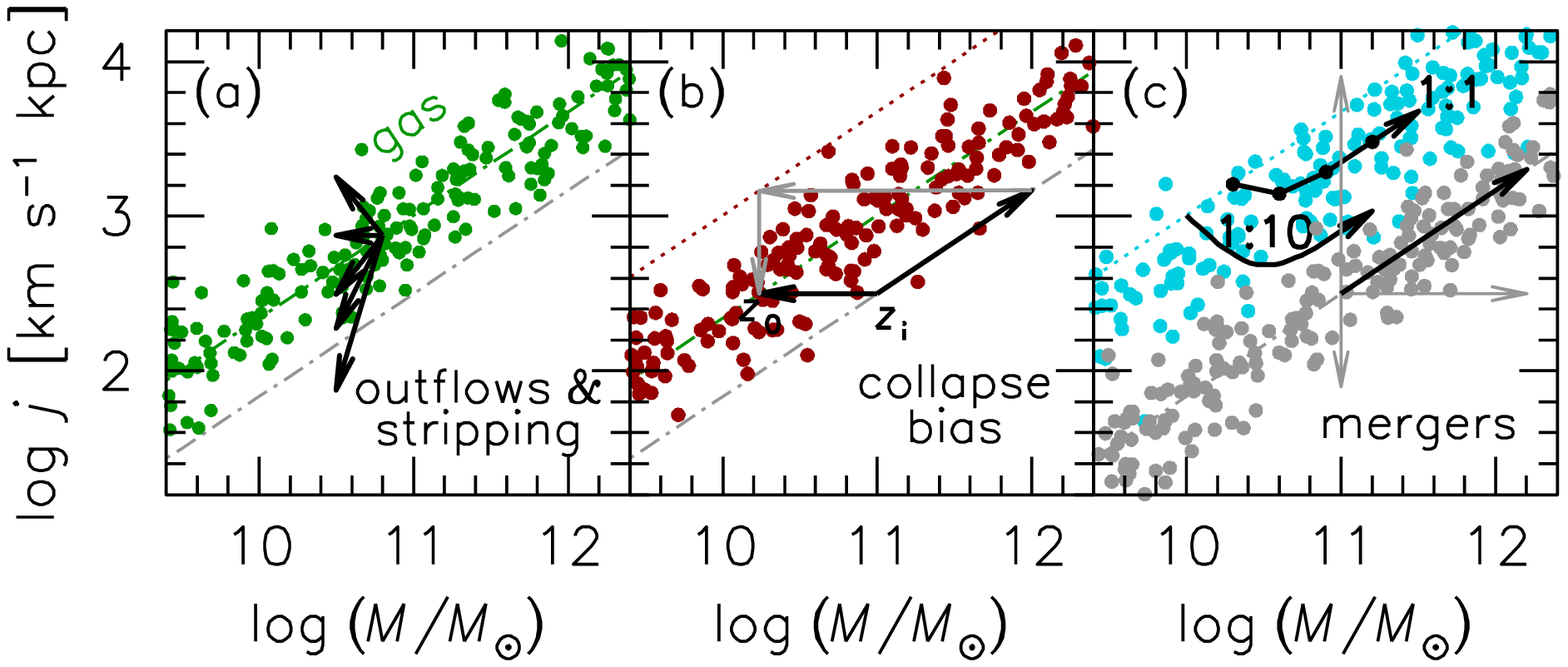} 
}
\caption{Schematic evolution of galaxies in specifical angular momentum and mass,
as in Figure~\ref{fig:schem2}, but now considering evolution through gas outflows,
stripping, and biased baryon collapse, and galaxy mergers.
Panel (a) show initial conditions for pre-collapse gas (dots), and
possible evolutionary vectors from outflows and stripping
(arrows; see text for details).
Panel (b) shows the collapse of gas and formation of 
stars at some initial redshift $z_{\rm i}$, preserving the \js--\Ms\
values until a final redshift $z_0$ (black arrow to the left, with dots illustrating
a population of galaxies).
The halo grows until redshift $z_0$ with no further star formation (black arrow to
upper right). At $z_0$, the expected trend with perfect $j$ conservation is the
dotted line, and net values for $f_\star$ and $f_j$ would be inferred using the
leftward and downward gray arrows, respectively.
Panel (c) shows initial conditions for DM halos as gray dots, and
schematic vectors of evolution through mergers (gray arrows):
mass growth (to the right), specific angular momentum decrease through
cancellation of the spin components (downwards), and increase through the
orbital component (upwards).  The net evolution is a black diagonal arrow to the upper
right.  
The upper dotted track marks the initial conditions for stellar disks, and 
the blue dots show disks
after having undergone four 1:1 mergers each.
The upper black curved vector illustrates the typical evolution of a galaxy,
with each black dot marking the beginning of a discrete merger event.
The lower black curved vector shows the same for a series of 1:10 mergers
(note that for clarity, the curved vectors are arbitrarily shifted relative to the $f_j=1$
starting point for the DM vector).
In both cases, after the mass has grown by a factor of $\sim$~2, the orbital \js\
dominates the evolution, moving merger remnants along a \js--\Ms\ track parallel to,
but lower than, the initial disk trend.
}
\label{fig:schem3}
\end{figure*}

Now that we have derived a comprehensive framework for connecting \js--\Ms\ observations
with simulated $\Lambda$CDM halos, and thereby derived generic constraints on specific
angular momentum retention, $f_j$ (Figure~\ref{fig:fvsf}), 
we will work through some 
case studies of plausible physical
processes in galaxy formation and evolution.
These cases are not meant to be exhaustive, nor to provide immediate
ammunition for current debates about galaxy formation,
but to serve as practical examples of how the $j$--$M$ diagram can be
used as a tool to furnish physical insight.
The models involved will treat $f_j$ and $f_\star$ as covariant parameters,
unlike in the previous sections where for simplicity they were independent.

A general constraint to keep in mind is
that for each galaxy type, $f_j$ is approximately constant
as a function of mass, including little additional scatter,
which accounts for the observed $j$--$M$
relations appearing so similar to those for theoretical DM halos.
{\it Any model for angular momentum evolution should
explain why galaxies appear to remember so faithfully the overall initial conditions of
their parent halos.}

The challenge of this $f_j$ constancy has been recognized previously for disk galaxies.
There are a variety of physical mechanisms during galaxy evolution that could involve $j$ 
transfer (e.g., gas cooling and feedback), but
unlike gravitational clustering,
these baryonic processes (and the resulting $f_j$ values)
are expected to depend strongly on mass,
which appears to require some degree of fine-tuning
to reconcile with the observations
(e.g., \citealt{2012MNRAS.421..608D}).
Our inclusion of early-type galaxies in this framework, with near-constant $f_j$,
deepens the mystery: there are now {\it two} fine-tuning conspiracies to explain.

Here we emphasize again a distinction from comparisons between
{\it internal} distributions with radius of $j$ for stars and DM halos
(e.g., \citealt{2001ApJ...555..240B,2001MNRAS.326.1205V,2002MNRAS.329..423M,2005ApJ...628...21S}).
As mentioned in Section~\ref{sec:intro}, there is ample reason to expect
redistribution of \js\ to occur within the baryonic component of a galaxy
and thereby violate strong $j$ conservation.
However, this does not affect our examination of weak conservation, where
the overall value of $j$ may remain roughly the same
(assuming negligible transfer of $j$ between baryons and DM).

We may reduce the potential explanations for the systematic difference in $f_j$
between spirals and ellipticals into two basic scenarios, which we will examine
before summarizing the overall picture.
One general scenario is an {\it internal} angular momentum bias, where 
high- and low-\js\ galaxies were formed from parts of their available gas supply
that had preferentially high or low $j$.
The other is that these galaxies experienced systematic differences in
angular momentum transport {\it after} star formation, and during 
subsequent galaxy assembly phases.

Below, 
Section~\ref{sec:outflow} discusses outflow and stripping scenarios, 
Section~\ref{sec:bias} considers biased collapse, and
Section~\ref{sec:merg} examines mergers.
Section~\ref{sec:eval} surveys 
the plausibility of these evolutionary modes in the light of the \js--\Ms\ observations.

\subsubsection{Outflows and stripping}\label{sec:outflow}

One example of the first scenario involves {\it gas outflows}, whether caused by
galactic winds or by some other mechanism.
Let us assume that the baryons in a galaxy collapse into a thin disk while
preserving the total specific angular momentum, i.e., $f_j=1$ (recall Figure~\ref{fig:schem2}(b)).
The local specific angular momentum within the disk, 
$j_{\rm g}(R) \propto R\,v_{\rm rot}(R)$,
is assumed to increase monotonically with galactocentric radius, 
which is unavoidable if the gas follows
co-rotating circular orbits (the rotation-velocity profile
cannot decrease any more rapidly than Keplerian, while the lever arm $R$ in the $j$ calculation
increases linearly).

Before many stars form, an outflow begins which we parameterize by a mass
loss that is proportional to the gas surface density to some unknown power $\beta$:
\begin{equation}
\Delta M_{\rm g} \propto \Sigma_{\rm g}^\beta.
\end{equation}
Because the gas is presumed to settle into a configuration where the density
increases toward the center (e.g., an exponential profile), the parameter
$\beta$ translates into a biased removal of gas from different disk {\it radii},
which in turn means depletion of gas parcels with systematically different $j_{\rm g}$.

To analyze this scenario further, we now introduce Figure~\ref{fig:schem3}, 
which like Figure~\ref{fig:schem2} illustrates schematic vectors
of mass and angular momentum evolution, but now extends to more specific,
physically-motivated processes.
In Figure~\ref{fig:schem3}(a), the horizontal arrow to the left illustrates
an outflow with $\beta=0$:
the gas everywhere in the disk is depleted by an equal fraction,
and its initial specific angular momentum is preserved, while its mass decreases.
If $\beta>0$, then the outflows occur preferentially in the high-density, central
regions that have relatively low $j_{\rm g}$, and so the overall $j_{\rm g}$ for the galaxy
increases (diagonal arrow toward upper left;
cf.\ \citealt{2001MNRAS.321..471B,2012ApJ...750..107S}).
If $\beta<0$, then the mass loss is preferentially from the outer regions, and the overall $j_{\rm g}$
decreases (diagonal arrows toward lower left).
Thus, outflows could in principle produce either a net increase or decrease in $f_j$.

It should be kept in mind that these outflows represent only material that is 
launched completely out of the galaxy, never to return.
Other types of outflows may also occur, where gas is expelled outward but remains
bound and falls inward again, as in a galactic fountain
(e.g., \citealt{2012MNRAS.419..771B}).
However, such internal processes might alter only the detailed distribution with radius
of $j$, and not affect the overall value which concerns us here
(see the discussion above of weak and strong $j$ conservation).
More complex scenarios could also be considered, where fountain material interacts with
halo gas and exchanges angular momentum
(e.g., \citealt{2009MNRAS.399.1089M,2011MNRAS.415.1534M}), 
leading to shifts in \js\ for the stellar disk that eventually forms.

A mechanism related to gas outflows is galaxy {\it stripping} through gravitational
interactions with other galaxies in a dense environment.  
Here the effects on \js\ and \Ms\ depend on whether the tidal stripping
occurs before or after the gas collapses.
If a galactic halo is tidally stripped {\it before} the gas collapses
(e.g., \citealt{1980ApJ...237..692L}),
then the reservoir of $M_{\rm g}$ and $j_{\rm g}$ available for collapse is depleted
in a manner that depends on the internal distribution of these quantities.
F83 adopted some plausible distributions and worked out the resulting $j$--$M$ changes:
we will not repeat the analysis here, but merely show the equivalent evolutionary vectors 
as the three arrows in Figure~\ref{fig:schem3}(a) pointing downwards to the left.

There are two key features to notice with the gaseous stripping arrows.  One is that unlike
outflows, this stripping can only {\it decrease} $f_j$ ($\beta<0$) since it acts solely
on the outer regions.
The second is that plausible $j$-loss vectors are accompanied by substantial mass loss,
which means that it is fairly difficult to move galaxies away from the initial
$j$--$M$ sequence.  
This conclusion is supported by $N$-body simulations of $\Lambda$CDM
halos, which find that the environmental dependencies of halo $\lambda$ are
fairly weak \citep{1988ApJ...330..519Z,1999MNRAS.302..111L,2005MNRAS.359.1537R}.

If instead the stripping occurs {\it after} the gas collapse, then $j$ and $M$
decrease for the DM but not for the baryons.  This leads to elevated
values of $f_j$ and $f_\star$, which could be investigated through observational
constraints on $M_{\rm vir}$ for field galaxies 
in comparison to satellite galaxies in massive groups.

\subsubsection{Biased collapse}\label{sec:bias}

There is another scenario that is functionally equivalent in the $j$--$M$ diagram
to outflow or stripping, but which merits special attention.
Here we consider a {\it spatially-biased} subcomponent of the initial gas 
which collapses and forms stars.  
Rather than our default assumption
of uniform efficiencies $f_\star$ and $f_j$ throughout the virial region,
we assume that stars form preferentially in the {\it inner regions} of the halo,
while the outer regions remain largely gaseous and form relatively few stars.

This scenario was introduced by \citet{2002ASPC..275..389F} and
is motivated by the higher densities, and thus overall gas dissipation rates
(through cooling and cloud collisions),
in the inner regions.
The consequent spatial bias in star formation can also be understood as a {\it temporal}
bias, if one considers an idealized onion-shell model wherein galaxies form
by inside-out collapse,
with virialization and star formation occurring first in the central regions
(cf.\ \citealt{1998ApJ...507..601V,1999ApJ...520...59K}).
Even in more realistic, hierarchical galaxy models, it is
uncontroversial that a large fraction of the baryons within a galaxy halo at any given
time will not yet have formed stars, and are
located preferentially at larger radii.
The stars observed in a galaxy at $z=0$ will have formed on average
at higher redshifts, and from gas that was more centrally confined than the 
$z=0$ virial volume.

Because $j$ for a $\Lambda$CDM halo is expected to increase systematically with
both internal radius and time, the above biasing scenario implies that $j_\star$ for
a galaxy will be lower than its total $j$ (including DM).
Such a biasing framework was used by \citet{2012MNRAS.424..502K} to connect
observed disk galaxies with simulated $\Lambda$CDM halos, and
thereby infer a radius of baryonic collapse.
Here we outline a generic toy model of collapse bias,
to understand its implications in the context of $j$--$M$ evolution vectors.

For simplicity, we adopt a step-function model where at an initial redshift $z_{\rm i}$,
all of the gas within the virial radius instantaneously collapses and forms stars
with perfect efficiency and angular momentum conservation ($f_\star=f_j=1$), and 
subsequently no star formation occurs ($f_\star=0$).
This scenario is illustrated by Figure~\ref{fig:schem3}(b), where $z_{\rm i}$
marks the initial halo parameters.  The leftward arrow shows the formation of the
stars, with \js--\Ms\ parameters that are preserved until $z_0=0$.
The diagonal arrow to the upper-right shows the subsequent evolution of the halo.
Because the halo continues to grow in $M$ and $j$, the net values of $f_\star$ and
$f_j$ for the stars will decrease with time, which is illustrated by the gray
arrows which are the inferences made by connecting the final conditions of the
halo and stars.

This biasing scenario might seem to provide a tidy alternative for understanding
galaxies that have {\it apparently} experienced baryonic angular momentum loss.
However, it is important to realize that such biasing cannot explain just any
arbitrary set of \js--\Ms\ observations.
For example, the vectors in Figure~\ref{fig:schem3}(b) were constructed to represent a
typical early-type galaxy with a net $f_\star=0.1$ at $z=0$, which turns out to
have a net $f_j=0.22$, i.e., not reproducing the apparent $\langle f_j\rangle\sim0.1$ 
from observations.
Note that this model had an initial $f_\star=1$, but in reality,
we expect an initial $f_\star < 1$, which would increase the discrepancy.
We will discuss this scenario further in Section~\ref{sec:eval};
for now, it serves as an important illustration of how
constructing physically-motivated vectors in the \js--\Ms\ diagram can provide
tight constraints on possible evolutionary scenarios.

\subsubsection{Mergers}\label{sec:merg}

We next consider galaxy {\it merging} following star formation,
which is likely to be more important for ellipticals than for spirals.
The mass of a galaxy increases through a merger, while
its final $j$ is determined by the vector sum of three initial $j$ components
(the internal $j$ for the two progenitor galaxies, and their relative orbital $j$),
as well as by any exchange of $j$ with the environment (e.g., between the stars
and their surrounding DM halos).
The random relative orientations of the first two components will cause them
to partially cancel out, which contributes a net {\it decrease} to $j$.
That is, after $N$ equal-mass mergers, there will be average trends for the remnant of
$J \propto N^{1/2}$ and $M \propto N$, and therefore $j \propto N^{-1/2}$
\citep{1979Natur.281..200F,1980ApJ...236...43A}.
The orbital $j$ and the $j$ exchange processes are more difficult
to model a priori.

The effects of mergers on DM halos have been studied extensively through
numerical simulations, resulting in a general picture where major mergers tend
to ``spin up'' the halos, while minor mergers and smooth accretion
tend to spin them down
(e.g., \citealt{2001ApJ...557..616G,2002MNRAS.329..423M,2002ApJ...581..799V,2004MNRAS.348..921P,2004ApJ...612L..13D,2006MNRAS.370.1905H}).
Given that the $j_{\rm vir}$--$M_{\rm vir}$ relation is scale-free and has a normalization
that is expected to change only gradually, if at all,
with time (e.g., \citealt{1997ApJ...478...13N}),
we conclude that for individual halos, the co-addition of the above processes must
amount to a random-walk that takes them on average {\it along} the $j_{\rm vir}$--$M_{\rm vir}$
sequence.

We illustrate this process in Figure~\ref{fig:schem3}(c) with a schematic evolutionary
vector for galaxy halos, broken down into subcomponents of $j_{\rm vir}$ and $M_{\rm vir}$
changes.\footnote{In the merging of DM halos, the resulting angular momentum and mass
are {\it not} the simple sum of those properties from the progenitors.
The combination of the two virial regions in a merger
increases the {\it density} within a fixed physical radius, but also increases the {\it volume}
of the virial region, so that more of the surrounding material falls under the gravitational
sway of the two galaxies together.
A 1:1 merger typically increases $M_{\rm vir}$ by a factor of $\sim$~2.3;
similar effects apply to $j_{\rm vir}$.}
Doubling the mass should typically increase \jv\ by a factor of $2^{2/3}=1.6$.

The effects of mergers on the stellar components of galaxies, which have collapsed
by large factors within their DM halos, are somewhat different.
Qualitatively speaking, it is a generic dynamical requirement that the stars
shed some of their orbital angular momentum, via tidal torques or dynamical friction,
in order to coalesce into a bound merger remnant
(e.g., \citealt{1985Natur.317..595F,1988ApJ...330..519Z,1988ApJ...331..699B,2006MNRAS.372.1525D}).

More quantitatively, we may make an initial, plausible guess that the ``final pass'' of the
merger before coalescence involves an impact parameter and relative velocity that are similar
to the stellar scalelength and circular velocity of the larger progenitor.
This would mean that the smaller progenitor would bring in an orbital $j_{\star,2}$ 
of a similar magnitude to internal $j_{\rm \star,1}$ of the larger progenitor
(i.e., $\Delta J_\star = j_{\star,2} \, M_{\star,2} \sim j_{\star,1} \, M_{\star, 2}$).

We sketch out some implications of this kind of merger evolution in Figure~\ref{fig:schem3}(c).
Starting with galaxy disks randomly selected along the median \js--\Ms\ trend as
in Figure~\ref{fig:schem2}(c) (adopting a simple $f_\star=0.1$ model
with scatter included for halo $\lambda$),
we apply a sequence of four mergers to each disk.
Each merger has a 1:1 mass ratio, and the relative vectors of internal \js\
and orbital \js\ are selected randomly
(this is similar in spirit to the orbital-merger model of 
\citealt{2002MNRAS.329..423M}).
The blue dots show the end result after the merger sequence, and the
upper arrow shows the median trend for a single galaxy, with black dots marking the
discrete merger events.
Note that at this point, the series of four 1:1 events is meant 
as a thought experiment and not necessarily as a likely merger history.

\begin{figure}
\centering{
\includegraphics[width=3.5in]{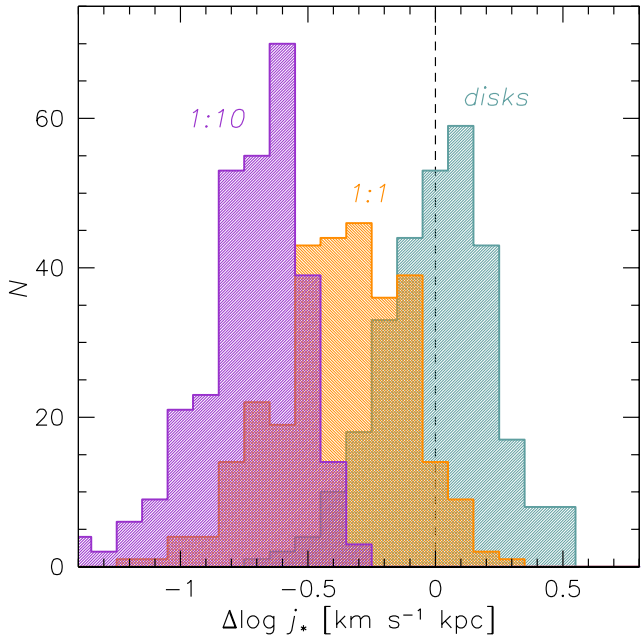} 
}
\caption{Distributions of specific angular momentum residuals, relative to
the mean trend for spiral disks, using the same analysis as in Figure~\ref{fig:schem3}(c).
The right histogram shows the disk initial conditions.
The middle and left histograms show merger remnants after having grown by 
a factor of 16 in mass, for 1:1 and 1:10 mergers, respectively.
The \js\ distribution has a smaller mean and dispersion for the
1:10 mergers than for the 1:1 mergers.
}
\label{fig:histdiff2}
\end{figure}

After an initial decrease of \js\ in the first merger from cancellation of the internal spin vectors,
the orbital \js\ dominates the evolution of the merger remnant
(e.g., \citealt{1980ApJ...236...43A,2006MNRAS.370.1905H};
this also means that the results hardly change if the ``accreted'' galaxies
are low-\js\ spheroids rather than disks as we have assumed here).
Because the orbital \js\ term is assumed to be similar to the disk \js--\Ms\ trend,
the final trend for the merger remnants parallels the disk trend, while being
offset to lower \js\ by a factor of $\sim$~2 ($\sim$~$-0.3$~dex).
Referring back to Figure~\ref{fig:schem2}, this corresponds to an effective
angular momentum loss term of $f_j \sim 0.5$.
The distribution of the offset is also shown by a histogram in Figure~\ref{fig:histdiff2}.

We have carried out the same exercise for a series of 1:10 mergers, with a median
trend shown by the lower vector in Figure~\ref{fig:schem3}(c).  The result is
similar to the 1:1 case, with orbital \js\ dominating the evolution after the
galaxy grows in mass by a factor of $\sim$~2.
However, the final \js\ trend is now lower
than the disks by a factor of $\sim$~6 ($\sim$~$-0.8$~dex; $f_j \sim$~0.15),
with less scatter than in the 1:1 case
(see  Figure~\ref{fig:histdiff2} again).
These differences arise because there is less stochasticity with the 1:10 mergers, where
random walk effects tend both to wash out variations and
to dilute the orbital contributions to \js.\footnote{This scenario has
some parallels to discussions in the literature about the systematic relations
between angular momentum and merger histories, and the implications for
the observed properties of galaxies 
(e.g., \citealt{2004ApJ...612L..13D,2002ApJ...581..799V,2005NewAR..49...25P,2007MNRAS.380L..58D,2012ApJ...750..107S}).
However, those studies did not always make a clear distinction between
the differing merger dynamics of DM halos and of their embedded stellar components.}
A more realistic mixture of multiple mergers with varying mass ratios
would presumably produce a \js\ distribution with a peak intermediate to our 1:1 and
1:10 scenarios, and with a larger scatter.

These calculations are laden with simplifying assumptions and could easily be
wrong by a factor of 2 in \js.  However, they are meant to illustrate some
possible implications of merger activity in a hierarchical context.
First of all, it is plausible that spheroids with a merger origin
would follow a \js--\Ms\ relation that is parallel to that of spiral disks,
but offset to lower \js\ by a factor of a few.\footnote{More generally,
a similar slope would presumably be driven by any merger history that
involves a scale-free mass spectrum of progenitors.
This is a basic property of $\Lambda$CDM halos, but is incorrect at
some level for stellar galaxies, owing to the strong break in their luminosity
function.}
Second, the scatter in \js\ introduced by random merging may be relatively small.

These two results in our toy model are both driven by the dominant
contributions of orbital \js.
Similar points were made by \citet{1979Natur.281..200F} and by \citet{1988ApJ...330..519Z}, 
in the latter case based on the prediction
that $\lambda$ would be fairly constant with radius inside DM halos.
The stars that condense at
the center of a halo, and then participate collisionlessly in its merger history,
would naturally follow the same $j$--$M$ scaling relations as the overall halos,
modulo a smaller scale-length in converting from $\lambda$ to $j$
(in Equation~(\ref{eqn:lambda}), $|E|$ is inversely proportional to the radius).

\subsubsection{Evaluating the possibilities}\label{sec:eval}

We now step back and consider how well the preceding evolutionary scenarios
(outflows, stripping, collapse bias, and mergers) 
mesh with the observational constraints
(Figures~\ref{fig:JMM0} and \ref{fig:fvsf}).  
The idea is to find a vector (or combination of vectors) that
connects up the well-established endpoints in the $j$--$M$ diagram:
the $\Lambda$CDM halo initial conditions and the $z=0$ galaxy observations.
It should however be remembered that the focus of this paper is not to
solve long-standing questions about galaxy evolution which may require
a detailed understanding of the physics involved.
Instead, our more modest goals are
to illustrate how the $j$--$M$ diagram can be used
in practical terms as a constraint on theory,
while looking for any hints as to the viability of various scenarios.

Recent work in numerical simulations of {\it disk} galaxy
formation has emphasized how outflows might remove low-$j_{\rm g}$ material,
which counteracts $j$-loss through tidal torques during galaxy collapse,
and maintains a high net level of $f_j$
(e.g., \citealt{2011MNRAS.415.1051B,2011ApJ...742...76G}).
We could then imagine that the differences between spiral and elliptical galaxies
originate from the spirals having much stronger outflows at early times.

This outflow scenario implies more mass loss in spirals and so
would initially seem to work the wrong way in explaining the $f_\star$ differences---but
there could be other factors besides gas-depletion that affect $f_\star$.
It is beyond the scope of this paper to explore this
scenario in detail, but we emphasize that the focus on reproducing $f_j$
and $f_\star$ for spirals needs to expand to include simultaneously
the constraints from ellipticals, beyond these being nuisance factors that
represent failed disks.

We have already discussed how stripping before baryonic collapse is not
expected to produce large changes in the observable \js--\Ms\ relations,
which may indeed be part of the reason that there is not more scatter
in these relations.\footnote{There is one case where
severe stripping has apparently led to a large reduction in \js:
NGC~4486B, which is a low-\js\ outlier in Figure~\ref{fig:JMM0}, and is discussed
in \citet{2012ApJ...748...29R}. This ``compact elliptical'' is a fairly rare
type of galaxy.}
There is also 
a more obvious constraint that both spirals and
ellipticals exist in the field and in clusters,
so present-day environment cannot be the unique driver of morphology and $j$ evolution.

Collapse bias is an appealing possibility because it would provide a natural explanation
for the positive correlation between $f_\star$ and $f_j$ as in Figure~\ref{fig:fvsf}.
In this scenario, elliptical galaxies would cease to build up both \Ms\ and \js\
at relatively early times, with
the remaining baryonic $M$ and $j$ at late times
either residing in a hot gas halo or having been blown out into intergalactic space.
Spiral galaxies would have more protracted star formation histories that
increase \Ms\ and \js\ monotonically with time.

Besides explaining the relative positions of ellipticals and spirals in the \js--\Ms\
diagram, this scenario also fits in naturally with the observation that the stars in spirals
are on average much younger than those in ellipticals.
There may be additional implications if one connects the {\it baryon} collapse to
the {\it overall halo} collapse, which has a well-understood theoretical underpinning.
At a given $z=0$ mass, some halos should have collapsed earlier than others,
leading to their DM distributions being more centrally concentrated.
Given a fixed $\lambda$, the central DM and associated stars would then have
relatively low $j$ values.  Since halo collapse time is correlated strongly with
environmental density, one would then expect the low-\js\ galaxies to reside preferentially
in high-density environments -- which is indeed what is found observationally
(through the traditional morphology-density relation).

A potential problem with this scenario is that it does not appear by itself
to be capable of explaining the apparent deficit of \js\ in ellipticals, as discussed in 
Section~\ref{sec:bias}.
More detailed analysis would be needed to see if halo concentration makes a difference,
and to understand the baryonic physics of 
why early-collapsing galaxies would also shut down
their star formation more drastically than late collapsers.
In addition to collapse bias, other effects may also need to be involved,
such as a bias to low spin for their halos, or a component of real $j$ loss.

The merger scenario is a common explanation for ellipticals, since it
accounts for spheroidal morphologies through violent relaxation
\citep{1977egsp.conf..401T}, and because there is strong observational evidence
for some elliptical galaxies actively forming through mergers
(e.g., \citealt{2006AJ....132..976R}).
Our toy model analysis suggests that the overall effect of mergers is
to {\it reduce} the \js\ of the remnant relative to an initial \js--\Ms\ trend for disks,
while the combination of {\it multiple} mergers may move the remnants parallel to that trend
(Figure~\ref{fig:schem3}(c)).  This might provide a natural explanation for the
observed \js--\Ms\ trend for ellipticals: the slope, scatter, and offset relative to disks.
Note that it is not entirely clear in this context
why the spiral bulges and the ellipticals would follow the same \js--\Ms\ trends.

A more quantitative comparison of our model to the observations allows us not only
to constrain the typical mass ratios in mergers (as Figure~\ref{fig:histdiff2}),
but also to infer the amount of mass growth in ellipticals since their
assumed primordial disk phase.
We do so by mapping our toy model vectors for mergers in the key $f_j$--$f_\star$
diagram (Figure~\ref{fig:fvsf}), starting from initial conditions similar to
present-day spirals ($f_\star=0.25,f_j=0.6$),
and requiring that they terminate at $(f_\star=0.1,f_j=0.1$).

Recalling that $M_{\rm vir}$ growth slightly outpaces \Ms\ growth
we find that reducing $f_\star$ by a factor of 2.5 requires a very long series
of mergers, with a final growth factor of $\sim$~100 in \Ms\ and $\sim$~300 in 
$M_{\rm vir}$.
Consideration of the $f_j$ constraint then 
suggests a typical merger mass ratio of $\sim$~1:3.
Such ``major mergers'' seem like a reasonable pathway to forming
elliptical galaxies, although recent work suggests a more
dominant role for {\it minor} mergers (e.g., $\sim$~1:10;
\citealt{2009ApJ...699L.178N,2009ApJ...697.1290B,2011MNRAS.417..845K,2012ApJ...744...63O,2012arXiv1202.3441J}),
which is motivated in part by explaining trends in size evolution, and
is also supported by the observed {\it shapes} of rotation-velocity profiles
(see Section~\ref{sec:obsresults} and
\citealt{2011ApJ...736L..26A}).\footnote{In more detail, the fast- and slow-rotator
subcategories of ellipticals (Section~\ref{sec:etgdata})
are often thought to originate in different merger histories,
such as binary versus multiple mergers 
(e.g., \citealt{2008ApJ...685..897B,2011MNRAS.416.1654B}).
Our discussion concerns primarily the fast-rotators, since these represent the vast
majority of ellipticals, and in addition, our \js\ constraints for the slow-rotators 
are less certain.
However, as discussed in Sections~\ref{sec:obsresults} and \ref{sec:replace},
we detect no systematic difference in \js--\Ms\ space between the two galaxy types,
suggesting that they may have relatively similar merger histories after all.
}
This apparent tension is not of great concern since our current results 
involve significant observational uncertainties and a crude model for the merging
vectors in Figure~\ref{fig:schem3}(c), 
while not taking proper account of the redshift-dependence of virial quantities.
However, they are intended to illustrate conceptually the kinds of constraints that are possible with more careful modeling.

A merger scenario may successfully explain the \js--\Ms\ properties
of ellipticals, but it should be remembered that in a cosmological context, all
galaxies including spirals should experience a continuous rain of accreting objects.
Even if spiral galaxies have systematically avoided the most extreme merger events,
they will have still experienced events in the $\sim$~1:10 range
(e.g., \citealt{1993MNRAS.261..921K,2008ApJ...683..597S,2010MNRAS.406.2267F}), 
which as shown in our toy models could significantly reduce \js.
A more detailed analysis of \js--\Ms\ evolution within a cosmological framework is
needed in order to investigate the quantitative differences that might arise between
spirals and ellipticals owing to varying merger histories.
In particular, an explanation for the observed bulge--disk \js\ bimodality is needed, 
since a spectrum of merger histories is more suggestive of a smooth distribution
of \js.
It should also be kept in mind that $\langle f_\star(M_\star)\rangle$ is observationally
constrained not only for present-day galaxies, but also at earlier
times  (e.g., \citealt{2009ApJ...696..620C,2012arXiv1205.5807M}), 
which introduces additional
``boundary conditions'' to $j$--$M$ evolution.

Synthesizing the scenarios above, it seems plausible that ellipticals might be explained
through a combination of collapse bias and multiple mergers--which
bears a notable resemblance to recent discussions of two-phase
galaxy formation \citep{2010ApJ...725.2312O}.  In this context,
an early burst of star formation would both imprint a relatively low initial \js,
and allow more opportunity for subsequent mergers to reduce \js\ further.
Spirals would be those systems where late gas infall both brings in higher $j$,
and avoids the most active merging period. 

There are of course other considerations besides angular momentum when
constructing models of galaxy evolution, which are beyond the scope
of this paper to evaluate.
We have also been able to cover only a subset of possible scenarios.

One significant omission is the disk-instability pathway for bulge formation
(e.g., \citealt{1964ApJ...139.1217T,1997ApJ...482..659D,1998ApJ...507..601V,2009MNRAS.396.1972P}), which is an internal
process where the bulge and disk either form from high- and low-$j$ material,
or else exchange $j$ through gravitational torques.
While this pathway is usually considered in connection with pseudo-bulges,
there are recent proposals that the special conditions in high-redshift
galaxy disks can lead to the massive, classical bulges of present-day spirals, lenticulars,
and ellipticals 
(e.g., \citealt{1999ApJ...514...77N,2004ApJ...611...20I,2008ApJ...688...67E,2009Natur.457..451D,2009ApJ...703..785D,2010MNRAS.404.2151C}).
The filamentary nature of mass and $j$ inflows at high redshift may also 
require significant revisions to standard spherical models
\citep{2012MNRAS.422.1732D,2012MNRAS.423.1544S,2012MNRAS.423.3616D,2011arXiv1106.0538K}.

Our overarching emphasis here is that whatever the mechanisms for galaxy formation,
they must reproduce the basic \js--\Ms\ scaling relations {\it observed for both
spiral and elliptical galaxies}.  A combination of all the processes mentioned above,
and more, could be operational in real galaxies, where each process must be associated
with a vector of \js--\Ms\ evolution that is not arbitrary but physically-motivated, 
as we have sketched in Figures~\ref{fig:fvsf}
and \ref{fig:schem2}.  
The sum of these vectors over the lifetime of the galaxy
must preserve the halo-like
scaling relations, {\it along with a relatively small scatter}.
These may be very challenging constraints to match in practice,
particularly if one includes boundary conditions
on $f_\star(M_\star)$ evolution with redshift,
and requires that the \js--\Ms\ relations hold for both bulge and disk
components simultaneously within the same galaxies.

Thus, a fresh approach to $j$--$M$ analysis appears to hold promise
for providing new, powerful constraints on galaxy evolution.
We would encourage numerical simulators to keep this
approach in mind as part of their toolkit, 
tracking the evolution of their simulated
galaxies in the $j$--$M$ diagram, 
while refining our schematic estimates of $\Delta j$--$\Delta M$ vectors,
and thereby gaining more insights
into the underlying physical processes in the simulations.

\section{Summary and conclusions}\label{sec:concl}

We have revisited the pioneering study of F83 which derived observational
estimates for the fundamental quantities \Ms\ and \js\ (stellar mass
and specific angular momentum) of spiral and elliptical galaxies, 
and compared these to theoretical expectations based on hierarchical assembly.
Although the amount and distribution of \js\ in late-type galaxies 
has been an intensively-studied
topic in the intervening years, even the most basic trends for early-types
have not been satisfactorily established.
We have capitalized on the advent of radially-extended kinematic data for a large sample
of early-type galaxies, to update and extend the analyses of F83.

We focus first on detailed analysis of a small sample of galaxies with data
extending to typically five effective radii, which is the distance one must reach
for a high degree of confidence in the \js\ estimates.
We derive various formulae for use in quantifying \js\ for pressure supported
systems, including deprojection effects.
In order to estimate \js\ for a larger sample of galaxies without 
requiring detailed modeling and data to very large radii, we test
a simple, heuristic \js-estimator.

Based on the shapes of observed rotation-velocity profiles for
the detailed sample of galaxies, we find that a convenient metric for
the characteristic rotation velocity $v_{\rm s}$ of a galaxy is provided by the observed
rotation at a semi-major-axis distance of two effective radii.
This approximation is accurate at the level of
$\sim$~0.1~dex, which is suitable for studying galaxy-to-galaxy variations in \js.

We next assemble a large sample of galaxies in the nearby universe with
adequate photometric and kinematic data for estimating \js\ and \Ms.
This sample covers the full spectrum of bright galaxy types from bulgeless-spiral to
diskless elliptical, as well as a wide range in \Ms,
centered approximately at the characteristic mass $M_\star^*$.
We use our simple formula for estimating \js, while adopting simple bulge+disk
models for the spiral galaxies.

Along the way, we also introduce an important new observational scaling relation for
galaxies of all types: $v_{\rm s}$ versus \Ms.  This relation is analogous
to the well-known Tully-Fisher relation for disk galaxies, but is more closely related to
angular momentum than to dynamical mass.
Unlike the generalized Tully-Fisher relation, the mass--rotation~velocity relation shows 
{\it near-perpendicular} rather than parallel trends for spiral and elliptical galaxies.
These rotation-velocity trends combine with size--mass trends to
trace the more fundamental \js--\Ms\ trends.

Our combined \js--\Ms\ estimates confirm the basic result of F83 that late-type spiral and
elliptical galaxies follow parallel sequences of roughly $\alpha \sim 2/3$ log-slope,
but with a large zeropoint difference (in our analysis,
the ellipticals have a factor of $\sim$~3--4 lower \js\ at a fixed \Ms).
Although this conclusion has already been used in some theoretical analyses,
now it has a much firmer observational basis.
In particular, the data do not support 
previous suggestions that major mergers have transported large
amounts of angular momentum into the outer regions of ellipticals.

We confirm for the first time that lenticular galaxies on average
lie intermediate to ellipticals and late-type spirals in the \js--\Ms\ plane,
with tentative indications for two families of lenticulars 
characterized by low and high \js.
We see no indication of systematic, overall
differences between centrally fast- and slow-rotator ellipticals.
We also find that spiral bulges
are consistent with following the \js--\Ms\ sequence for ellipticals, despite
having very different relations between mass, size, and rotation.
Thus, as far as the fundamental parameters \js\ and \Ms\ are concerned,
spiral bulges are essentially like mini-ellipticals.

We examine the residuals of the combined galaxy \js--\Ms\ data with respect to the 
disk-only trend, and find that these correlate better with disk-to-bulge ratio 
than with Hubble type.
They also deviate from a lognormal distribution, possibly suggesting instead
a bimodality in \js.  Considering all of these results together, we propose an
alternative framework to the Hubble sequence, based on more physically motivated parameters.
In this picture, all galaxies are a combination of a bulge and a disk, which are
distinct subcomponents with different characteristic amounts of \js.
Galaxy morphology may then be seen as a secondary manifestation of the
mix of high- and low-$j$ material, or equivalently, the position of a galaxy in 
\js--\Ms\ parameter space is a reflection of its bulge-to-disk ratio.

We next connect our observational results to a theoretical framework based on
the hierarchical assembly of galaxy halos in a $\Lambda$CDM cosmology.
We use numerically-informed analytic methods that are much simpler than
hydrodynamical simulations,
but less susceptible to the large, lingering uncertainties about 
baryonic recipes, resolution effects, and other numerical issues.
We find that the predictions for universal mean values of halo spin translate into 
$j_{\rm vir}$--$M_{\rm vir}$ relations with an $\alpha=2/3$ log-slope, which
is remarkably similar to the observed \js--\Ms\ relations.
The zeropoint differences among these relations provide valuable clues
to the formation processes of different galaxy types.

Mapping between halo and stellar quantities involves two basic parameters:
the net fraction of baryons turned into stars, $f_\star$, and the fraction of specific $j$
retained, $f_j$.  We find that realistic variations of $f_\star$ with mass
produce surprisingly mild deviations of the \js--\Ms\ relation from a simple $\alpha=2/3$
power-law.  The most noticeable correction is a slightly shallower predicted slope for
the spirals, which turns out to agree well with the observations.

We explore two simplified alternative scenarios for explaining the spiral-elliptical 
dichotomy in the \js--\Ms\ plane:  the formation of spiral and elliptical
galaxies in low- and high-spin halos, respectively (spin-bias scenario); 
and a difference in $j$ retention (variable-$f_j$ scenario).
We find that spin-bias does not explain the tails of the observed
\js\ distribution, nor does it agree with the observed trend as a function of mass
for the elliptical galaxies.
The variable-$f_j$ scenario, on the other hand, matches the data well and
suggests universal values of $f_j\sim0.55$ and $f_j\sim0.1$ for spirals and ellipticals,
or for disks and bulges, respectively.
The near-constancy of these values is intriguing, 
and means that 
all the complexities of galaxy evolution 
somehow effectively reduce to a simple model, where
galactic stars have
preserved the ``initial'' conditions of their host halos, including
the $j_{\rm vir}$--$M_{\rm vir}$ slope and scatter.
This interpretation may be useful for semi-analytically
populating DM halos with both spiral and elliptical galaxies
(cf.\ \citealt{1998MNRAS.295..319M}).

Our $f_j$ result for spirals confirms similar conclusions going back decades,
that these galaxies have retained most of their primordial specific angular momentum. 
This is an unavoidable conclusion from basic comparisons between observational constraints
and theoretical expectations, which for many years presented a major challenge
to numerical simulations of galaxy formation, as these apparently predicted very low values 
for $f_j$.  It has gradually been realized that such simulations 
overpredicted
angular momentum transport (e.g., \citealt{2011arXiv1109.4638K}), 
with major uncertainties still lingering in the baryonic physics included in the
simulations.

Our consolidation of the $f_j$ result for elliptical galaxies provides a new benchmark
for models of galaxy evolution, which to be fully credible must
reproduce the observed $f_j$ (and $f_\star$) trends for both spirals and ellipticals 
{\it in the same simulation}.  For example, any feedback processes that are invoked
to prevent overcooling and $j$ loss in spirals should be much less effective for
ellipticals.

We have explored a few toy models for galaxy evolution that exploit the basic
constraints provided by the parameter space of \js\ and \Ms, or equivalently of
$f_j$ and $f_\star$.  Galaxies cannot evolve arbitrarily in this parameter space,
requiring physically-motivated diagonal vectors of change 
($\Delta j_\star, \Delta M_\star$).
Thus we suggest the $j$--$M$ diagram as a key tool for assessing and interpreting
any model of galaxy formation.

Our simplified models suggest that
a combination of early baryonic collapse and
repeated galaxy merging (major or minor)
might account for the parallel-but-offset
trend of ellipticals relative to spirals.  We have provided illustrative constraints
on the numbers and mass-ratios of the mergers, which after refinement with more
detailed modeling could be compared with cosmologically-based
predictions for mass growth and merging.

In summary, we have established a new synthesis of \js--\Ms\ trends from observations,
whose general resemblance to halos in $\Lambda$CDM cosmology
provides important support for that theory, and in turn 
furnishes a valuable framework for
constraining baryonic processes as discussed above.  
Our course, the observations presented here must be relevant to any model
of galaxy formation, even if $\Lambda$CDM theory eventually needs revision.
More generally, we propose that the morphologies of galaxies are
closely tied to the evolution of angular momentum during their assembly,
with late-types being very efficient at retaining $j$, and
early-types proficient at shedding $j$.

In this context, there are several areas ripe for additional progress.
First, clear predictions from high-resolution cosmological simulations
are needed for \js\ versus \Ms\ as a function of galaxy type, to 
explore whether the dichotomy between spirals and ellipticals, or
disks and bulges, can be explained by differences in their assembly histories.
Second, the observational work on nearby galaxies should be extended to the
next level, via a volume-limited, homogeneous survey of all types of galaxies
including full two-dimensional, wide-field photometric-kinematic bulge--disk
decompositions, 
with attention to stellar mass-to-light ratio variations.
This would allow for more robust analysis of the deviations of the \js\ residuals
from lognormality, and for more secure treatments of S0/Sa galaxies and of
bulge and disk subcomponents.
Third, the extensions of our study to lower-mass
(e.g., \citealt{2012ApJS..198....2K})
as well as to higher-redshift galaxies
(cf.\ \citealt{2007A&A...466...83P,2010ApJ...725.2324B}),
to freshly accreted material within galaxies
(cf.\ \citealt{2011ApJ...738...39S}), 
and to the orientations of the ${\bf j}_\star$ vectors
(e.g., \citealt{2012arXiv1201.5794C,2012arXiv1207.0068T}),
would provide additional, valuable diagnostics.

\vspace{0.3cm}

\noindent
\acknowledgements
We thank Brad Whitmore for assistance with the spiral galaxy data compilation,
and the referee for a constructive review.
We thank Frank van den Bosch, Andi Burkert, Roger Davies, Aaron Dutton, George Efstathiou,
Eric Emsellem, Ken Freeman, Marcel Haas, Phil Hopkins, Koen Kuijken, Surhud More,
Joel Primack, and Mike Williams for helpful comments and discussions.
This research was supported in part by the National Science Foundation under
Grants No. AST-0507729, AST-0808099, AST-0909237, 
and PHY05-51164.
This research has made use of the NASA/IPAC Extragalactic Database (NED) which is operated by the Jet Propulsion Laboratory, California Institute of Technology, under contract with the National Aeronautics and Space Administration. 
This publication makes use of data products from the Two Micron All Sky Survey, which is a joint project of the University of Massachusetts and the Infrared Processing and Analysis Center/California Institute of Technology, funded by the National Aeronautics and Space Administration and the National Science Foundation.

\bigskip

\noindent
\vspace{0.2cm}\\

\newpage

\appendix

\section{Appendix A: Angular momentum formulae}\label{sec:form}

\subsection{A.1. General formulae}

We begin with the description of a galaxy as the six-dimensional
phase-space distribution function of its particles (gas and stars)  $f({\bf r},{\bf v})$, 
where {\bf r} and {\bf v}
are the (vector) position and velocity coordinates relative to the galactic center,
and $f$ is normalized to unity when integrated over all positions and velocities.
Given a total mass $M$, the three-dimensional spatial mass density at position {\bf r} is
\begin{equation}
\rho({\bf r}) = M \int f({\bf r},{\bf v}) d^3{\bf v} ,
\end{equation}
and the mean velocity at that position is
\begin{equation}
\bar{\bf v}({\bf r}) = \frac{M}{\rho({\bf r})}\int {\bf v} \, f({\bf r},{\bf v}) \, d^3{\bf v} .
\end{equation}
The true (i.e., not projected) specific angular momentum is then
\begin{equation}\label{eqn:A4}
{\bf j}_{\rm t} \equiv \frac{{\bf J}_{\rm t}}{M} = \frac{\int {\bf r}\times \bar{\bf v}({\bf r}) \, \rho({\bf r}) \, d^3{\bf r}}{\int \rho({\bf r}) \, d^3{\bf r}} .
\end{equation}

Given the loss of information in observed galaxies (one positional dimension lost in
projection, and two velocity dimensions usually unmeasurable as proper motions),
one must adopt some simplifying assumptions in order to recover ${\bf j}_{\rm t}$
from observations.
Our main assumptions here are that galaxies are transparent, 
have {\it axisymmetric} density distributions, and
{\it rotate on cylinders} that are aligned with the symmetry axis of the density
(with no other net velocity component such as expansion or contraction).

Adopting cylindrical galactic coordinates $(R,z,\phi)$, our modeling assumptions imply that $\bar{\bf v}({\bf r})$
becomes a simple rotation-velocity profile $v_{\rm rot,t}(R) \, \hat{\mbox{\boldmath$\phi$}}$,
independent of $z$ and $\phi$ 
(where $\hat{\mbox{\boldmath$\phi$}}$ is the unit vector in the azimuthal direction), 
and that the density $\rho(R,z)$ is independent of $\phi$.
Equation~(\ref{eqn:A4}) then reduces to a one-dimensional integral:
\begin{equation}\label{eqn:A5}
j_{\rm t} = \frac{\int v_{\rm rot,t}(R) \, \Sigma(R) \, R^2 \, dR}{\int \Sigma(R) \, R \, dR} ,
\end{equation}
where
\begin{equation}\label{eqn:Sigma}
\Sigma(R) = \int \rho(R,z) \, dz
\end{equation}
is the surface mass density  when the galaxy is viewed pole-on.
With these assumptions, all
galaxies with the same $v_{\rm rot,t}(R)$ rotational profile and the same pole-on $\Sigma(R)$
have the same ${\bf j}_{\rm t}$, e.g., whether they are thin disks or
round spheroids.

Even under these fairly restrictive assumptions,
recovering the true angular momentum of an observed galaxy is
a difficult inverse problem.
Fortunately, there is a way to structure the problem that makes it
conceptually and computationally simpler.
We separate the calculation for \jt\ into two factors:
\begin{equation}\label{eqn:jCp}
j_{\rm t} = C_i \, j_{\rm p} .
\end{equation}
Here the second factor on the right is the analogue of Equation~(\ref{eqn:A5})
constructed {\it purely from observations} along the projected semi-major axis $x$:
\begin{equation}\label{eqn:jest}
j_{\rm p} \equiv \frac{\int v_{\rm rot,p}(x) \, \Sigma(x) \, x^2 \, dx}{\int \Sigma(x) \, x \, dx} .
\end{equation}
Note that this ``projected specific angular momentum'' is {\it not} literally the projection
of ${\bf j}_t$ on the plane of the sky 
(which we will discuss at the end of this subsection).
Also, the two denominators in Equations~(\ref{eqn:A5}) and (\ref{eqn:jest})
are closely related.  In the case of spherical symmetry for the density, 
they are identical, and
for now we will adopt this assumption for simplicity, returning to the more
general axisymmetric case later.

The first factor in Equation (\ref{eqn:jCp}), $C_i$, is a numerical coefficient
incorporating all of the additional deprojection effects that
depend on inclination and on the shapes (but not the amplitudes) of
the surface density profile and of the rotation-velocity curve.
Substituting Equations~(\ref{eqn:A5}) and (\ref{eqn:jest}) into Equation~(\ref{eqn:jCp}),
we have
\begin{equation}\label{eqn:Cj0}
C_i = \frac{\int v_{\rm rot,t}(R) \, \Sigma(R) \, R^2 \, dR}{\int v_{\rm rot,p}(x) \, \Sigma(x) \, x^2 \, dx} .
\end{equation}

Before manipulating Equation~(\ref{eqn:Cj0}) further, we point out that
the advantage of this formulation of the angular momentum problem is that the complicated expression $C_i$ 
need not be evaluated for every individual galaxy---provided that it is not
sensitive to the details of the density and rotation-velocity profiles, and
instead depends primarily on the inclination.
Using simple models below, we will verify that this is the case, so that we can
treat $C_i$ as a numerical coefficient (calibrated
by models) that we combine with the observables in Equation~(\ref{eqn:jest}),
and thereby reconstruct \jt\ using Equation~(\ref{eqn:jCp}).

\begin{figure*}
\centering{
\includegraphics[width=5in]{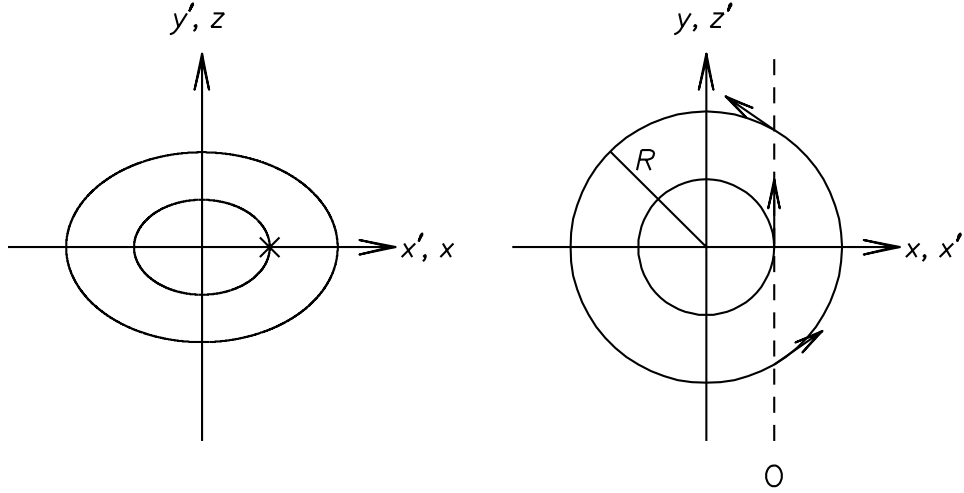} 
}
\caption{Geometry and coordinate systems for the calculation of the
deprojection factor $C_i$.
Here the galaxy is spheroidal and axisymmetric, with an inclination of $i=\pi/2$
(edge-on).  The left-hand diagram shows the observer reference frame 
$(x^\prime,y^\prime,z^\prime$), while the right-hand diagram shows the galaxy
reference frame $(x,y,z)$.
An example line-of-sight observation is illustrated by the $\times$ symbol on the left, and by
the dashed line on the right leading to the O symbol.
The arrows tangential to the circles show sample rotation-velocity vectors.
More general inclinations are similar, but with the line-of-sight components of
the rotation reduced by factors of $\sin i$.
}
\label{fig:coords}
\end{figure*}

Next, to derive a more specific expression for $C_i$, we begin by
considering a galaxy observed from an arbitrary direction, which we characterize
by the usual inclination $i$, the angle between the rotation axis $z$ and the line-of-sight.
The observer's reference frame is ($x^\prime,y^\prime,z^\prime$), 
where $z^\prime$ is the line-of-sight (measured away from the observer),
and $x^\prime$ is aligned with $x$ for convenience (see Figure~\ref{fig:coords}).
Since we are adopting an approach of modeling rotation-velocity profiles along
the observed semi-major axis, we simplify the problem by doing calculations only
for the case $y^\prime=0$.
The coordinate transformations between galaxy and observer frames are then
\begin{equation}\label{eqn:xx}
x=x^\prime
\end{equation}
\begin{equation}\label{eqn:yy}
y = - z^\prime \sin i
\end{equation}
\begin{equation}\label{eqn:zz}
z = z^\prime \cos i
\end{equation}
(see \citealt{1997MNRAS.287...35R} for more general expressions).
The azimuthal unit vector in the galaxy frame can be expressed as
\begin{equation}
\hat{\mbox{\boldmath$\phi$}} = \sin \phi \, \hat{\bf x} + \cos \phi \, \hat{\bf y} ,
\end{equation}
which after using unit-vector relations analogous to Equations~(\ref{eqn:xx}) and (\ref{eqn:yy}), becomes
\begin{equation}\label{eqn:phip}
\hat{\mbox{\boldmath$\phi$}} = \sin \phi \, \hat{\bf x}^\prime - \cos \phi \, \sin i \, \hat{\bf z}^\prime .
\end{equation}

For a given parcel of material,
the observer can measure only the projection of the mean 
velocity vector onto the line-of-sight, which we express as a dot product:
\begin{equation}\label{eqn:vp}
v_{\rm p} = \bar{\bf v}({\bf r}) \cdot \hat{\bf z}^\prime =  v_{\rm rot,t}(R) \, \hat{\mbox{\boldmath$\phi$}}\cdot\hat{\bf z}^\prime .
\end{equation}
From Equation~(\ref{eqn:phip}), we then find
\begin{equation}\label{eqn:vp2}
v_{\rm p} = - \, v_{\rm rot,t}(R) \, \cos \phi \, \sin i  = 
- \, v_{\rm rot,t}(R) \, \frac{x}{R} \, \sin i  ,
\end{equation}
where we use
Equations~(\ref{eqn:xx}) and (\ref{eqn:yy}) to make the substitution
$R \equiv \left(x^2+{z^\prime}^2\sin^2 i\right)^{1/2}$.
We thus define a projected profile of rotation velocity along the semi-major axis:
\begin{equation}
v_{\rm rot,p}(x) \equiv - \, v_{\rm p}(x) ,
\end{equation}
which we calculate by integrating Equation~(\ref{eqn:vp2})
along the line-of-sight $z^\prime$ while weighting by the density $\rho({\bf r})$.
Recalling also that for now, we are assuming spherical symmetry for the density,
we find the projected rotation-velocity profile
\begin{equation}\label{eqn:vrotp2}
v_{\rm rot,p}(x) = \frac{x \sin i \int \rho(r) \, v_{\rm rot,t}(R) \, R^{-1} dz^\prime}{\int \rho(r) \, dz^\prime} ,
\end{equation}
where we can also substitute $r=\left(x^2+{z^\prime}^2\right)^{1/2}$.
The denominator is the surface density [Equation~(\ref{eqn:Sigma})], and we have
\begin{equation}\label{eqn:vrotp1}
v_{\rm rot,p}(x) = \frac{x \sin i}{\Sigma(x)} \int \rho(r) \, v_{\rm rot,t}(R) \frac{dz^\prime}{R} .
\end{equation}

\begin{figure*}
\centering{
\includegraphics[width=3.5in]{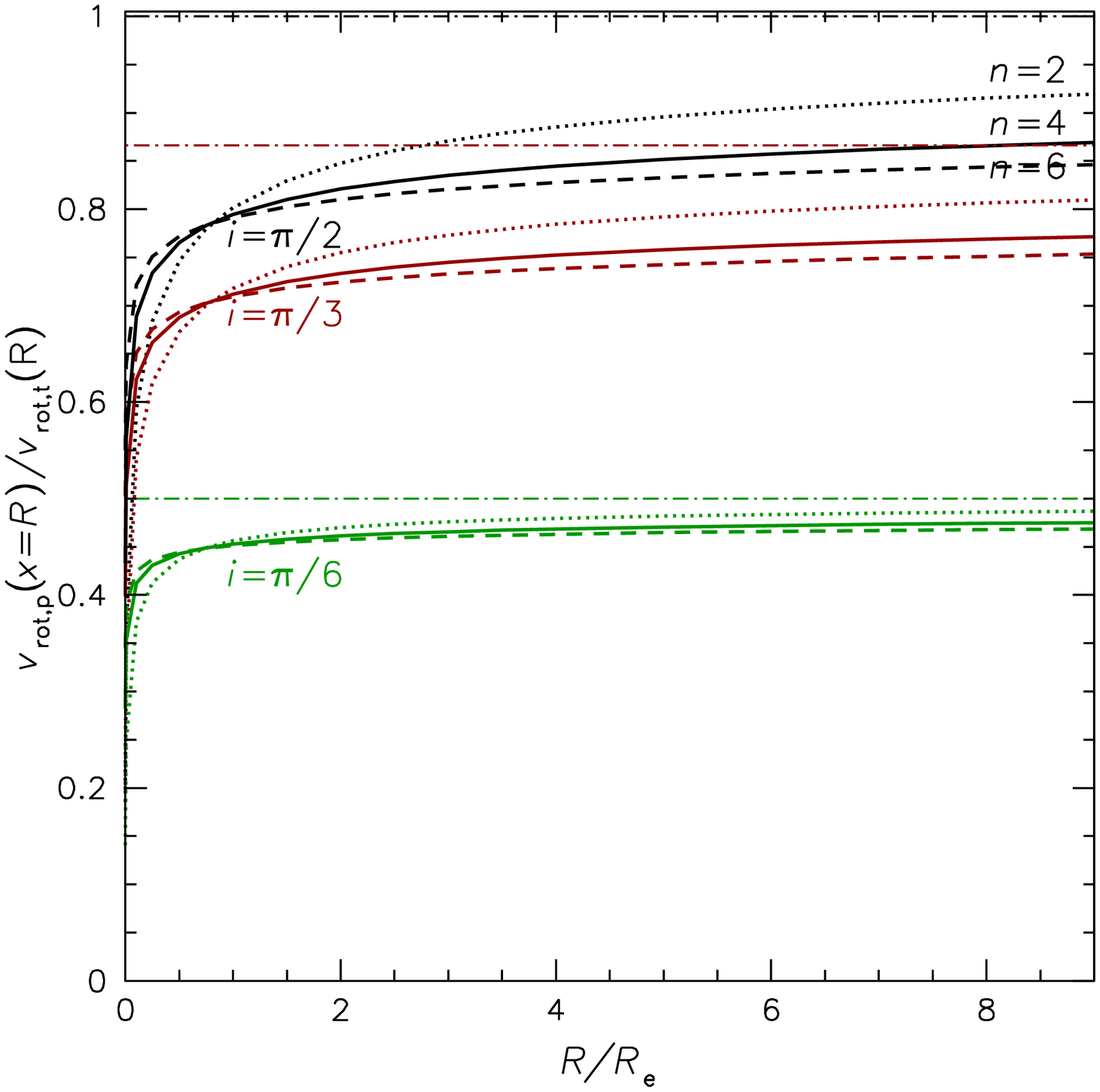} 
\includegraphics[width=3.5in]{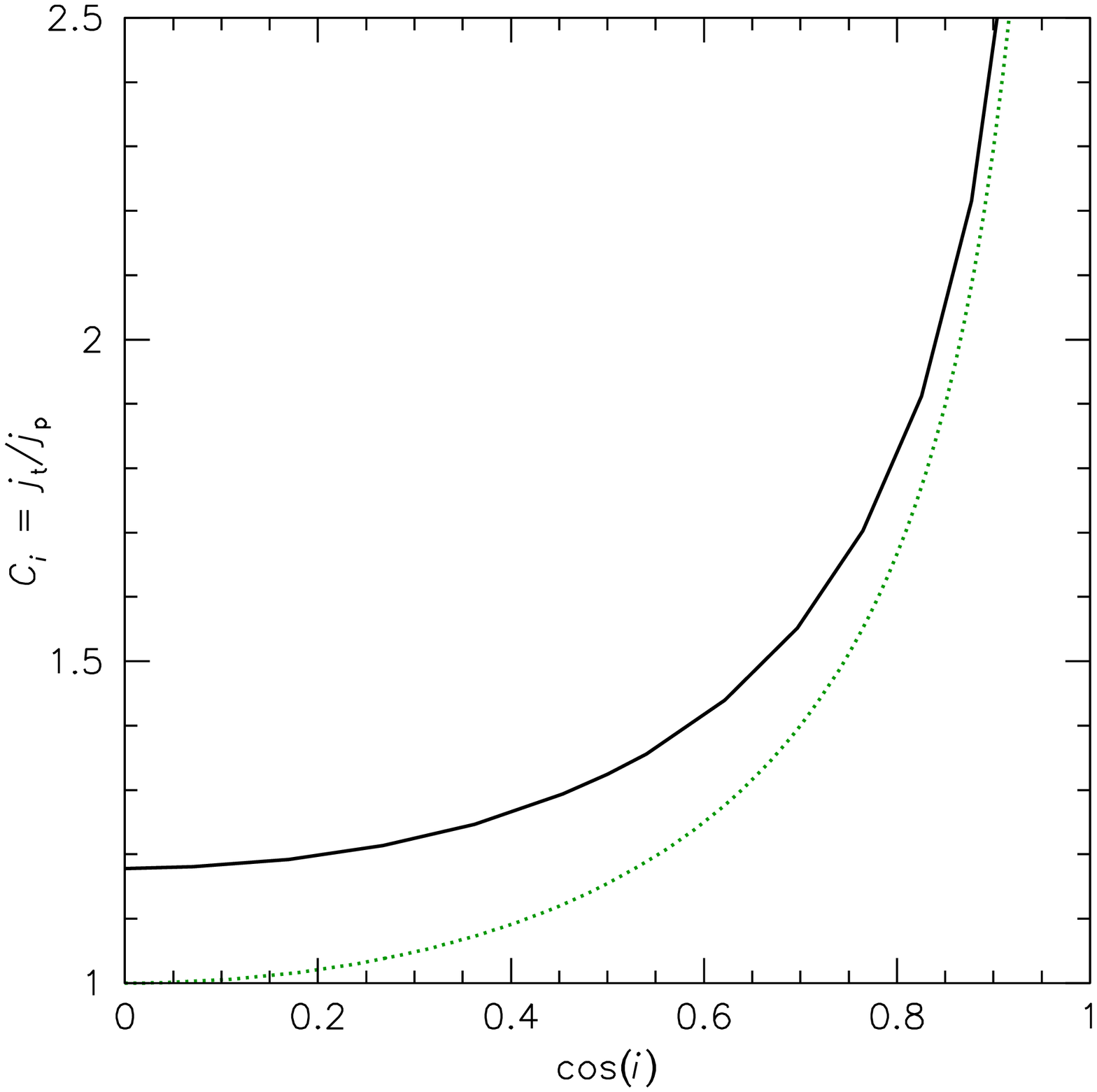} 
}
\caption{Relations between true and observed properties of
idealized spherical galaxies with flat intrinsic rotation-velocity profiles.
{\it Left:}
Ratio of observed to true rotation velocity, vs.\ galactocentric distance,
in units of the effective radius.
Three different density profiles are used, with different line-styles and their
S\'ersic $n$ indices labeled in the plot.
Three inclinations are also used, with different colors, as labeled in the plot.
The horizontal dot-dashed lines show the results if the dilution effect is ignored.
We have verified that we reproduce related results from the literature
\citep{1978MNRAS.183..501B,1978ApJ...222..450Y}.
{\it Right}: 
Ratio of the true specific angular momentum to the value estimated by using the observed
rotation-velocity profile [Equation~(\ref{eqn:jest})], as a function of the cosine of the inclination.
The black solid curve includes the dilution effect, 
while the green dotted one does not.
\vskip 0.2cm
}
\label{fig:Cr}
\end{figure*}

There are a couple of notable features about Equations~(\ref{eqn:vp2}) and (\ref{eqn:vrotp1}).  
One is that the difference between true and observed rotation velocity is more than a simple matter of
the ``$\sin i$'' inclination effect for a galaxy tilted away from edge-on.  
For a system of finite thickness,
there is an additional $\cos\phi$ term that represents a ``dilution'' effect
(cf.\ \citealt{1978MNRAS.183..501B,1978ApJ...222..450Y,1986ApJ...302..208F,1999AJ....117.2666N}),
corresponding to the projection of a circular orbit seen at varying azimuth $\phi$
(see Figure~\ref{fig:coords}).
Even for an edge-on case,
the rotation vector is not in the line-of-sight except at the
tangent point for a given semi-major axis distance $x$.
This effect also implies that even if the true rotation velocity ($v_{\rm rot,t}$)
is constant with radius, the projected rotation velocity ($v_{\rm rot,p}$) 
is generally {\it not}, and goes to zero toward the
center of the galaxy (see left-hand panel of Figure~\ref{fig:Cr} for examples).

Next, using Equation~(\ref{eqn:vrotp1}), we find our final value for
the numerical coefficient,
\begin{equation}\label{eqn:Cj2}
C_i = \frac{\int v_{\rm rot,t}(R) \, \Sigma(R) \, R^2 \, dR}{\sin i \int x^3 \, dx \int \rho(r) \, v_{\rm rot,t}(R) \, R^{-1} dz^\prime} ,
\end{equation}
where $\rho(r)$ can in principle also be written in terms of the
directly observed profile $\Sigma(x)$ via an Abel integral.

Having arrived at our general expressions for calculating \jt, we note that there
is another, related approach found in the literature.
This is to construct an observational analogue to ${\bf j}_{\rm t}$, using 
the data directly to calculate a two-dimensional cross-product in
an expression similar to (\ref{eqn:A4})
\citep{1988skeg.book.....F,2007MNRAS.379..401E,2009MNRAS.397.1202J}.
This technique has the merit of potentially providing a general characterization of the data,
including non-axisymmetric rotation fields.
We have experimented with this approach and found that the fairly small
improvement in accuracy (assuming the availability of full two-dimensional data)
 comes with the price of added complexity, 
while it also becomes unclear how to connect the results to 
the true three-dimensional angular momentum ${\bf j}_{\rm t}$,
 which is our concern here.

\subsection{A.2. Thin disks}\label{sec:thin}

Following the general framework above, we continue with more specific galaxy models.
For context, we begin with the simple case of an infinitely thin disk that has
a constant intrinsic rotation-velocity profile, which is the same as the
circular velocity, $v_{\rm rot}(R)=v_{\rm c}(R)$. 
It also has an exponential pole-on surface density profile:
\begin{equation}
\Sigma(R) \propto \exp\left(-R/\Rd\right) ,
\end{equation}
where $\Rd$ is the disk scale-length.
Using Equation~(\ref{eqn:A5}), we find
\begin{equation}\label{eqn:jd}
j_{\rm t} = 2 \, v_{\rm c} \, \Rd\ = 1.19 \, v_{\rm c} \, R_{\rm e} ,
\end{equation}
where \Reff\ is the radius enclosing half of the light of a face-on disk.
and we have used the relation \Reff$=1.68 \Rd$ for an exponential profile
(note that for other disk inclinations, the observed half-light radius is smaller).
Because the disk is infinitesimally thick, there is no dilution, except for a
perfectly edge-on case, and we have simply 
$C_i = (\sin i)^{-1}$.
This means that the observed (constant) rotation velocity
is $\vs = v_{\rm c} \sin i$, and
we can map between observations and true specific angular momentum
by the relation
\begin{equation}
j_{\rm t} = \frac{2 \, \vs \, \Rd}{\sin i} .
\label{eqn:simpform}
\end{equation}
Using the notation of Equation~(\ref{eqn:jCp}), we may say alternatively that
$j_{\rm p}=2 \, \vs \, \Rd$ and $C_i=(\sin i)^{-1}$.
The exception is the perfectly edge-on case ($i=\pi/2$), where we find by constructing
the appropriate analogues to Equations~(\ref{eqn:A5}) and (\ref{eqn:jest}) that $C_i=2$.

Real disks do have finite thickness and consequently an appreciable
amount of observed rotation-velocity dilution toward their centers,
even for $i < 90^\circ$.
Still, detailed formulae like (\ref{eqn:jCp})--(\ref{eqn:Cj2})
are not really necessary, and as we verify in the main text,
one can instead arrive at a very good approximation to \jt\ by taking
the maximum or asymptotic rotation velocity observed at
large radii to be $\vs$, which is then used in Equation~(\ref{eqn:simpform}).

For a disk galaxy it is fairly straightforward to estimate the
inclination $i$ based on the apparent axial ratio $q$, using the formula
\begin{equation}\label{eqn:diski}
\sin i = \left(\frac{1-q^2}{1-q_i^2}\right)^{1/2} ,
\end{equation}
where we adopt an intrinsic axis ratio of $q_i=0.11$ (e.g., \citealt{1992A&AS...93..255G}).
However, in order to tie in with spheroidal galaxies, 
we first treat $i$ as an unknown for individual galaxies,
so that a ``typical'' \jt\ is recovered by statistical deprojection of an ensemble of galaxies.  One approach is to consider the
{\it median} case $C_{\rm med}$, where for randomly oriented galaxies,
half of them have $C_i < C_{\rm med}$;
this is equivalent to identifying the median value for $i$.
Since random inclinations in an axisymmetric geometry
are distributed uniformly in $\cos i$, the median is $i=\pi/3$, which
corresponds to $C_{\rm med}=2/\sqrt{3} \simeq 1.15$.

One may ask instead what is the {\it average} inclination correction, $C_{\rm avg}$,
which if applied uniformly to all observed galaxies would provide the correct ensemble average for \jt.
Interestingly, the answer appears to depend on the exact question being posed.
One may start in the reference frame of a galaxy (or planetary system) and ask what 
the average projection is (cf.\ \citealt{2008PASP..120..531C,2009ApJ...697.2057L}).
Or one may start with the {\it observations} and estimate the average 
{\it deprojection} (e.g., \citealt{2006ApJ...646..107E}).
In the spirit of mapping from projected to intrinsic quantities, we take
the latter approach as follows.

The average value of \jt\ is calculated as an expectation value of
a probability distribution:
\begin{equation}
\langle j_{\rm t} \rangle = \int j_{\rm t} \, \frac{dP}{dj_{\rm t}} \, dj_{\rm t} .
\end{equation}
We consider a single observation $j_{\rm p}$ with infinitesimally small
uncertainties, and since the probability $P$ is uniform in $\cos i$, 
we recast this equation (see Equation~(\ref{eqn:jCp})) as
\begin{equation}\label{eqn:Cavg}
\langle C_i \rangle = \int C_i \sin i \, di .
\end{equation}
Given $C_i=(\sin i)^{-1}$ for a thin disk, 
we then find $C_{\rm avg}=\langle C_i \rangle = \pi/2 \simeq 1.57$.
The equivalent calculation using the average {\it projection} yields
$C_{\rm avg}=\langle C_i^{-1} \rangle^{-1} = 4/\pi \simeq 1.27$.
These two alternative correction factors differ by 23\% or 0.09 dex,
which will be a relatively minor concern in our overall \jt\ analysis.
In the main text, we analyze a sample of spiral galaxies where the deprojections are known;
because this sample is chosen to be avoid galaxies that are near face-on, 
the different options for averaging or medianing imply very similar values for $C$, and
we cannot distinguish a best choice.
In any case, the important point
is to carry out {\it relative} comparisons of disk and
spheroidal galaxies using the {\it same} deprojection approach,
which we choose to do using median inclinations.

\subsection{A.3. Spheroids}\label{sec:spheroid}

We next consider rotating spheroids, with the goals of 
(1) calculating the coefficient $C_i$ to be used in combination with
observational estimates of \jp\ in Equation~(\ref{eqn:jCp});
(2) deriving an even simpler, more approximate expression for \jp\ that avoids
the detailed calculation of Equation~(\ref{eqn:jest}).
We assume spherical symmetry for now, and adopt the general
\citet{1968adga.book.....S} law for surface density profiles,
\begin{equation}
\Sigma(R) \propto \exp\left[-b_n(R/R_{\rm e})^{1/n}\right] ,
\label{eqn:ser1}
\end{equation}
where the shape index $n$ determines the steepness of the outer density profile
(higher values are shallower: $n=1$ is an exponential profile, $n=4$ is a de Vaucouleurs law),
and the numerical coefficient is
\begin{equation}\label{eqn:bn}
b_n \simeq 2n -1/3+0.009876/n
\end{equation}
\citep{2000A&A...353..873M}.
Approximate analytical deprojection formulae for $\rho(r)$ are also provided in the preceding reference.

If we somehow knew the intrinsic rotation-velocity profile $v_{\rm rot,t}(R)$,
it would be straightforward to evaluate Equation~(\ref{eqn:A5}) and find \jt.
For example, a de Vaucouleurs ($n=4$) model with constant intrinsic rotation velocity
$v_{\rm t}$ would yield:
\begin{equation}\label{eqn:exact}
j_{\rm t} = 2.29 \, v_{\rm t} \, R_{\rm e} 
\end{equation}
(cf.\ \citealt{1985SvAL...11..277Z}).
To deal with the projection effects, we begin by adopting a constant intrinsic rotation velocity,
and calculate $C_i$ using Equation~(\ref{eqn:Cj2}).
We find that $C_i$ depends only on $i$, and not on $n$
(Figure~\ref{fig:Cr}, right).
The inclination dependence is weaker than $C_i \propto (\sin i)^{-1}$, which means
it is partially counteracted by the dilution effect.
An example is the edge-on case ($i=\pi/2$), for which we find $C_i=1.18$, as compared
with $C_i=1$ if dilution were neglected.\footnote{That is,
given an observed rotation velocity $\vs$, we would have 
$j_{\rm t}=2.70 \, v_{\rm s} R_{\rm e}$, which may be contrasted with the expression
$j_{\rm t}=1.03 \, v_{\rm s} R_{\rm e}$ from \citet{1995A&A...293...20S},
who took dilution and non-cylindrical rotation into account but neglected all of
the angular momentum outside of 5~\Reff.}
We performed the same tests for rotation-velocity profiles that vary smoothly with radius
and found very similar results.
For example, $C_i$ has only a $\sim$~10\% dependence on the details of the rotation-velocity
profile in Equation~(\ref{eqn:Cj2}).

The implication is that given observations of spherical galaxies, we can recover an accurate
estimate of \jt\ using Equation~(\ref{eqn:jCp}) and (\ref{eqn:jest}), where $C_i$ is a numerical
factor that depends only on inclination.
A simple expression for $C_i$ that is accurate to better than 3\% everywhere is:
\begin{equation}\label{eqn:Cform}
C_i \simeq \frac{0.99+0.14\,i}{\sin i} ,
\end{equation}
where $i$ in the numerator is in radians.
Note that this expression for $C_i$ differs from the undiluted value by no more than 20\%,
which might suggest that we ignore the dilution effect in our studies of \jt, 
but we include it because we want to avoid collecting multiple systematic
errors of this level.

In practice, we do not usually know the individual inclinations of spheroidal galaxies,
and instead need to do a statistical deprojection as introduced
above for disk galaxies.
Considering the median case ($i=\pi/3$), we find $C_{\rm med}=1.32$.
For the inclination averaged value, we again use Equation~(\ref{eqn:Cavg})
but with the spheroidal expression for $C_i$ [Equation~(\ref{eqn:Cj2}), where in
practice only the denominator is affected by the inclination averaging].
We find 
$C_{\rm avg} = \langle C_i^{-1}\rangle^{-1} = 1.45$;
the alternative value is 
$C_{\rm avg}=\langle C_i \rangle = 1.73$.

When modeling real early-type galaxies as in the main part of this paper,
we do not use any of these choices for $C$.
This is because our galaxy sample is not randomly selected in sub-type (E or S0),
and these sub-types are known to have a systematic connection with inclination.
Lenticular galaxies are difficult to identify when close to face-on,
so the samples of lenticular and elliptical galaxies will be biased to 
high and low inclinations, respectively.
To correct for this bias, we adopt a simplified picture where there is 
only {\it one} species of early-galaxy, which gets classified as E or S0 depending on whether 
its inclination is below or above a boundary of $i=\pi/3$ (cf.\ \citealt{1994ApJ...433..553J}).
Therefore we apply median deprojection factors of
$C_{\rm med}=1.21$ ($i=1.32$) for the lenticulars,
and $C_{\rm med}=1.65$ ($i=0.72$) for the ellipticals.

When applying this approach for \jt\ estimation to real data, one has to make allowances
for the limited radial extent of the available $v_{\rm rot,p}(R)$ profiles.
One can still use Equation~(\ref{eqn:jest}) and extrapolate beyond the data,
while taking care to quantify the uncertainties that this entails.
We will provide examples of this procedure in the main text.
However, a much easier approach is also possible---sacrificing some
accuracy for the sake of speed and simplicity---which we outline below
and test with detailed calculations.

\subsection{A.4. Simple angular momentum estimator}\label{sec:simpapp}

The general idea for simple \jp\ estimation is that every galaxy can be
characterized by a single observed rotation velocity $\vs$, which if substituted
as a constant value in Equation~(\ref{eqn:jest}) would give the same answer as
using the full $v_{\rm rot,p}(x)$ profile.
The rotation-velocity and surface brightness components in Equation~(\ref{eqn:jest})
are then separable, and we can reduce the calculation to a 
product of a numerical coefficient, a velocity scale, and a scale-length
\begin{equation}\label{eqn:K}
\tilde{j_{\rm p}} = k_n \, \vs \, R_{\rm e} ,
\end{equation}
which is analogous to Equation~(\ref{eqn:simpform}).
Here \tjp\ is a general approximation of \jp,
and the coefficient $k_n$ is a spatial weighting factor calculated using 
Equation~(\ref{eqn:jest}) with $v_{\rm rot,p}(x)$ set to unity
(i.e., corresponding to a weighting derived from the stellar density profile combined
with a radius $x$ lever-arm).
$k_n$ is a function of $n$, with the following handy approximation,
accurate to better than 4\%:
\begin{equation}\label{eqn:kn}
k_n \simeq 1.15 + 0.029 \, n + 0.062 \, n^2 .
\end{equation}
Thus, $k_1=1.19$ and $k_4=2.29$, as in Equations~(\ref{eqn:jd})
and (\ref{eqn:exact}).

The crux of this approach is estimating $\vs$ from observations without having to 
evaluate the full integral in Equation~(\ref{eqn:jest}), 
nor requiring that the rotation-velocity profile be known to the outermost radii.
The trick comes from realizing that every galaxy has at least one radius $x_s$
where the local projected rotation velocity is equal to $\vs$.
This radius varies from one galaxy to another, but we might expect the variation to
be modest if real galaxies have rotation-velocity profiles that are not too dissimilar.
The implication then is that we can simply measure the rotation velocity at the same
radius $x_s/R_{\rm e}$ for all galaxies and adopt $\vs =v_{\rm rot,p}(x_s)$ 
for use in Equation~(\ref{eqn:K}).

There are two potentially different goals here:  one is to pick the
radius $x_s$ that yields {\it on average} the correct answer for \jp\ for a
variety of galaxies; the other is to pick a radius where the {\it scatter} in 
the \jp\ approximation is minimized.  
The anticipated origin of this scatter is the variety of radial behaviors of
early-type galaxy rotation-velocity profiles, e.g., rising or falling from central
to outer regions, such that the radius $x_s$ which applies to one galaxy might not
work well for another.
It is the issue of scatter that is most important, since a systematic
offset in \tjp\ can be calibrated out and subsumed in the value of $k_n$.

We test this heuristic concept via some simple model galaxies below and via some real
observations in the main text. We find that $x_s$ values anywhere in the range
of $\sim$~(2--5)~\Reff\ appear to work reasonably well, and in order to maximize
the number of suitable galaxies, we adopt $x_s \simeq$~2~\Reff.

To arrive at this point,
we first experiment with galaxy models having simple power-law
rotation-velocity profiles $v_{\rm rot,t}(R)$
that range between $\propto R^{-1/2}$ and $\propto R^{+1/2}$.
For each S\'ersic index $n$, 
and rotation-velocity profile, we
compute \jp\ via Equation~(\ref{eqn:jest}), next determine the equivalent $\vs$ value,
and then find the radial location $x_s$ where the local projected rotation velocity is equal to $\vs$.
Finally, we examine the ratio $\tilde{j_{\rm t}}/j_{\rm t}$,
[where \jt\ is known exactly from Equation~(\ref{eqn:A5})] for a range of $x_s$,
which allows us to diagnose the best values of $x_s$ to use in general.
We make \tjt\ and \jt\ comparisons here, rather than \tjp\ and \jp, in order to 
incorporate the impact of neglecting the mild dependence of $C_i$ on rotation-velocity profile
[Equation~(\ref{eqn:Cj2})].

\begin{figure*}
\centering{
\includegraphics[width=3.5in]{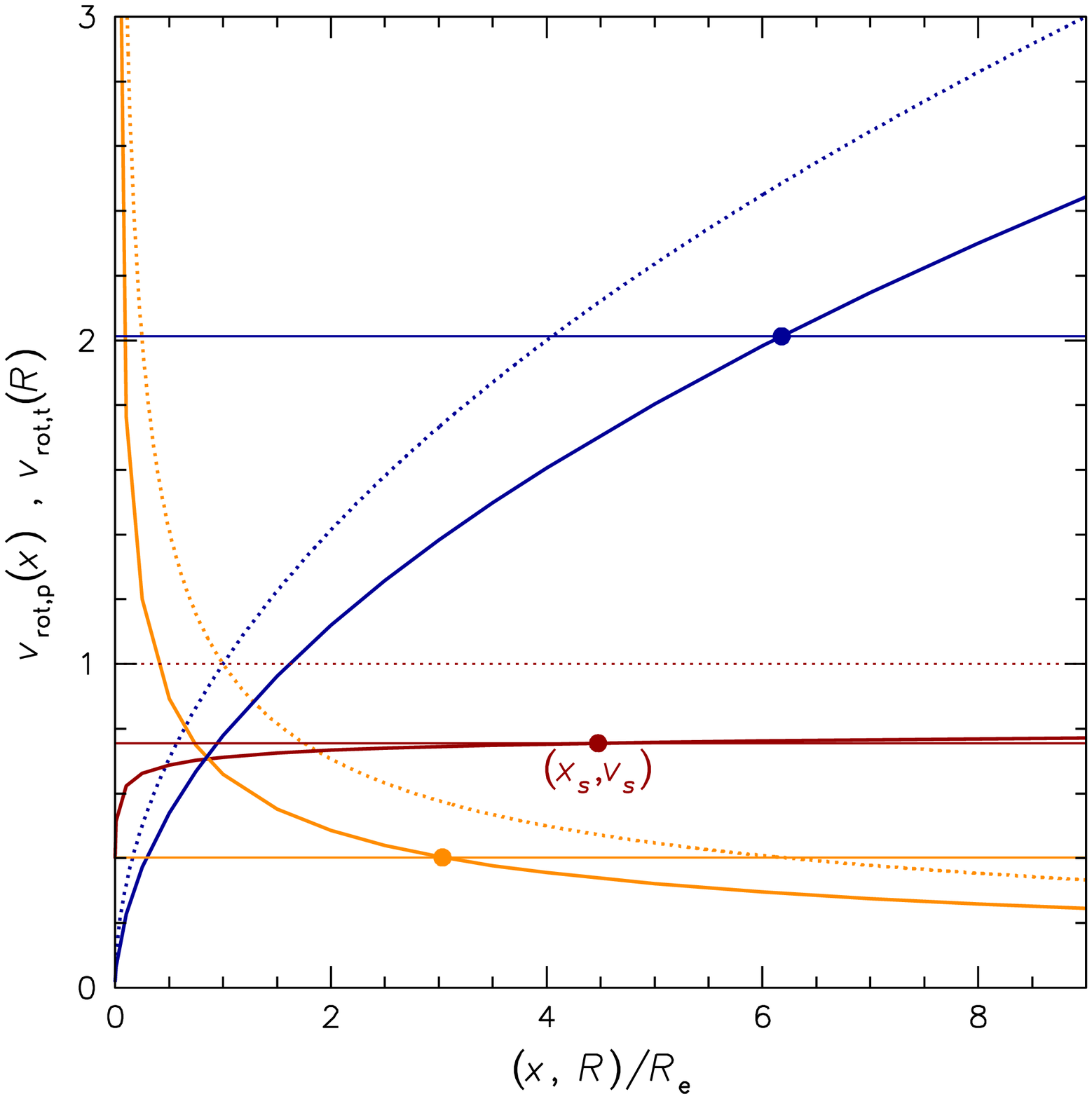} 
\includegraphics[width=3.5in]{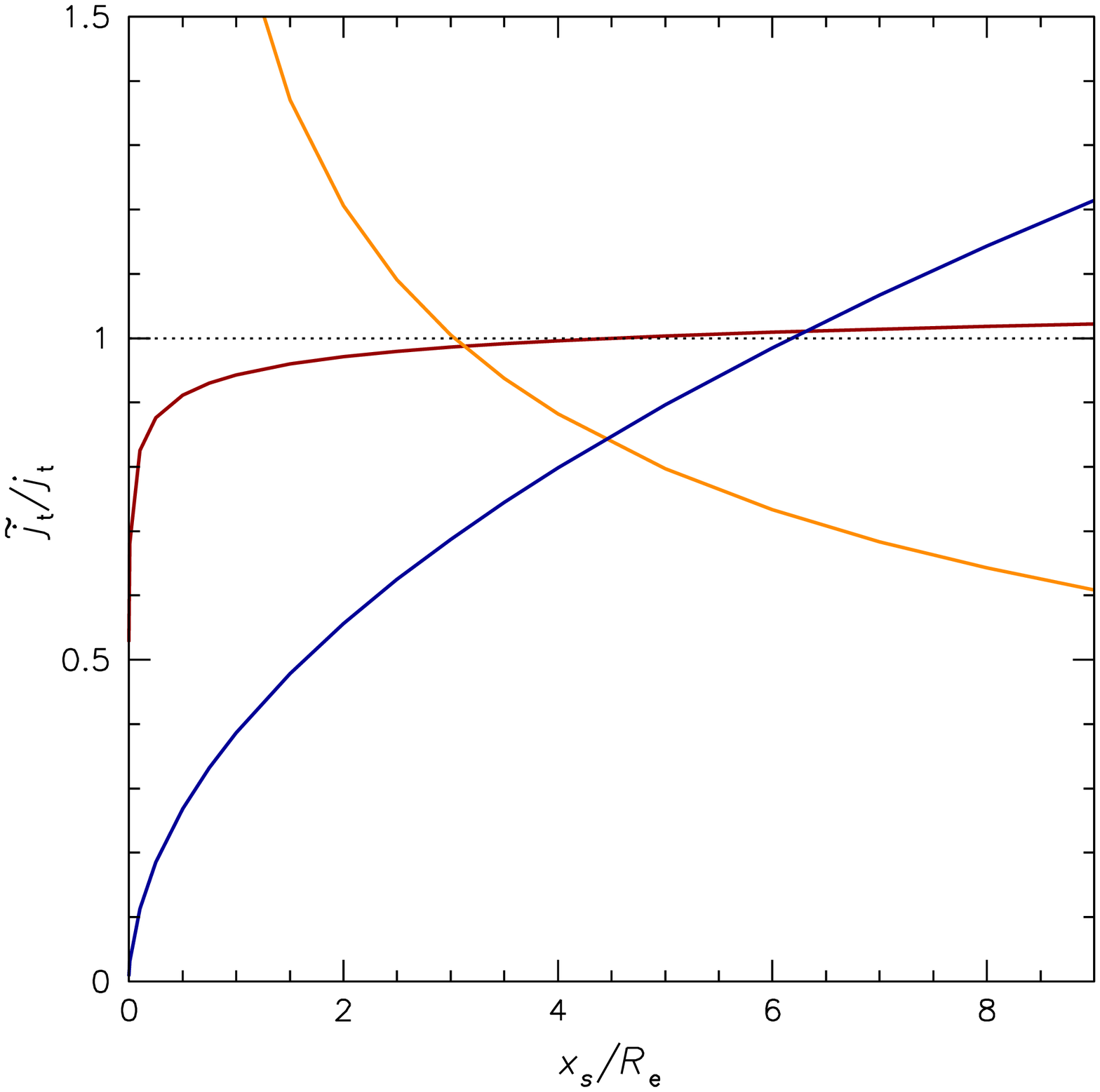} 
}
\caption{Relations between true and observed properties of
idealized spherical galaxies with de Vaucouleurs profiles ($n=4$) and inclination $i=\pi/3$.
Different colors indicate models with distinct rotation-velocity profiles
(red: constant; blue: $v_{\rm rot,t}(R) \propto R^{ 1/2}$,
orange: $v_{\rm rot,t}(R) \propto R^{ -1/2}$).
{\it Left:}
True (dotted curves) and projected (solid curves) rotation velocities vs.\ galactocentric distance,
in units of the effective radius.
The horizontal dashed lines show the characteristic global rotation velocities $\vs$,
with the large dots marking the radii $x_s$ where $\vs = v_{\rm rot,p}(x_s)$.
{\it Right}: 
Ratio of estimated to true specific angular momentum, vs.\ the chosen measurement location
$x_s$ (see text for details).
\vskip 0.2cm
}
\label{fig:models}
\end{figure*}

This procedure is illustrated for one case ($n=4$, $i=\pi/3$) in Figure~\ref{fig:models}.
The left-hand panel shows the intrinsic and projected rotation-velocity profiles for three different
models.  The horizontal lines show the values of $\vs$ which would yield the 
correct $\tilde{j_{\rm p}}=j_{\rm p}$ when substituting $k_4=2.29$ in Equation~(\ref{eqn:K}).
The intercept of the value with the corresponding $v_{\rm rot,p}(x)$ profile gives
the appropriate radius $x_s$.
In the right-hand panel, the accuracy of \tjt\ is plotted versus the chosen $x_s$
for each model.  For the constant-rotation-velocity model, $\tilde{j_{\rm t}} \simeq j_{\rm t}$
for a wide range of $x_s$, i.e., the results are insensitive to $x_s$.
For the other rotation-velocity models, the choice of $x_s$ is more critical and ranges from
$\simeq 3 R_{\rm e}$ to $\simeq 6 R_{\rm e}$.
A compromise radius that works reasonably well is
$x_s \simeq 4.5 \, R_{\rm e}$, which provides accurate \tjt\ estimates
at the 15\% level or better for all three models.

Exploring other values of $n$ and $i$, 
we find that inclination is not important, 
and that some aspects of the $n=4$ example are generally true for other $n$.
For each $n$, the radius $x_s$ that minimizes the errors in \tjt\ among the different rotation-velocity
profiles is close to the exact value of $x_s$ for the constant rotation velocity case.
This ``sweet spot'' also depends systematically on the density profile:
for $n=$~2, 4, and 6, it occurs at 
$x_s \sim 2$~\Reff, 4.5~\Reff, and 10~\Reff\footnote{These
locations are very similar to the corresponding radius of gyration (where
a point mass would have the same moment of inertia as the full mass profile;
see \citealt{2009MNRAS.394.1249C}).}.

This $x_s$ trend demonstrates that
the extended envelopes of higher-$n$ galaxies contribute relatively
more to \jt, and require correspondingly extended kinematics observations.
Given that the \Reff\ values for high-$n$ galaxies are already large, the
observations to $\sim$~10~\Reff\ become extremely challenging, and we will in general
omit such galaxies from our current analysis.
Observations at $x_s\sim$~4.5~\Reff\ as needed for $n\sim4$ galaxies
are also relatively rare, but fortunately the systematic bias in \tjt\
when using smaller $x_s$ is small, and the scatter is still only $\sim$~$\pm$~$0.1$~dex
when $x_s\sim$~2.5~\Reff.

The foregoing tests involve model galaxies whose intrinsic properties we fully understand,
but we also need to carry out checks with real galaxies.
In the main text (Section~\ref{sec:examp}),
we present more detailed models of a half-dozen real galaxies, using Equation~(\ref{eqn:jest})
along with the full observed rotation-velocity profiles to estimate \jp.
There we show that the constant $\vs$ approach works fairly well for a broad range of radii,
and we adopt $x_s\simeq2~R_{\rm e}$; at this radius,
the systematic offset in \tjp\ appears to be small, and 
the galaxy-to-galaxy scatter is $\sim \pm$~0.1~dex.\footnote{For
the real galaxies, we do not attempt a full \tjt\ and \jt\ comparison because we do not know the
detailed intrinsic rotation-velocity profiles needed to calculate $C_i$ exactly.}

We have thus arrived at our adopted procedure for simplified \jp\ estimation
using Equation~(\ref{eqn:K}).  
This \tjp\ is then multiplied by the deprojection factor $C_i$ to yield
an estimate of the intrinsic \tjt.
Of special interest is the de Vaucouleurs profile ($n=4$), with the 
approximate result for a median inclination
\begin{equation}
\tilde{j_{\rm t}} = 3.03 \, \vs \, R_{\rm e} , 
\label{eqn:j1}
\end{equation}
which is related to Equation~(\ref{eqn:exact}) by a factor of $C_{\rm med}$
in order to account for the rotation-velocity dilution 
($\vs < v_{\rm t}$;
note that as discussed in the previous subsection, in practice,
$C_{\rm med}$ may need to be corrected for additional inclination-selection effects).
This can be compared to the expression used by F83:
\begin{equation}
j_{\rm t} = 2.5 \, v_{\rm m} \, R_{\rm e} ,
\end{equation}
where $v_{\rm m}$ was the maximum value of the projected rotation velocity within
the (central) observed region.
This formula is now superseded by the more accurate expressions above.

\subsection{A.5. Flattening corrections}

Real galaxies of course are not spheres.
We could in principle set up a standard axisymmetric density model by making an
ellipsoidal substitution for $r$ as needed with $\rho(r)$:
\begin{equation}
m^2 = R^2+\frac{z^2}{q^2} = {x^\prime}^2+{z^\prime}^2\sin^2 i + \frac{{z^\prime}^2\cos^2i}{q^2} ,
\end{equation}
where $q$ is the intrinsic axis ratio, and
again we are measuring rotation velocity only along the projected semi-major axis $(y^\prime=0)$.
Substituting this expression into Equation~(\ref{eqn:vrotp2}) along with
additional modifications of Equation~(\ref{eqn:Cj2}),
we find from some test calculations that even strong flattening
makes only a very small difference to $v_{\rm rot,p}$ and to $C_i$,
which typically increase and decrease (respectively)
by $\sim$~5--10\% relative to the spherical case for $q=0.3$.
Given the mildness of these effects, and the unknown systematics
of the cylindrical-rotation assumption, we will not attempt to make any
correction based on the flattening.  The extrapolation of the kinematic
data to larger radii is in any case probably the dominant uncertainty
for our final results.

The other potential concern here is
the effective radius.
We have already mentioned for disk galaxies that the value of $\Rd$ used in
Equation~(\ref{eqn:jd}) is not the observed but instead the deprojected value.
Similarly, for flattened spheroidal galaxies, we should 
not use in Equation~(\ref{eqn:K}) the circularized value \Reff\ normally tabulated 
in catalogs, but the equivalent distance along the semi-major axis,
$a_{\rm e}\equiv R_{\rm e}/\sqrt{q^\prime}$, where $q^\prime$ is the observed axis ratio
(in the rest of the paper we simplify this to $q$).
This is because even though we may not know the inclination of a galaxy, we do know (for 
an axisymmetric case) that if it were face-on, the true \Reff\ needed to calculate
\jt\ would be roughly the same as \aeff.
Since early-type galaxies can be as flattened as $q^\prime\sim0.3$, this is a significant
correction.

A caveat here is that the observed flattening is correlated with the inclination:
e.g., a galaxy with $q^\prime=0.3$ is probably nearly edge-on, and adopting the
median inclination for random orientations would then cause us to overestimate its \jt.
Also, the intrinsic \Reff\ value for a spheroidal galaxy 
is slightly smaller than $a_{\rm e}/\sqrt{q^\prime}$
(see also \citealt{2009MNRAS.400.1665W}).
However, these two effects are much weaker in general than neglecting the
$q^\prime$ correction entirely.
Therefore in the main text we carry out all of the calculations using
densities and kinematics along the semi-major axis: e.g., 
$\Sigma(x)$ and \aeff\ rather than $\Sigma(R)$ and \Reff.

\newpage

\section{Appendix B: Detailed observational results for individual galaxies}\label{sec:obsapp}

Here we present in some detail the data and methods used to derive
angular momentum profiles for individual early-type galaxies.
Our general approach is to map all of the data to an equivalent semi-major axis
rotation-velocity profile, as though making a standard long-slit observation.
In some cases, the data are already available in this format while in others
that involve discrete and semi-discrete data at various position angles,
we must fit a two-dimensional kinematic model before mapping onto the semi-major axis.

Our modeling method was recently developed for use with 
sparsely sampled two-dimensional data
\citep{2009MNRAS.398...91P,2011ApJ...736L..26A,2011MNRAS.415.3393F,2011ApJS..197...33S,Pota12}.
It  is based on flattened sinusoidal curves of mean line-of-sight
velocity versus azimuth $\phi$:
\begin{equation}
\bar{v}_{\rm p}(\phi,x) = \pm v_{\rm rot,p}(x)\left[1+\frac{\tan^2 (\phi-\phi_0)}{q^2}\right]^{-1/2} ,
\label{eqn:quasi}
\end{equation}
where \vrotpx\ is the rotation velocity versus the semi-major axis distance $x$,
and $\bar{v}_{\rm p}$ is evaluated along an ellipse with axis ratio $q$.
The on-sky coordinates are the same as introduced in Appendix~\ref{sec:form}, but without
primes (${}^\prime$) for the sake of simplicity; the position angle of maximum receding rotation velocity is
at $\phi=\phi_0$.
This model is exactly equivalent both to a $\cos \phi$ rotation velocity multiplied by the
equation for an ellipse in polar coordinates, and to
the classic tilted-ring approach used for gas disks (e.g., \citealt{1988AJ.....96..851G,1989A&A...223...47B}).
Therefore it is a natural observational model to use in conjunction with our \jt\ modeling
scheme, which assumes a cylindrical rotation field in the reference frame 
of the galaxy (Appendix~\ref{sec:form}).
There is also a connection here to the ``kinemetric'' methods of modeling 
data from integral-field spectrographs, where the two-dimensional kinematics
are expanded in a Fourier series \citep{2006MNRAS.366..787K,2008MNRAS.390...93K}.
Our model is equivalent to the first-order term of this expansion.

In principle, the parameters $q$ and $\phi_0$ should be determined by
fitting to the kinematic data, but in practice, the azimuthal sampling is often 
too sparse for such constraints.
For the sake of a uniform treatment of the data,
we adopt the {\it photometric} values for $q$ and $\phi_0$ in every case;
these are generally held constant at a global value,
e.g., neglecting kinematic twists with radius.
The errors introduced through these simplifications
are generally much smaller than the
uncertainties in extrapolating \vrotpx\ outwards to radii beyond the data.

\begin{figure*}
\centering{
\includegraphics[width=3.5in]{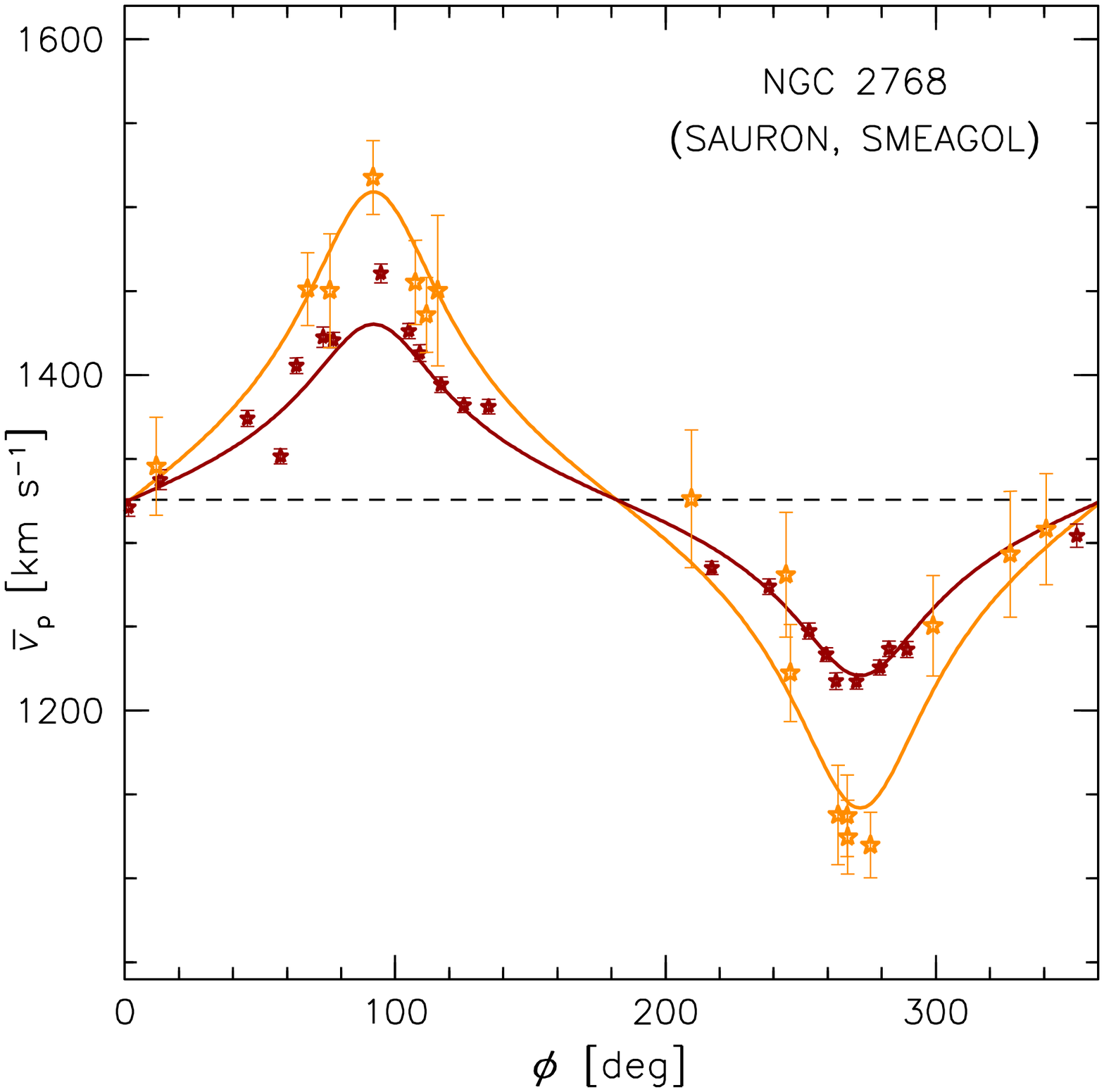} 
\includegraphics[width=3.5in]{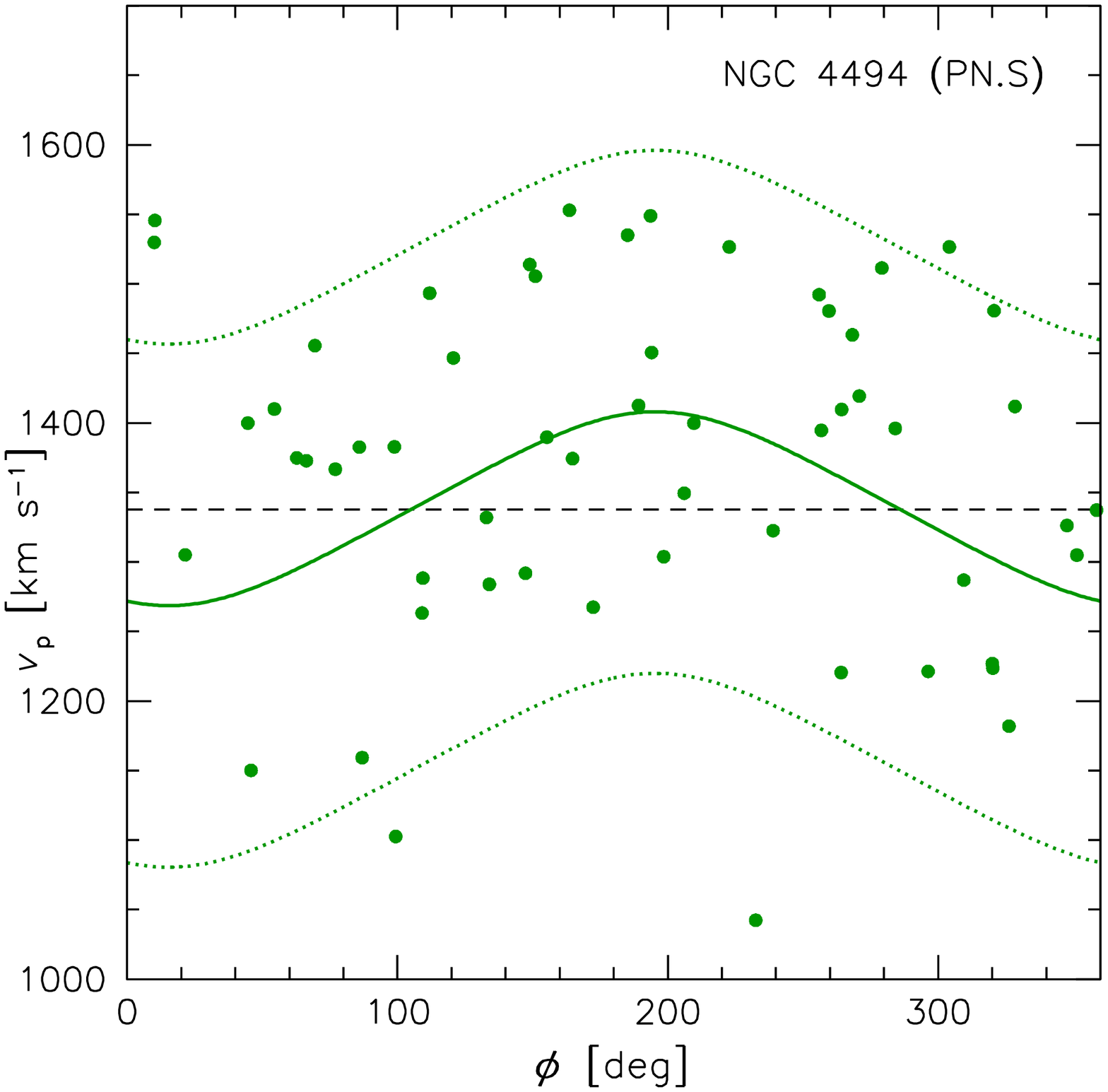} 
}
\caption{Examples of kinematic model fits to two-dimensional velocity data in early-type galaxies.
On the {\it left} are subsets of stellar kinematics measurements in the galaxy
NGC~2768 from SAURON using integral field spectroscopy \citep{2004MNRAS.352..721E}
and from SMEAGOL using the SKiMS technique \citep{2009MNRAS.398...91P}.
These data sets are marked with small and large points, respectively, and
are taken from radii
$x \sim 35''$ ($\sim$~0.3~\aeff) and $x \sim 125''$ ($\sim$~1.3~\aeff), respectively.
The horizontal dashed line marks the systemic velocity (the SAURON data have
been offset to match up with SMEAGOL), and
the curves show flattened sinusoidal model fits to the data [see Equation~(\ref{eqn:quasi})].
The increase of rotation velocity with radius is apparent here by eye.
The {\it right} panel shows PN data for NGC~4494 \citep{2009MNRAS.393..329N}
at a radius of $x \sim 60''$ ($\sim$~1.3~\aeff).
The solid curve shows the mean velocity model fit, while dotted curves show the
intrinsic dispersion $\sigma$ of the model, marking $\pm 2\sigma$ boundaries.
For clarity, the measurement uncertainties of the individual data points (typically
20~\kms) are not shown.
\label{fig:N4494kin1}
}
\end{figure*}

Figure~\ref{fig:N4494kin1} illustrates the kinematic fitting procedure
for several types of data.
One type that may be unfamiliar is the 
``stellar kinematics with multiple slits'' (SKiMS) technique from the SMEAGOL project
(\citealt{2009MNRAS.398...91P,2011ApJ...736L..26A,2011MNRAS.415.3393F,Forbes12}), which uses a wide-field multi-slit spectrograph
to provide sparse sampling of
a two-dimensional field as needed for reconstructing the basic kinematical structure.
Equation~(\ref{eqn:quasi}) is fitted to the data (left-hand panel), while occasionally 
disregarding outlying data points so as not to skew the results.
The uncertainties on the rotation velocities are estimated via Monte Carlo simulations of mock
data sets. 

This procedure is modified slightly for discrete velocity data such as from PNe and GCs,
which are sparsely sampled not only in position but also in velocity.
Here, each data point is not a measure of the local mean velocity, but is a random {\it discrete}
velocity drawn from an underlying distribution.  The kinematic modeling of rotation
therefore involves simultaneous fitting for the velocity {\it dispersion}
using a maximum likelihood method
(which represents a refinement, along with the rotation-field flattening,
of the commonly-used least-squares method;
e.g., \citealt{1997ApJ...486..230C}).
Note also that when using data from the PN.Spectrograph
\citep{2009MNRAS.393..329N,2011MNRAS.411.2035N,2009MNRAS.394.1249C}, 
we ignore the results inside $\sim$~\Reff\ 
since there are known systematic biases that can occur in the central regions,
where PN detection may be velocity-dependent.

To combine different data sets on the same galaxy, 
we could in principle fit them all simultaneously to a single kinematic model.
However, for the sake of simplicity and transparency, we create separate
one-dimensional rotation-velocity profiles from each data set separately, and then
combine these into a single profile.
For this final combination, we set up a new grid of data bins that are smoothly
distributed in radius, and in each bin average the weighted contribution of
each data point according to its uncertainties and its fractional overlap with that bin.
The price of generating these well-behaved profiles is that the values in
each bin are not fully independent of those in adjacent bins.

A significant complication in our analysis comes in extrapolating the rotation-velocity profile beyond
the outermost measured data points.
In a spiral galaxy, this can be done assuming a profile that becomes asymptotically
constant, but we cannot assume any such model for an elliptical or lenticular.
Instead, we fit a power-law model of rotation velocity with radius to the final few data points:
$v_{\rm rot,p}(x) \propto x^\gamma$.

The optimum range of points to use in this fit is not straightforward to determine:
use too few, and the extrapolation becomes merely an amplification of 
noise in the outermost measurements;
use too many, and the inner measurements with their small uncertainties dominate
the fit, which may not even agree with the outer measurements.
Therefore some degree of subjectivity is involved, where the radial region used in the fit
is determined by the widest range that still appears consistent with a smooth power-law trend,
which typically occurs outside $\sim$~2.5~\Reff\ for our sample.
The permitted range of power-law exponents $\gamma$ is then determined by a $\Delta \chi^2=1$ criterion.
In some cases the formal range of $\gamma$ is large and we adopt a plausibility prior
that the rotation velocity increases no more rapidly than linearly with radius
(i.e. $\gamma < 1$).
Note that although there may be lingering concerns about the robustness of these
extrapolation procedures, half of the galaxies in our detailed sample have data out to
$\sim$~8--10~\Reff, in which case the extrapolation is relatively unimportant 
for the \js\ calculation.

\begin{figure}
\centering{
\includegraphics[width=2.85in]{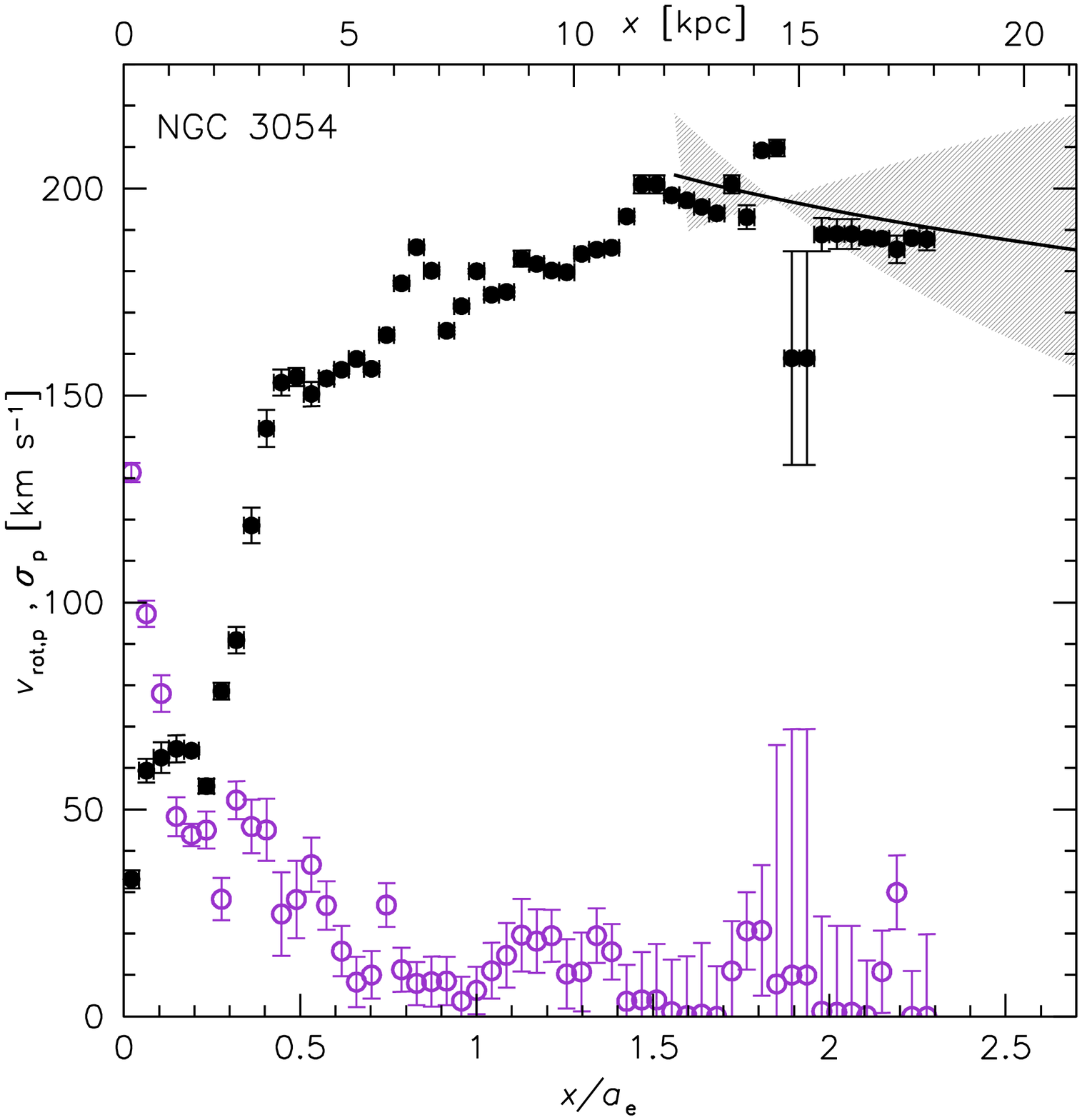} 
\includegraphics[width=2.85in]{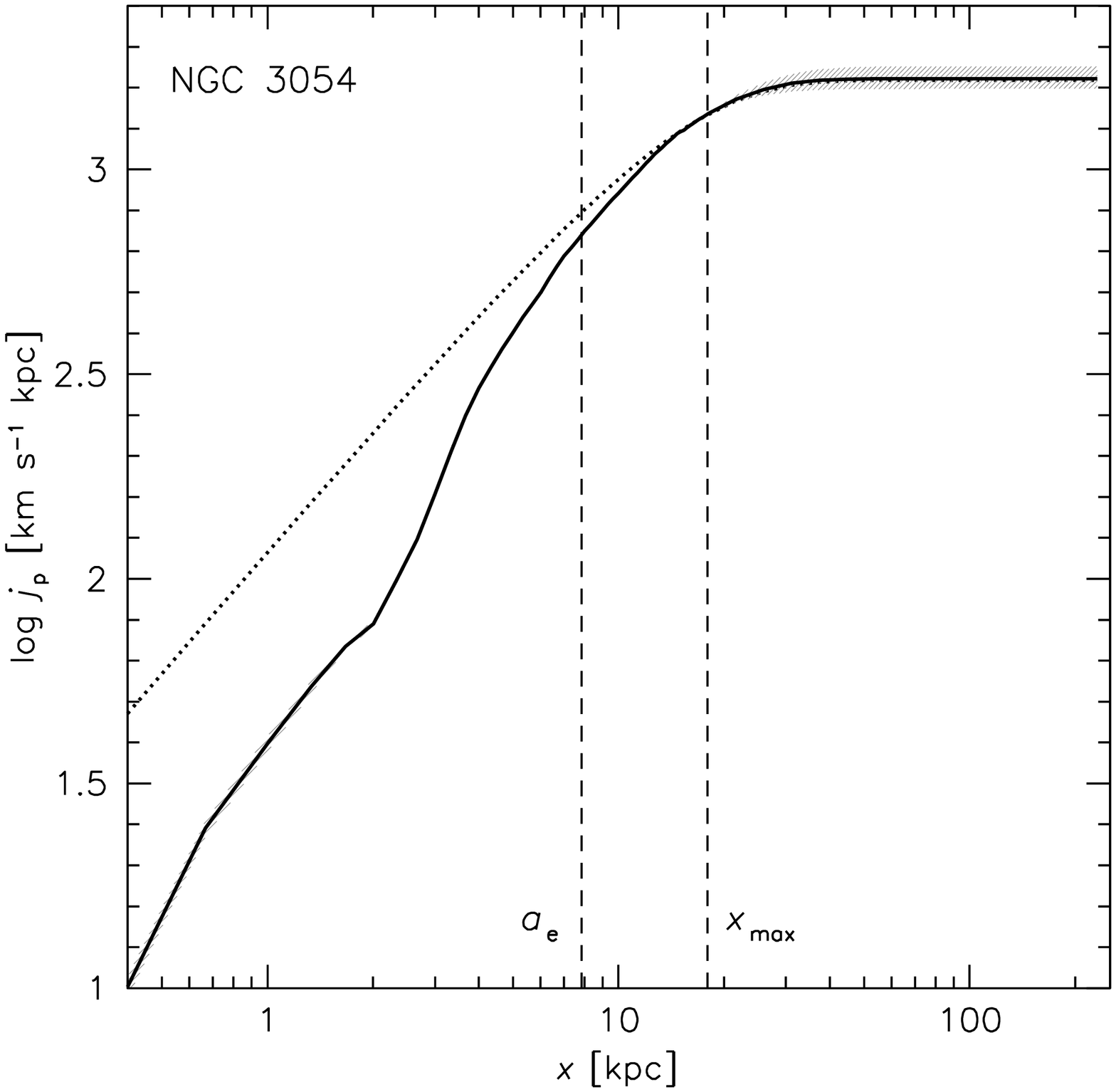} 
\vskip 20pt
\includegraphics[width=2.85in]{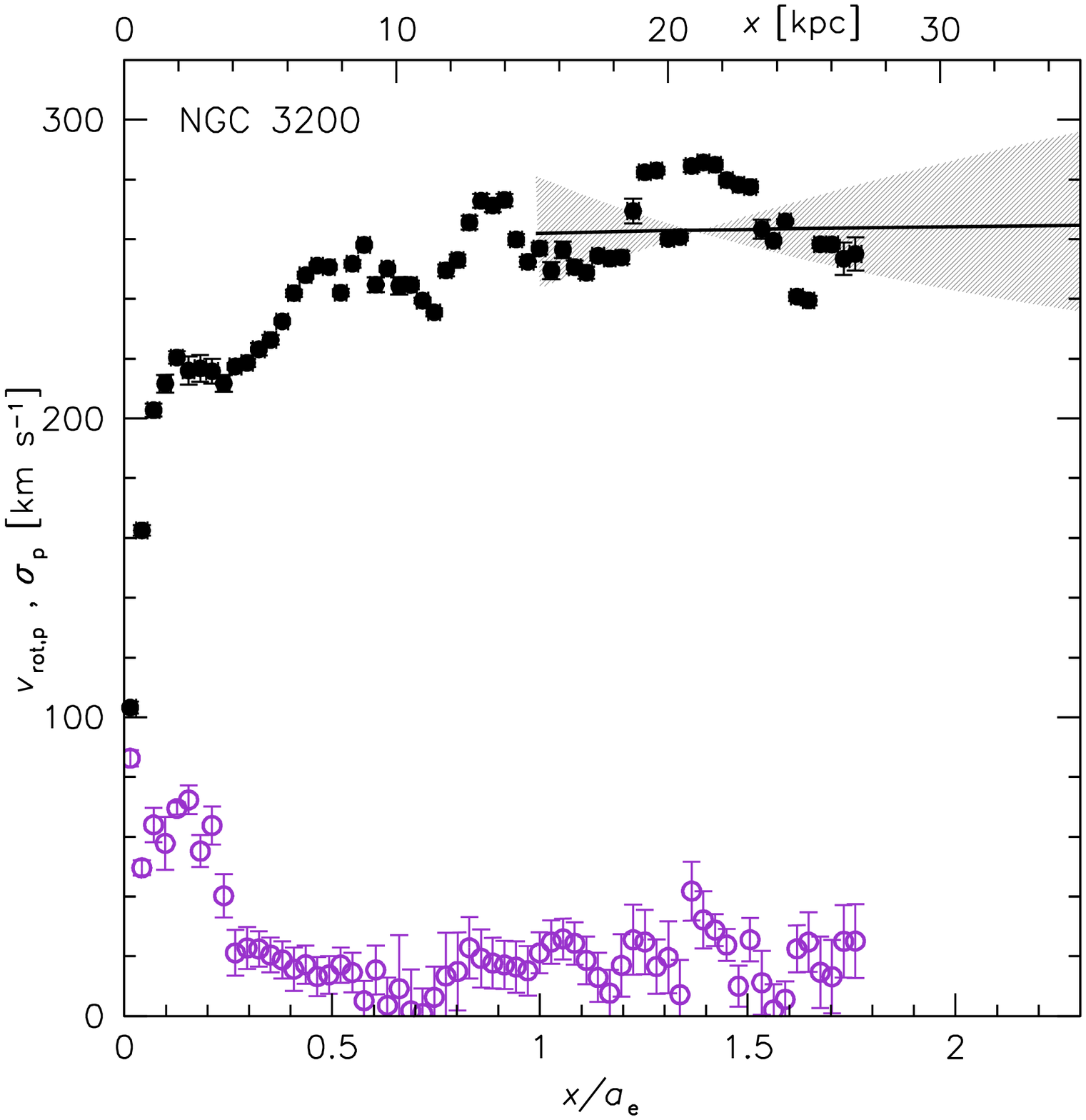} 
\includegraphics[width=2.85in]{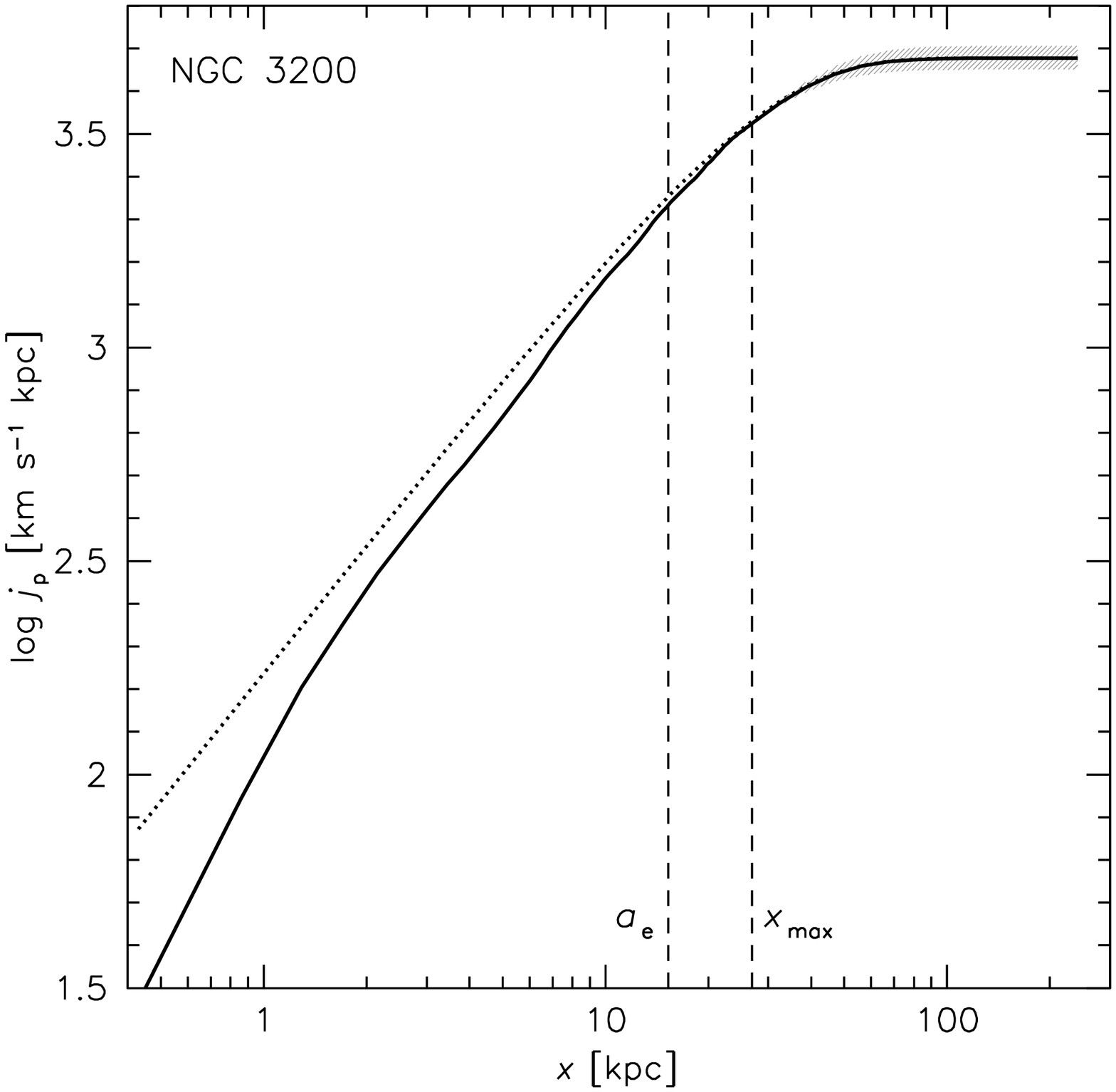} 
}
\caption{Kinematic results for the spiral galaxies NGC~3054 ({\it top panels})
and NGC~3200 ({\it bottom panels}).
The {\it left-hand} panels show data for the profiles of projected rotation velocity
 (black filled circles)
and velocity dispersion (purple open circles) vs.\ semi-major axis radius.
The black curve with shaded region shows the outer power-law extrapolation of
the rotation velocity and its estimated uncertainty.
The {\it right-hand} panels show the data converted into a profile of cumulative projected specific
angular momentum vs.\ radius (solid curve),
with the shaded regions showing the uncertainties
(barely visible for this galaxy).
The dotted curve shows the expected distribution of $j_{\rm p}(<x)$ 
for a constant rotation-velocity profile.
Vertical dashed lines show the effective radius of the galaxy and the maximum
extent probed by the kinematic data.
\label{fig:grid3054}
}
\end{figure}

\begin{figure}
\centering{
\includegraphics[width=2.85in]{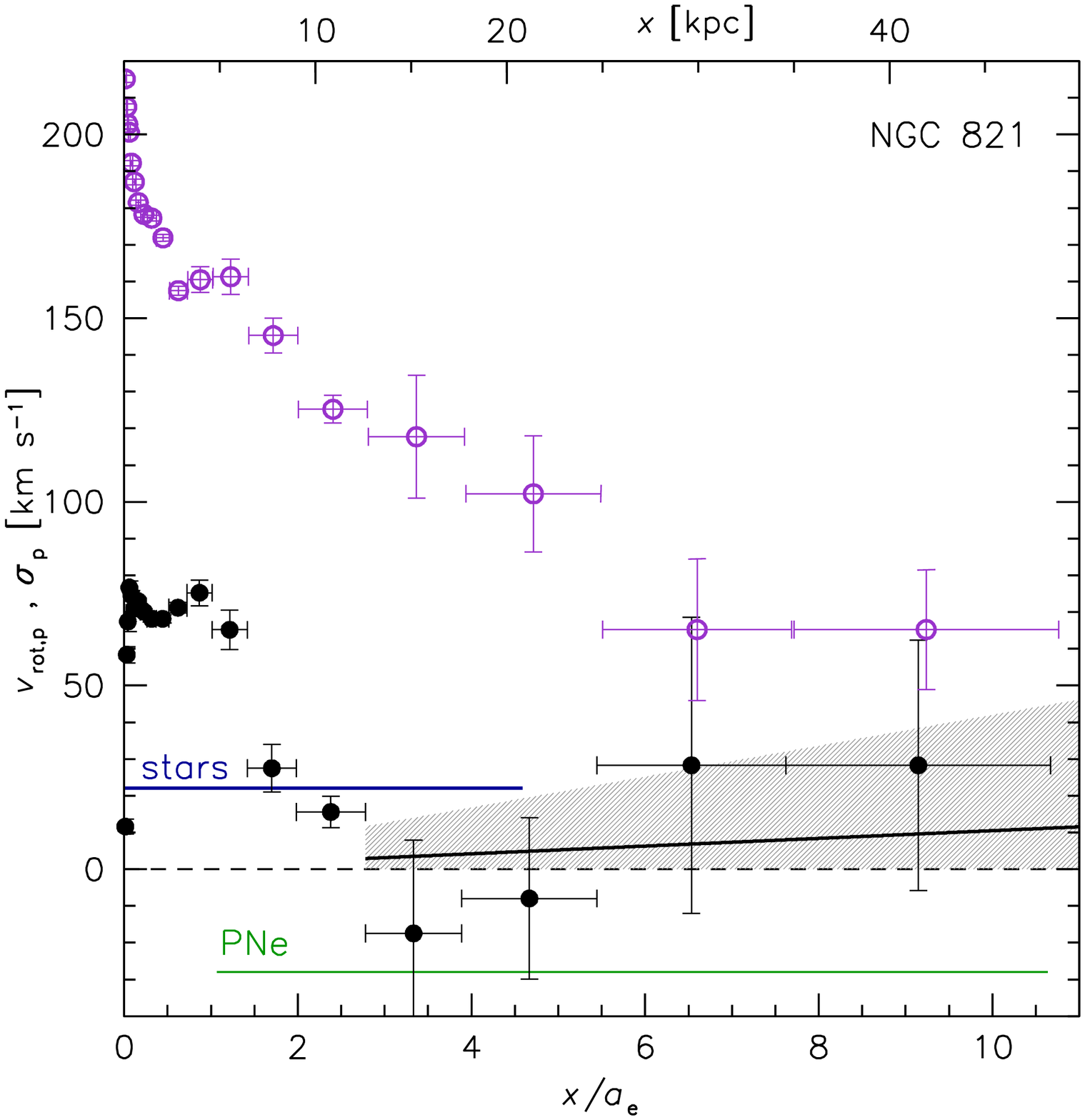} 
\includegraphics[width=2.85in]{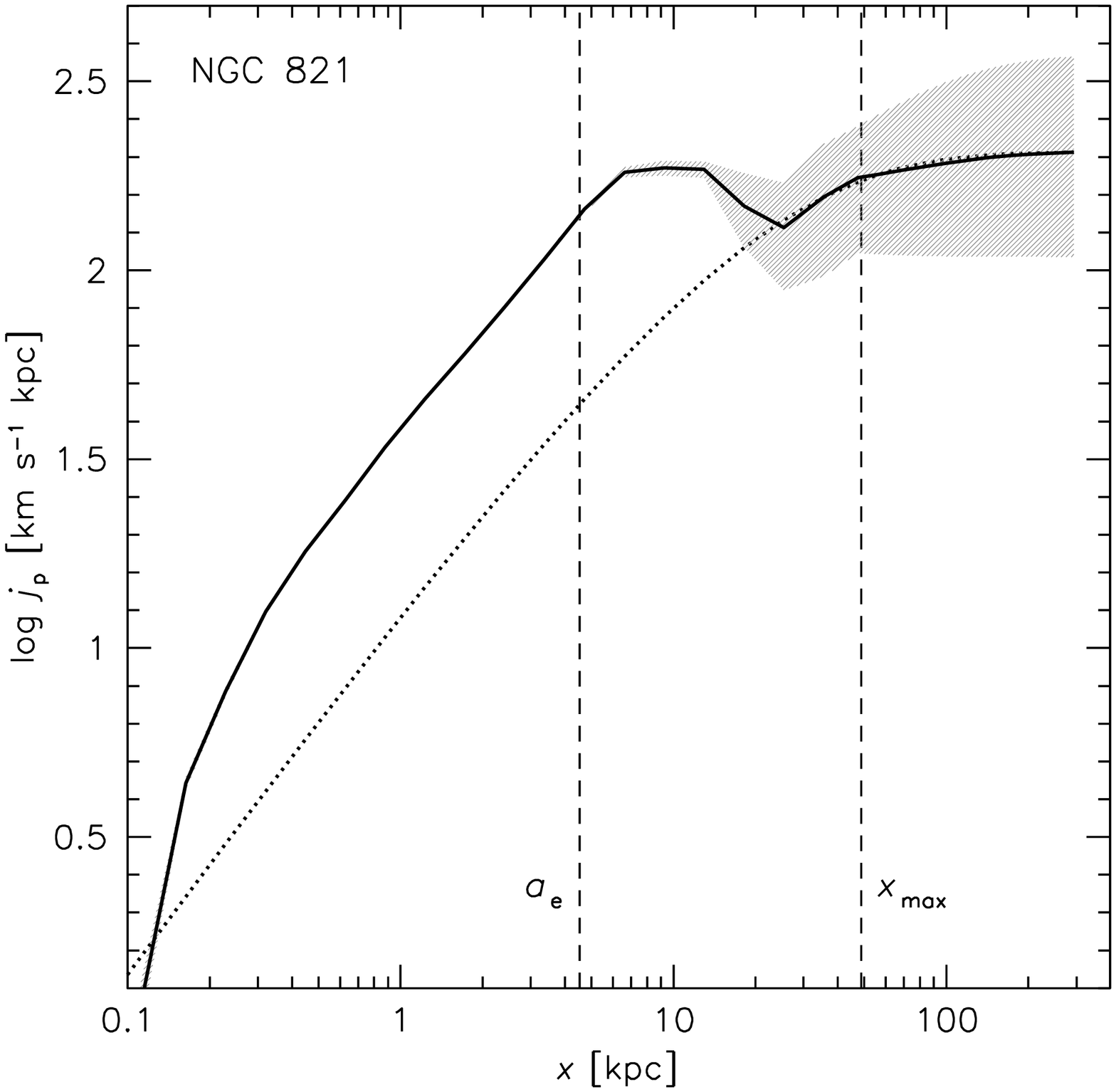} 
\vskip 20pt
\includegraphics[width=2.85in]{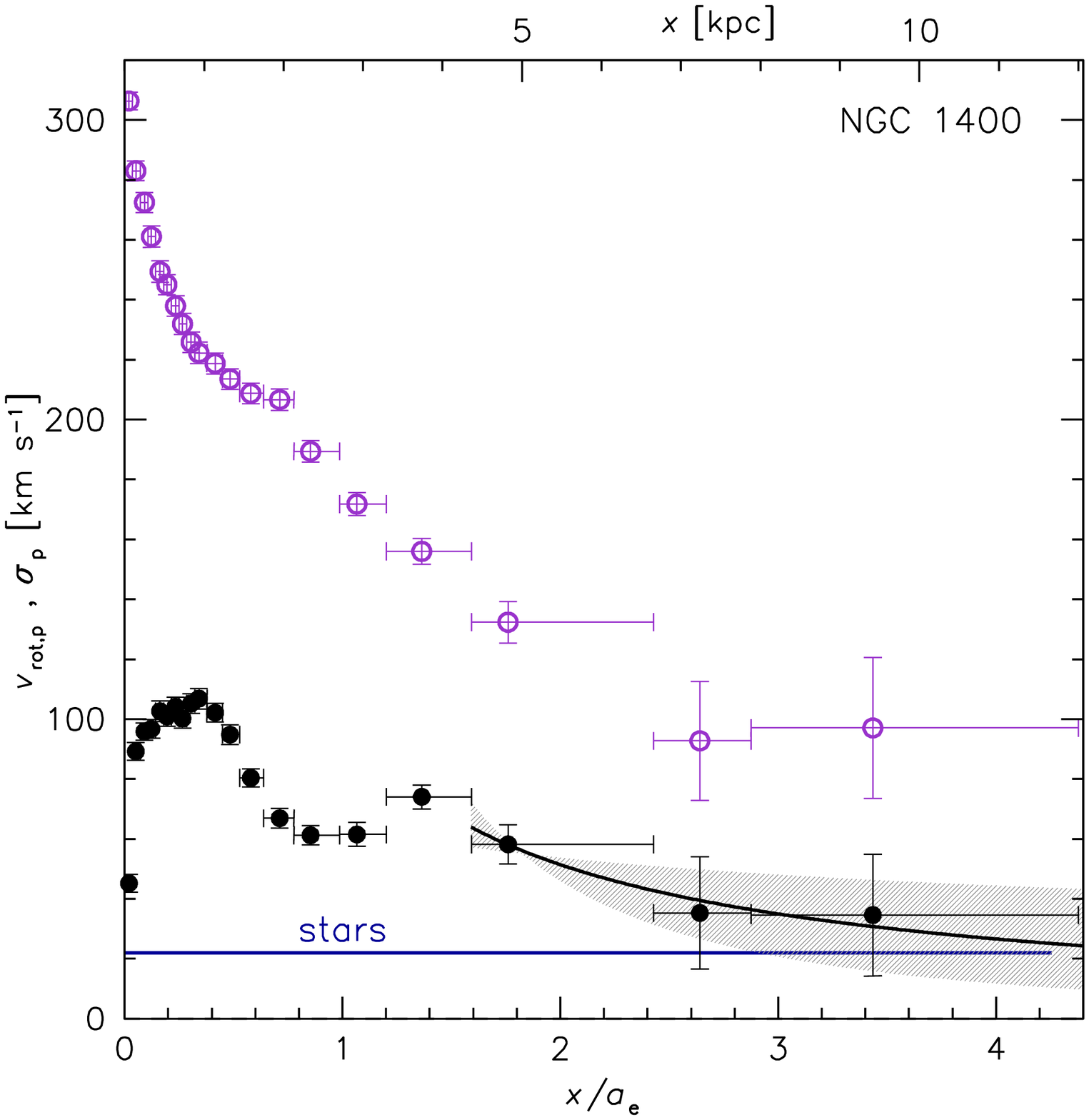} 
\includegraphics[width=2.85in]{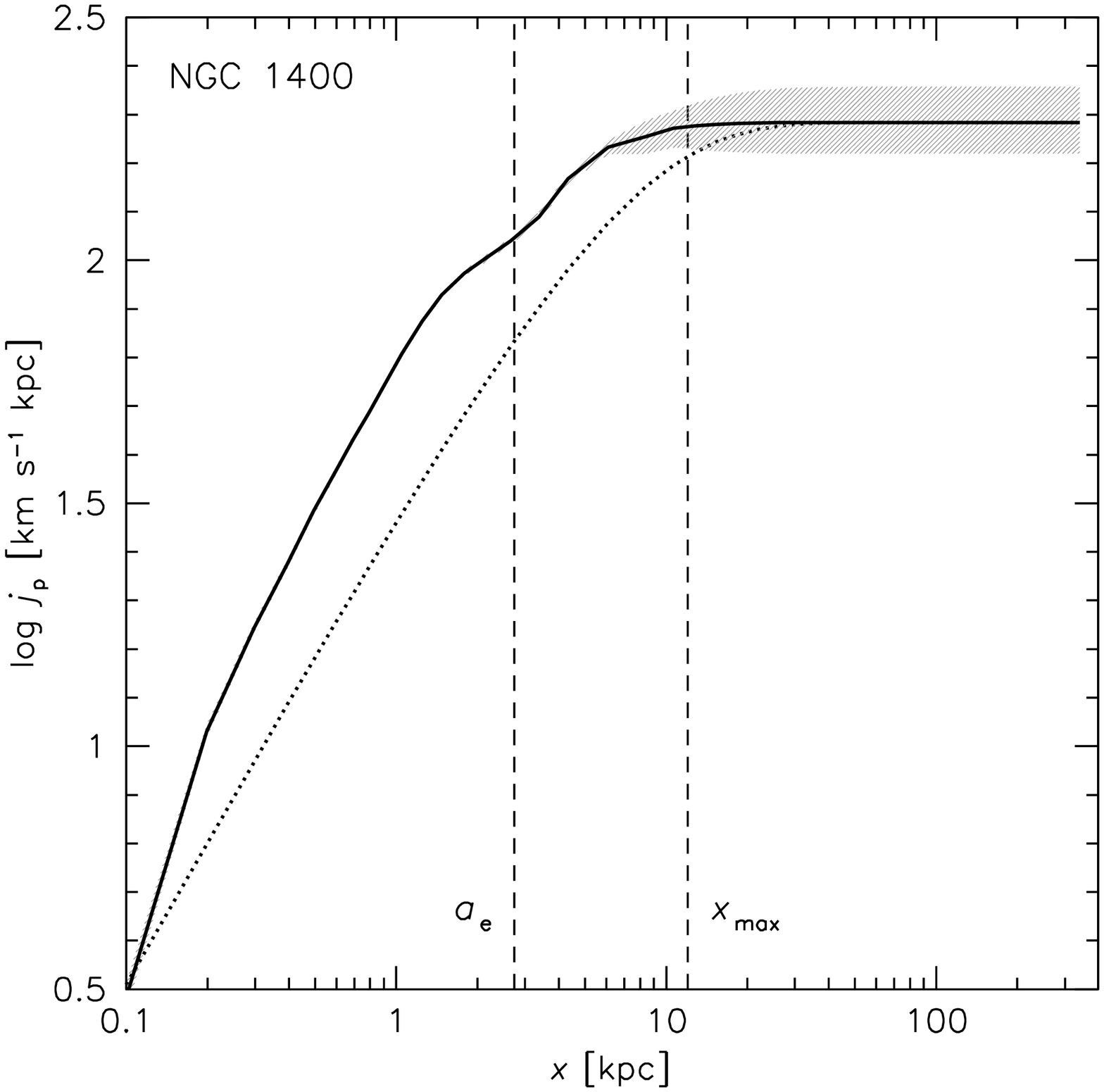} 
}
\caption{Kinematic results for the early-type galaxies NGC~821
({\it top panels}) and NGC~1400 ({\it bottom panels}).
See Figure~\ref{fig:grid3054} for explanations.
In addition, horizontal lines in the left-hand panels
indicate the radial ranges spanned by the different data sets.
Note also the dashed line at zero rotation velocity in the case of NGC~821.
\label{fig:grid821}
}
\end{figure}

\begin{figure}
\centering{
\includegraphics[width=2.85in]{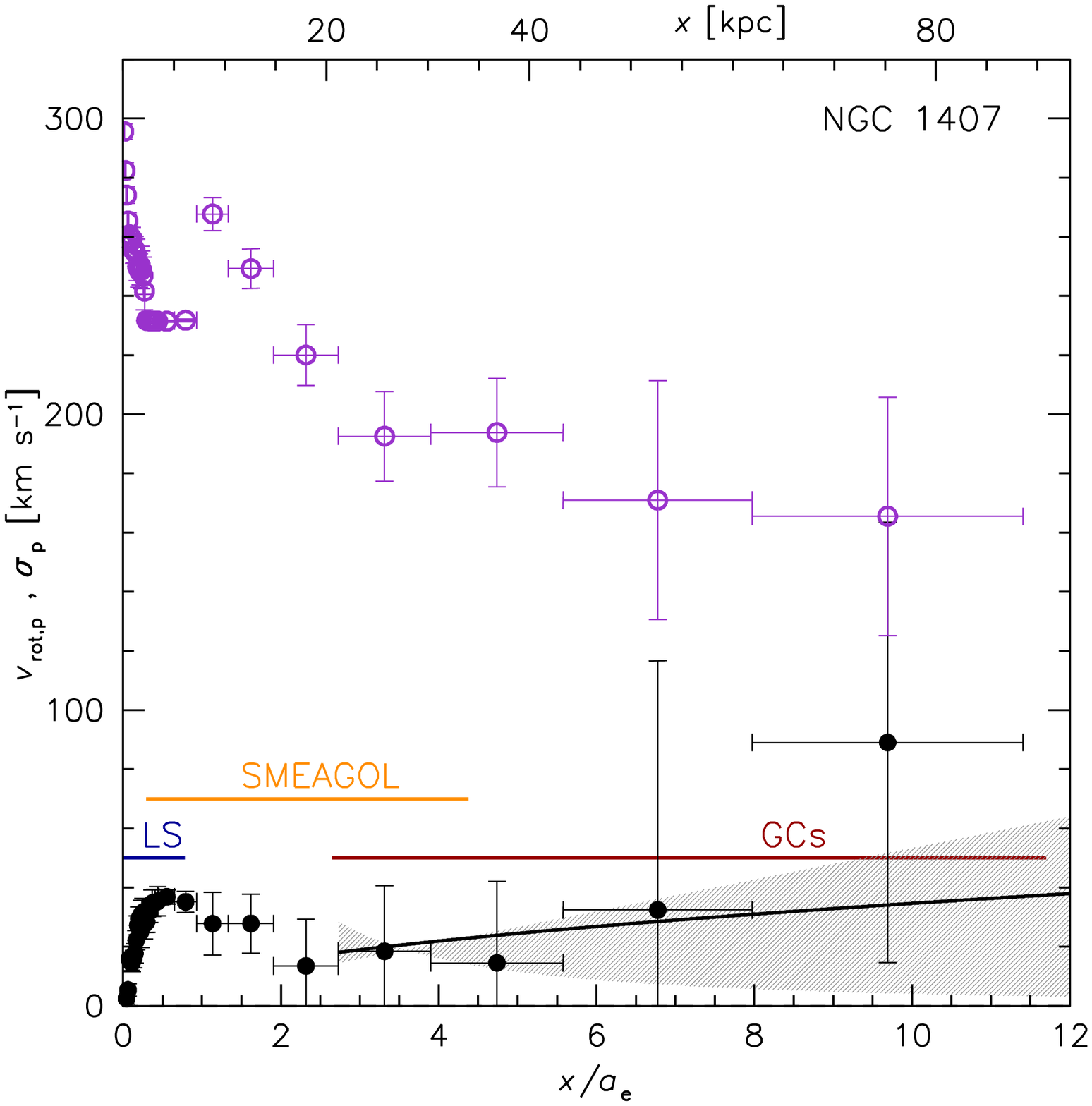} 
\includegraphics[width=2.85in]{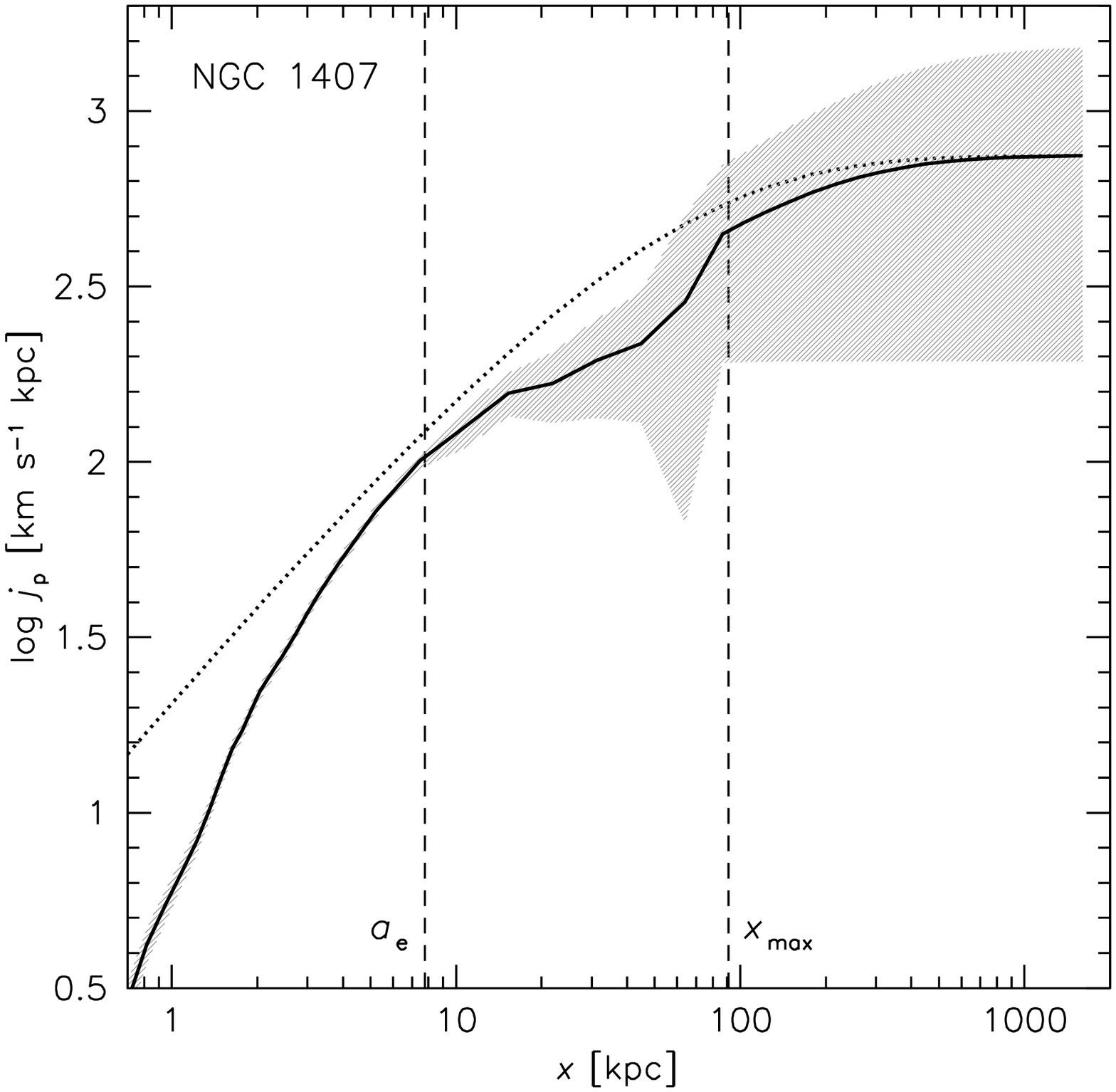} 
\vskip 20pt
\includegraphics[width=2.85in]{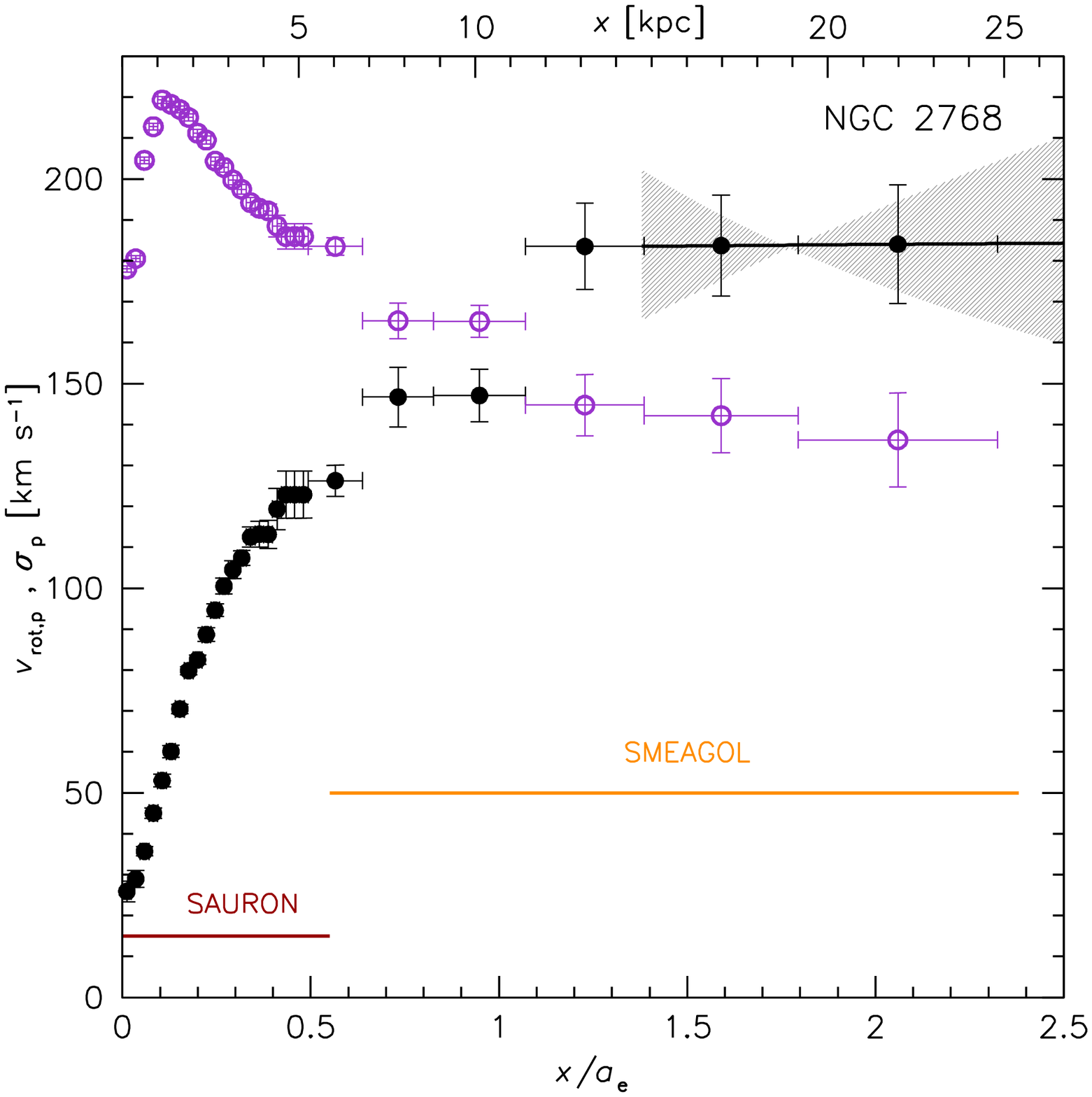} 
\includegraphics[width=2.85in]{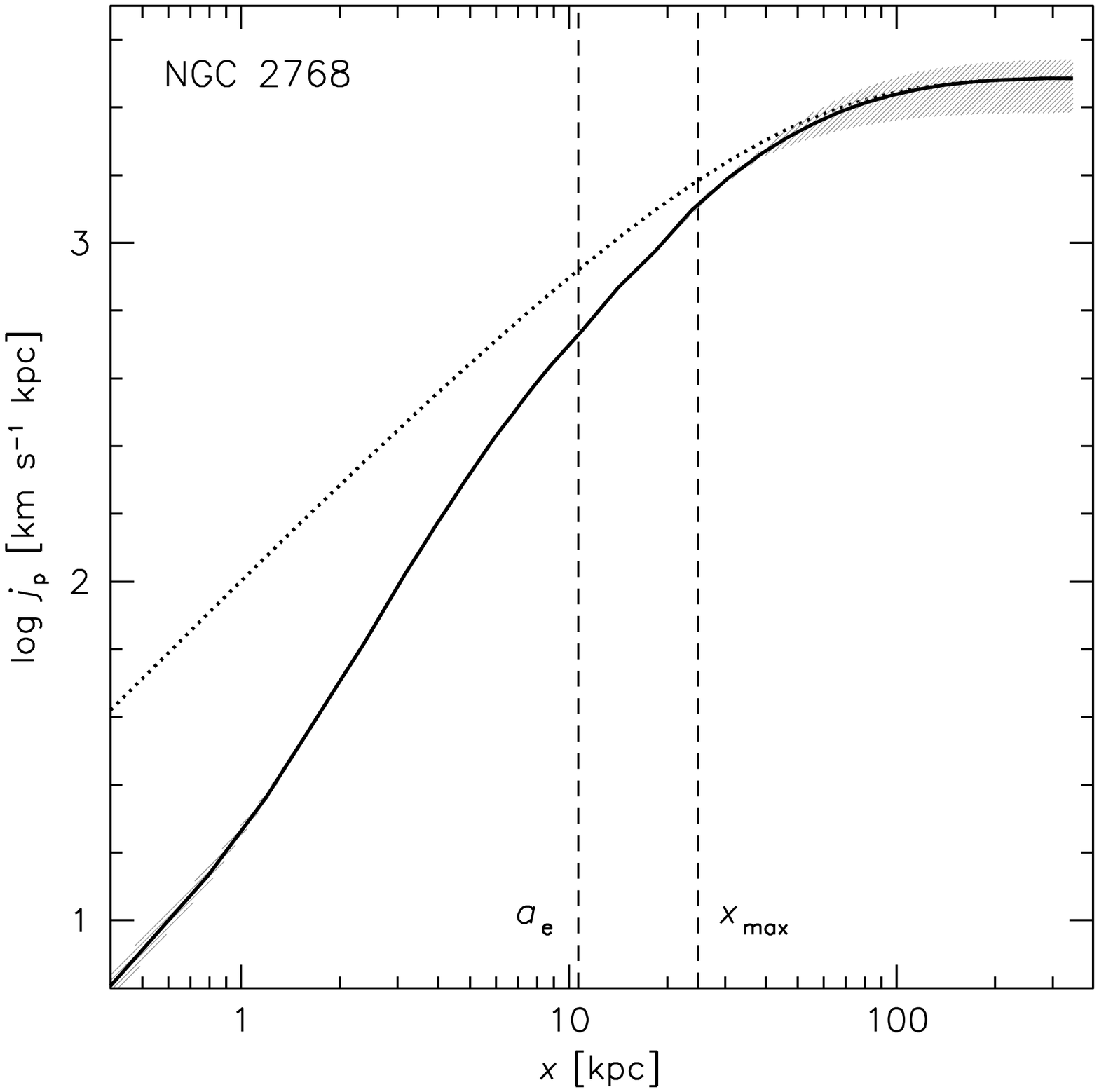} 
}
\caption{Kinematic results for the early-type galaxies NGC~1407
({\it top panels}) and NGC~2768 ({\it bottom panels}).
See Figures~\ref{fig:grid3054} and \ref{fig:grid821} for explanations.
\label{fig:grid1407}
}
\end{figure}

\begin{figure}
\centering{
\includegraphics[width=2.85in]{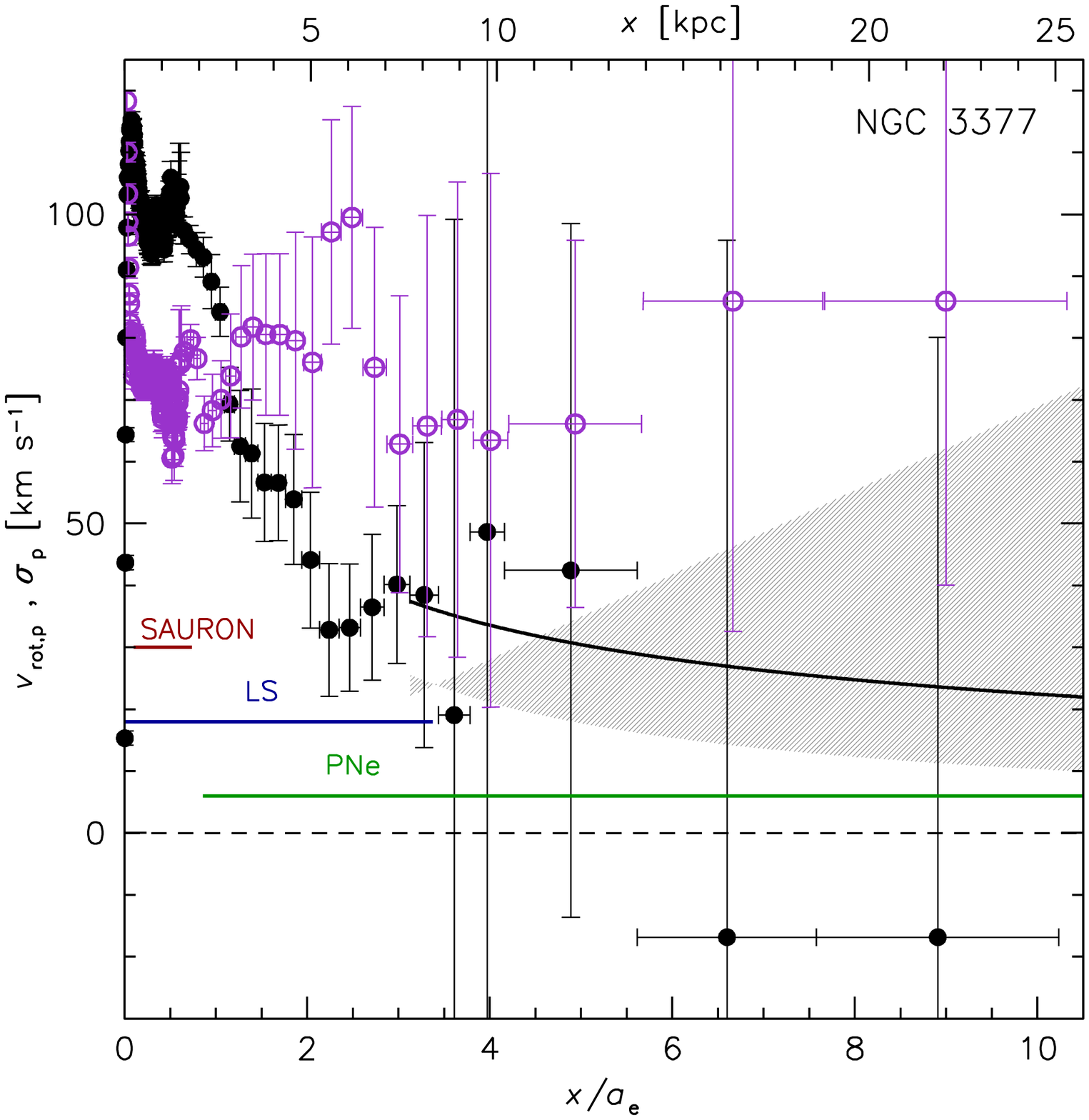} 
\hskip 15pt
\includegraphics[width=2.85in]{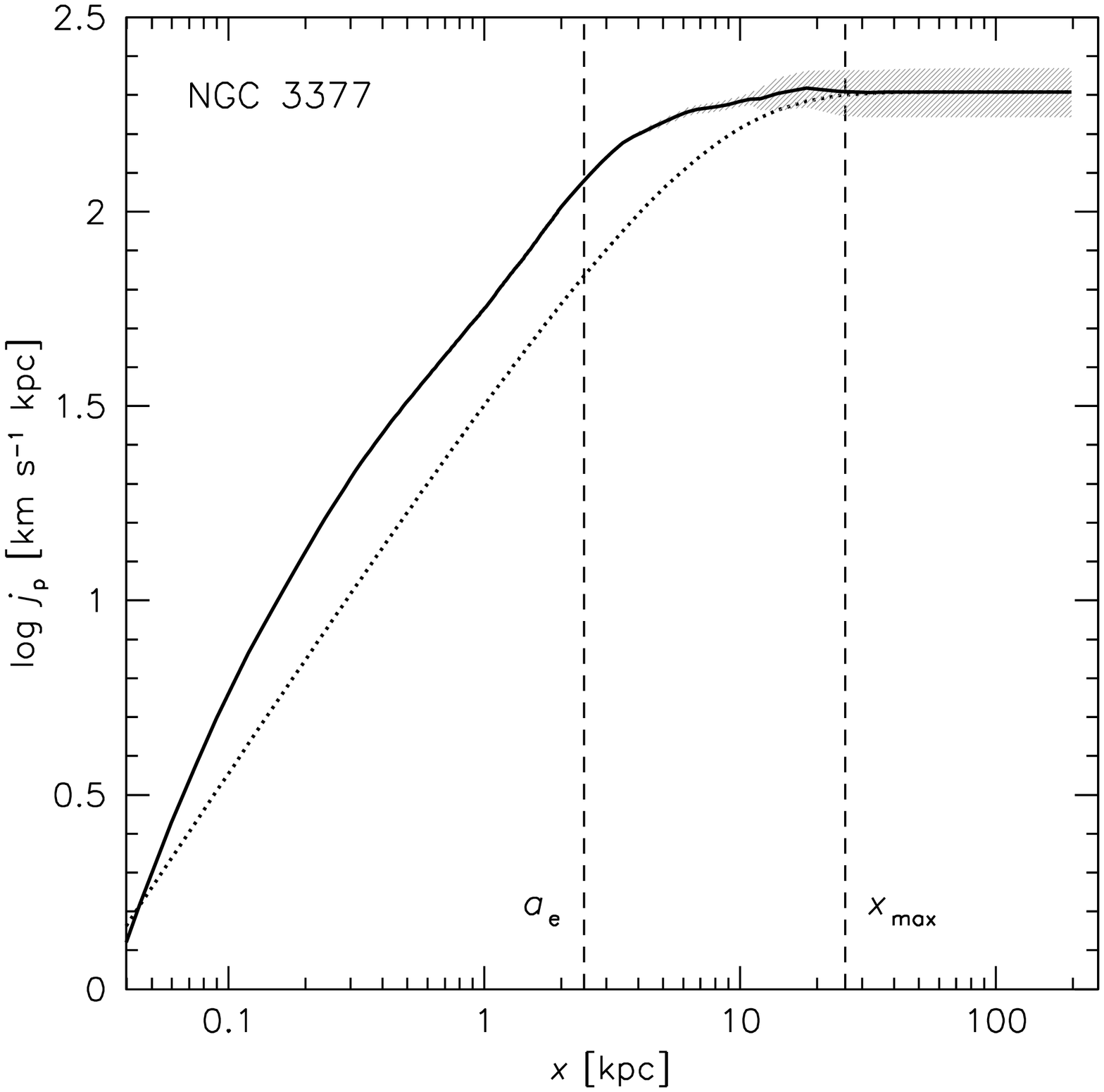} 
\vskip 20pt
\includegraphics[width=2.85in]{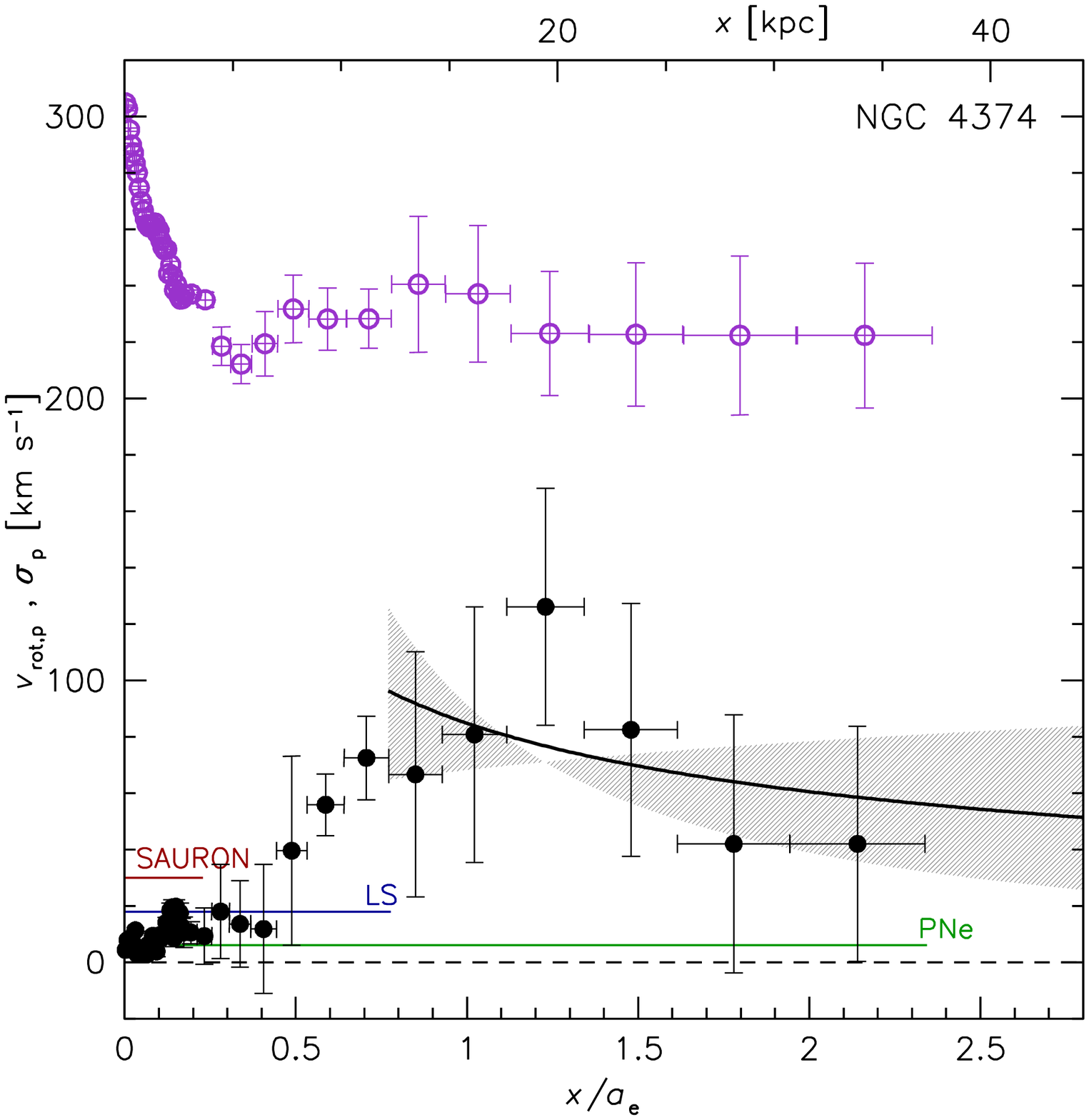} 
\hskip 15pt
\includegraphics[width=2.85in]{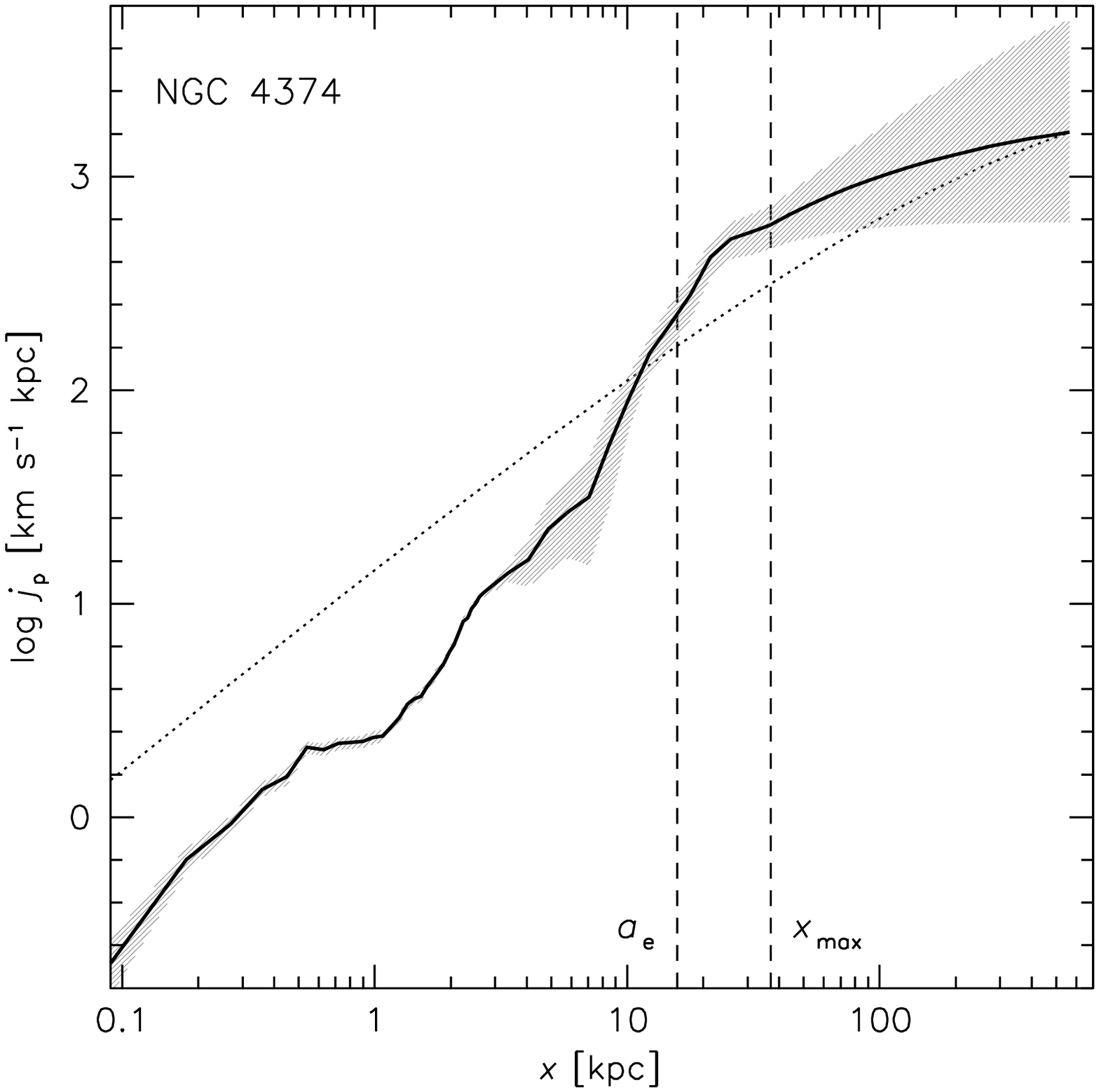} 
}
\caption{Kinematic results for the early-type galaxies NGC~3377
({\it top panels}) and NGC~4374 (M84; {\it bottom panels}).
See Figures~\ref{fig:grid3054} and \ref{fig:grid821} for explanations.
\label{fig:grid3377}
}
\end{figure}

\begin{figure}
\centering{
\includegraphics[width=2.85in]{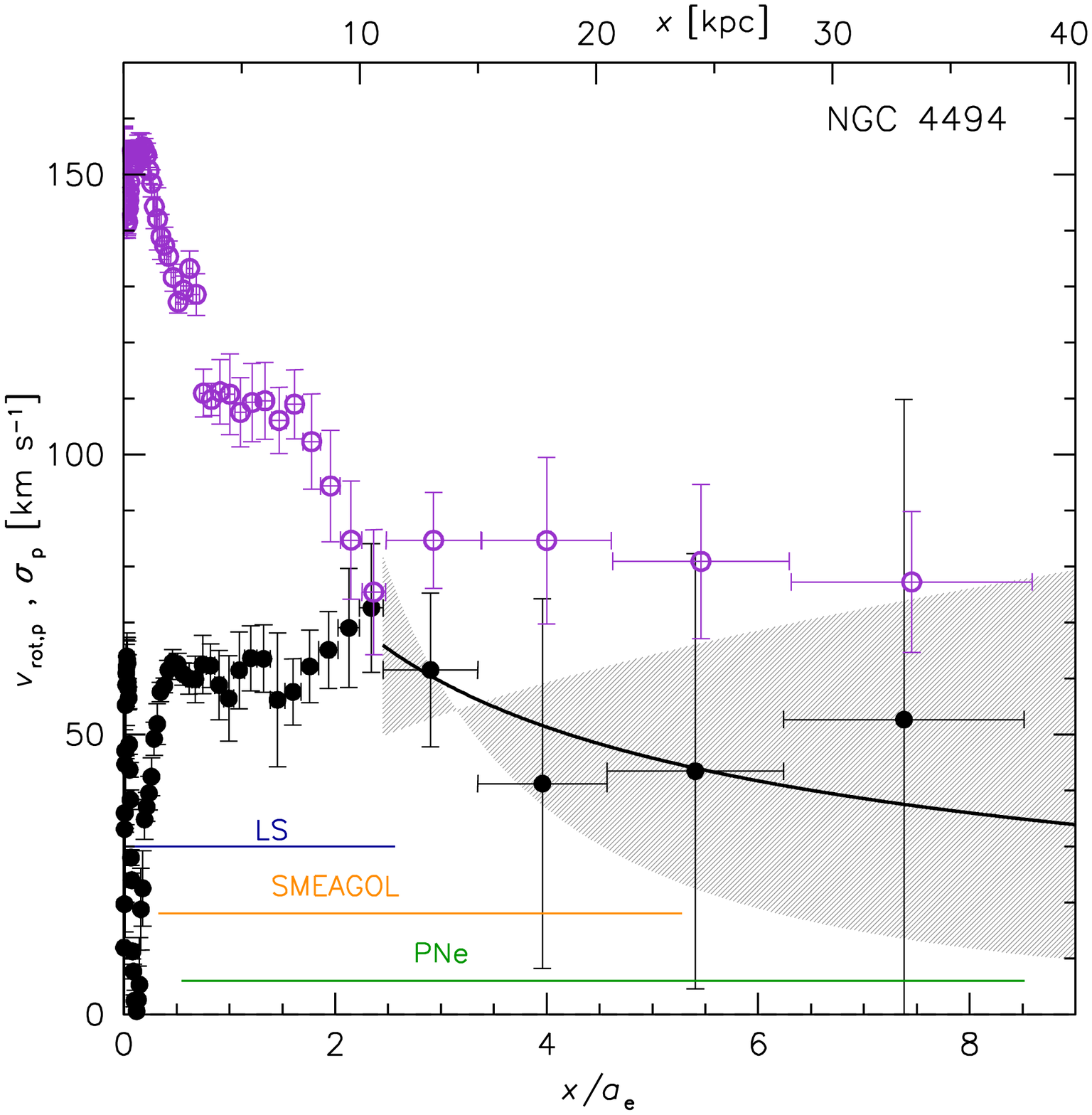} 
\hskip 15pt
\includegraphics[width=2.85in]{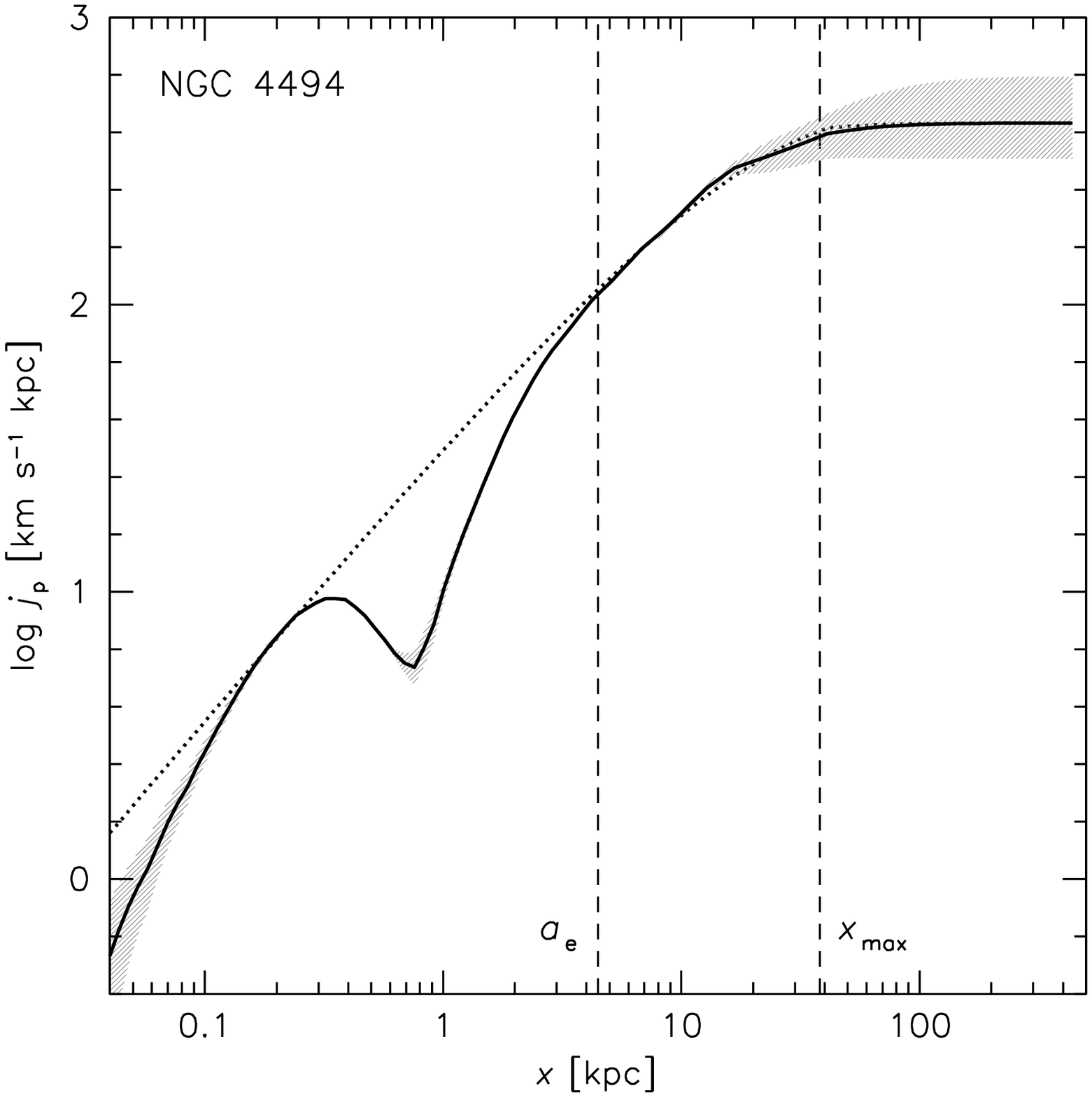} 
\vskip 20pt
\includegraphics[width=2.85in]{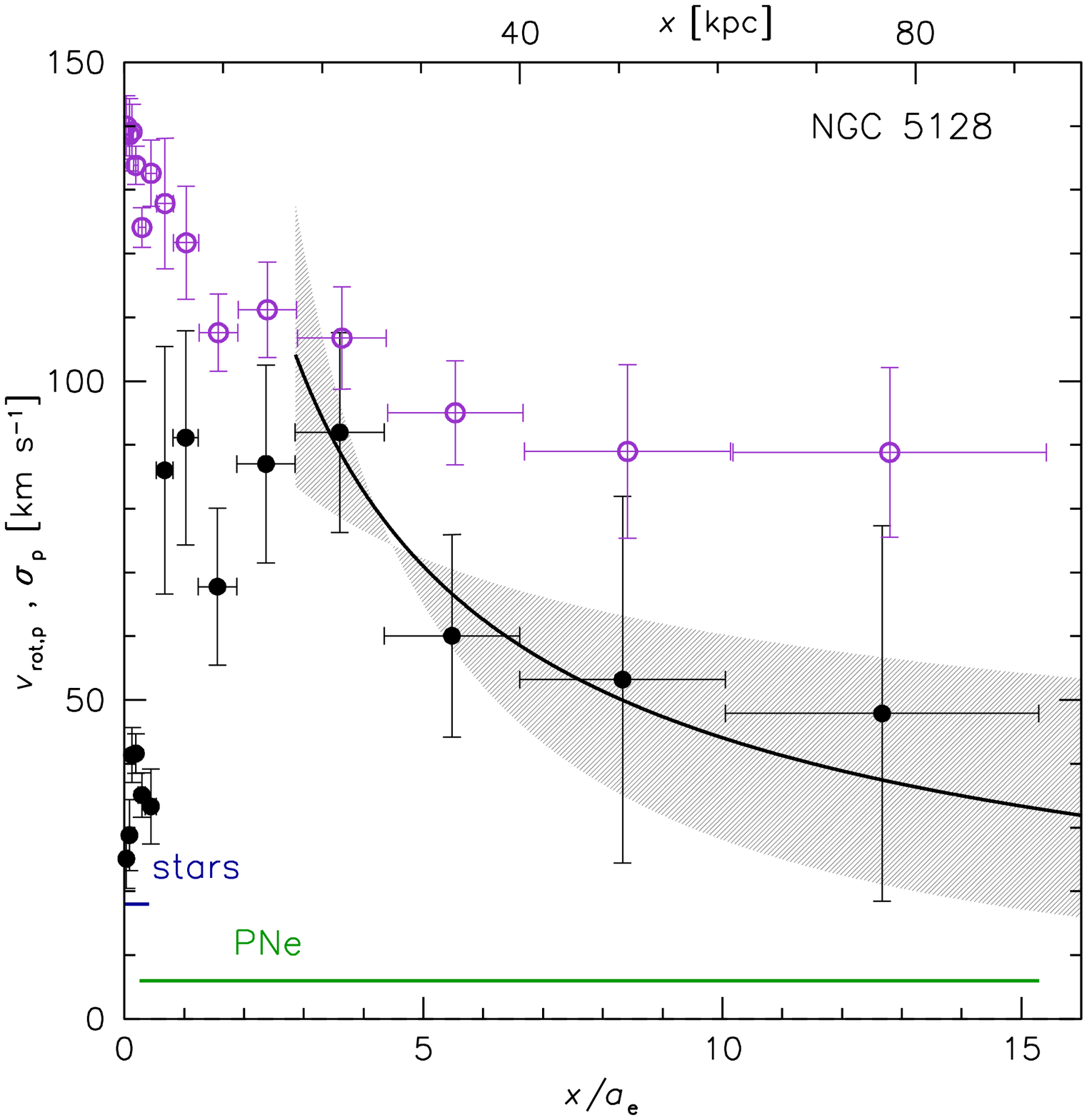} 
\hskip 15pt
\includegraphics[width=2.85in]{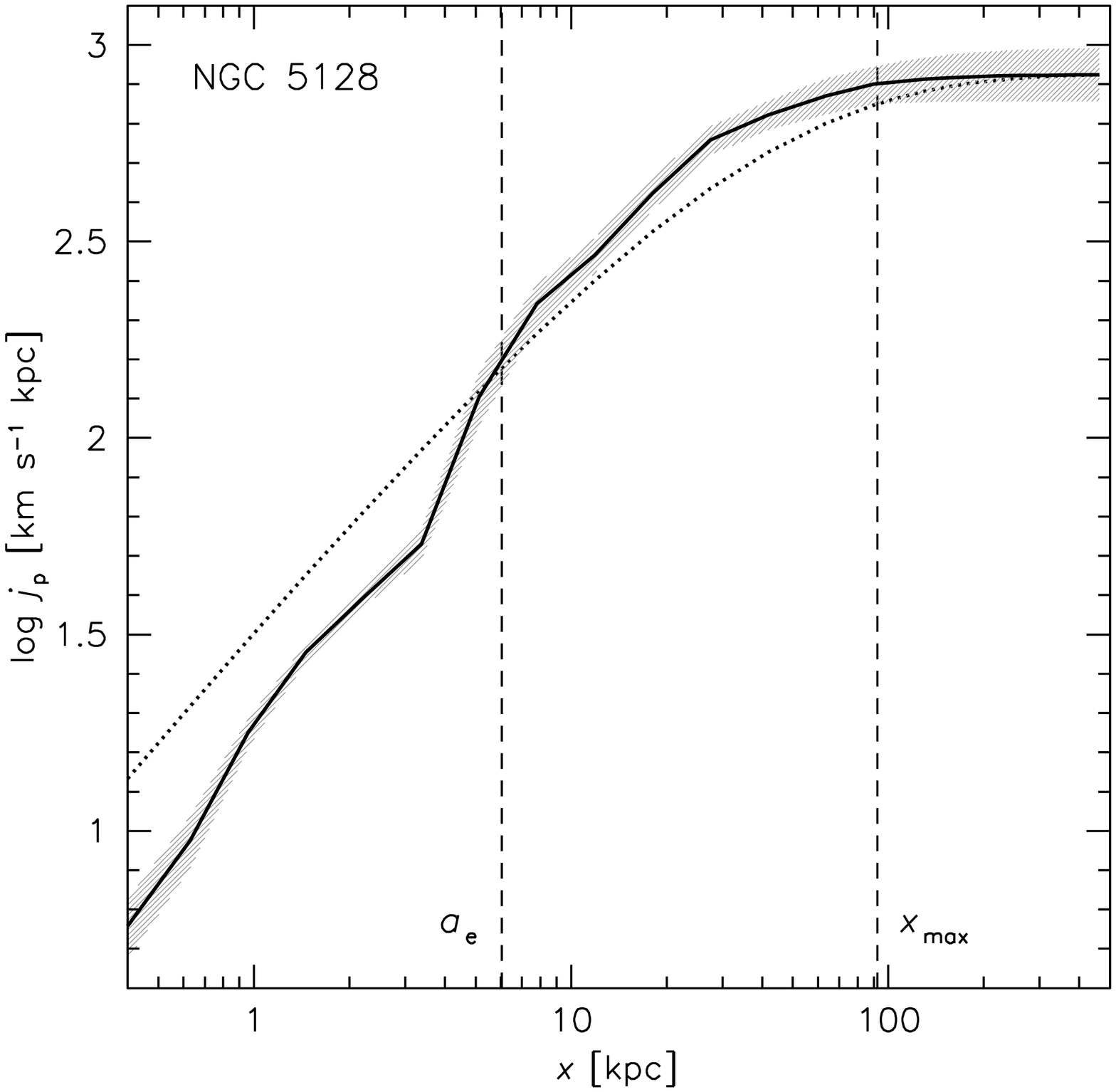} 
}
\caption{Kinematic results for the early-type galaxies NGC~4494
({\it top panels}) and NGC~5128 ({\it bottom panels}).
See Figures~\ref{fig:grid3054} and \ref{fig:grid821} for explanations.
\label{fig:grid4494}
}
\end{figure}

The kinematic data for the eight early-type galaxies that we model in detail are described
in the remainder of this section, followed by plots of the kinematic profiles in the left
panels of Figures~\ref{fig:grid821}--\ref{fig:grid4494}. Here the velocity
dispersion profiles are included for comparison with the rotation-velocity profiles,
but are not otherwise used in this paper.
For completeness, we also include the two spiral galaxies that we model in detail
as described in Section~\ref{sec:disk}
(Figure~\ref{fig:grid3054}).

We convert the rotation velocity data 
into profiles of projected specific angular momentum \jp\
using the methods of Section~\ref{sec:gen} and Appendix~\ref{sec:form}.
These are also presented
in Figures~\ref{fig:grid3054}--\ref{fig:grid4494} (right-hand panels).
We summarize the relevant observational and modeling parameters for these
galaxies in Table~B\ref{tab:gal}.
Note that in many cases from this sample, new GC kinematics data were recently obtained by
\citet{Pota12}, but are not incorporated here.

\vskip 0.3cm

\noindent
{\it NGC 821}---Isolated, disky elliptical.
Stellar kinematics data to $\sim$~4~\aeff\ are combined in a single kinematic model,
drawing on a number of sources including SAURON, SMEAGOL, HET, GMOS and MDM
\citep{2003ApJ...596..903P,2005MNRAS.362..857P,2009MNRAS.398..561W,2009MNRAS.398...91P,2010ApJ...716..370F}.
PN data are taken from \citet{2009MNRAS.394.1249C}, who used non-parametric kinematic mapping techniques
and found a strong kinematic misalignment relative to the stellar kinematics.
We find the same result, such that the PNe between $\sim$~1 and 2~\aeff\
are strongly {\it counter}-rotating with respect to the stars.
At larger radii, the PNe and stars agree, and we suspect some kind of
contamination in the intermediate-radius PN sample (cf.\ \citealt{2006AJ....131..837S}).
However, the PN rotation velocity along the major axis is still marginally consistent
(given the errors) with the stellar rotation velocity outside 1~\aeff, so we include these PN
data anyway since the overall results are dominated by the stellar data in these regions.
The final composite profile has a remarkably low outer rotation velocity,
as emphasized by \citet{2009MNRAS.398...91P}.

\vspace{0.2cm}

\noindent
{\it NGC 1400}---Apparently round early-type galaxy, probably a member of the NGC~1407 group.
It may be a lenticular seen near face-on.  The data include major-axis long-slit spectra
out to $\sim$~1~\aeff, and SKiMS covering $\sim$~1--3~\aeff\
from SMEAGOL \citep{2009MNRAS.398...91P}.
As Proctor et al. discussed, there is an unexplained velocity dispersion discrepancy 
between these two stellar-light data sets, but the rotation-velocity profiles are consistent.

\vspace{0.2cm}

\noindent
{\it NGC 1407}---Bright, round elliptical at the center of the Eridanus~A group.
Stellar kinematics data out to $\sim$~3~\aeff\ come from major-axis long-slit spectra
and from SMEAGOL \citep{2009MNRAS.398...91P}.
Kinematics data for 55 metal-rich GCs extend to $\sim$~12~\aeff\
\citep{2009AJ....137.4956R}, where we have removed objects brighter than $i=21.3$
owing to their peculiar kinematics, which may imply contamination by ultra compact dwarfs
(cf.\ \citealt{2011AJ....142..199B}).
There is also a mismatch between the stellar and GC velocity dispersions around
$\sim$~1~\aeff\ that could be caused by additional contamination from substructure or
from the metal-poor GC subpopulation, but the overall rotation-velocity profile is not affected.
We adopt $q=1$ and a fixed $\phi_0=250^\circ$ for all of the kinematic modeling.

\vspace{0.2cm}

\noindent
{\it NGC 2768}---Flattened E/S0 in the field;
we have overruled the RC3 classification of E6 in favor of other
classifications from the literature, including the RSA, as S0.
Stellar kinematics data out to $\sim$~2~\aeff\
are drawn from SAURON and SMEAGOL (see left-hand panel of Figure~\ref{fig:N4494kin1}).
From our detailed early-type galaxy sample, this is the case with the most strongly
increasing $v_{\rm rot,p}(x)$ profile.
However, the kinematic data are still somewhat limited, and 
there are preliminary indications from more radially extended data for a 
decreasing rotation-velocity profile (A. Romanowsky et al., in preparation).
Note that the extrapolation of the data outwards can reach an unphysical
 rotation velocity, exceeding a plausible circular velocity of
$\sim$~300~\kms, but restricting the maximum rotation velocity does not substantially
alter the permitted range of \jp.

\vspace{0.2cm}

\noindent
{\it NGC 3377}---Disky, low-luminosity elliptical in a loose group (Leo I).
The kinematic data are from SAURON inside 0.7~\aeff, 
deep long-slit VLT/FORS2 major axis spectra to 3.4~\aeff\ \citep{2009MNRAS.394.1249C},
and 112 PN velocities between 1 and 12~\aeff\ \citep{2009MNRAS.394.1249C}.
In comparing data-sets in their overlap regions (including other long-slit
kinematics from the literature: 
Bender et al. 1994; Kormendy et al. 1998; Halliday et al. 2001; Prugniel \& Simien 2002)
we discovered some fairly large discrepancies, particularly in the velocity dispersion.
Assuming that the higher spectral resolution of the FORS2 results makes them more
reliable, we therefore subtract 59~\kms\ in quadrature from the SAURON dispersions,
and multiply the SAURON rotation velocities by a factor of 1.13.
The impact of this issue on rotation velocity and \jp\ is not large, but we note that
characterizations of low-mass galaxies in general using quantities such as 
$v_{\rm rot}/\sigma_{\rm p}$
should be viewed with caution since they could be strongly affected by 
systematic problems, as in NGC~3377.
There is some indication of a $\sim 20^\circ$ kinematic twist in the outer
parts of this galaxy, where the true rotation velocity could be $\sim 60$~\kms\
rather than $\sim 20$~\kms.

\vspace{0.2cm}

\noindent
{\it NGC 4374 (M84)}---Bright, slow-rotator elliptical, apparently in a
subgroup along with M86 falling in to the Virgo cluster.
The kinematic data include SAURON inside 0.2~\aeff, deep long-slit VLT/FORS2
major and minor axis spectra to 0.8~\aeff\ \citep{2009MNRAS.394.1249C}, and 450 PN velocities 
between 0.1 and 2.3~\aeff\ \citep{2011MNRAS.411.2035N}.
As with NGC~3377 (above), we found discrepancies between the SAURON and FORS2
data, and subtracted 85~\kms\ in quadrature from the SAURON velocity dispersions.
Both the FORS2 and PN data show a sudden kinematic twist from near major axis alignment
inside 65\arcsec\ to near minor axis alignment at larger radii. (We recovered this
information from the FORS2 data by fitting a two-dimensional kinematic model to the major and
minor axis data.)  This minor-axis rotation was also apparent in the PN analysis of
\citet{2009MNRAS.394.1249C},
although the isophotes also twist so that by $\sim$~200\arcsec, the rotation is
actually along the major axis.
There may be further kinematic twisting outside $\sim$~250\arcsec\ such that
the rotation is again along the minor axis, with a velocity  of $\sim$~40~\kms.
Because of the strong twisting in this triaxial galaxy, and the fact that
$q=$~0.9--1.0 in the outer regions, we adopt the circularized \Reff{}~$=176^{\prime\prime}$
value for \aeff.
There is one other complication with a high-$n$ galaxy like NGC~4374,
as discussed in Section~\ref{sec:bulge}:
the cumulative \jp\ converges slowly with radius,
and so we choose the virial radius $r_{\rm vir} \sim 6095^{\prime\prime}$ of the galaxy
as the boundary for defining ``total'' enclosed \jp.

\vspace{0.2cm}

\noindent
{\it NGC 4494}---Round elliptical in the Coma I cloud.
The kinematic data include stellar long-slit spectroscopy to $\sim$~2~\aeff{}
\citep{2009MNRAS.393..329N},
stellar kinematics with multiple slits (SKiMS) to $\sim$~4~\aeff{} 
\citep{2009MNRAS.398...91P},
and PN velocities to $\sim$~9~\aeff{} \citep{2009MNRAS.393..329N,2009MNRAS.394.1249C}.
The newest SKiMS dataset from \citet{2011MNRAS.415.3393F} was not used, but would
yield essentially the same rotation-velocity profile;
note that this paper provided an inclination estimate of $i \sim \pi/4$.
An example of the PN data is shown in the right-hand panel of Figure~\ref{fig:N4494kin1}.

\vspace{0.2cm}

\noindent
{\it NGC 5128}---Peculiar early-type galaxy at only 4~Mpc distance.
The kinematic data include pioneering two-dimensional spectroscopy out to $\sim$~0.5~\aeff\ from
\citet{1986MNRAS.218..297W} and 780~PN velocities to $\sim$~15~\aeff\ from \citet{2004ApJ...602..685P}.
Analysis of the PN data does not show any strong kinematic twists with radius,
so for simplicity we use a fixed $\phi_0=259^\circ$ for the entire galaxy.
Also, because of the lack of detailed photometry for the galaxy, we set $q=1$.

\begin{table}
\begin{center}
\caption{Angular momenta of galaxies modeled in detail.\label{tab:gal}}
\noindent{\smallskip}\\
\begin{tabular}{l c c c c c c c c c c c}
\hline
\hline
NGC & Type & $D$ & $n$ & \aeff\ & $a_{\rm max}$ & $q$ & $\vs$ & \jp\ & $\log\left(\frac{M_\star}{M_\odot}\right)$ & Tracers & Ref. \\
& & (Mpc) & & (kpc) & (\aeff) & & (\kms) & (\kms~kpc)  & \\
\hline
\noindent{\smallskip}\\
3054 & Sb & 34.4 & 1 & 7.8 & 2.3 & 0.62 & 177 & $1670^{+110}_{-100}$ & 11.12 & H$\alpha$ & P+04a \\
3200 & Sc & 52.2 & 1 & 15.3 & 1.8 & 0.28 & 261 & $4750^{+330}_{-280}$ & 11.37 & H$\alpha$ & P+04a \\
821 & E6 & 23.4 & 3.4 & 4.5 & 10.7 & 0.60 & 23 & $210^{+160}_{-100}$ & 11.02 & IFU, LS, MS, PN & E+04, C+09, P+09,\\ & & & & & & & & & & & W+09a, FG10 \\
1400 & S0 & 25.7 & 1.9 & 2.7 & 4.4 & 0.89 & 48 & $190^{+40}_{-20}$ & 11.05 & LS, MS & P+09 \\
1407 & E0 & 28.1 & 4.3 & 7.8 & 11.7 & 1.0 & 39 & $750^{+760}_{-560}$ & 11.66 & LS, MS, GC & P+09, R+09 \\
2768 & S0 & 21.8 & 2.6 & 10.7 & 2.3 & 0.40 & 169 & $3060^{+410}_{-640}$ & 11.33 & IFU, MS & E+04, P+09 \\ 
3377 & E5 & 10.9 & 2.0 & 2.5 & 10.5 & 0.67 & 56 & $200\pm30$ & 10.50 & IFU, LS, PN & E+04, C+09 \\
4374 & E1 & 18.5 & 8.3 & 15.8 & 2.3 & 1.0 & 32 & $1610^{+3860}_{-1000}$ & 11.68 & IFU, LS, PN & E+04, C+09 \\
4494 & E1 & 16.6 & 3.2 & 4.5 & 8.5 & 0.84 & 58 & $430^{+190}_{-110}$ & 11.08 & LS, MS, PN & N+09, C+09, P+09 \\
5128 & S0 &  4.1 & 4 & 6.0 & 15.3 & 1.0 & 61 & $840^{+140}_{-120}$ & 11.28 & MS, PN & W+86, P+04b \\
\hline
\noindent{\smallskip}\\
\end{tabular}
\tablecomments{
Morphological Types are generally taken from the RC3 catalog \citep{1991trcb.book.....D}.
$\vs$ and \jp\ are calculated from the
full, detailed treatment of Equations~(\ref{eqn:JMp}) and (\ref{eqn:proxy}).
The distances $D$ are taken from surface brightness fluctuation analyses
\citep{2001ApJ...546..681T,2009ApJ...694..556B} where available, and otherwise from redshifts.
The S\'ersic parameters $n$ and \aeff\ are taken from various literature sources.
The tracers include H$\alpha$ gas-emission rotation-velocity curves;
long-slit stellar kinematics (LS);
multi-slit stellar kinematics (MS);
integral-field stellar kinematics (IFU);
planetary nebulae (PN); and metal-rich globular clusters (GC).
The references are
\citealt{1986MNRAS.218..297W} (W+86);
\citealt{2004MNRAS.352..721E} (E+04);
\citealt{2004A&A...424..447P} (P+04a);
\citealt{2004ApJ...602..685P} (P+04b);
\citealt{2009MNRAS.394.1249C} (C+09);
\citealt{2009MNRAS.393..329N} (N+09);
\citealt{2009MNRAS.398...91P} (P+09);
\citealt{2009AJ....137.4956R} (R+09);
\citealt{2009MNRAS.398..561W} (W+09a);
\citealt{2010ApJ...716..370F} (FG10).
}
\end{center}
\end{table}

\newpage
\clearpage

\newpage
\clearpage

\newpage
\clearpage

\newpage
\clearpage

\newpage
\clearpage

\clearpage
\newpage

\section{Appendix C: Supplemental observational information for full sample}\label{sec:obsfull}

Here we describe some details of the full sample of observational data,
with results reported 
in Tables~C\ref{tab:spirals} and C\ref{tab:etg}.
The methods for the spiral galaxies are
described in Section~\ref{sec:ltgdata} and do not require further elaboration.
For the early-types, in addition to the generalities in Section~\ref{sec:etgdata}, 
here we discuss details including
estimates of the S\'ersic index $n$, an evaluation of bias
in our galaxy sample selection, and some nuances of the size--mass relation.

\begin{figure}
\centering{
\includegraphics[width=3.5in]{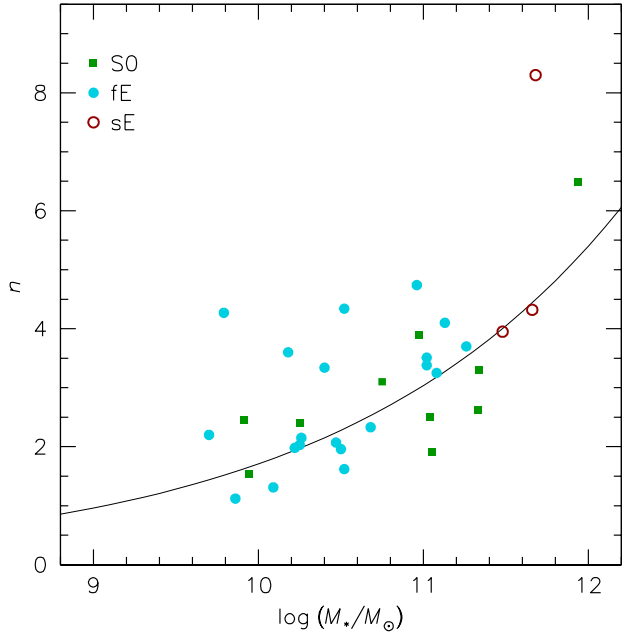} 
}
\caption{Relation between S\'ersic index $n$ and stellar mass \Ms\ for early-type galaxies.
Points show results from the literature for individual galaxies in our sample.
Symbol and colors indicate different galaxy types as in the legend;
fE and sE mean `fast rotator' and `slow rotator' ellipticals based on 
their central properties (see text for details).
The black curve shows our adopted mean relation, Equation~(\ref{eqn:nM}).
}
\label{fig:sersic}
\end{figure}

The issue with $n$ is that, as discussed in the main text, its value
is a significant factor in accurate $j$ calculation, but is not available for
some of the galaxies in our sample.  Instead, we use a statistical estimate based
on trends among other galaxies.
Figure~\ref{fig:sersic} shows the $n$--\Ms\ correlation
for the early-type galaxies in our sample that {\it do} have S\'ersic fits available.
There is a correlation between \Ms\ and $n$ that echos other trends
in the literature, and in particular matches the power-law slope of $\sim0.3$ found
by \citet{2003AJ....125.2936G}.
The mean relation that we adopt is
\begin{equation}\label{eqn:nM}
n = 5.4\times10^{-3} \times (M_\star/M_\odot)^{0.25}.
\end{equation}
There is a scatter of $\sigma_n \sim 1$ about this relation,
which is also similar to variations found in the literature for different S\'ersic fits
of the same galaxies.
This uncertainty in $n$ translates to an uncertainty of $\sim$~25\% in $k_n$ 
[see Equation~(\ref{eqn:kn})], and thus $\sim$~0.1~dex uncertainty in $j$
[Equation~(\ref{eqn:jCK0})].

Next, to investigate 
whether or not our galaxy sample provides an unbiased, fair representation
of the nearby universe,
we could check a variety of different galaxy parameters, but
given our focus on angular momentum, the most relevant parameter is the
readily-measured central stellar rotation velocity.

Our point of comparison is the \atlas3d\ survey, which presented
photometric and kinematic properties for a complete, volume-limited sample of 260 nearby 
early-type galaxies, selected only by $K$-band luminosity \citep{2011MNRAS.413..813C}.
We use their tabulated central rotation metric, $(v/\sigma)_{\rm e/2}$, 
which is flux-weighted in two dimensions within an aperture of $R_{\rm e}/2$ 
\citep{2011MNRAS.414..888E}.
From the galaxies in common between our samples, we calibrate this metric to
our own major-axis kinematic data measured locally at $a_{\rm e}/2$.
We find that the \atlas3d\ values of $(v/\sigma)_{\rm e/2}$ should be
multiplied by a factor of $\sim$~2.2 to match our data
(which, as expected, is a large difference because of the two-dimensional versus major-axis
measurement locations).

We next divide $(v/\sigma)$ from both data sets
 by a factor of $[\epsilon/(1-\epsilon)]^{1/2}$ to derive
a rotation-dominance parameter $(v/\sigma)^*$.
This is analogous to the standard parameter in Equation~(\ref{eqn:binney}),
but is constructed differently for the sake of convenient comparisons
between the two data sets, and should not be used to make detailed
inferences about the properties of these galaxies.

\begin{figure}
\centering{
\includegraphics[width=3.5in]{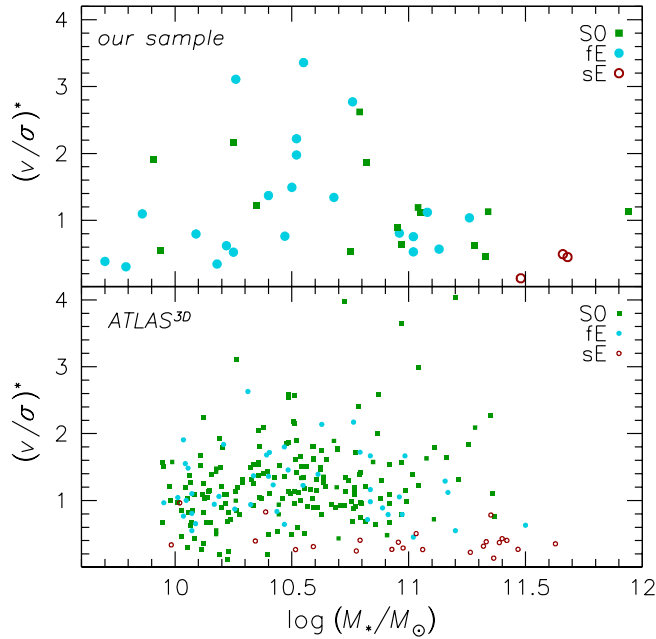} 
}
\caption{Rotation dominance parameter vs.\ stellar mass for early-type
galaxies (see text for details).  
Our galaxy sample is shown in the top panel, and the \atlas3d\ sample in the
bottom panel. Symbols show different galaxy types 
as in the legend.
Our sample is biased toward ellipticals and more massive galaxies but otherwise appears
to provide a fair representation of the central kinematics of nearby
early-types.
Note that the fraction of ``slow rotators'' is highest for the 
most massive systems.
} \label{fig:atlas1}
\end{figure}

We plot $(v/\sigma)^*$ versus stellar mass in Figure~\ref{fig:atlas1}.
Our sample follows a similar trend to the unbiased sample,
with central rotation becoming systematically less important for the
more massive galaxies.  Both samples span a similar range in mass.

Another important issue here is the inclusion of different sub-classes
of early-type galaxies.  Our sample consists of 15 lenticulars and 25 ellipticals, which
implies a strong selection bias for ellipticals, since
$\sim$~75\% of nearby early-types are S0s \citep{2011MNRAS.413..813C}.

In Section~\ref{sec:scale} we compare
the size--mass relation of our sample (both the early- and late-types) with 
the \atlas3d results.  For \atlas3d, we adopt the power-law relations
provided by \citet{2011MNRAS.413..813C}, after adapting them
to the parameters that we plot for our sample.
For the spirals, we convert
from circularized \Reff\ to \aeff\ by assuming a median inclination of $60^\circ$.
For the fast-rotator E/S0s, we use a median axis ratio of $q=0.63$.

Comparison with our sample (in Figure~\ref{fig:size}) 
is not straightforward since we have used {\it disk}
sizes rather than total galaxy sizes (bulge+disk) for the spirals.
In any case, we find that our results are consistent overall with \atlas3d\, but
there appears to be a difference for the lower-mass ellipticals;
at $\log\,(M_\star/M_\odot)\sim$~10--10.5, these galaxies 
are $\sim$~30\% more compact in our sample than the trend found by \atlas3d.
Examining several galaxies in common between the samples, we find
that the discrepancy persists for cases where we have made use of high-quality 
photometric analyses \citep{2009ApJS..182..216K}.   Figure~8 of \citet{2011MNRAS.413..813C}
suggests that for the smaller galaxies, the \atlas3d\ sizes, which are partially 
based on NIR measurements, are overestimated relative to optically-based sizes.
This illustrates the type of systematic uncertainties remaining
in our analysis, and also of the comfortably low impact that they have 
on our \js--\Ms\ analyses (only 
$\sim \pm$~0.05~dex or $\sim \pm$~15\% in \js).

One additional issue is that our sample of spirals was selected to avoid
strong bars which might in principle introduce a bias in \js.
Inclusion of barred spirals could be an improvement made in a next generation,
volume-limited \js--\Ms\ survey.

\clearpage
\newpage

\LongTables

\begin{deluxetable}{l l r r c r l l c r r r r r r}
\tablewidth{0pt}
\tabletypesize{\footnotesize}
\tablecaption{Angular momenta of spiral galaxies\label{tab:spirals}}
\tablehead{
\noalign{\smallskip}
\noalign{\smallskip}
\noalign{\smallskip}
Name & Type & $T_{\rm Hubble}$ &  $D$ & $B/T$ & $R_{\rm d}$ & $a_{\rm e,b}$ & $q_{\rm b}$ & $i$ & $v_{s,{\rm d}}$ & $v_{s,{\rm b}}$ & \jp\ & \jt\ & $\log\left(\frac{M_\star}{M_\odot}\right)$  \\
& & & (Mpc) & & (kpc) & (kpc) &  & (deg) & (\kms) & (\kms) &  (\kms\ & (\kms\ & & \\
& & & & & & & & & & & kpc) & kpc) & & }
\noalign{\smallskip}
\startdata
\noalign{\smallskip}
\noalign{\smallskip}
\noalign{\smallskip}
NGC 224 & Sb & 3.0 & 0.70 & 0.19 & 5.9 & 0.17 & 0.70 & 77 & 234 & 73 & 2230 & 2290 & 11.06 \\
\noalign{\smallskip}
NGC 247 & Sd & 6.9 & 3.8 & 0.00 & 4.1 & \nodata & \nodata & 76 & 92 & \nodata & 770 & 790 & 9.84 \\
\noalign{\smallskip}
NGC 300 & Sd & 6.9 & 1.8 & 0.00 & 1.6 & \nodata & \nodata & 47 & 60 & \nodata & 190 & 260 & 9.23 \\
\noalign{\smallskip}
NGC 701 & Sc & 5.0 & 24.3 & 0.00 & 1.9 & \nodata & \nodata & 61 & 127 & \nodata & 480 & 550 & 10.46 \\
\noalign{\smallskip}
NGC 753 & Sbc & 4.9 & 65.8 & 0.04 & 4.8 & 0.02 & 0.86 & 50 & 161 & 31 & 1480 & 1930 & 11.31 \\
\noalign{\smallskip}
NGC 801 & Sc & 5.3 & 77.3 & 0.21 & 13.1 & 1.1 & 0.75 & 86 & 205 & 56 & 4270 & 4280 & 11.50 \\
\noalign{\smallskip}
NGC 1024 & Sab & 2.4 & 46.8 & 0.34 & 6.2 & 0.09 & 0.56 & 68 & 252 & 101 & 2070 & 2230 & 11.34 \\
\noalign{\smallskip}
NGC 1035 & Sc & 5.0 & 16.4 & 0.00 & 1.3 & \nodata & \nodata & 71 & 126 & \nodata & 340 & 360 & 10.13 \\
\noalign{\smallskip}
NGC 1085 & Sbc & 3.6 & 89.9 & 0.22 & 6.0 & 0.2 & 0.90 & 39 & 194 & 68 & 1810 & 2900 & 11.42 \\
\noalign{\smallskip}
NGC 1087 & Sc & 5.2 & 20.1 & 0.00 & 3.0 & \nodata & \nodata & 50 & 103 & \nodata & 630 & 820 & 10.62 \\
\noalign{\smallskip}
NGC 1325 & Sbc & 4.0 & 20.9 & 0.03 & 4.2 & 0.3 & 0.82 & 70 & 169 & 56 & 1360 & 1450 & 10.61 \\
\noalign{\smallskip}
NGC 1353 & Sb & 3.3 & 20.0 & 0.15 & 2.8 & 1.0 & 0.74 & 65 & 205 & 34 & 990 & 1090 & 10.74 \\
\noalign{\smallskip}
NGC 1357 & Sab & 1.9 & 26.4 & 0.34 & 4.7 & 2.3 & 0.85 & 45 & 184 & 34 & 1210 & 1720 & 11.03 \\
\noalign{\smallskip}
NGC 1417 & Sb & 3.3 & 54.7 & 0.07 & 6.4 & 0.2 & 1.0 & 50 & 218 & 0 & 2580 & 3370 & 11.31 \\
\noalign{\smallskip}
NGC 1421 & Sbc & 4.1 & 27.7 & 0.05 & 6.0 & 4.0 & 0.73 & 78 & 192 & 79 & 2230 & 2280 & 10.93 \\
\noalign{\smallskip}
NGC 1620 & Sbc & 4.5 & 47.2 & 0.07 & 7.6 & 0.7 & 0.85 & 69 & 224 & 34 & 3150 & 3380 & 11.32 \\
\noalign{\smallskip}
NGC 2403 & Scd & 6.0 & 3.2 & 0.00 & 2.0 & \nodata & \nodata & 69 & 116 & \nodata & 470 & 500 & 10.01 \\
\noalign{\smallskip}
NGC 2590 & Sbc & 3.6 & 71.8 & 0.25 & 5.9 & 0.5 & 0.54 & 75 & 246 & 110 & 2220 & 2300 & 11.34 \\
\noalign{\smallskip}
NGC 2608 & Sb & 3.3 & 33.1 & 0.07 & 2.9 & 0.4 & 0.86 & 50 & 107 & 37 & 580 & 760 & 10.67 \\
\noalign{\smallskip}
NGC 2639 & Sa & 0.9 & 49.7 & 0.61 & 2.8 & 4.8 & 0.79 & 65 & 294 & 69 & 1110 & 1300 & 11.33 \\
\noalign{\smallskip}
NGC 2708 & Sb & 3.1 & 30.2 & 0.08 & 3.4 & 0.3 & 1.0 & 61 & 201 & 0 & 1250 & 1430 & 10.81 \\
\noalign{\smallskip}
NGC 2715 & Sc & 5.2 & 23.2 & 0.01 & 5.6 & 0.03 & 0.87 & 70 & 136 & 37 & 1520 & 1620 & 10.83 \\
\noalign{\smallskip}
NGC 2742 & Sc & 5.3 & 22.6 & 0.01 & 3.5 & 0.14 & 0.89 & 58 & 148 & 15 & 1030 & 1210 & 10.65 \\
\noalign{\smallskip}
NGC 2775 & Sab & 1.7 & 21.5 & 0.20 & 4.5 & 0.4 & 0.89 & 44 & 195 & 41 & 1400 & 2020 & 11.31 \\
\noalign{\smallskip}
NGC 2815 & Sb & 2.9 & 36.5 & 0.32 & 5.7 & 1.1 & 0.80 & 72 & 262 & 67 & 2070 & 2190 & 11.23 \\
\noalign{\smallskip}
NGC 2841 & Sb & 3.0 & 12.2 & 0.36 & 3.1 & 0.12 & 0.71 & 65 & 263 & 87 & 1070 & 1180 & 11.10 \\
\noalign{\smallskip}
NGC 2844 & Sa & 0.6 & 25.4 & 0.19 & 1.8 & 0.13 & 0.90 & 74 & 154 & 24 & 460 & 480 & 10.32 \\
\noalign{\smallskip}
NGC 2903 & Sbc & 4.0 & 7.4 & 0.00 & 2.4 & \nodata & \nodata & 65 & 185 & \nodata & 900 & 1000 & 10.74 \\
\noalign{\smallskip}
NGC 2998 & Sc & 5.2 & 70.1 & 0.02 & 5.8 & 0.05 & 0.81 & 62 & 185 & 29 & 2100 & 2380 & 11.13 \\
\noalign{\smallskip}
NGC 3031 & Sab & 2.4 & 3.6 & 0.15 & 2.7 & 0.05 & 0.82 & 59 & 197 & 50 & 920 & 1070 & 10.95 \\
\noalign{\smallskip}
NGC 3054 & Sb & 3.5 & 34.4 & 0.04 & 4.7 & 0.2 & 0.90 & 50 & 188 & 31 & 1680 & 2190 & 11.12 \\
\noalign{\smallskip}
NGC 3067 & Sab & 2.1 & 25.7 & 0.03 & 2.1 & 0.2 & 0.62 & 68 & 134 & 41 & 550 & 600 & 10.61 \\
\noalign{\smallskip}
NGC 3109 & Sm & 9.1 & 1.3 & 0.00 & 1.3 & \nodata & \nodata & 82 & 47 & \nodata & 120 & 120 & 8.62 \\
\noalign{\smallskip}
NGC 3198 & Sc & 5.2 & 12.2 & 0.00 & 3.4 & \nodata & \nodata & 71 & 147 & \nodata & 990 & 1040 & 10.49 \\
\noalign{\smallskip}
NGC 3200 & Sc & 4.5 & 52.2 & 0.11 & 9.1 & 0.2 & 0.87 & 72 & 262 & 45 & 4250 & 4470 & 11.37 \\
\noalign{\smallskip}
NGC 3495 & Sd & 6.3 & 17.8 & 0.02 & 3.9 & 0.30 & 0.63 & 78 & 172 & 89 & 1320 & 1350 & 10.39 \\
\noalign{\smallskip}
NGC 3593 & S0/a & $-0.4$ & 5.5 & 0.07 & 0.9 & 0.03 & 0.59 & 67 & 99 & 30 & 160 & 180 & 9.86 \\
\noalign{\smallskip}
NGC 3898 & Sab & 1.7 & 22.7 & 0.66 & 3.7 & 2.1 & 0.62 & 67 & 230 & 107 & 930 & 1070 & 11.07 \\
\noalign{\smallskip}
NGC 4062 & Sc & 5.3 & 10.8 & 0.03 & 2.0 & 0.50 & 0.96 & 64 & 139 & 13 & 550 & 610 & 10.22 \\
\noalign{\smallskip}
NGC 4236 & Sdm & 8.0 & 4.4 & 0.00 & 3.7 & \nodata & \nodata & 72 & 79 & \nodata & 580 & 610 & 9.43 \\
\noalign{\smallskip}
NGC 4258 & Sbc & 4.0 & 7.3 & 0.00 & 6.0 & \nodata & \nodata & 69 & 186 & \nodata & 2240 & 2410 & 11.08 \\
\noalign{\smallskip}
NGC 4378 & Sa & 1.0 & 41.1 & 0.52 & 6.0 & 2.8 & 1.0 & 35 & 185 & 0 & 1060 & 1850 & 11.42 \\
\noalign{\smallskip}
NGC 4419 & Sa & 1.1 & 12.5 & 0.10 & 1.2 & 0.07 & 0.92 & 71 & 180 & 19 & 370 & 400 & 10.43 \\
\noalign{\smallskip}
NGC 4448 & Sab & 1.8 & 7.0 & 0.22 & 1.0 & 0.03 & 0.90 & 69 & 182 & 38 & 300 & 320 & 10.03 \\
\noalign{\smallskip}
NGC 4594 & Sa & 1.1 & 12.7 & 0.86 & 3.1 & 8.4 & 0.70 & 84 & 348 & 104 & 2030 & 2380 & 11.56 \\
\noalign{\smallskip}
NGC 4605 & Sc & 5.1 & 4.5 & 0.01 & 0.8 & \nodata & \nodata & 68 & 79 & \nodata & 130 & 140 & 9.58 \\
\noalign{\smallskip}
NGC 4682 & Scd & 5.9 & 34.7 & 0.02 & 4.4 & 0.04 & 0.77 & 57 & 143 & 40 & 1220 & 1460 & 10.73 \\
\noalign{\smallskip}
NGC 4698 & Sab & 1.4 & 13.7 & 0.55 & 3.6 & 1.6 & 0.96 & 70 & 233 & 18 & 790 & 850 & 10.96 \\
\noalign{\smallskip}
NGC 4736 & Sab & 2.4 & 4.8 & 0.35 & 1.5 & 0.14 & 0.90 & 30 & 89 & 23 & 180 & 360 & 10.76 \\
\noalign{\smallskip}
NGC 4800 & Sb & 3.0 & 18.4 & 0.09 & 1.1 & 0.01 & 1.0 & 53 & 136 & 0 & 260 & 330 & 10.54 \\
\noalign{\smallskip}
NGC 4845 & Sab & 2.3 & 24.9 & 0.07 & 5.2 & 0.3 & 0.80 & 72 & 171 & 25 & 1650 & 1740 & 11.10 \\
\noalign{\smallskip}
NGC 5033 & Sc & 5.1 & 17.2 & 0.10 & 6.9 & 0.10 & 0.70 & 69 & 195 & 57 & 2440 & 2620 & 11.31 \\
\noalign{\smallskip}
NGC 5055 & Sbc & 4.0 & 8.7 & 0.05 & 4.0 & 0.2 & 0.86 & 54 & 169 & 27 & 1280 & 1590 & 11.11 \\
\noalign{\smallskip}
NGC 6314 & Sa & 1.1 & 98.2 & 0.56 & 6.2 & 2.4 & 0.70 & 70 & 215 & 83 & 1430 & 1560 & 11.50 \\
\noalign{\smallskip}
NGC 7171 & Sb & 3.1 & 39.3 & 0.04 & 4.0 & 0.04 & 0.72 & 52 & 169 & 35 & 1300 & 1650 & 10.84 \\
\noalign{\smallskip}
NGC 7217 & Sab & 2.5 & 16.4 & 0.25 & 2.5 & 2.9 & 0.93 & 35 & 149 & 23 & 590 & 1030 & 11.07 \\
\noalign{\smallskip}
NGC 7331 & Sb & 3.9 & 14.4 & 0.26 & 4.8 & 0.2 & 0.40 & 67 & 202 & 111 & 1450 & 1580 & 11.34 \\
\noalign{\smallskip}
NGC 7537 & Sbc & 3.6 & 37.3 & 0.18 & 2.5 & 0.2 & 0.70 & 79 & 137 & 34 & 570 & 580 & 10.47 \\
\noalign{\smallskip}
NGC 7541 & Sbc & 4.7 & 37.5 & 0.01 & 4.7 & 0.3 & 0.62 & 72 & 219 & 34 & 2050 & 2150 & 11.23 \\
\noalign{\smallskip}
NGC 7606 & Sb & 3.0 & 31.3 & 0.08 & 5.8 & 1.22 & 0.60 & 66 & 238 & 79 & 2540 & 2780 & 11.34 \\
\noalign{\smallskip}
NGC 7664 & Sc & 5.0 & 47.9 & 0.03 & 2.3 & 0.2 & 0.80 & 58 & 157 & 36 & 710 & 830 & 10.86 \\
\noalign{\smallskip}
IC 467 & Sc & 5.2 & 32.1 & 0.00 & 3.7 & \nodata & \nodata & 67 & 129 & \nodata & 960 & 1040 & 10.44 \\
\noalign{\smallskip}
IC 724 & Sa & 1.0 & 89.5 & 0.41 & 8.2 & 1.0 & 0.78 & 55 & 233 & 86 & 2360 & 2890 & 11.51 \\
\noalign{\smallskip}
UGC 2259 & Sdm & 7.8 & 9.8 & 0.00 & 1.1 & \nodata & \nodata & 22 & 31 & \nodata & 70 & 190 & 9.14 \\
\noalign{\smallskip}
UGC 2885 & Sc & 5.2 & 78.0 & 0.14 & 15.1 & 10.4 & 0.88 & 65 & 281 & 76 & 7560 & 8380 & 11.74 \\
\noalign{\smallskip}
UGC 11810 & Sbc & 3.7 & 66.4 & 0.05 & 5.8 & 0.4 & 0.76 & 74 & 183 & 22 & 2010 & 2090 & 10.82 \\
\noalign{\smallskip}
UGC 12810 & Sbc & 3.8 & 109.6 & 0.07 & 8.0 & 0.7 & 0.83 & 70 & 211 & 67 & 3140 & 3350 & 11.26 \\
\noalign{\smallskip}
\noalign{\smallskip}
\noalign{\smallskip}
\enddata
\tablecomments{
Morphological Types are taken from the RC3 catalog \citep{1991trcb.book.....D}.
The Hubble stage $T_{\rm Hubble}$ is from HyperLeda \citep{2003A&A...412...45P}.
Distances are from NED, using the redshifts with a Hubble constant of $h=0.73$ and correcting the Hubble flow for Virgo, Shapley, and the Great Attractor.
$B/T$ is the $r$-band bulge-to-total luminosity ratio from \citet{1986AJ.....91.1301K,1987AJ.....93..816K,1988AJ.....96..514K};
the $B/T$ value for NGC~4736 was not tabulated, and we have estimated it from the
rotation curve decomposition.
The semi-major axis disk and bulge scale-lengths ($\Rd$ and $a_{\rm e,b}$) are from
parametric fits to the Kent surface brightness profiles.
The bulge axis ratio $q_{\rm b}$ is also from Kent,
except for NGC~4736 \citep{2008A&A...487..555M}.
The inclination $i$ is from \citet{1980ApJ...238..471R,1982ApJ...261..439R,1985ApJ...289...81R} 
or \citet{1984ApJ...287...66W}, except in some cases where it is derived directly
from the observed disk flattening.
The characteristic disk rotation velocity $v_{s, {\rm d}}$ is based on the (2--3)~$\Rd$ regions of
the Kent deprojected rotation curves, reprojected to the observed values.
The bulge rotation velocity $v_{s, {\rm b}}$ is estimated using 
a simplified model based on observed ellipticity
(see main text).
The projected and intrinsic specific angular momenta \jp\ and \jt\ are derived
using Equations~(\ref{eqn:jCK0}) and (\ref{eqn:F83eq1}), respectively.
The stellar mass \Ms\ is based on 2MASS $K$-band photometry (see main text).
}
\end{deluxetable}

\newpage

\begin{deluxetable}{l c c r l c l r r r c l  }
\tablewidth{0pt}
\tabletypesize{\footnotesize}
\tablecaption{Approximate angular momenta of early-type galaxies\label{tab:etg}}
\tablehead{
\noalign{\smallskip}
\noalign{\smallskip}
\noalign{\smallskip}
Name & Type & $T_{\rm Hubble}$ & $D$ & $n$ & \aeff\ & $q$ & $\vs$ & \tjp\ & $\log\left(\frac{M_\star}{M_\odot}\right)$ & Tracers & Ref. \\
& & & (Mpc) & & (kpc) & & (\kms) & (\kms\  & \\
 & & & & & & & & kpc) & }
\noalign{\smallskip}
\noalign{\smallskip}
\startdata
\noalign{\smallskip}
NGC 821 & E6 & $-4.8$ & 23.4 & 3.4 & 4.5 & 0.60 & 20 & 180 & 11.02 & LS, MS, PN & C+09, P+09, \\
 & & & & & & & & & & & W+09a, FG10\\
\noalign{\smallskip}
NGC 1023 & S0 & $-2.7$ & 11.1 & 3.9 & 3.2 & 0.34 & 175 & 1270 & 10.97 & LS, PN & C+09 \\
\noalign{\smallskip}
NGC 1316 & S0 & $-1.8$ & 21.0 & 6.5 & 13.5 & 0.68 & 70 & 3640 & 11.94 & PN & M+12 \\
\noalign{\smallskip}
NGC 1339 & E3 & $-4.3$ & 19.7 & 4.3 & 1.5 & 0.71 & 105 & 400 & 10.52 & LS & G+98 \\
\noalign{\smallskip}
NGC 1344 & E4 & $-4.0$ & 20.9 & 4.1 & 5.1 & 0.58 & 20 & 240 & 11.13 & PN & C+09 \\
\noalign{\smallskip}
NGC 1373 & E2 & $-3.9$ & 19.3 & 4.3 & 1.0 & 0.77 & 10 & 25 & 9.79 & LS & G+98 \\
\noalign{\smallskip}
NGC 1379 & E0 & $-4.8$ & 19.5 & 2.3 & 2.3 & 0.97 & 20 & 70 & 10.68 & LS & G+98 \\
\noalign{\smallskip}
NGC 1380 & S0 & $-2.3$ & 21.2 & 3.3 & 5.4 & 0.51 & 220 & 2330 & 11.34 & LS & B+06 \\
\noalign{\smallskip}
NGC 1381 & S0 & $-2.2$ & 21.9 & 3.1 & 3.6 & 0.26 & 155 & 1040 & 10.75 & LS & W+09b \\
\noalign{\smallskip}
NGC 1400 & S0 & $-3.7$ & 25.7 & 1.9 & 2.7 & 0.89 & 50 & 3100 & 11.05 & LS, MS & P+09 \\
\noalign{\smallskip}
NGC 1404 & E1 & $-4.8$ & 20.2 & 3.7 & 2.6 & 0.88 & 115 & 630 & 11.26 & LS & G+98 \\
\noalign{\smallskip}
NGC 1407 & E0 & $-4.5$ & 28.1 & 4.3 & 7.8 & 1.0 & 20 & 680 & 11.66 & MS & P+09 \\
\noalign{\smallskip}
NGC 1419 & E0 & $-4.8$ & 22.9 & 3.6 & 1.1 & 0.99 & 3 & 7 & 10.18 & LS & G+98 \\
\noalign{\smallskip}
NGC 1428 & S0 & $-3.0$ & 20.7 & 1.5 & 1.4 & 0.63 & 65 & 130 & 9.97 & LS & G+98 \\
\noalign{\smallskip}
NGC 2310 & S0 & $-2.0$ & 14.9 & 2.1* & 3.4 & 0.19 & 135 & 680 & 10.35 & LS & W+09b \\
\noalign{\smallskip}
NGC 2577 & S0 & $-3.0$ & 31.7 & 2.7$^*$ & 2.8 & 0.62 & 210 & 1000 & 10.79 & LS & R+99 \\
\noalign{\smallskip}
NGC 2592 & E2 & $-4.8$ & 31.7 & 2.6$^*$ & 2.2 & 0.83 & 160 & 610 & 10.76 & LS & R+99 \\
\noalign{\smallskip}
NGC 2699 & E1 & $-5.0$ & 27.9 & 2.4$^*$ & 1.7 & 0.93 & 90 & 240 & 10.55 & LS & R+99 \\
\noalign{\smallskip}
NGC 2768 & S0 & $-4.4$ & 21.8 & 2.6 & 10.7 & 0.40 & 185 & 3360 & 11.33 & MS & P+09 \\
\noalign{\smallskip}
NGC 2778 & E2 & $-4.8$ & 22.3 & 2.2 & 1.5 & 0.80 & 120 & 270 & 10.26 & LS & R+99, H+01 \\
\noalign{\smallskip}
NGC 3115 & S0 & $-2.9$ & 9.4 & 2.5 & 3.7 & 0.50 & 190 & 1150 & 11.04 & LS, MS, GC & A+11 \\
\noalign{\smallskip}
NGC 3156 & S0 & $-2.5$ & 21.8 & 2.4 & 2.0 & 0.58 & 75 & 240 & 10.24 & LS & R+99\\
\noalign{\smallskip}
NGC 3203 & S0 & $-1.2$ & 34.0 & 2.9 & 5.7 & 0.20 & 165 & 1700 & 10.95 & LS & W+09b \\
\noalign{\smallskip}
NGC 3377 & E5 & $-4.8$ & 10.9 & 2.0 & 2.5 & 0.67 & 45 & 160 & 10.50 & LS, PN & C+09 \\
\noalign{\smallskip}
NGC 3379 & E2 & $-4.8$ & 10.3 & 4.7 & 2.6 & 0.84 & 30 & 210 & 10.96 & SA, LS, PN & D+07, C+09, W+09a \\
\noalign{\smallskip}
NGC 3605 & E3 & $-4.5$ & 20.1 & 1.3 & 1.4 & 0.66 & 65 & 110 & 10.09 & LS & R+99 \\
\noalign{\smallskip}
NGC 4318 & E3 & $-5.0$ & 22.0 & 1.1 & 0.8 & 0.66 & 85 & 80 & 9.86 & LS & S+10 \\
\noalign{\smallskip}
NGC 4374 & E1 & $-4.3$ & 18.5 & 8.3 & 15.8 & 1.0 & 40 & 2040 & 11.68 & PN & C+09 \\
\noalign{\smallskip}
NGC 4387 & E4 & $-4.9$ & 18.0 & 2.0 & 1.3 & 0.57 & 50 & 90 & 10.25 & LS & H+01 \\
\noalign{\smallskip}
NGC 4434 & E1 & $-4.8$ & 22.5 & 3.3 & 1.2 & 0.92 & 45 & 110 & 10.40 & LS & S+10 \\
\noalign{\smallskip}
NGC 4464 & S0 & $-2.1$ & 15.8 & 2.5 & 0.6 & 0.70 & 50 & 50 & 9.91 & LS & SP97b, S+97, \\
& & & & & & & & & & & H+01, S+10 \\
\noalign{\smallskip}
NGC 4478 & E2 & $-4.8$ & 17.1 & 2.1 & 1.1 & 0.81 & 65 & 110 & 10.47 & LS & D+83, SP97a, H+01 \\
\noalign{\smallskip}
NGC 4486B & E4 & $-5.0$ & 16.3 & 2.2 & 0.2 & 0.60 & 35 & 10 & 9.70 & LS & S+10 \\
\noalign{\smallskip}
NGC 4494 & E1 & $-4.8$ & 16.6 & 3.2 & 4.5 & 0.84 & 65 & 570 & 11.08 & LS, MS, PN & N+09, C+09, P+09 \\
\noalign{\smallskip}
NGC 4551 & E3 & $-4.9$ & 16.2 & 2.0 & 1.2 & 0.68 & 45 & 80 & 10.22 & LS & S+07, S+10 \\
\noalign{\smallskip}
NGC 4564 & E5 & $-4.8$ & 14.6 & 1.6 & 2.2 & 0.55 & 120 & 370 & 10.52 & LS, PN & C+09 \\
\noalign{\smallskip}
NGC 4697 & E4 & $-4.4$ & 11.4 & 3.5 & 4.4 & 0.59 & 35 & 320 & 11.02 & PN & C+09 \\
\noalign{\smallskip}
NGC 5128 & S0 & $-2.1$ & 4.1 & 4.0$^*$ & 6.0 & 1.0 & 80 & 1110 & 11.28 & PN & P+04b \\
\noalign{\smallskip}
NGC 5846 & E1 & $-4.7$ & 24.2 & 4.0 & 8.4 & 0.93 & 25 & 480 & 11.48 & PN & M+05, C+09\\
\noalign{\smallskip}
NGC 7617 & S0 & $-1.9$ & 57.1 & 2.7$^*$ & 3.3 & 0.78 & 80 & 470 & 10.82 & LS & R+99\\
\noalign{\smallskip}
\noalign{\smallskip}
\noalign{\smallskip}
\enddata
\tablecomments{
Morphological Types are generally taken from the RC3 catalog \citep{1991trcb.book.....D}.
The Hubble stage $T_{\rm Hubble}$ is from HyperLeda \citep{2003A&A...412...45P}.
The distances $D$ are taken from the surface brightness fluctuation analyses
\citep{2001ApJ...546..681T,2009ApJ...694..556B} where available, and otherwise from redshifts.
The parameters of the stellar mass profile $(n, a_{\rm e}, q)$ are taken
from various literature sources as described in Section~\ref{sec:etgdata}.
Galaxies with a `$*$' by their S\'ersic index $n$ have this value predicted
from their stellar masses, rather than being measured directly (see text for details).
$\vs$ is estimated from the observations at $x_s=2\,a_{\rm e}$,
and \tjp\ is calculated using Equation~(\ref{eqn:jCK0}).
For the subset of 8 galaxies with detailed \jp\ calculations listed in Table~\ref{tab:gal},
the \tjp\ values are listed here for reference, but
are not used in our main analyses.
The observational tracers are:
long-slit stellar kinematics (LS);
multi-slit stellar kinematics (MS);
integral-field stellar kinematics (IFU);
planetary nebulae (PN); and metal-rich globular clusters (GC).
The references are:
\citealt{1983ApJ...266...41D} (D+83);
\citealt{1997A&AS..122..521S,1997A&AS..126...15S} (SP97a,b); 
\citealt{2007MNRAS.377..759S} (S+97);
\citealt{1998A&AS..133..325G} (G+98);
\citealt{1999ApJ...513L..25R} (R+99);
\citealt{2001MNRAS.326..473H} (H+01);
\citealt{2004ApJ...602..685P} (P+04b);
\citealt{2006MNRAS.371.1912B} (B+06);
\citealt{2007ApJ...664..257D} (D+07);
\citealt{2009MNRAS.393..329N} (N+09);
\citealt{2009MNRAS.394.1249C} (C+09);
\citealt{2009MNRAS.398...91P} (P+09);
\citealt{2009MNRAS.398..561W} (W+09a);
\citealt{2009MNRAS.400.1665W} (W+09b);
\citealt{2010ApJ...716..370F} (FG10);
\citealt{2010MNRAS.408..254S} (S+10);
\citealt{2011ApJ...736L..26A} (A+11);
\citealt{2012A&A...539A..11M} (M+12).
Other photometric references are:
\citealt{2005AJ....130.1502M} (M+05).
}
\end{deluxetable}

\newpage

\section{Appendix D: Decomposing early-type galaxies}\label{sec:decomp}

In Section~\ref{sec:obsresults}, we sought to understand the \js--\Ms\ trends for
spiral galaxies by considering their bulge and disk components separately. 
Many early-type galaxies (both lenticulars and fast-rotating ellipticals) 
are also thought to consist of such subcomponents,
but the decompositions tend to be more difficult.
Here we develop a novel method to estimate the bulge fraction, 
$f_{\rm b} = B/T$, 
in order to study correlations involving this parameter, while we
also analyze a small set of ellipticals with decompositions from the literature.

First, we suppose for simplicity
that the relative dominance of rotation observed in an early-type
galaxy indicates the disk fraction, i.e., the bulge is
assumed to be non-rotating, with all of the observed rotation attributable to the disk.
This disk is assumed to rotate at the circular velocity $v_{\rm c}$, which we relate to
the stellar rotation velocity and velocity dispersion, $v$ and $\sigma$, by
$v_{\rm c} \simeq [v^2+2\sigma^2]^{1/2}$.
The observed rotation is then $v = C^\prime (1-f_{\rm b}) \times v_{\rm c}$, where
the parameter $C^\prime$ accounts for projection and aperture effects 
(e.g., a central aperture probably encloses a larger disk fraction than the
global value). After rearranging terms, we can solve for the bulge fraction
based on the observed $(v/\sigma)$ parameter:
\begin{equation}\label{eqn:fb}
f_{\rm b} \simeq 1-\frac{C^\prime\left(v/\sigma\right)}{\left[2+{C^\prime}^2\left(v/\sigma\right)^2\right]^{1/2}}
\end{equation}

\begin{figure*}
\centering{
\includegraphics[width=3.5in]{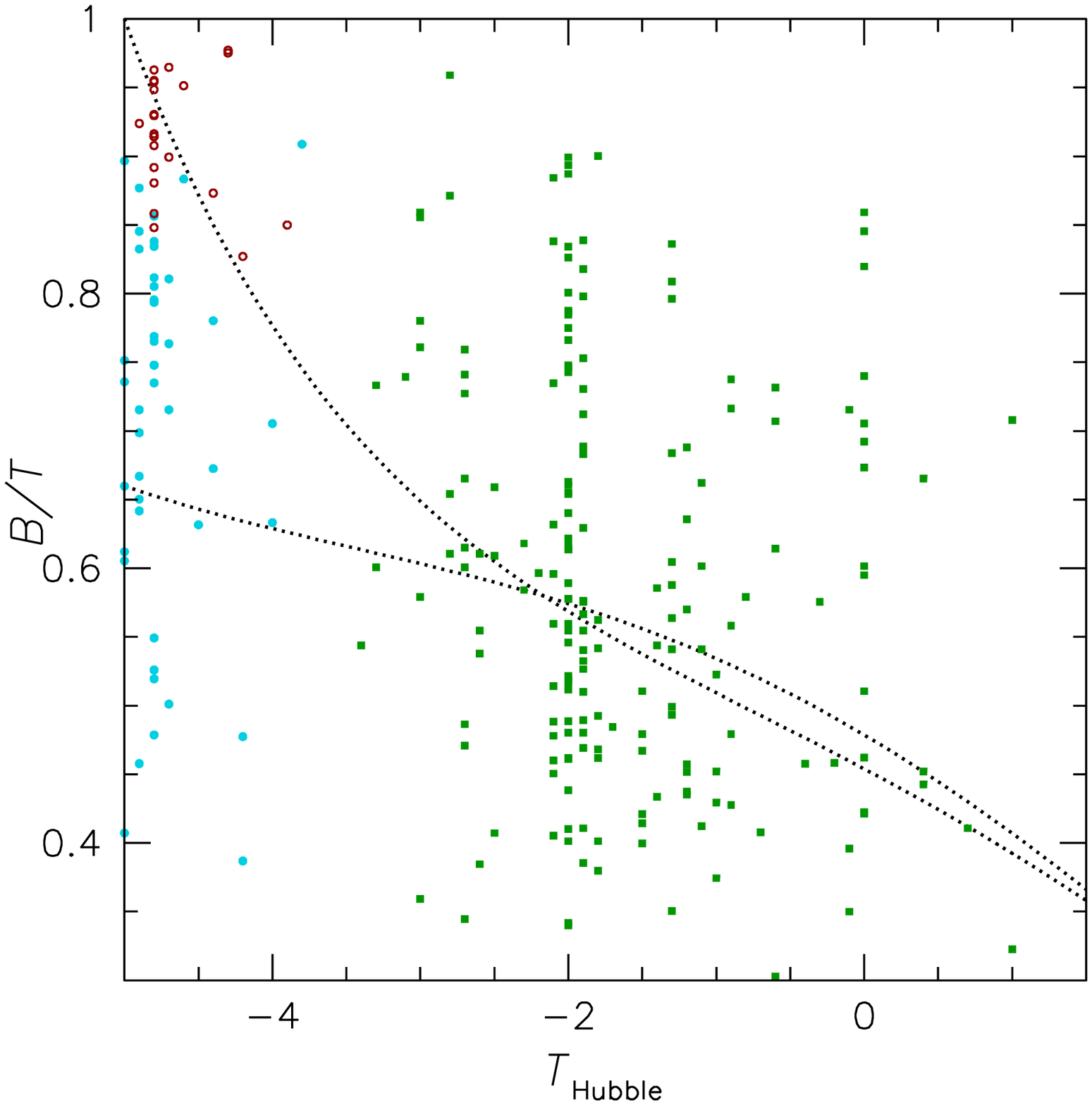} 
\includegraphics[width=3.5in]{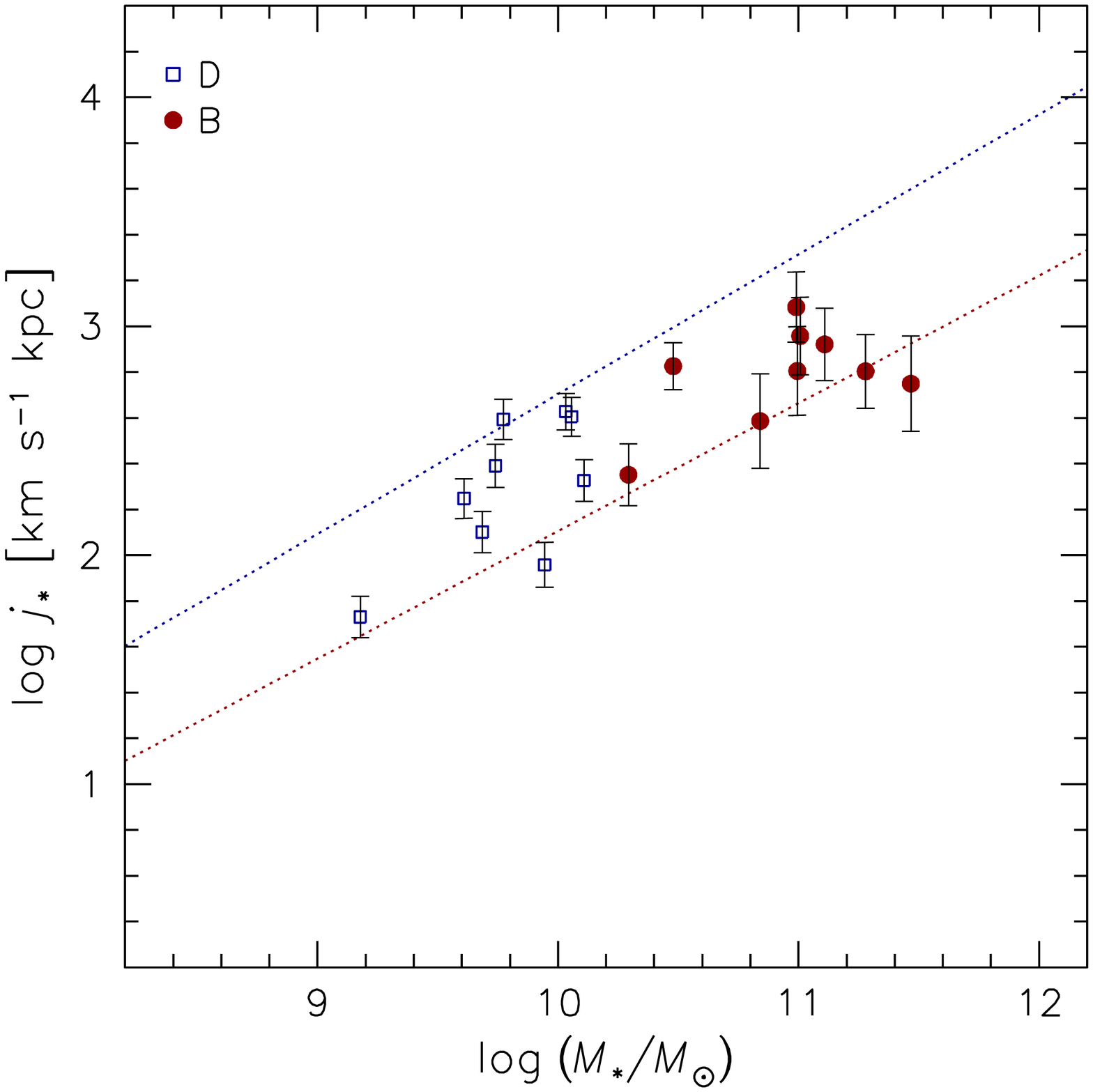} 
}
\caption{Bulge-fraction vs.\ morphological type for early-type galaxies
({\it left panel}), and  specific angular momentum vs.\ stellar mass
for the disk and bulge components of disky ellipticals ({\it right panel}).
On the left, dotted curves show two alternative observational trends from
\citet{1986ApJ...302..564S}.  
The points show data from \atlas3d\ galaxies, using Equation~(\ref{eqn:fb})
to estimate $B/T$ with $C^\prime=1.35$.
The symbol types are as in Figure~\ref{fig:atlas1}.
The dotted lines at right
show the best-fit relations for spiral disks and ellipticals
from the right-hand panel of Figure~\ref{fig:JMM0}.
}
\label{fig:scorza}
\end{figure*}

Next, to estimate the numerical factor $C^\prime$, we calibrate against the
relation between morphology and bulge fraction derived from photometric data
by \citet{1986ApJ...302..564S}. 
As shown in the left-hand panel of Figure~\ref{fig:scorza},
these authors derived two alternative relations with different extrapolations
for the elliptical galaxies, and we select the version that extends to $B/T=1$,
i.e., with some ellipticals being truly disk-less.
We then plot the \atlas3d\ data set, using their morphological types and
$(v/\sigma)_e$ values, and estimating $B/T$ using Equation~(\ref{eqn:fb}).
Based on the lenticular galaxy data ($T_{\rm Hubble}\sim -2$), 
we judge $C^\prime\simeq1.35$ to provide a reasonably good match;
this is similar to our adopted model of $C_{\rm med}=1.21$ for these galaxies
(Appendix~\ref{sec:spheroid}).\footnote{\citet{2011MNRAS.417..845K} adopted a similar convention for
connecting early-type galaxy rotation with $B/T$ but did not quantify a relation
except to estimate $f_{\rm b}\sim$~0.85--0.9 as the division between fast-
and slow-rotators, based on semi-analytical modeling of galaxy formation.
This is consistent with our relation, since the slow/fast boundary is
typically at $(v/\sigma)_{\rm e}\sim$~0.15, which our relation translates to
$f_{\rm b}\sim$~0.85.
On the other hand, the observational study of \citet{2011AdAst2011E..18L}
does {\it not} support a strong correlation between $(v/\sigma)_{\rm e}$ and $B/T$.
}
As in Appendix~\ref{sec:obsfull}, there is then an additional
factor (of $1/2.4$ at 1~\aeff) to convert our $(v/\sigma)$ observations to the \atlas3d\ values
when those are not available.
We estimate our final $B/T$ values to be generally accurate at the $\sim \pm 0.2$ level.

Next we investigate the bulge and disk \js--\Ms\ trends for a small sample of
elliptical galaxies, making use of the pioneering work of \citet{1995A&A...293...20S}, 
who carried out decompositions {\it including the kinematics of the 
bulge and disk subcomponents}.
These authors reported \js\ for the subcomponents, and
concluded that while the surface brightnesses
and scale-lengths of these disks
were consistent with an extension of scaling relations for lenticular
and spiral galaxies, the same may not be true for the disk \js\
values, with those of the ellipticals systematically
lower than those of the disk galaxies.  
These data were further used by \citet{1998ApJ...507..601V} to support a scenario where
disks in ellipticals are not related to spiral disks, but such conclusions neglected
the mass dependence naturally expected for \js.

Revisiting the \citet{1995A&A...293...20S} data set, we take their size and
rotation decompositions as given, and update the distances, stellar masses, and
\js\ calculations according to our current methodology.
From a few galaxies in common, we find that these \js\ values tend to be higher than
what we would derive from scratch
(e.g., because of decreasing rotation velocity outside the central regions).
Nevertheless, we plot the results in Figure~\ref{fig:scorza} (right-hand panel), comparing them to the 
\js--\Ms\ relations that we have found for spiral disks, and for ellipticals overall
(which we recall is roughly similar to the trend for the massive bulges of spirals).

There does appear to be a systematic separation between the relations for
the bulges and disks of these ellipticals, but not as large as we would expect based on
the results for spirals.
Given the uncertainties in these decompositions, we conclude that it is not yet clear
whether or not the disk and bulge subcomponents of ellipticals follow the same \js--\Ms\
relations as in spirals. If they do, it would be natural to expect the total trend for
ellipticals to be offset to slightly higher \js\ values from the bulge trend (owing to the disk
contributions), which from inspection of Figure~\ref{fig:JMM0} (right-hand panel) may indeed be true.

\end{document}